%% file: PhD_Thesis_Main_File.tex
\documentclass[11pt,a4paper,twoside,openright]{report}

\usepackage{amsfonts,amsmath,amssymb,caption,enumerate,float,graphicx,setspace}
\usepackage{epsfig}
\usepackage{siunitx}
\usepackage{lipsum,blindtext} 
\usepackage{fancyhdr}
\usepackage{color}
\usepackage{subfigure}
\usepackage[nottoc]{tocbibind}
\graphicspath{{figures/}}

\usepackage{url,array,csquotes,multirow,tabularx}
\usepackage{bbm}
\usepackage{enumerate}
\usepackage{slashed}
\usepackage{lscape}
\usepackage{hhline}
\usepackage{xcolor}
\usepackage{enumitem}
\usepackage[acronym,nomain,toc,nopostdot,nonumberlist]{glossaries}
\let\oldgls\gls
\renewcommand{\gls}[1]{\textcolor{black}{\oldgls{#1}}}   
\makeglossaries
\usepackage{tcolorbox}
\definecolor{bred}{rgb}{1.0,0.0,0.0}
\definecolor{bgreen}{rgb}{0.0,1.0,0.0}
\newtcbox{\rb}{on line,
  colframe=bred,colback=bred!15!white,
  boxrule=0.0pt,arc=3pt,boxsep=0pt,left=2pt,right=2pt,top=2pt,bottom=2pt}
\newtcbox{\gb}{on line,
  colframe=bgreen,colback=bgreen!15!white,
  boxrule=0.0pt,arc=3pt,boxsep=0pt,left=2pt,right=2pt,top=2pt,bottom=2pt}

\newcommand{\beqa}{\begin{eqnarray}}
\newcommand{\eeqa}{\end{eqnarray}}
\newcommand{\be}{\begin{equation}}
\newcommand{\ee}{\end{equation}}
\newcommand{\ba}{\begin{array}} 
\newcommand{\ea}{\end{array}}

\usepackage{subcaption} 

\usepackage{slashed} 	
\usepackage{empheq}

\usepackage[sc,noBBpl]{mathpazo}
\usepackage[T1]{fontenc}
\usepackage{kpfonts}

\usepackage{geometry}
\usepackage[times]{quotchap}

\makeatletter
\@ifundefined{c@lofdepth}{}{\let\c@lofdepth\relax}
\makeatother
\makeatletter
\@ifundefined{c@lotdepth}{}{\let\c@lotdepth\relax}
\makeatother

\usepackage[nottoc]{tocbibind}
\usepackage[titles]{tocloft}

\makeatletter
\newlength{\@chapterlength} 
\newcommand\@chapterheadsmark{1}
\ifnum \@chapterheadsmark = 1
\fi

\def\l@section{\@dottedtocline{1}{\@chapterlength}{2.3em}}

\newlength{\@subsectionlength}
\def\l@subsection{\@dottedtocline{2}{\@subsectionlength}{3.2em}}

\newlength{\@subsubsectionlength}
\def\l@subsubsection{\@dottedtocline{3}{\@subsubsectionlength}{0em}}

\makeatother

\renewcommand\cftchappresnum{\chaptername~}
\newlength\mylength
\makeglossaries
\usepackage[intoc]{nomencl}
\makenomenclature

\usepackage[toc,page]{appendix}

\usepackage{datetime}
\newdateformat{thisdate}{%
	\twodigit{\THEDAY}}
\newdateformat{thismonthdigit}{%
	\twodigit{\THEMONTH}}
\newdateformat{thismonth}{%
	\monthname[\THEMONTH]}
\newdateformat{thisyear}{%
	\THEYEAR}
 
\providecommand{\keywords}[1]
{
    \large
    \textbf{\textit{Keywords---}} #1
}

\usepackage[backend=biber, sorting=none,
            defernumbers=false, style=numeric-comp, doi=true]{biblatex}
\usepackage{hyperref}
\hypersetup{linktoc=all, colorlinks=true, linkcolor=blue, citecolor=magenta, urlcolor=magenta}

\begin{document}

\pagenumbering{gobble}
\input{00_Cover/cover_page}

\pagenumbering{roman}
\include{01_Dedication/Dedication}
\include{05_Acknowledgement/Acknowledgment}

\setstretch{1.5}  
\let\cleardoublepage\clearpage  
\input{06_Abstract/Abstract}
\input{110_Publications/Publications}

\newpage
\thispagestyle{empty}
\null\newpage
\input{08_Symbols/symbols}

\setstretch{1.5}

\input{09_Contents/Contents}

\setstretch{1.5}
\pagestyle{fancy}
\renewcommand{\sectionmark}[1]{\markright{\thesection~#1}{}}
\renewcommand{\chaptermark}[1]{\markboth{\thechapter~-~#1}{}}
\fancyhf{}
\fancyhead[RE]{\leftmark}
\fancyhead[LO]{\rightmark}
\fancyhead[LE,RO]{\thepage}

\newpage
\thispagestyle{empty}
\null\newpage
\pagenumbering{arabic}
\input{10_Chapter_1/Introduction}

\clearpage
\input{20_Chapter_2/Literature_review}

\clearpage
\input{30_Chapter_3/chapter_3}

\clearpage
\newpage\null\newpage
\input{40_Chapter_4/chapter_4}

\clearpage
\newpage\null\newpage
\input{50_Chapter_5/chapter_5}

\clearpage

\input{60_Chapter_6/chapter_6}

\clearpage
\newpage
\thispagestyle{empty}
\null\newpage
\input{70_Conclusion/Conclusion}
\clearpage

\appendix

\renewcommand{\chaptername}{Appendix}
\addtocontents{toc}{\protect\renewcommand{\protect\cftchappresnum}{\chaptername~}}

\input{100_Appendices/appendix_1}
 \input{100_Appendices/appendix_2}

\setstretch{1}
\newpage
\thispagestyle{empty}
\null\newpage

\renewcommand{\bibname}{References}
\printbibliography[heading=bibintoc]

\pagestyle{fancy}
\fancyhf{}
\fancyhead[RE]{\bfseries \bibname}
\fancyhead[LO]{\bfseries \bibname}
\fancyhead[LE,RO]{\thepage}

\end{document}

%% file: 00_Cover/cover_page.tex
\begin{titlepage}
\begin{center}

{\huge \bfseries Radiative Mass Generation in \\ \vspace{0.1cm}Gauged Theories of Flavour : \\ \vspace{0.15cm}A Path to Fermion Mass Hierarchies}\\
\vspace*{2cm}

{\large \bfseries \itshape A thesis submitted to the\\
Indian Institute of Technology Gandhinagar\\
for the award of the degree}\\
\vspace*{0.5cm}
{\large \itshape {of}}\\
\vspace*{0.5cm}

{\fontencoding{T1}\fontfamily{pzc}\fontseries{m}\fontshape{n}\selectfont
\huge \bfseries Doctor of Philosophy}\\
\vspace*{0.5cm}

{\large \itshape {by}}\\
\vspace*{0.5cm}

{\Large \bfseries Gurucharan Mohanta \\}
\vspace*{0.5cm}

{\large Under the guidance of\\}
\vspace*{0.2cm}

{\Large \bfseries Dr. Ketan M. Patel\\}


\vspace*{3.0cm}

\includegraphics[scale=0.07]{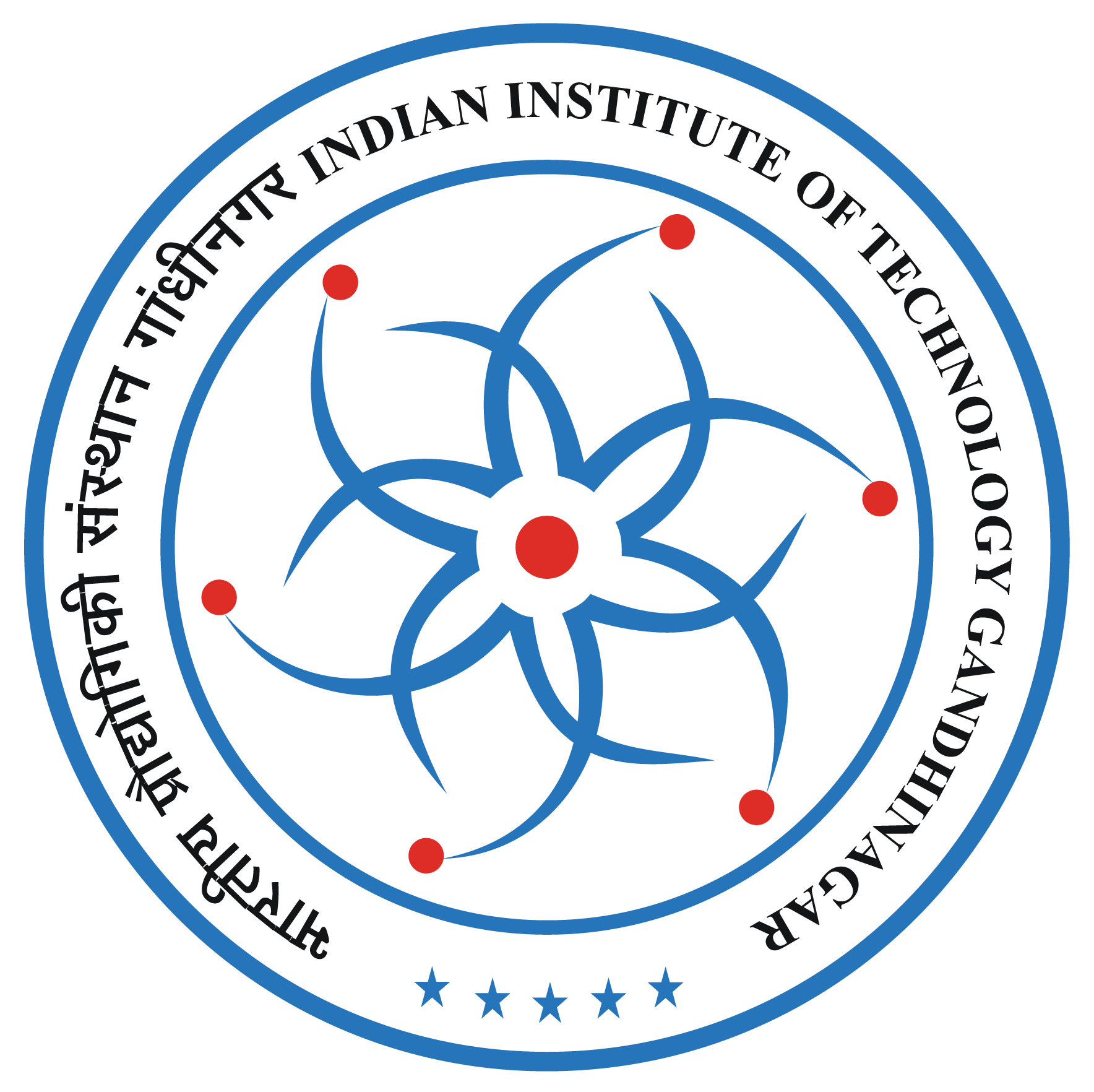}
\vspace*{0.5cm}

\textsc{\Large \bfseries{Department of Physics\\
Indian Institute of Technology Gandhinagar\\
Gujarat, 382055, India\\}}
{\Large \bfseries {\thisyear\today}\\}
\vspace*{0.1cm}

\end{center}
\end{titlepage}

\newpage\null\newpage

%% file: 01_Dedication/Dedication.tex

\chapter*{}

\begin{flushleft}
\textit{\textbf{\Large Dedicated to,}}
\end{flushleft}
\vspace*{1em}
\begin{flushright}
\textit{\textbf{\Large My beloved Maa, Bapa, and friends without whose endless love and support, I could not achieve this. }}
\end{flushright}
\vspace*{\fill}

\addcontentsline{toc}{chapter}{Dedication}

\newpage
\thispagestyle{empty}
\null\newpage

%% file: 05_Acknowledgement/Acknowledgment.tex
\chapter*{\centering Acknowledgments}
\addcontentsline{toc}{chapter}{Acknowledgments}

\textit{The journey of Ph.D. is truly remarkable, marked by moments of excitement, passion, and at times, frustration and overwhelming challenges, all of which shape and transform an individual into a researcher. The successful completion of this path would not have been possible without the unwavering support and contributions of so many wonderful people. I am deeply grateful to each and every one of you who has stood by me, offering help, encouragement, and guidance throughout this journey.}

\textit{ I consider it a privilege to express my deepest and most heartfelt gratitude to my PhD supervisor, Dr. Ketan M. Patel. His exceptional vision, expertise, and invaluable guidance have been fundamental in shaping the direction of my research. His unwavering support has not only inspired me but has also been a continuous source of motivation throughout my entire graduate journey. He has truly been a defining factor in my academic growth, and I will forever be grateful for his encouragement and belief in my abilities.}

\textit{I am sincerely grateful to my DSC members, Prof. Namit Mahajan, Dr. Satyajit Seth and Prof. Ramitendranath Bhattacharyya. Their insightful suggestions and constructive feedback during the DSC seminars have significantly enriched my research, enabling me to approach my work from a broader and more comprehensive perspective. Their input has been instrumental in refining my ideas and expanding my horizons.}

\textit{I extend my heartfelt thanks to Prof. Srubabati Goswami, Prof. Navinder Singh, Prof. Partha Konar, Dr. Naveen Chauhan, Prof. Varun Sheel, and Dr. Rajesh Kumar Kushawaha for their teachings and the enlightening discussions during the coursework and projects. Their invaluable teachings have played a crucial role in enhancing my understanding of key concepts and advancing my research. I also deeply appreciate Dr. Paramita Dutta for her kind and supportive behaviour.}

\textit{I would like to express my thanks to Prof. D. PallamRaju, Dean PRL, and Prof. Bijaya Sahoo, Chair of the Academic Committee, for their valuable contributions and guidance during my academic pursuits. A special acknowledgment is owed to Dr. Bhushit Vaishnav, Head of PRL Academic Services, whose steadfast assistance and encouragement from my very first day at PRL have been indispensable. Furthermore, I am thankful to various other administrative divisions at PRL, including accounts, purchase, library, administration, computer center, canteen, dispensary, transport, and housekeeping sections, for providing a conducive working environment and ensuring a smooth research experience.}

\textit{ Words cannot fully express my gratitude to my M.Sc. teachers from the Department of Physics, Sambalpur University, Odisha, for inspiring me to pursue research after completing my M.Sc. Their guidance and support served as a strong foundation throughout my Ph.D. journey, and they have consistently encouraged me to stay focused on my research. Their unwavering belief in my potential has been a continuous source of motivation.}

\textit{ A Ph.D. journey would never feel complete without encountering truly unforgettable colleagues and friends. Over the past five years, I have had the privilege of staying at the PRL Navrangpura Hostel, where I was fortunate to meet and connect with many remarkable individuals. I want to express my heartfelt gratitude to the B-lobby gang—Debashis, Kamran, and Dharmendra. You guys were always by my side, supporting me through every high and low of my personal life. The memories we created, especially during our TRIPs across different parts of India, will forever hold a special place in my heart. The only person who can come close to the legendary B-lobby gang is none other than Arup Bhai. Even though we did not stay in the same hostel, I have always treasured his companionship. His witty remarks and dark humour have never failed to bring a smile to my face. Thank you for being there for me, not just as a friend but as a constant source of joy throughout this journey. I am also grateful to Wafikul aka W-Bhai, Sanjay, Arup Mailty, Aditya, Trinesh, Sandeep, Mansi, Malika, Ananya, Akansha and Chandrima for their friendship and positive support throughout my Ph.D. journey. A big thank you to my amazing seniors—Saurabh Da, Supriya Da, and Bharathi Bhai. You guys are my go-to person for literally anything. The memories of participating in volleyball, football, and cricket tournaments with you will always stay vivid in my mind. A huge shoutout to Anshika Di, Monica Di, Meghna Di, Monal Bhai,  Sovan Da, Chandrima Di, Pravin Bhai, Anupam Da, Tanmoy Da and Ramanuj Da for creating a delightful and enjoyable atmosphere in the PRL. I have also met with amazing juniors, Deepanshu, Suvendu, Namit bhai, Achintya and Bhavya.}

\textit{A special shoutout to my college crew-Chandru-Debu (CDG members), Anitya, Pujarini, Ladli, Bijay, Debashish, Dharmu, Ranjan, Gouri, Gouranga, Panda, and Ajay-you guys made this journey so much more fun! I also want to give a special mention to the masters HEP group members; your interactions and discussions made my master’s days truly enjoyable.}  

\textit{Additionally, I’d like to thank some of my incredible juniors-Krishna, Saurabh, Manas, Kak, Gitanjali, Babulu, Chandan, Bikash, Truptilata, Anagha, Arpita and Ranjan-for their engaging discussions on physics problems and solutions. Your curiosity and insights were truly inspiring!} 

\textit{It is special for me to mention my childhood friends-Thakura, Goutam, Anshika, Riku and Chanda-Apu-Rudra-Raju- who hold a special place in my heart for their unwavering belief in me and constant support.}

\textit{I would like to express my profound gratitude to my beloved parents and family. Their unwavering love, constant encouragement, tireless support, and countless sacrifices have been the foundation of my strength throughout this entire journey. Their prayers have been my rock, providing me with the strength to overcome every obstacle. It is through their hard work, dedication, and continuous prayers that I have reached this moment, writing my doctoral thesis.} 

\textit{Last but certainly not least, I would like to express my sincere gratitude to all those individuals whom I may have unintentionally overlooked in this acknowledgement. Your contributions, support, and positive influence have been invaluable to me. Please know that your involvement and assistance have played a crucial role in the successful completion of my Ph.D. thesis, and I deeply appreciate everything you have done.}

\textit{Thank you all for your invaluable contributions and for being part of this significant milestone in my academic career.}

\noindent
\begin{minipage}[t]{0.35\textwidth}
    Date: \\
    Place:
\end{minipage}%
\hfill
\noindent
\begin{minipage}[t]{0.4\textwidth}
\begin{flushright}
    \bfseries
    \rule{\textwidth}{1pt}\\
	Gurucharan Mohanta\\
	Roll No.: 20330009
\end{flushright}
\end{minipage}


%% file: 06_Abstract/Abstract.tex
\chapter*{\centering Abstract}

\pagestyle{fancy}
\fancyhf{}
\fancyhead[LO,RE]{Abstract}
\fancyhead[LE,RO]{\thepage}

\addcontentsline{toc}{chapter}{Abstract}
A class of models based on gauged flavour symmetries is proposed to explain the observed hierarchical structure of fermion masses in the Standard Model (SM). These frameworks introduce a mechanism where only the third-generation fermions acquire masses at the tree level, while the first and second generations receive their masses through radiative corrections involving new gauge bosons. A pair of vector-like fermions for each sector is introduced, which play a key role in giving seesaw masses to third-generation fermions only and also take part in the radiative mass generation mechanism. It is explicitly shown that the Abelian theories that can incorporate the radiative mass generation mechanism have to be flavour non-universal in nature.

We construct a renormalisable model by extending the SM gauge symmetry by two additional \(U(1)\) gauge groups. The extended two abelian symmetries are all fermion generalisations of the well-known leptonic symmetries such as \(L_\mu - L_\tau\) and \(L_e - L_\mu\). In this setup, both the first and second-generation fermion masses arise from 1-loop corrections induced by the new gauge bosons, with the mass gap between them being dictated by a little hierarchy in the masses of the associated vector bosons.  This approach provides a natural explanation for the mass hierarchy among the three generations of charged fermions and results in interesting phenomenological consequences. Phenomenological implications, including constraints on the new gauge boson masses, are discussed. It is shown that the phenomenologically viable solutions require a new physics scale of the order of ${\cal O}(10^{5})$ TeV. The latter's large separation from the electroweak scale poses a challenge from the naturalness point of view.

Next, we discuss a scenario based on a single Abelian symmetry \(G_F\) as the gauged flavour symmetry in which gauge charges are optimised to suppress the contribution to the flavour-violating processes involving the lighter generations, leading to improved constraints on new physics. In this setup, a tiny first-generation fermion mass is induced at the 2-loop level. It is shown that there exists a strong correlation between the flavour violations in the $1-2$ sector and 2-loop masses such that in the vanishing limit of the earlier, the latter vanishes completely. An anomaly-free implementation of this analysis within the SM framework is demonstrated, and constraints from flavour violation place a lower bound on the new physics scale, estimated to be around \(10^3\) TeV which is nearly two orders of magnitude smaller than the previous framework. This framework may allow further improvements in the constraints at the expense of deviations in the light quark masses and, hence, can be tested from precise measurement of light quark masses.

The Abelian frameworks usually contain a large number of free parameters; thus, they are unpleasant from the computability point of view. Therefore, we discuss non-Abelian extensions in which three generations of fermions are horizontally unified in the three-dimensional representations, which leads to a comparatively smaller number of parameters in the theory. This, along with the feature that the non-Abelian gauge bosons have non-diagonal couplings, leads to a more predictive framework for the radiative mass mechanism. The aspects of radiative generation of fermion masses in this class of theories are demonstrated by considering the $SU(3)$ as the extended symmetry acting in flavour space. The specific choice of gauge charges ensures that only third-generation fermions obtain tree-level masses, while the first-order corrections induce mass terms for both second and first-generation fermions. The mass hierarchy between the first two generations is attributed to the sequential breaking of $SU(3)_F$ with an intermediate \(SU(2)\) breaking step, which determines the ordering of the gauge boson masses. The minimal setup contains less parameters than the previous setup and thus becomes much more constrained and more predictive such that it computes the strange quark mass slightly deviating from its current central value by $3 \, \sigma$.

We also study the radiative mass mechanism in the context of left-right (L-R) symmetric theories. We show that an abelian extension of the parity invariant LR framework, which implements the mass generation mechanism, has the potential to explain the smallness of the strong CP phase. The explicit model is based on $U(1)_{2-3}$ abelian flavour symmetry, and the parity is softly broken in the scalar sector. We explicitly show that new gauge boson-induced 1-loop and 2-loop corrections to the fermion masses and the scalar-induced one-loop correction do not generate nonzero $\bar{\theta}$ as they take a Hermitian form. It is also shown that the gauge boson-induced corrections don't induce a strong CP phase in any order of perturbation theory. However, since the scalar sector breaks the parity invariance, scalar-induced quantum corrections contribute to the non-vanishing $\bar{\theta}$ at the 2-loop level. The minimal realisation of this framework predicts a correlation between the $U(1)_{2-3}$ symmetry breaking scale and the \(SU(2)_R\) breaking scale. As the flavour violation constraints require a new physics scale around \(10^3\) TeV or higher, thereby it predicts the strong CP phase to be \(\bar{\theta} \lesssim 10^{-14}\).

Overall, these approaches highlight the role of gauged flavour symmetries in viably addressing the fermion mass hierarchy problem while yielding rich phenomenological predictions. 



\keywords{fermion mass hierarchy, flavour symmetry, flavour violation, radiative mass generation,  vector-like fermions, neutrinos.}

\newpage
\thispagestyle{empty}
\null\newpage

%% file: 110_Publications/Publications.tex
\pagestyle{fancy}
\fancyhf{}
\fancyhead[RE]{\bfseries References}
\fancyhead[LO]{\bfseries References}
\fancyhead[LE,RO]{\thepage}

\chapter*{Publications}
\addcontentsline{toc}{chapter}{Publications}

    
    
    
\section*{Research Articles (All included in the thesis)}
\begin{enumerate}
    \item  Gurucharan Mohanta.\ 
\enquote{Radiative Mass Mechanism: Addressing the Flavour Hierarchy and Strong CP Puzzle}.
In: \emph{JHEP 04 (2025) 170}, 
\textsc{doi}: \href{https://doi.org/10.1007/JHEP04(2025)170}{10.1007/ JHEP04(2025)170}. arXiv:\href{https://doi.org/10.48550/arXiv.2411.13385}{2411.13385[hep-ph]};\.\, 

\item Gurucharan Mohanta and \ and Ketan M. Patel.\ 
\enquote{Loop-induced masses for the first two generations with optimum flavour violation}.
In: \emph{JHEP 12 (2024) 158},
\textsc{doi}: \href{https://doi.org/10.1007/JHEP12(2024)158}{10.1007/JHEP12(2024)158}. arXiv:\href{https://doi.org/10.48550/arXiv.2406.19179}{2406.19179 [hep-ph]}.

\item Gurucharan Mohanta and \ and Ketan M. Patel.\ 
\enquote{Gauged $SU(3)_F$ and loop induced quark and lepton masses}.
In: \emph{JHEP 10 (2023) 128},
\textsc{doi}: \\ \href{https://doi.org/10.1007/JHEP10(2023)128}{10.1007/JHEP10(2023)128}. arXiv:\href{https://doi.org/10.48550/arXiv.2308.05642}{ 2308.05642
[hep-ph]}.

\item Gurucharan Mohanta and \ and Ketan M. Patel.\ 
\enquote{Radiatively generated fermion mass hierarchy from flavor nonuniversal gauge symmetries}. In:
\emph{Phys. Rev. D 106 (2022)7, 075020}, \textsc{doi}: \href{https://doi.org/10.1103/PhysRevD.106.075020}{10.1103/PhysRevD.106.075020}. \\ arXiv:\href{https://doi.org/10.48550/arXiv.2207.10407}{ 2207.10407
[hep-ph]}.
\end{enumerate}

\section*{Conference proceedings}
\begin{enumerate}
    \item Gurucharan Mohanta.\ 
\enquote{Radiatively Generated Quarks and Lepton Masses in Extended Gauge Theories}.
In: \emph{Springer Proc.Phys. 304 (2024) 282-285},
DOI: \href{https://doi.org/10.1007/978-981-97-0289-3_61}{https://doi.org/10.1007/978-981-97-0289-3${}_{-}$61}.
\item Gurucharan Mohanta. \enquote{Minimal $Z^\prime$ for radiative generation of fermion masses}.
In: \textit{17th International Conference on Interconnections between Particle Physics and Cosmology (\textbf{PPC 2024})}, $14-18$th October, IIT Hyderbad, India;\ \url{Under}\, \url{ review}.
\end{enumerate}

%% file: 08_Symbols/Symbols.tex
\nomenclature{$\nu(\Bar{\nu})$}{Neutrino (Anti-neutrino)}
\nomenclature{$P_{L,R}$}{Left handed, right handed chirality}
\nomenclature{$\nu_e$}{Electron neutrino}
\nomenclature{$\nu_\mu$}{Muon Neutrino}
\nomenclature{$\nu_\tau$}{Tau Neutrino}
\nomenclature{$\gamma_5$}{Chirality operator}
\nomenclature{$e^{\pm}$}{Electron, Positron}
\nomenclature{$\mu^{\pm}$}{Positively, negatively charged muon}
\nomenclature{$\tau$}{Tau}
\nomenclature{$Y$}{Hyper charge}
\nomenclature{$Q$}{Electro Magnetic Charge}
\nomenclature{$T_3$}{Third component of Isospin}
\nomenclature{$\theta_{ij}$}{Mixing angles}

%% file: 09_Contents/Contents.tex
\renewcommand{\contentsname}{Table of Contents}
\tableofcontents
\addcontentsline{toc}{chapter}{\contentsname}
\pagestyle{fancy}
\fancyhf{}
\fancyhead[LO,RE]{\contentsname}
\fancyhead[LE,RO]{\thepage}
\newpage
\thispagestyle{empty}
\null\newpage
\listoffigures
\pagestyle{fancy}
\fancyhf{}
\fancyhead[LO,RE]{List of Figures}
\fancyhead[LE,RO]{\thepage}
\newpage
\thispagestyle{empty}
\listoftables
\pagestyle{fancy}
\fancyhf{}
\fancyhead[LO,RE]{List of Tables}
\fancyhead[LE,RO]{\thepage}
\newpage
\thispagestyle{empty}
\null\newpage
\printglossary[type=\acronymtype, title=Abbreviations, toctitle=List of Abbreviations
]
\pagestyle{fancy}
\fancyhf{}
\fancyhead[LO,RE]{List of Abbreviations}
\fancyhead[LE,RO]{\thepage}
\input{07_Abbreviations/Abbreviations}



%% file: 07_Abbreviations/Abbreviations.tex

\newacronym{SM}{SM}{Standard Model}
\newacronym{BSM}{BSM}{Beyond the Standard Model}
\newacronym{CKM}{CKM}{Cabibbo-Kobayashi-Maskawa} 
\newacronym{VL}{VL}{vector-like}
\newacronym{VEV}{VEV}{vacuum expectation value}
\newacronym{FCNC}{FCNC}{flavour changing neutral current}
\newacronym{RGE}{RGE}{renormalisation group evolution}
\newacronym{L-R}{L-R}{left-right symmetry}
\newacronym{WCs}{WCs}{Wilson coefficients}
\newacronym{LH}{LH}{Left-handed}
\newacronym{RH}{RH}{Right-handed}
\newacronym{FV}{FV}{flavour violation}
\newacronym{LFV}{LFV}{lepton flavour violation}

%% file: 10_Chapter_1/Introduction.tex
\chapter{Introduction}\label{chap:intro}
\graphicspath{{Chapter_1/Vector/}{Chapter_1/}}
The ultimate goal of particle physics is to uncover the fundamental properties of elementary particles and the interactions governing them. These serve as building blocks for deeper principles that describe the underlying structure of nature. Over time, it has become clear that the interactions of elementary particles are governed by symmetry principles, guiding our understanding of fundamental forces. The theories aimed in this direction are the gauge theories of particle physics. Among the many open questions in this field, one of the most intriguing is the ambiguity surrounding the origin of elementary particle masses, especially fermion masses. Addressing this issue is the central objective of this thesis.  

The \gls{SM} of particle physics \cite{Glashow:1961tr,Salam:1964ry,Weinberg:1967tq} provides a comprehensive framework for describing these fundamental fermions and their interactions via the strong, weak, and electromagnetic forces. It classifies elementary particles into quarks and leptons, which constitute matter, and bosons, which mediate fundamental forces. The quarks and leptons together form the flavour sector of the SM, each arranged into three generations. While the SM successfully unifies the electroweak and strong interactions into a common theoretical framework of the gauge theory and introduces the  Brout-Englert-Higgs (B-E-H) mechanism \cite{Higgs:1964pj,Higgs:1964ia,Higgs:1966ev} to explain how particles acquire mass, it does not explain the vast differences in fermion masses, known as the fermion mass hierarchy.  Giving a basic introduction to the issue of the hierarchy is the main topic of this chapter.
 
 In the next section, we begin our discussion by giving a very brief overview of the symmetry-breaking mechanism and the origin of the masses of different particles in the SM. We also point out the number of free parameters associated with the mass sector of fermions.  Next, we outline some unexplained issues of the flavour sector of the SM in section \ref{intro:issues}. With the fermion mass hierarchy issue in mind, in section \ref{intro:attempts}, we briefly discuss the attempts that have been made to explain it. Finally, the outline of the rest of the chapters in the thesis will be given in the last section.

\section{Fermion masses in the Standard Model}
In the SM, \gls{LH} and \gls{RH} components of fermions transform differently under the SM gauge symmetry, \( G_{SM} \). Consequently, gauge invariance forbids any bare mass terms for these fermions. Instead, their masses arise dynamically through the spontaneous breaking of \( G_{SM} \) via the B-E-H mechanism.  

At the core of this mechanism is the Higgs field, \( \Phi \), a scalar field that interacts with SM fermions through Yukawa couplings. These interactions enable fermions to acquire mass after electroweak symmetry breaking. The transformation properties of the SM fermions and the Higgs field under \( G_{SM} \) are summarised in Table \ref{tab:sm_particles}.  

\begin{table}[t]
    \centering
    \renewcommand{\arraystretch}{1.5}
    \begin{tabular}{ccc}
        \hline
        \textbf{Field} & & \textbf{\( SU(3)_C \times SU(2)_L \times U(1)_Y \)} \\  
        \hline
        \hline
        Quark Doublet &\( Q_{L_i} = \begin{pmatrix} u_{L_i} \\ d_{L_i} \end{pmatrix} \) & \( (\mathbf{3}, \mathbf{2}, \frac{1}{3}) \) \\  
        Up-type Quark Singlet &\( u_{R_i} \) & \( (\mathbf{3}, \mathbf{1}, \frac{4}{3}) \) \\  
        Down-type Quark Singlet &\( d_{R_i} \) & \( (\mathbf{3}, \mathbf{1}, -\frac{2}{3}) \) \\  
        \hline
        Lepton Doublet &\( \Psi_{L_i} = \begin{pmatrix} \nu_{L_i} \\ e_{L_i} \end{pmatrix} \) & \( (\mathbf{1}, \mathbf{2}, -{1}{}) \) \\  
        Charged Lepton Singlet &\( e_{R_i} \) & \( (\mathbf{1}, \mathbf{1}, -2) \) \\  
        \hline
        Higgs Doublet &\( \Phi = \begin{pmatrix} \phi^+ \\ \phi^0 \end{pmatrix} \) & \( (\mathbf{1}, \mathbf{2}, {1}{}) \) \\  
        \hline
        Gluon &\( g \) & \( (\mathbf{8}, \mathbf{1}, 0) \) \\  
        Weak Bosons &\( W^\pm, W^0 \) & \( (\mathbf{1}, \mathbf{3}, 0) \) \\  
        Hypercharge Boson &\( B \) & \( (\mathbf{1}, \mathbf{1}, 0) \) \\  
        \hline
    \end{tabular}
    \caption{Standard Model particle content and their transformation properties under \( SU(3)_C \times SU(2)_L \times U(1)_Y \). The index \( i = 1,2,3 \) represents the three generations of quarks and leptons. Here we follow the convention in which electromagnetic charges are normalised as \(Q = T_3 + \frac{Y}{2}\).}
    \label{tab:sm_particles}
\end{table}


The renormalisable Yukawa couplings responsible for the charged fermion mass generation take the form:  
\be \label{SM:yukawa}
-{\cal L}_Y = {Y^d}_{ij}\,\overline{Q}_{L}{}_i\, {\Phi}\, d_{Rj} + {Y^u}_{ij}\,\overline{Q}_{L}{}_i\, \tilde{\Phi}\, u_{Rj} + {Y^e}_{ij}\,\overline{\Psi}_{L}{}_i\, {\Phi}\, e_{Rj} + H.c.
\ee  
where \( \tilde{\Phi } \) is the conjugate Higgs field, defined as \( \tilde{\Phi }= i \sigma_2 \Phi^* \). Also, the sum over the repeated indices is implied. 

In general, the Yukawa couplings \( Y^{u,d,e}_{ij} \) are non-diagonal, implying that fermion mass eigenstates are not necessarily aligned with their interaction eigenstates. However, by performing a suitable basis transformation, \( Y^u \) and \( Y^e \) can be diagonalised without loss of generality, while \( Y^d \) remains a general \( 3\times3 \) matrix.  

The electroweak interaction between gauge bosons and fermions in the Glashow-Weinberg-Salam theory is governed by the gauge-invariant Lagrangian:  
\be\label{ch1:Lgauge}
{\cal L} \supset \sum_{f=Q,L} \bar{f_L}_i i \gamma^\mu D_\mu\,{f_L}_i + \sum_{f=u,d,e} \bar{f_R}_i i \gamma^\mu D_\mu\,{f_R}_i\,,
\ee  
where the gauge covariant derivative \( D_\mu \) is given by:  
\be
D_\mu f_{Li} = \left(\partial_\mu + igW^a_\mu T^a + i g^\prime \frac{Y}{2} B_\mu \right)f_{Li}\,,
\ee  
\be
D_\mu f_{Ri} = \left(\partial_\mu + i g^\prime \frac{Y}{2} B_\mu \right)f_{Ri}\,.
\ee  
Here, the gauge bosons \( W^a_\mu \) (\( a = 1,2,3 \)) correspond to the three generators of the \( SU(2)_L \) symmetry, while \( B_\mu \) is associated with the \( U(1)_Y \) hypercharge symmetry. The generators of these symmetries are denoted by \( T^a \) for \( SU(2)_L \) and \( \frac{Y}{2} \) for \( U(1)_Y \).  

From eq. (\ref{ch1:Lgauge}), it is evident that the gauge Lagrangian remains invariant under the following global unitary transformations:  
\beqa \label{ch1:U(3)trans}  
Q_{Li} &\to & \left(U^Q\right)_{ ij} Q_{Lj}\,,~~u_{Ri} \to \left( U^u\right)_{ ij} u_{Rj}\,,~d_{Ri} \to \left(U^d\right)_{ ij} d_{Rj}\,,\nonumber\\
\Psi_{Li} &\to & \left(U^\Psi\right)_{ ij} \Psi_{Lj}\,,~~~~~e_{Ri} \to \left(U^e\right)_{ ij} e_{Rj}\,.  
\eeqa  
where \( U^{Q,\Psi,u,d,e} \) are \( 3\times3 \) unitary matrices denoting arbitrary rotations in the flavour space. This implies that the gauge Lagrangian possesses a global symmetry of \( U(3)^5 \), which commutes with the SM gauge group. If the entire Lagrangian were invariant under these transformations, fermions would remain massless. However, this symmetry is explicitly broken by the Yukawa interactions.  

Under the transformations in eq. (\ref{ch1:U(3)trans}), the Yukawa Lagrangian, eq. (\ref{SM:yukawa}), transforms as:  
\beqa \label{ch1:LY2}  
-{\cal L}_Y &=& \tilde{Y}^d_{ij}\,\overline{Q}_{L}{}_i\, {\Phi}\, d_{Rj}\, + \tilde{Y}^u_{ij}\,\overline{Q}_{L}{}_i\, \tilde{\Phi}\, u_{Rj}\,+\,\tilde{Y}^e_{ij}\,\overline{L}_{L}{}_i\, {\Phi}\, e_{Rj}\,+\,H.c\,.\eeqa  
The transformed Yukawa coupling matrices, \( \tilde{Y}^{u,d,e} \), are given by:  
\be\label{ch1:tildeY}  
\tilde{Y}^d = (U^Q)^\dagger Y^d U^d\,,~~\tilde{Y}^u = (U^Q)^\dagger Y^u U^u\,,~~\tilde{Y}^e = (U^\Psi)^\dagger Y^e U^e\,.  
\ee  
From eq. (\ref{ch1:tildeY}), it follows that an appropriate choice of \( U^\Psi \) and \( U^e \) allows the matrix \( \tilde{Y}^e \) to be diagonalised. This procedure, known as \textit{biunitary diagonalisation}, is achieved by selecting \( U^\Psi \) and \( U^e \) such that:  
\be  
\tilde{Y}^e{}^2 = (U^\Psi)^\dagger Y^e Y^{e \dagger}U^\Psi\,,~~(\tilde{Y}^e)^2 = (U^e)^\dagger  Y^{e \dagger}Y^e\,U^e\,,  
\ee  
where  
\be  
\tilde{Y}^e= {\rm Diag.}(y^e_1,y^e_2,y^e_3).  
\ee  
Here, \( y^e_1, y^e_2, y^e_3 \) corresponds to the real Yukawa couplings of the charged leptons-electron, muon, and tau, respectively, in the diagonal basis.  

A similar procedure applies to the up-type and down-type quark Yukawa matrices. Since the transformation matrix \( U^Q \) appears in both \( \tilde{Y}^u \) and \( \tilde{Y}^d \), one of them can always be chosen diagonal. By convention, choosing  
\be  
\tilde{Y}^u= {\rm Diag.}(y^u_1,y^u_2,y^u_3)  
\,,\ee  
allows the Yukawa Lagrangian in eq. (\ref{ch1:LY2}) to be rewritten as  
\beqa \label{ch1:LY3}  
-{\cal L}_Y &=& {y}^d_{ij}\,\overline{Q}_{L}{}_i\, {\Phi}\, d_{Rj}\, + {y}^u_{i}\,\overline{Q}_{L}{}_i\, \tilde{\Phi}\, u_{Ri}\,+\,{y}^e_{i}\,\overline{L}_{L}{}_i\, {\Phi}\, e_{Ri}\,+\,H.c\,.\eeqa  
Here, \( \tilde{Y}^d_{ij}={y}^d_{ij} \). The \( 3\times3 \) matrix \( {y}^d \) remains complex, containing a total of 18 parameters. Since some phases can still be absorbed, the total number of independent phases in \( y^d \) can further be reduced. Specifically, for \( n = 3 \) here, $2n-1$ number of phases can be absorbed, and therefore there are 13 independent real parameters in the down-quark Yukawa sector (9 real elements and 4 phases).  

Thus, in the unbroken phase of the SM, the total number of independent parameters in the Yukawa Lagrangian is 19: 3 real parameters in \( y^u_i \) (up-type quarks), 3 real parameters in \( y^e_i \) (charged leptons), and 13 real parameters in \( y^d_{ij} \) (down-type quarks).  When the symmetry is broken, some of these parameters remain unphysical as the mixing in the right-handed sector is not observable in the SM.

Since the Higgs field \( \Phi \) resides in the fundamental representation of \( SU(2)_L \) and carries hypercharge \( Y = 1 \), it spontaneously breaks the electroweak symmetry,  
\be  
SU(2)_L \times U(1)_Y \longrightarrow U(1)_{EM}\,,  
\ee  
when it acquires a nonzero \gls{VEV}. The unbroken \( U(1)_{EM} \)  must remain as the symmetry of the vacuum since the electric charge is found to be conserved. Thus, the generator of the unbroken \( U(1)_{EM} \) symmetry, denoted as \( Q \), must satisfy the condition \( Q \langle \Phi \rangle = 0 \). It is given by:  
\be  
Q = T_3 + \frac{Y}{2}\,.  
\ee  
A vacuum configuration that preserves this condition is:  
\be \label{ch1:vev}  
\langle \Phi \rangle = \begin{pmatrix} 0 \\ v/\sqrt{2} \end{pmatrix} \,.  
\ee  
Substituting eq. (\ref{ch1:vev}) into the Yukawa Lagrangian (\ref{ch1:LY3}), the mass terms for the charged fermions are generated as:  
\beqa \label{ch1:LY4}  
-{\cal L}_Y &=& \frac{{y}^d_{ij}v}{\sqrt{2}}\,\overline{d}_{L}{}_i\,  d_{Rj}\, + \frac{{y}^u_{i}v}{\sqrt{2}}\,\overline{u}_{L}{}_i\, \, u_{Ri}\,+\,\frac{{y}^e_{i}v}{\sqrt{2}}\,\overline{e}_{L}{}_i\,  e_{Ri}\,+\,H.c\,.\eeqa  

Defining the mass matrices as:  
\be \label{ch1:massmatrix}  
m^d_{ij} =\frac{{y}^d_{ij}v}{\sqrt{2}}\,,~~m^u_i=\frac{{y}^u_{i}v}{\sqrt{2}}\,,~~m^e_i=\frac{{y}^e_{i}v}{\sqrt{2}}\,,  
\ee  
it follows that \( m^u_i \) and \( m^e_i \) represent the physical masses of the three generations of up-type quarks and charged leptons, respectively. However, in the down-quark sector, the mass matrix \( m^d_{ij} \) is not diagonal in general.   

In the \( SU(3)_C \times U(1)_{EM} \) interaction basis, up-type and down-type quarks mix in charged current (CC) interactions. Consequently, the unitary transformations that diagonalise the down-quark mass matrix also appear in these interactions. The resulting mixing matrix is known as \gls{CKM} matrix: \(V_{CKM}\) \cite{Kobayashi:1973fv}, and  is defined via the charged current Lagrangian:  
\be  
{\cal L}_{CC} \sim W^+_\mu \, \overline{u}_{L}{}_i\,\gamma^\mu\, \left(V_{CKM}\right)_{ij}\,d_{Lj}\,.  
\ee  
The explicit form of \( V_{CKM} \) is obtained from the relation:  
\be  
{\rm Diag. }(m^2_d,m^2_s,m^2_b) \,=\, \left(V_{CKM}\right)^\dagger\, m^d (m^d)^\dagger \,V_{CKM}\,,  
\ee  
where \( m_d, m_s, m_b \) are the physical masses of the three generations of down-type quarks. In general, the elements of $V_{CKM}$ are written as:
\be V_{CKM} = \left(\ba{ccc} V_{ud} & V_{us} & V_{ub} \\V_{cd} & V_{cs} & V_{cb} \\V_{td} & V_{ts} & V_{tb}  \ea\right)\,. \ee
\(V_{CKM}\) is unitary in the SM and can be parameterised in terms of 3 angles and 1 phase.
These 4 CKM parameters, as well as the 9 charged fermion masses, are complicated functions of the 19 independent input Yukawa couplings.  The details of parameter counting and biunitary diagonalisation can be found in \cite{Grossman:2023wrq}.

These masses and mixing angles are experimentally measurable and are therefore considered as observables. Thus, the fermion sector of the SM contains 13 observables (nine charged fermion masses and four CKM parameters). Among these observables, the lepton masses and the quark mixings are measured experimentally, with uncertainties at the few per cent level. The heavy quark masses are determined up to \(10\%\) uncertainty, whereas the light quark masses have nearly \(30\%\) uncertainty as they are extracted from lattice simulation. The success of the SM in explaining fermion mass generation was solidified with the discovery of the Higgs boson at 125 GeV by the ATLAS and CMS collaborations in 2012 \cite{ATLAS:2015yey}. Beyond this, the SM has consistently passed all experimental tests, confirming its predictive power.

Despite its remarkable success, the SM faces several significant challenges that point towards the need for physics beyond its framework. Observational issues like neutrino masses, dark matter, and anomalies in certain SM predictions like $(g-2)_\mu$ indicate that the SM is an incomplete theory. However, there are also structural issues like gauge hierarchy, flavour puzzle, strong CP problem, etc. We focus on the flavour puzzle in the next section.

\section{Flavour puzzle}
\label{intro:issues}
The flavour puzzle in the Standard Model refers to the lack of an underlying explanation for the observed pattern of fermion generations, mass hierarchy, mixing angles, and CP violation. One fundamental question in this context is why nature exhibits exactly three generations of quarks and leptons. So far, there is no deeper theoretical justification for this duplication of fermion families in the SM. However, the fact that both the quark and lepton sectors contain the same number of generations suggests that an undiscovered fundamental mechanism might be responsible. On the other hand, it is also possible that this structure is merely a coincidence. The number of generations is also closely tied to the origin of CP violation \cite{Kobayashi:1973fv}. In particular, three generations of quarks are the minimum requirement for CP violation to occur in weak interactions, a phenomenon that has been experimentally confirmed \cite{Christenson:1964fg}. However, CP violation has not been observed in strong interactions, raising the question of why strong interactions appear to preserve CP symmetry. This unresolved issue, known as the strong CP problem, will be explored in detail in the next subsection.

Additionally, the SM does not provide an explanation for the vast differences in fermion masses across three generations, although each transforming identically under the gauge group \( SU(3)_C \times SU(2)_L \times U(1)_Y \). The Higgs mechanism allows fermions to acquire mass through their Yukawa couplings, but the specific values and hierarchy of these masses remain arbitrary and incalculable within the framework of the SM. While neutrinos, initially thought to be massless, have extremely small but nonzero masses as inferred from oscillation experiments \cite{Super-Kamiokande:1998kpq}.  However, in the SM, the neutrinos remain massless at the renormalisable level, and their mass terms can be generated through the famous dim. 5 Weinberg operator \cite{Weinberg:1979sa}. From experimental observations the mass ordering of neutrinos is still not determined, also the observed lepton mixing has almost all entries of ${\cal O}(1)$ unlike in the quark sector where mixings are small \cite{ParticleDataGroup:2022pth}. For a comprehensive review of the aforementioned puzzles and potential explanations, refer to \cite{Feruglio:2015jfa}.

\subsection{Quantifying the flavour hierarchies}
Quark masses span several orders of magnitude, from the up quark (2–3 MeV) to the top quark (173 GeV) \cite{ParticleDataGroup:2022pth}. Similarly, among leptons, the electron (0.511 MeV) is much lighter than the tau (1.78 GeV). For convenience, we give the relative strength of lighter fermion masses with respect to the third-generation fermion masses: 
\begin{align}
     \frac{m_u}{m_t} \simeq 1.4\times10^{-5}, \ ~& ~~~~~~~~~~~\ \frac{m_c}{m_t} \simeq 7.4\times 10^{-3}\, ,\nonumber   \\
    \frac{m_d}{m_b}  \simeq 1.14\times 10^{-5}, \ & ~~~~~~~~~~~\ \frac{m_s}{m_b}  \simeq 2.4 \times 10^{-2} \, ,\nonumber \\
    \frac{m_e}{m_{\tau}}  \simeq 2.8\times 10^{-4}, ~~ & ~~~~~~~~~~~\ \frac{m_{\mu}}{m_{\tau}} \simeq5.9\times 10^{-2}\, ,
\end{align}
Apparently, there is a several orders of magnitude difference between the masses of charged fermions. For example, the top quark is about five orders of magnitude heavier than the up quark. Also, it can be noticed that the intergenerational hierarchy is nearly more or less the same for all types of charged fermions and is proportional to $10^{-2}$. In a similar fashion, the elements of the CKM matrix governing quark mixing exhibit a hierarchical structure as\cite{ParticleDataGroup:2022pth}:
\be |V_{ud}| \sim 1,~~|V_{us}|\sim \lambda,~~|V_{ub}| \sim \lambda^3\,, \ee
with all the diagonal entries are close to $1$, and $\lambda$ is a small parameter and close to $\sim 0.2$. The reason for this hierarchical structure of masses and mixing remains largely unknown. Some of the popular solutions to this puzzle will be briefly discussed in section \ref{intro:attempts}. 

\subsection{The anarchy}
Neutrinos, once thought to be massless, have extremely small but nonzero masses inferred from oscillation experiments \cite{Super-Kamiokande:1998kpq}. Unlike the charged fermion sector, where hierarchical masses and mixings follow distinct patterns, neutrino masses and mixings appear to be less constrained and anarchical in nature. For example, considering the lightest neutrino mass is zero: \( m_1 = 0 \), the other two active neutrino masses assuming normal ordering (NO) are given by:
\begin{equation}
    m_2 = \sqrt{\Delta m_{sol}^2} \sim 0.01 \,{\rm eV}, ~~ \quad m_3 = \sqrt{\Delta m_{atm}^2}\,\sim\, 0.05 \,{\rm eV}
\end{equation}
Here, $\Delta m_{sol}^2$ and $\Delta m_{atm}^2$ are solar and atmospheric neutrino oscillation mass differences, which are extracted from the global fit to neutrino oscillation data \cite{Esteban:2024eli}. From above, we see that the neutrino masses are less hierarchical in nature. However, this is the widest range, $ 0$ to $0.05$ eV, that can be spanned by the three active neutrino masses. Alternatively, all neutrinos could have a similar magnitude of masses, following anarchic solutions \cite{Hall:1999sn,deGouvea:2012ac}, while still satisfying the cosmological bound \(\sum_{\nu_i} m_{\nu_i} = 0.12\) eV \cite{Planck:2018vyg}.
 
 The lepton mixing matrix, known as the Pontecorvo-Maki-Nakagawa-Sakata (PMNS) matrix, has large mixing angles in contrast with the small mixing angles in the quark sector. This corresponds to almost ${\cal O}(1)$ entries for the mixing elements. For convenience, the PMNS matrix for normal ordering can be written as:
 \begin{equation}
|U_{\text{PMNS}}| \approx
\begin{bmatrix}
0.814 & 0.580 & 0.148 \\
0.363 & 0.692 & 0.623 \\
0.453 & 0.620 & 0.640
\end{bmatrix}\, ,
\end{equation}
where the best-fit values used are: $
\theta_{12} = 33.68^\circ, 
\theta_{23} = 43.3^\circ, 
\theta_{13} = 8.52^\circ, 
\delta_{\text{CP}} = 177^\circ.$ These values are taken from the NuFIT 6.0 (2024) global fit \cite{Esteban:2024eli}.

\section{Strong CP problem}
\label{strong CP}
Another unresolved issue in the SM is the Strong CP problem. In Quantum Chromodynamics (QCD), the physical CP-violating parameter is given by:  
\be \label{theta-bar}\bar{\theta } =\, \theta_{QCD}\, +\, \arg (\det(M_u M_d))\,,\ee  
where \( M_u \) and \( M_d \) are the mass matrices for the up and down quark sectors, respectively, and \( \theta_{QCD} \) is defined through the term:  
\be {\cal L}_{\theta_{QCD}} \,=\, \frac{\theta_{QCD}\, g_s^2}{32\pi^2} \, G^a_{\mu \nu}\tilde{G^a}{}^{\mu \nu}\, .\ee  
This term, permitted by SM gauge symmetry, violates both parity (P) and time-reversal symmetry (T), resulting in CP violation in the strong interaction. However, this CP-violating term is not uniquely defined, as it can be shifted into the second term of eq. (\ref{theta-bar}) (or vice versa) through a redefinition of the quark fields. The physical parameter \( \bar{\theta} \) has observable consequences, particularly in the neutron's electric dipole moment (EDM). The non-observation of a neutron EDM places a constraint \cite{Abel:2020pzs}:  
\be \bar{\theta} < 10^{-10} \ee  
This is an extremely small number, which is puzzling because, from a theoretical standpoint, one would expect \( \bar{\theta} \) to be of order \( \mathcal{O}(1) \), similar to the CP-violating phase in weak interactions. This unexpected smallness is what constitutes the Strong CP problem.  

Several mechanisms have been proposed to explain why \( \bar{\theta} \) is so small. One possible solution is the massless quark scenario \cite{Georgi:1981be,Choi:1988sy}, where if one of the light quarks (up or down) were exactly massless, the strong CP phase could be rotated away, making \( \bar{\theta} \) unobservable. However, lattice QCD studies and experimental data indicate that all SM quarks have nonzero masses, ruling out this possibility. Another well-known solution is the Peccei-Quinn (PQ) mechanism \cite{Peccei:1977hh,Peccei:1977ur,Wilczek:1977pj,Weinberg:1977ma,Kim:1979if,Shifman:1979if,Dine:1981rt}, which introduces a new global \( U(1)_{PQ} \) symmetry that is spontaneously broken, leading to the emergence of a new light scalar particle called the axion. The axion dynamically adjusts itself to cancel out the CP-violating term, naturally driving \( \bar{\theta} \) to zero. 

Another type of approach is the discrete symmetry-based solution to the puzzle.  The leading order contribution to the strong CP-violating phase is forbidden by either CP or P as the symmetry of the ultraviolet theory, and a tiny value is generated when higher order corrections are taken into account. The complete theory which assumes CP as the symmetry of the Lagrangian at high energy is known as the Nelson-Barr mechanism \cite{Nelson:1983zb,Barr:1984qx}. It introduces new heavy quarks or vector-like fermions with a carefully arranged flavour structure to ensure that CP violation appears only in weak interactions, leaving \( \bar{\theta} \) naturally small. The other kind, parity-based solutions \cite{Mohapatra:1978fy,Babu:1988mw,Babu:1989rb}, impose \gls{L-R} at high energies, which forces \( \bar{\theta} = 0 \) at tree level, with small corrections arising only through radiative effects. In chapter \ref{chap6}, we will show that our proposed mechanism which explains the mass hierarchy also solves the strong CP puzzle in a very effective manner.

\section{Attempts to generate hierarchy}
\label{intro:attempts}
Several \gls{BSM} frameworks have been proposed to explain the fermion mass hierarchy and flavour structure, including the Froggatt-Nielsen (FN) mechanism, Clockwork mechanism, Extra-dimensional models, and Radiative mass generation.

The FN mechanism \cite{Froggatt:1978nt} introduces a horizontal symmetry, typically a global or gauged \( U(1) \), under which different generations of fermions carry distinct charges. This symmetry is spontaneously broken by the VEV of a flavon field, leading to an effective suppression of Yukawa couplings by powers of a small parameter \( \epsilon \sim \frac{\langle \phi \rangle}{\Lambda} \), naturally generating the observed mass hierarchy. For example, if we obtain the mass matrices of the following form \cite{Shaikh:2024ufv}:
\be M_u \sim \left(\ba{ccc}\epsilon^3 &\epsilon &1\\ \epsilon^3 &\epsilon &1\\\epsilon^3 &\epsilon &1\ea\right)\,, \quad M_d \sim \left(\ba{ccc}\epsilon^3 &\epsilon^2 &\epsilon\\ \epsilon^3 &\epsilon^2 &\epsilon \\ \epsilon^3 &\epsilon^2 &\epsilon\ea\right)\,, \quad M_e \sim \left(\ba{ccc}\epsilon^3 &\epsilon^2 &\epsilon\\ \epsilon^3 &\epsilon^2 &\epsilon \\ \epsilon^3 &\epsilon^2 &\epsilon\ea\right)\,,\ee
then it lead to following pattern of masses:
\beqa m_t &\sim& 1\,,~~~~~ m_{b,c,\tau} \sim \epsilon\,, \nonumber\\
m_{s,\mu} &\sim& \epsilon^2\, , ~~~~~m_{u,d,e} \sim \epsilon^3\,.\eeqa
For $\epsilon\sim 0.02-0.03$, the above structure reproduces the realistic spectrum of hierarchical masses.

The clockwork mechanism \cite{Giudice:2017suc}, originally developed in the context of axions and massive gauge bosons, has been applied to flavour physics as well. In this framework, fermions couple sequentially across a series of sites in a discrete or continuous "clockwork chain," with the mass hierarchy emerging due to exponential localization of the effective zero mode in the chain. This mechanism can explain why some Yukawa couplings are naturally small without requiring finely tuned parameters. 

Extra-dimensional models, such as those based on warped geometries (e.g., Randall-Sundrum models) \cite{Randall:1999ee}, provide an alternative perspective by localizing fermions at different positions along an extra spatial dimension. In these scenarios, the overlap of fermion wavefunctions with the Higgs field determines their effective Yukawa couplings. This leads to an exponential mass suppression for lighter generations. 

Another intriguing possibility is the radiative mass generation mechanism \cite{tHooftRenormalizableLF,Weinberg:1972ws,Georgi:1972hy,Georgi:1972mc,Mohapatra:1974wk,Barr:1976bk,Barr:1978rv}, where the masses of certain fermions arise at the loop level rather than at the tree level. In these models, new heavy particles running in quantum loops generate effective Yukawa couplings, naturally explaining the smallness of certain fermion masses. Since quantum corrections generate masses for some of the fermions, these frameworks make masses calculable parameters of the theory, which is a distinctive feature compared to the previously discussed mechanisms. One more advantage of this framework is that the loop suppression factor $\frac{1}{16\pi^2}$ is of the order of intergenerational mass hierarchy.  Again, this mechanism naturally leads to a small number of unknown parameters. All these features will be discussed in detail in the next subsection.

\subsection{Radiative mass generation}

An elegant approach to understand the peculiar hierarchical structure of charged fermion masses is to allow only the third-generation fermions to acquire mass at leading order, while the masses of the first two generations emerge through quantum corrections. This mechanism makes the masses of the first- and second-generation fermions fully or partially calculable within the theory. Such an approach was first explored in \cite{tHooftRenormalizableLF,Weinberg:1972ws} shortly after the development of the SM, aiming to treat mass differences as computable parameters of the theory. Later on, in \cite{Georgi:1972hy,Georgi:1972mc,Mohapatra:1974wk,Barr:1976bk,Barr:1978rv}, the mechanism is explicitly used to explain the electron-muon mass ratio; in other words, the electron mass is computed in terms of muon mass. In these approaches, it was realised that the successful implementation of the mechanism requires an extension of the SM through the inclusion of massive fields.

The necessity of extending the SM gauge sector can be understood as follows. Assigning nonzero masses only to third-generation fermions at the zeroth order leads to an accidental \( U(2)^5 \) global symmetry in the mass Lagrangian. Since this symmetry is respected by the full Lagrangian, quantum corrections from other SM fields do not contribute to the masses of the lighter-generation fermions. This follows from the principle of technical naturalness \cite{tHooft:1979rat}, which states that corrections to a mass term in the SM are proportional to the mass itself. Consequently, if certain fermion masses are initially set to zero, they will remain zero at all orders in perturbation theory. Therefore, in order to successfully incorporate the radiative mass generation mechanism within the SM, the field content must be extended.  

A more complete realisation of the radiative mass mechanism was introduced by Balakrishna in \cite{Balakrishna:1987qd}, successfully reproducing the observed hierarchical mass spectrum of quarks and charged leptons. This work proposed a L-R symmetric model with an extended gauge group, \( SU(3)_C \times SU(2)_L \times SU(2)_R \times U(1)_{B-L} \), and an expanded fermion sector incorporating an additional generation of massive  \gls{VL} states. These VL fermions played a crucial role in generating tree-level masses exclusively for third-generation fermions through a seesaw-like mechanism. The model’s scalar sector was minimally extended to spontaneously break L-R symmetry and enable the radiative generation of lighter-generation fermion masses via one-loop and two-loop diagrams. Subsequent studies \cite{Balakrishna:1988ks,Balakrishna:1988xg,Balakrishna:1988bn,Babu:1988fn,Babu:1989tv,Ma:1989ys,Babu:1990fr1} further developed this framework, reinforcing the viability of radiative mass generation as a mechanism for explaining fermion mass hierarchies.

However, after initial interest, research in this direction significantly declined, with only a handful of works published over the next two decades \cite{Arkani-Hamed:1996kxn,Ma:1988qc,HernandezGaleana:2004cm,Appelquist:2006ag,Barr:2007ma,Dobrescu:2008sz,Graham:2009gr}. Interest in the field was later revived by Weinberg \cite{Weinberg:2020zba}, leading to a resurgence of studies on radiative mass generation \cite{Jana:2021tlx,Mohanta:2022seo,Chiang:2021pma,Chiang:2022axu,Baker:2020vkh,Baker:2021yli,Yin:2021yqy}. Notably, Weinberg proposed a model based on horizontal gauge symmetry \( SO(3)_L \times SO(3)_R \), where the three generations of left-handed quarks and leptons, as electroweak doublets, form a \( (3,1) \) representation under \( SO(3)_L \times SO(3)_R \), while the right-handed quarks and charged leptons occupy separate \( (1,3) \) representations. In this framework, the masses of first- and second-generation fermions arise through the exchange of \( SO(3)_{L,R} \) gauge bosons at the two-loop and one-loop levels, respectively. However, the model predicts several unrealistic correlations among various quarks and lepton masses, which are proven wrong by the present experimental observations. For example, the model predicts that the ratio of second-generation to third-generation fermion masses is the same for up quarks, down quarks, and charged leptons, which is not even approximately true of observed masses. Another unrealistic feature of the framework is that they do not exhibit any quark mixings in theory.

Extended models of radiative fermion masses introduce self-energy corrections through loops involving heavy fermions, with contributions mediated by either scalars or vector bosons. Based on the dominant contribution, these models can generally be classified into two categories:  
\begin{enumerate}[label= (\roman*)]
    \item \textbf{Spin-0 mediated models}: These involve scalar particles in the fermion self-energy loops (see \cite{Balakrishna:1987qd,Balakrishna:1988ks,Balakrishna:1988xg,Balakrishna:1988bn,Babu:1988fn,Babu:1989tv,Ma:1989ys,Babu:1990fr1,Arkani-Hamed:1996kxn,Ma:1988qc,Barr:2007ma,Graham:2009gr,Dobrescu:2008sz,Crivellin:2010ty,Crivellin:2011sj,Chiang:2021pma,Chiang:2022axu,Baker:2020vkh,Baker:2021yli,Yin:2021yqy,Chang:2022pue,Greljo:2023bix,Chang:2024snt} for examples), and require the extension of the SM Yukawa sector. 
    \item \textbf{Spin-1 mediated models}: These involve vector bosons in the self-energy loops \cite{HernandezGaleana:2004cm,Appelquist:2006ag,Reig:2018ocz,Weinberg:2020zba,Jana:2021tlx,Mohanta:2022seo,Mohanta:2023soi,Kuchimanchi:2024nkt,Mohanta:2024wcr,Jana:2024icm,Mohanta:2024wmh,Klett:2024ubx}. In these models, extending the SM gauge symmetry to include a gauged flavour symmetry group \( G_F \) is necessary as we show in the next chapter. The radiative corrections in this framework are dictated by the gauge couplings and the masses of the new gauge bosons.  
\end{enumerate}
It is worth pointing out that Case (ii) frameworks have some more advantages over Case (i) in the following regime:  
 \begin{itemize}
     \item \textit{Predictive Power}: The quantum corrections to fermion masses depend on independently measurable gauge couplings and gauge boson masses, making the loop-corrected masses, in principle, calculable quantities. 
     \item \textit{Fewer Free Parameters}: Symmetry-based extensions typically introduce fewer arbitrary parameters compared to scalar-based models; thus are potentially more economical. This also enhances theoretical predictability.
     \item \textit{Natural Gauge Symmetry Extension}: These models arise naturally from fundamental principles of gauge theories, aligning well with established ideas in particle physics.  
 \end{itemize}
Given these advantages, radiative models where a new gauge sector plays a primary role in generating the masses of lighter quarks and leptons are particularly promising and ask for systematic investigation. At the time I began my doctoral research, no realistic gauged extension of radiative mass generation frameworks had been developed, motivating the choice of this topic for my thesis.

\section{The objectives and Thesis structure}
The central objective of the thesis is to explain the fermion mass hierarchy issue of the SM through the radiative mechanism in a calculable and predictive manner. Some specific objectives aligned in this direction are outlined below:
\begin{enumerate}[label= (\roman*)]
\item  Systematic investigation of the possibility of radiative generation of fermion masses in gauged extensions of the SM.\\
 - We study these aspects by considering an Abelian theory and outline general conditions under which such extensions can viably and effectively accommodate the observed flavour hierarchies.
\item Implementing radiative mass generation mechanism in extended Abelian theories. \\
   - We discuss a class of ultraviolet complete renormalisable theories and qualitatively evaluate the spectrum of masses, scales and couplings. Additionally, we point out inherent complexities in an Abelian extension and mention some possible way-outs. 
%
\item Studying the implementation of the radiative mechanism in non-Abelian extensions and identifying the improvements over Abelian counterparts.\\
   - We outline various advantages of non-Abelian gauge groups to incorporate the radiative mechanism and discuss a complete model based on $SU(3)_F$ extension, which leads to a more predictive framework.
   
\item Left-Right extension and solving the strong CP puzzle.\\
- We discuss the implementation of the radiative mass mechanism in left-right symmetric theory and show that when parity symmetry is imposed on such a framework, it naturally explains the strong CP puzzle.
\end{enumerate}

All the above-mentioned objectives are incorporated in the thesis with great detail and organized in the following manner. In the next chapter, we systematically outline the steps of the radiative mass generation mechanism and explicitly compute the self-energy correction loop diagram. For simplicity, the above details are carried out by considering a toy framework based on an Abelian gauge theory.  The mass-generating loop diagrams in this framework are formed by the emission and absorption of vector boson of the Abelian symmetry. We also outline the limitations of the framework at 1-loop and point out possible way-outs on which subsequent chapters are based.

Chapter \ref{chap3} integrates the radiative mass mechanism within the SM extensions by utilising the findings of the toy framework analysis. The explicit renormalisable model constructed is based on two flavour non-universal Abelian gauge symmetries. The first and second-generation fermion masses, both, are induced at 1-loop, while the hierarchy between these two generations results from a gap between the masses of two vector bosons of the extended gauge symmetries. Although it reproduces the observed hierarchical pattern of fermion masses, phenomenologically, the lowest new physics predicted turns out to be $\sim 10^8$ GeV. Chapter \ref{chap4} discusses a framework in which a slightly lower allowed scale is obtained, as well as the first generation fermion masses are induced in the two-loop. It is shown that the couplings, which dominantly constrain the lower bound on flavour symmetry breaking scale and the mass of the first generation fermions, are related. In the limit of vanishing of one leads to the vanishing of the other.

Next, in chapter \ref{chap5}, we discuss the additive mass generation mechanism using non-Abelian gauge symmetries, i.e., $SU(3)_F$. We also point out some of the advantages over the Abelian theories of the radiative mechanism. Most importantly, we show that the $SU(3)_F$ framework typically leads to fewer parameters compared to the Abelian theories discussed in previous chapters. We also discuss a pattern of $SU(3)_F$ breaking, which leads to hierarchical masses for the first two generation fermions, although they are induced at the 1-loop level.

In Chapter \ref{chap6}, we explore the radiative generation of fermion masses within the Left-Right symmetric framework. Furthermore, we demonstrate that imposing parity symmetry naturally provides a resolution to the strong CP problem. Chapter \ref{chap7} concludes our study, where we also outline potential future research directions. Lastly, two appendices are included to clarify key technical aspects discussed in the main text.

%% file: 20_Chapter_2/Literature_review.tex
\chapter{Radiative generation of fermion masses: General aspects}\label{chap:2}

As discussed in the previous chapter, gauge extension models provide a more predictive framework compared to other types of radiative mass models. In this chapter, we explore the simplest gauge theory, an Abelian framework, which serves as a toy model for studying the fundamental aspects of the radiative mass generation mechanism. This framework involves three generations of chiral fermions, \( f'_{L i} \) and \( f'_{R i} \) (\( i=1, 2, 3 \)), which are non-trivially charged under the \(G_F= U(1)_F \) gauge symmetry, making them suitable for incorporating the radiative mechanism. An additional pair of vector-like states is postulated, which plays a crucial role in obtaining the rank-1 structure of the mass matrix as well as in the underlying mechanism. In this framework, radiative masses arise exclusively from the interaction between the gauge boson of the Abelian symmetry and the fermion multiplets. It is shown that the abelian symmetries which can viably implement the mechanism must be flavour non-universal in nature.

In the next section, we analyze the conditions under which an Abelian theory can generate fermion masses for different generations at various orders of perturbation theory. We start with obtaining the rank 1 structure of the mass matrix for the chiral fermions, and then we compute the leading-order corrections to it. The rank of a matrix is determined from the number of non-vanishing eigenvalues it corresponds. 

\section{The Toy model and Loop-induced  masses}
\label{sec:general}
Consider $q_{Li}$ and $q_{Ri}$ as the charges of the three generations of the aforementioned chiral fermions, $f^\prime_{L i}$ and $f^\prime_{R i}$, respectively, under the abelian symmetry. It is to be noted that under a chiral gauge symmetry, such as the electroweak symmetry of the SM, \( f'_{Li} \) and \( f'_{Ri} \) must transform differently.  We also introduce a pair of VL fermions, \( F'_{L} \) and \( F'_{R} \), which may or may not be charged under the \( U(1)_F \) gauge symmetry but transform identically under a chiral symmetry. Unlike the chiral fermions, the vector-like fermions possess a symmetry-preserving mass. The rank 1 structure for the mass matrix of the chiral fermions ($f^\prime_{L i}$ and $f^\prime_{R i}$) can be obtained by allowing Yukawa interaction between the SM fermions and vector-like fermions only. Any direct coupling among the chiral fermions are prohibited through a careful selection of \( U(1)_F \) charges. 

The interaction term of chiral and VL fermions with the gauge boson of $U(1)_F$ can be written as:
\beqa \label{L_gauge}
-{\cal L}_{\rm gauge} &=& g_X X_\mu \left({ q}_{L \alpha}\, \overline{f}^\prime_{L \alpha} \gamma^\mu f^\prime_{L \alpha} + { q}_{R \alpha}\, \overline{f}^\prime_{R \alpha} \gamma^\mu f^\prime_{R \alpha} \right)\, \nonumber\\
&\equiv& g_X X_\mu \left( \overline{f}^\prime_{L }{\mathbb Q}_{L} \gamma^\mu f^\prime_{L } +  \overline{f}^\prime_{R } {\mathbb Q}_{R}\gamma^\mu f^\prime_{R} \right)\,.\eeqa
where $\alpha=1,...,4$, $f^\prime_{L \alpha} = (f^\prime_{L i},F^\prime_L)$, and  $f^\prime_{R\alpha} = (f^\prime_{R i},F^\prime_R)$ . $q_{L,R i}$ is the  charge of  chiral fermion $f^\prime_{L,Ri}$ under $G_F$ and $q_{L4} = q_{R4} $. The  $4\times4$ charge matrix is defined by:
\be {\mathbb Q}_{L,R} = \left(\ba{cc}q_{L,R} & 0\\0&q_{L,R4}\ea\right) \, ,\ee
with 
\be \label{q_LR}
q_{L} = {\rm Diag.} \left(q_{L1},\,q_{L2},\,q_{L3}\right)\,,~~q_{R} = {\rm Diag.} \left(q_{R1},\,q_{R2},\,q_{R3}\right)\,.\ee

\subsection{The tree level}
\label{toy:treelevel}
After the breaking of \( U(1)_F \) and chiral symmetries, the mass Lagrangian of the fermions at leading order can be arranged to take the following form:
\beqa \label{L_mass}
-{\cal L}_{m} &=& \mu_{L i}\, \overline{f}^\prime_{Li} F_R^\prime + \mu_{R i}\, \overline{F}^\prime_L f_{Ri}^\prime + m_F\, \overline{F}^\prime_L F^\prime_R + {\rm h.c.}\,, \nonumber \\
&\equiv & \overline{f}^\prime_{L \alpha}\, {\cal M}^{(0)}_{\alpha \beta}\, f^\prime_{R \beta} + {\rm h.c.}\,, \eeqa
with $f^\prime_{L(R)4} = F^\prime_{L(R)}$ and $\alpha = i,4$. The $4 \times 4$  mass matrix ${\cal M}^{(0)}$ has the following form:
\be \label{M0}
{\cal M}^{(0)} = \left(\ba{cc} 0_{3 \times 3} & (\mu_L)_{3 \times 1} \\ (\mu_R)_{1 \times 3} & m_F \ea \right)\,,\ee
where $\mu_L = (\mu_{L1},\mu_{L2},\mu_{L3})^T$ and $\mu_R = (\mu_{R1},\mu_{R2},\mu_{R3})$. The specific form of interactions in \({\cal L}_{m}\) can be derived by treating \(\mu_{L,R}\) as spurions and assigning them appropriate charges under the full symmetry of the theory. Both \(\mu_L\) and \(\mu_R\) break \(U(1)_F\) symmetry, while at least one of them also breaks chiral symmetry, depending on the gauge charge assignments of the VL fermions.

The structure of the tree-level mass matrix, given in eq. (\ref{M0}), closely aligns with the form introduced in the framework of the universal seesaw mechanism \cite{Berezhiani:1983hm,Berezhiani:1985in,Chang:1986bp,Rajpoot:1986nv,Davidson:1987mh}, where all the three-generation chiral fermions obtain masses through the seesaw mechanism. However, in our case, due to the specific structure of \({\cal M}^{(0)}\), only one of the chiral generations acquires a nonzero mass. In the so-called seesaw approximation, when VL fermions mass scale is higher from the other scales of the theory, i.e., $  m_F \gg \mu_{L,R}$ \cite{Minkowski:1977sc,Yanagida:1979gs,GellMann:1979vob}, the vector-like states can be integrated out which leads following effective mass matrix for chiral states:
\be \label{M0_eff}
M^{(0)}_{ij} = -\frac{1}{m_F}\, \mu_{L i}\, \mu_{R j}\,.\ee
As can be seen, $M^{(0)}$ is a direct product of two vectors and, therefore, is a rank-1 matrix. The non-vanishing eigenvalue corresponding to $M^{(0)}$ is proportional to the trace of this matrix. As $\mu_{L,R} \ll m_F$, the mass of this state is suppressed compared to that of the VL fermion. This state is identified as the third-generation fermion and, in this way, only one pair of chiral fermions is arranged to acquire a tree-level mass in the underlying framework.

The fermion fields in eq. (\ref{L_mass}) are primed to separate them from the physical basis. The physical basis is obtained by considering unitary transformations as $f_{L,R} = {\cal U}^{(0) \dagger}_{L,R} f^\prime_{L,R}$. Since the VL fermions and third generation chiral fermions are the only massive states at the tree level,  the biunitary diagonalisation can be written as:
\be \label{M0_diag}
{\cal U}_L^{(0) \dagger}\,{\cal M}^{(0)}\,{\cal U}_R^{(0)} \equiv {\cal D}^{(0)} = {\rm Diag}.\left( 0,0,m_3^{(0)},m_4^{(0)}\right)\,.\ee
In the seesaw approximation, the unitary matrices can have a simplified analytical expression and can be written as \cite{Joshipura:2019qxz}:
\be \label{U0_ss}
{\cal U}^{(0)}_{L,R} = \left(\ba{cc} U_{L,R}^{(0)} & - \rho_{L,R}^{(0)} \\
 \rho_{L,R}^{(0) \dagger} U_{L,R}^{(0)} & 1 \ea\right) + {\cal O}(\rho^2)\,, \ee
where
\be \label{rho0}
\rho^{(0)}_L = -\frac{1}{m_F} \mu_L\,,~~\rho_R^{(0) \dagger} = - \frac{1}{m_F} \mu_R\,,\ee
are three-dimensional column and row vectors, respectively, commonly known as seesaw expansion parameters. $U_{L,R}^{(0)}$ are $3 \times 3$ unitary matrices defined from the biunitary diagonalisation:
\be \label{M0_eff_diag}
U_L^{(0) \dagger}\, M^{(0)}\, U_R^{(0)} = {\rm Diag.}\left(0,0,m_3^{(0)}\right)\,.\ee

The simple form of $M^{(0)}$ can be used to determine $U_{L,R}^{(0)}$ analytically. The third column of this unitary matrix can be obtained by deriving the eigenvector corresponding to the only non-zero eigenvalue of $M^{(0)}$. The other two eigenvectors can be derived from the orthogonality. We find:
\be \label{U0_ana}
U_L^{(0)} = \left( \ba{ccc} -\frac{\mu_{L2}^*}{\sqrt{N_1}} & -\frac{\mu_{L1} \mu_{L 3}^*}{\sqrt{N_2}} & \frac{\mu_{L1}}{\sqrt{N_3}} \\ 
\frac{\mu_{L1}^*}{\sqrt{N_1}} & -\frac{\mu_{L2} \mu_{L 3}^*}{\sqrt{N_2}} & \frac{\mu_{L2}}{\sqrt{N_3}} \\
0 & \frac{|\mu_{L1}|^2 + |\mu_{L2}|^2}{\sqrt{N_2}} & \frac{\mu_{L3}}{\sqrt{N_3}} \\
 \ea\right)\, V^{[12]}_L,\ee
with $N_{1,2,3}$ as normalisation constants that can be obtained by normalising each of the three columns. $V^{[12]}_L$ denotes an arbitrary unitary rotation in the $1$-$2$ plane, which remains unspecified due to degeneracy among the first two generation masses. A similar expression applies for $U_R^{(0)}$, with substitution $\mu_L \to \mu_R^{*}$.

\subsection{At 1-loop}
\label{subsec:1loop}
The higher-order corrections to the tree-level mass matrix can be straightforwardly calculated in the physical basis, where the \( U(1)_F \) gauge interactions are expressed as:
\be \label{L_gauge_mass}
-{\cal L}_X = g_X X_\mu\, \left(({\cal Q}^{(0)}_L)_{\alpha \beta}\, \overline{f}_{L \alpha} \gamma^\mu f_{L \beta} + ({\cal Q}^{(0)}_R)_{\alpha \beta}\, \overline{f}_{R \alpha} \gamma^\mu f_{R \beta} \right)\,, \ee 
where
\be \label{Q0}
{\cal Q}_{L,R}^{(0)} ={\cal U}_{L,R}^{(0) \dagger}\,{\mathbb Q}_{L,R}\,{\cal U}_{L,R}^{(0)}\,,= {\cal U}_{L,R}^{(0) \dagger}\,\left(\ba{cc} q_{L,R} & 0\\0 & q_{L,R4} \ea \right)\,{\cal U}_{L,R}^{(0)}\,.\ee
For non-universal charges $q_{L,R\alpha}$, the gauge interactions are not flavour diagonal in the physical basis. This plays a crucial role in inducing the masses of the remaining massless fermions at higher orders in perturbation theory. To compute such effects, it is convenient to go to the Dirac fermion basis, where the gauge interactions in eq. (\ref{L_gauge_mass}) can be written as:
\beqa \label{a11}
-{\cal L}_{\rm gauge} &=& g_X\, X_\mu  \overline{f}_{ \alpha} \gamma^\mu \,{\cal C}_{\alpha \beta}\,  f_{\beta},\eeqa 
with the coupling defined as:
\be \label{a12}
{\cal C}_{\alpha \beta} = ({\cal Q}^{(0)}_L)_{\alpha \beta}\, P_L + ({\cal Q}^{(0)}_R)_{\alpha \beta} P_R \ . \ee
Here $P_L$ and $P_R$ are the usual chiral projection operators:
\be P_L=\frac{1-\gamma_5}{2} \,,~~~P_R= \frac{1+\gamma_5}{2}\ee

\begin{figure}[t]
\centering
\includegraphics[width=0.6\textwidth]{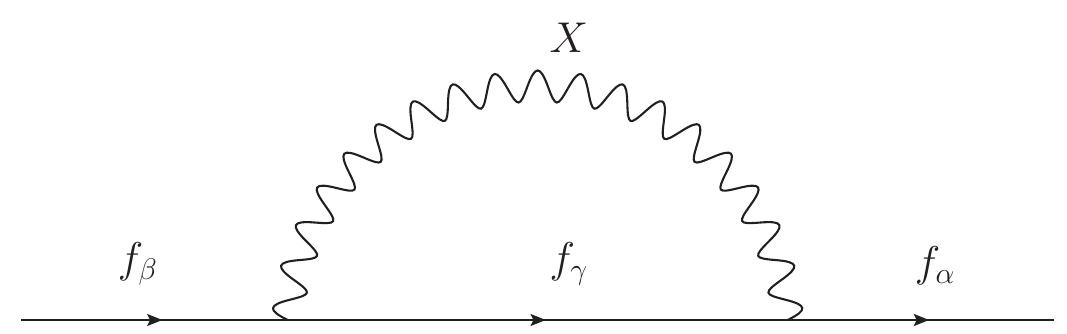}
\caption{Gauge boson induced fermion self-energy correction at 1-loop.}
\label{fig:1loop}
\end{figure}

At the 1-loop level, the first and second-generation fermions can acquire masses via diagrams that include the $U(1)_F$ gauge boson $X_\mu$ and massive fermions ($f_{3}, f_{4}$) within the loop, as shown in Fig. {\ref{fig:1loop}}. The one-particle-irreducible (1PI) two-point function can be written using this diagram is given by
\beqa \label{a13}
-i\Sigma_{\alpha \beta}(p) &=& (-ig_X)^2\sum_{\gamma}\int \frac{d^4 k}{(2\pi)^4}\frac{(\gamma^\mu {\cal C}^{\dagger}_{\alpha \gamma})\cdot i(\slashed{k}+\slashed{p}+m_{\gamma})\cdot (\gamma^\nu {\cal C}_{\gamma \beta})}{[(k+p)^2-m_{\gamma}^2 + i\epsilon]}\, \nonumber \\
&& \hspace{4cm}\times~~ \Delta_{\mu \nu}(k)\,,\eeqa
with 
\be \label{a14}
\Delta_{\mu \nu}(k) = \frac{-i}{k^2-{M_{X}}^2 + i\epsilon} \left( \eta_{\mu \nu}-(1-\zeta)\frac{k_\mu k_\nu}{k^2-\zeta M_X^2}\right)\,.\ee

We set $p=0$ in order to go to the on-shell condition for the massless fermion and calculate the loop contribution in the Feynman-'t Hooft gauge ($\zeta = 1$), using the dimensional regularization scheme. Since the denominator is an even function of $k$ when $p$ approaches zero, terms in the numerator containing an odd power of $k$ vanish. Also,  ${\cal Q}^{(0)}_{L,R}{}^\dagger = {\cal Q}^{(0)}_{L,R}$ (see eq. (\ref{Q0})). Therefore,
\begin{align}
\Sigma_{\alpha \beta}(0) =\ -i g_X^2 \mu^\epsilon\ & \sum_{\gamma}\left[({{\cal Q}^{(0)}_L})_{\alpha \gamma} P_R + ({{\cal Q}^{(0)}_R})_{\alpha \gamma} P_L \right]  \nonumber \\
& \times \int \frac{d^dk}{(2\pi)^d} \frac{d \ m_{\gamma} }{k^2-m_{\gamma}^2 + i\epsilon}\frac{1}{k^2-{M_{X}}^2 + i\epsilon}\ {{\cal C}}_{\gamma \beta}\,, \label{a15}
\end{align} 
where, $d=4-\epsilon$ and $\mu$ as the renormalisation scale. Also, we have used the identities $\gamma^\mu {\cal C}_{\alpha \gamma} = \left[({{\cal Q}^{(0)}_L})_{\alpha \gamma} P_R + ({{\cal Q}^{(0)}_R})_{\alpha \gamma} P_L \right] \gamma^\mu\,$ and $\gamma^\mu \gamma_\mu = d$ in order to obtain eq. (\ref{a15}). After some algebraic simplifications, it can be written as
\begin{align}
\Sigma_{\alpha \beta}(0) =\ \frac{d g_X^2}{16\pi^2}\ & \sum_{\gamma}\ m_\gamma \ \left[({{\cal Q}^{(0)}_L})_{\alpha \gamma} ({{\cal Q}^{(0)}_R})_{\gamma \beta} P_R + ({{\cal Q}^{(0)}_R})_{\alpha \gamma} ({{\cal Q}^{(0)}_L})_{\gamma \beta} P_L \right]  \nonumber \\
& \times \frac{({2 \pi \mu})^\epsilon}{i \pi^2} \int d^dk \frac{ 1 }{k^2-m_{\gamma}^2 + i\epsilon}\frac{1}{k^2-{M_{X}}^2 + i\epsilon}
\end{align}
The integration in the above can be computed and expressed in terms of Passarino-Veltmann function $B_0$ \cite{Passarino:1978jh} leading to a final expression
\be \label{a19}
\Sigma_{\alpha \beta}(0) =\  \frac{ g_X^2}{4\pi^2}\  \sum_{\gamma}\ m_\gamma \ \left[({{\cal Q}_L})_{\alpha \gamma} ({{\cal Q}_R})_{\gamma \beta} P_R + ({{\cal Q}_R})_{\alpha \gamma} ({{\cal Q}_L})_{\gamma \beta} P_L \right]\ B_0[M_X,m_\gamma ]   \ee 
with 
\beqa\label{B0}
B_0[M,m ] &=& \frac{({2 \pi \mu})^\epsilon}{i \pi^2} \int d^dk\, \frac{ 1 }{k^2-m^2 + i\epsilon}\frac{1}{k^2-{M}^2 + i\epsilon} \, \nonumber \\
& = &  \Delta_\epsilon + 1 - \frac{M^2 \ln \frac{M^2}{\mu^2} - m^2 \ln \frac{m^2}{\mu^2}}{M^2 - m^2}\,. \eeqa
Here, $\Delta_\epsilon$  the singular part of loop integration for $\epsilon \to 0$. Explicitly,
\be \label{Depsilon}
\Delta_\epsilon = \frac{2}{\epsilon} - \gamma + \ln 4\pi\,.\ee
From, eq. (\ref{a19}) the 1PI selfenergy function can be parametrised as following:
\be \label{Sigma}
\Sigma_{\alpha \beta}(p=0) = \sigma^L_{\alpha \beta}\, P_L + \sigma^R_{\alpha \beta}\, P_R\,,\ee
with
\beqa \label{sigma_LR}
\sigma^L_{\alpha \beta} & = &   \frac{g_X^2}{4 \pi^2}\, \sum_\gamma ({\cal Q}^{(0)}_R)_{\alpha \gamma}\, ({\cal Q}^{(0)}_L)_{\gamma \beta}\, m^{(0)}_\gamma\, B_0[M_X,m_\gamma^{(0)}{}]\,, \nonumber \\
\sigma^R_{\alpha \beta} & = &   \frac{g_X^2}{4 \pi^2}\, \sum_\gamma ({\cal Q}^{(0)}_L)_{\alpha \gamma}\,({\cal Q}^{(0)}_R)_{\gamma \beta}\,m^{(0)}_\gamma \, B_0[M_X,m^{(0)}_\gamma{}]\,,\eeqa
The 1-loop corrected $4\times4$ fermion mass matrix can be expressed in the form:
\be \label{M_loopcorr}
{\cal M}^{(1)} = {\cal M}^{(0)} + \delta{\cal M}^{(0)}\,,\ee
where
\be \label{del_calM}
\delta{\cal M}^{(0)} = {\cal U}^{(0)}_L\,\sigma^R\,{\cal U}_R^{(0)\dagger}\,. \ee
The explicit form of $\sigma^R$ can be obtained from eq. (\ref{sigma_LR}). In general, the loop contributions $\delta{\cal M}^{(0)}$ contain divergent terms proportional to $\Delta_\epsilon$. Renormalisability demands that the $3 \times 3$ upper-left block of $\delta{\cal M}^{(0)}$ must be finite, as no corresponding counterterms are available in tree-level Lagrangian to cancel these divergences. Representing the divergent part of $\delta{\cal M}^{(0)}$ as $\delta{\cal M}^{(0)}_{\rm div}$, we obtain
\be \label{DM_div}
\delta{\cal M}^{(0)}_{\rm div} \propto {\cal U}^{(0)}_L\, {\cal Q}^{(0)}_L\, {\cal D}^{(0)}\, {\cal Q}^{(0)}_R\, {\cal U}_R^{(0)\dagger} = {\mathbb Q}_L\,{\cal M}^{(0)}\,{\mathbb Q}_R\,, \ee
where the last equality follows from eqs. (\ref{Q0},\ref{M0_diag}). Using the diagonal nature of ${\mathbb Q}_{L,R}$ and the form of the mass matrix ${\cal M}^{(0)}$ given in eq. (\ref{M0}), one finds
\be \label{DM_div2}
\left(\delta{\cal M}^{(0)}_{\rm div}\right)_{ij} = 0\,.\ee
Therefore, the $3 \times 3$ upper-left block of $\delta{\cal M}^{(0)}$ is finite, as expected from the renormalisability \cite{Barr:1978rv,Weinberg:1972ws}.

The finite part of $\delta {\cal M}^{(0)}$ can be simplified to
\be \label{del_calM_2}
\left(\delta{\cal M}^{(0)} \right)_{\alpha \beta} = \frac{g_X^2}{4 \pi^2} q_{L \alpha} q_{R \beta}\, \sum_\gamma \left({\cal U}^{(0)}_L\right)_{\alpha \gamma} \left({\cal U}_R^{(0)*}\right)_{\beta \gamma}\, m^{(0)}_\gamma\,b_0[M_X,{m^{(0)}_\gamma}]\,, \ee 
where $b_0$ denotes the finite part of the loop function $B_0$ as defined
in the $\overline {\rm MS} $ scheme. Explicitly,
\be \label{b0}
b_0[M,m] \equiv 1  - \frac{M^2 \ln \frac{M^2}{\mu^2} - m^2 \ln \frac{m^2}{\mu^2}}{M^2 - m^2}\,,.\ee
Further simplifications is achieved in the seesaw approximation. Substituting eqs. (\ref{U0_ss},\ref{M0_diag}) in eq. (\ref{del_calM_2}), we find
\beqa \label{del_calM_33}
\left(\delta{\cal M}^{(0)} \right)_{ij} &\simeq& \frac{g_X^2}{4 \pi^2} q_{L i} q_{R j}\, \left(U^{(0)}_L\right)_{i3} \left(U^{(0)*}_R\right)_{j3}\, m^{(0)}_3\, \nonumber \\
&& \hspace{1em} \times ~\left(b_0[M_X,{m^{(0)}_3}] - b_0[M_X,m_F] \right)\,.\eeqa
From eq. (\ref{M0_eff_diag}), it can be shown that:
\be \label{Meff_diag_comp}
\left(U^{(0)}_L\right)_{i3} \left(U^{(0)*}_R\right)_{j3}\, m^{(0)}_3\, = M^{(0)}_{ij} = -\frac{1}{m_F}\mu_{Li} \mu_{Rj}\,.\ee 
Thus, the simplified eq. (\ref{del_calM_33}) is:
\beqa  \label{dM0_eff}\left(\delta{\cal M}^{(0)} \right)_{ij} & \simeq & \frac{g_X^2}{4 \pi^2} q_{L i} q_{R j}\, M^{(0)}_{ij}\left(b_0[M_X,{m^{(0)}_3}] - b_0[M_X,m_F] \right)\,,\eeqa
with $i,j=1,2,3$ and the repeated indices are not summed over. After performing some straightforward algebraic simplifications, the corrections to the remaining components of the mass matrix are derived as follows:
\beqa \label{del_calM_44}
\left(\delta{\cal M}^{(0)} \right)_{i4} &\simeq & \frac{g_X^2}{4 \pi^2} q_{L i} q_{R 4}\, \mu_{Li}\, \left(b_0[M_X,m_F] + \sum_j \frac{|\mu^\prime_j|^2}{m_F^2}\, b_0[M_X,m_3^{(0)}]\right)\,,\nonumber \\
\left(\delta{\cal M}^{(0)} \right)_{4i} &\simeq & \frac{g_X^2}{4 \pi^2} q_{L 4} q_{R i}\, \mu_{Ri}\, \left(b_0[M_X,m_F] + \sum_j \frac{|\mu_j|^2}{m_F^2}\, b_0[M_X,m_3^{(0)}]\right)\,,\nonumber \\
\left(\delta{\cal M}^{(0)} \right)_{44} &\simeq & \frac{g_X^2}{4 \pi^2} q_{L 4} q_{R 4}\,m_F\, \left(b_0[M_X,m_F] -  \frac{{m^{(0)}_3}^2}{m_F^2}\, b_0[M_X,m_3^{(0)}]\right)\,. \eeqa
Based on the aforementioned findings, the fermion mass matrix with 1-loop corrections, as given by eq. (\ref{M_loopcorr}), can be expressed as:
\be \label{M_corr}
{\cal M}^{(1)} = \left( \ba{cc} (\delta M)^{(0)}_{3 \times 3} & (\tilde{\mu})_{3 \times 1} \\ (\tilde{\mu}^\prime)_{1 \times 3} & \tilde{m}_F \ea \right)\,,\ee
with 
\beqa \label{DeltaM}
\left(\delta M^{(0)} \right)_{ij} &=& \left(\delta{\cal M}^{(0)} \right)_{ij}\,,~~~~~~~~~~~\tilde{\mu}_i \,= \,\mu_i + \left(\delta{\cal M}^{(0)} \right)_{i4}\,,\nonumber\\
\tilde{\mu}^\prime_i &=& \mu^\prime_i + \left(\delta{\cal M}^{(0)} \right)_{4i}\,,~\tilde{m}_F \,=\, m_F +\left(\delta{\cal M}^{(0)} \right)_{44}\,.\eeqa
Since, $\delta M^{(0)}_{ij} \ll \tilde{\mu}_i, \tilde{\mu}^\prime_i \ll \tilde{m}_F$, the effective $3 \times 3$ mass matrix for the chiral fermions can then be written as:
\be \label{M_eff_corr}
M^{(1)} = \delta M^{(0)}  - \frac{1}{\tilde{m}_F}\tilde{\mu_L} \tilde{\mu}_R\,.\ee
By comparing the above with eq. (\ref{M0_eff}), it is seen that the second term has the form similar to $M^{(0)}$ with the original elements being swapped by their 1-loop corrected values. This term remains of rank 1 and contributes only to the masses of the third-generation fermions.

\section{Consequences of 1-loop result}
Two significant characteristic features can be identified from the above 1-loop  corrected fermion mass matrix given in eq. (\ref{M_eff_corr}).
\begin{itemize}
    \item The flavour universal $U(1)$ symmetry cannot induce radiative masses for the lighter generations. This can be understood as follows. For $q_{L1}=q_{L2}=q_{L3}$ and $q_{R1}=q_{R2}=q_{R3}$, we find $\delta M \propto M^{(0)}$ from eq. (\ref{dM0_eff}) and $\tilde{\mu} \propto \mu$, $\tilde{\mu^\prime} \propto \mu^\prime$, $\tilde{m_F} \propto m_F$ from eq. (\ref{del_calM_44}). The latter implies: 
    \be - \frac{1}{\tilde{m}_F}\tilde{\mu_L} \tilde{\mu}_R\,\propto - \frac{1}{{m}_F}{\mu}_L {\mu}_R\,\propto M^{(0)}\,. \ee 
    In summary, this suggests that $M^{(1)} \propto M^{(0)}$, indicating that the rank of the 1-loop corrected mass matrix stays at one. Thus, a flavour non-universal $U(1)$ is essential for generating masses for the lighter fermion generations. Our finding contradicts the results presented in \cite{Jana:2021tlx}, which employ a flavour universal $U(1)_{B-L}$ to produce radiative masses for the first-generation fermions.
    \item The generic choices of charges $q_{L,Ri}$, the mass matrix $M^{(1)}$ always leads to one vanishing eigenvalue. 
\end{itemize}

The second point can be proved in the following manner. To simplify subsequent analyses, we assume vector-like fermions are neutral with respect to the $U(1)_F$ symmetry. This choice also doesn't lead to a loss of generality, as from various gauge anomaly cancellation constraints, one of the gauge charges can be fixed. Also, from the first bullet point, it can be seen that even for arbitrary charges, it doesn't induce loop masses for otherwise massless fermions. Consequently, the expression for ${\cal M}^{(1)}$ can be represented as:
\be \label{M1_1}
{\cal M}^{(1)} = \left( \ba{cc} \left(\delta M^{(0)}\right)_{3 \times 3} & \mu_L\\ \mu_R & m_F \ea \right)\,.\ee

In the seesaw limit, the unitary matrices that perform the block diagonalisation of ${\cal M}^{(1)} $ given in eq. (\ref{M1_1}), can be approximated by:
\be \label{U1_ss}
{\cal U}^{(1)}_{L,R} \approx \left(\ba{cc} U_{L,R}^{(1)} & - \rho_{L,R}^{(1)} \\
 \rho_{L,R}^{(1) \dagger} U_{L,R}^{(1)} & 1 \ea\right)\,, \ee
with $\rho^{(1)}_{L,R}=\rho^{(0)}_{L,R}$ and $U_{L,R}^{(1)}$ are the unitary $3\times3$ matrices which diagonalises the effective 1-loop corrected $3\times 3$ mass matrix
\be \label{M1_eff}
M^{(1)}_{ij} = M^{(0)}_{ij} + \delta M^{(0)}_{ij}\,.\ee
 By substituting eq. (\ref{dM0_eff}) into eq. (\ref{M1_eff}), we can simplify it to
\be \label{M1_eff_2}
M^{(1)}_{ij} = M^{(0)}_{ij}\,\left(1+ C\, q_{Li}\, q_{Rj} \right)\,, \ee
where $C= \frac{g_X^2}{4 \pi^2} (b_0[M_X,m_3^{(0)}] - b_0[M_X,m_F])$. It can then be seen that the above mass matrix has a vanishing determinant. This results from the fact that one of the columns of $M^{(1)}$ is not independent. For example,
\be \label{}
M^{(1)}_{i1} = \frac{q_{R1}-q_{R3}}{q_{R2}-q_{R3}}\, \frac{\mu_{R1}}{\mu_{R2}}\, M^{(1)}_{i2} + \frac{q_{R2}-q_{R1}}{q_{R2}-q_{R3}}\, \frac{\mu_{R1}}{\mu_{R3}}\, M^{(1)}_{i3}\,,\ee
for $i=1,2,3$. An analogous relation is observed for the rows of $M^{(1)}$ when $q_R$ and $\mu_R$ are substituted with $q_L$ and $\mu_L$, respectively. This indicates that for the most general choices of $q_R$ and $q_L$, the 1-loop corrected effective mass matrix is of rank-2, resulting in one state being massless. Consequently, the diagonalisation of $M^{(1)}$ presented in eq. (\ref{M1_eff_2}), can be written as:
\be \label{M1_eff_diag}
U_L^{(1) \dagger}\, M^{(1)}\, U_R^{(1)} = {\rm Diag.}\left(0,m_2^{(1)},m_3^{(1)}\right)\,.\ee

Deriving analytical expressions for $U_{L,R}^{(1)}$ can often be a challenging task; however, the first column of $U_{L,R}^{(1)}$ can be obtained with relative ease. This process involves determining the eigenvector associated with the zero eigenvalue of both $M^{(1)} M^{(1) \dagger}$ and $M^{(1) \dagger} M^{(1)}$. The former provides the first column for $U_L^{(1)}$, while the latter corresponds to that of $U_R^{(1)}$. We find that,
\be \label{ev_m0}
\left(\ba{c} U^{(1)}_{L 11} \\ U^{(1)}_{L 21} \\ U^{(1)}_{L 31} \ea \right) = \frac{1}{\sqrt{N}} \left(\ba{c} 1 \\ \frac{\mu_{L1}^*}{\mu_{L2}^*} \frac{q_{L3} - q_{L1}}{q_{L2}-q_{L3}}\\ -\frac{\mu_{L1}^*}{\mu_{L3}^*} \frac{q_{L2} - q_{L1}}{q_{L2}-q_{L3}} \ea \right)\,.\ee
Similar expression can be obtained for $U_{R}^{(1)}$ by the replacement $L \to R$ in the above expression.

Since the lightest generation fermion remains massless in this setup at the 1-loop level. There are two possible ways to induce radiative masses for the otherwise massless fermion, thus establishing a hierarchical mass spectrum for the chiral fermions. These are:
\begin{enumerate}
    \item \textit{Extending the Gauge Sector}: To generate masses for both the first and second generations at the 1-loop level, the gauge sector must be extended beyond a single \( U(1) \). The minimal extensions include either a \( U(1) \times U(1) \) structure or an \( SU(3) \) flavour symmetry.

\item  \textit{Higher-Order Contributions}: The massless state is generally a non-trivial combination of all three fermion flavours. As a result, it can acquire mass at a higher order in perturbation theory, following a mechanism similar to how a massive state emerges at the 1-loop level. 
\end{enumerate}

We outline the simplest approach to generating first-generation masses using a \( U(1) \times U(1) \) framework. In this setup, the first \( U(1) \) should be responsible for inducing mass exclusively for the second generation, which can be achieved by setting \( q_{L1} = q_{R1} = 0 \). As shown in eq. (\ref{del_calM_33}), this leads to a massless first-generation fermion and loop-suppressed mass for the second generation. The masses of the first-generation fermions are then introduced through corrections mediated by the gauge boson of the second \( U(1) \), under which they are nontrivially charged. The mass hierarchy between the first and second generations can be orchestrated by implementing hierarchical masses for the respective gauge bosons of these two $U(1)$ symmetries. This can be understood as follows:  Incorporating the corrections from both \( U(1) \) gauge bosons, the total 1-loop corrected mass matrix takes the form:
\be \label{M1_eff_u1u1}
\left( M^{(1)}\right)_{ij} = \left( M^{(0)}\right)_{ij} + \left(\delta M_1^{(0)}\right)_{ij}\,+ \left(\delta M_2^{(0)}\right)_{ij}\,,\ee
Like eq. (\ref{M1_eff_2}), for VL states neutral under both the $U(1)$s, it can be written as 
\be \label{M1_eff_2_u1u1}
M^{(1)}_{ij} = M^{(0)}_{ij}\,\left(1+ C_1\, q^{(1)}_{Li}\, q^{(1)}_{Rj} +C_2\, q^{(2)}_{Li}\, q^{(2)}_{Rj}\right)\,, \ee
where $C_n= \frac{g_X^2}{4 \pi^2} (b_0[M_{Xn},m_3^{(0)}] - b_0[M_{Xn},m_F])$ for $n=1,2$. $q^{(n)}_{L,Ri}$ is charge of the chiral fermion $f^\prime_{L,Ri}$ under $U(1)_n$ symmetry. As mentioned earlier $q^{(1)}_{L1,R1}=0$ and $q^{(2)}_{L1,R1}\neq 0$, to induce radiative masses for second and first-generation fermions using gauge corrections due to $U(1)_1$ and $U(1)_2$ respectively. The suppression factor for first-generation fermion masses to second-generation is $\frac{C_2}{C_1}$. 
For instance, the loop integration factor with the condition $M_X \gg m_F \gg m^{(0)}_3$ can be expressed as
    \be \label{b0_limit}
    b_0[M_X,m_3^{(0)}] - b_0[M_X,m_F] \simeq -\frac{m_F^2}{M_X^2}\,\ln\frac{m_F^2}{M_X^2}\,.\ee
The suppression factor will be
\be\frac{C_2}{C_1} \sim \frac{g^2_{X2}}{g^2_{X1}} \frac{M^2_{X1}}{M^2_{X2}} \ee
    Thus, two $U(1)$s with $M_{X_2} \gg M_{X_1}$ can lead to hierarchical masses of first and second generations despite both being generated at 1-loop. The hierarchy between them will be proportional to $\frac{M_{X1}^2}{M_{X2}^2}$. A suitable choice of $U(1)_1 \times U(1)_2$ charges and an explicit model that can achieve this is discussed in chapter \ref{chap3}. 
    
    It was straightforward to obtain hierarchical loop-induced masses for first and second-family fermions for two $U(1)$ case. However, for the aforementioned $SU(3)$ framework, obtaining hierarchy between the first two generation fermion masses is a non-trivial task and will be explored in detail in chapter \ref{chap5}.

Another possibility, the second way, requires a framework in which second-generation fermion masses arise at the 1-loop level, and first-generation gets mass through higher-order corrections. Such a scenario, in which relatively suppressed first family masses can be obtained, is discussed in chapter \ref{chap4}.
\section{Symmetry analysis}
\label{sec:sym_dec}
Examining the symmetries involved provides valuable insights into how the assumed structure of the mass Lagrangian, combined with gauge interactions, results in mass generation for three generations of chiral fermions at subsequent orders in the perturbation theory. This section demonstrates how this approach aids in determining the precise characteristics of gauge interactions and the parameters within the mass Lagrangian necessary for obtaining a specific flavour spectrum.

Initially, in the absence of ${\cal L}_m$ and ${\cal L}_X$, the kinetic terms and any flavor-universal gauge interactions, like those found in the SM, exhibit invariance under a global $U(3)_L \times U(3)_R$ symmetry. This symmetry is broken by nonzero values of $\mu_L$ and $\mu_R$, as demonstrated by ${\cal L}_m$ in eq. (\ref{L_mass}). Nevertheless, it is possible to perform an $U(3)_{L,R}$ transformations that transform $\mu_{L,R}$ into the configuration $(0,0,\times)$, with ``$\times$'' representing a nonzero value. Consequently, the mass terms in eq. (\ref{L_mass}) leads to:
\be \label{tree_symm}
U(3)_L \times U(3)_R\, \xrightarrow{\mu_{L} \neq 0,\, \mu_{R} \neq 0}\, U(2)_L \times U(2)_R\,.\ee
This accidental $U(2)_L \times U(2)_R$ symmetry is responsible for the emergence of two massless states, corresponding to the fermions of the first and second-generation fermions. In the physical basis, denoted as $f_{L,R} = {\cal U}^{(0) \dagger}_{L,R} f^\prime_{L,R}$, the symmetry of ${\cal L}_m$ becomes more apparent.

Subsequently, should the complete theory maintain invariance under the $U(2)_L \times U(2)_R$ symmetry, the perturbative approach cannot produce nonzero masses for the lighter generation fermions. In this context, the symmetry is broken using the gauge interaction Lagrangian ${\cal L}_X$ as specified in eq. (\ref{L_gauge}). At the leading order, ${\cal L}_X$ is:
\be \label{L_guage_33}
-{\cal L}^{(0)}_X = g_X X_\mu\, \left(({ Q}^{(0)}_L)_{i j}\, \overline{f}_{L i} \gamma^\mu f_{L j} + ({ Q}^{(0)}_R)_{i j}\, \overline{f}_{R i} \gamma^\mu f_{R j} \right)\,+\, {\cal O}(\rho), \ee 
with $Q_{L,R}^{(0)}$ is upper-left $3\times3$ block of the matrix ${\cal Q}_{L,R}^{(0)}$ given in eq. (\ref{Q0}) and have the form:
\be \label{Q0_33}
Q_{L,R}^{(0)} = U_{L,R}^{(0) \dagger}\, q_{L,R}\, U^{(0)}_{L,R}\,.  \ee
Only for specific choices of $q_{L,R}$ leading to $Q_{L,R}^{(0)} = {\rm Diag.}\left(q,q,q^\prime\right) $, the gauge interactions in ${\cal L}_X$ do not break $U(2)_L \times U(2)_R$ symmetry. However, for generic choices of $q_{L,R}$ this symmetry is broken by ${\cal L}_X$. 
The expression for \( U^{(0)}_{L,R} \) in eq.~(\ref{U0_ana}) includes an arbitrary unitary rotation matrix \( V^{[12]}_{L,R} \), which mixes the first two generations of fermions. As a result, when substituted into eq.~(\ref{Q0_33}), the elements \( \left(Q_{L,R}^{(0)}\right)_{12} \) depend on these arbitrary mixing parameters, making them unphysical.

In the mass Lagrangian at the $1$-loop level, the breaking of $U(2)_L \times U(2)_R$ is evident. Nonetheless, the corrected mass matrix $M^{(1)}_{ij}$ reinstates an incidental $U(1)_L \times U(1)_R$ subgroup from the initial symmetry,
\be \label{1loop_symm}
U(2)_L \times U(2)_R\, \xrightarrow{\text{at 1-loop}}\, U(1)_L \times U(1)_R\,,\ee
leading to a massless state. Once more, this symmetry can be more readily articulated within the physical basis. It is defined by the transformations:
\be \label{U1U1_trans}
f_{L 1} \to e^{i \alpha_L} f_{L 1}\,,~~f_{R 1} \to e^{i \alpha_R} f_{R 1}\,.\ee
The subsequent generation of non-vanishing mass for the first family fermions at higher loops involving $X$-boson requires the breaking of this symmetry in ${\cal L}_X$. Following  1-loop correction, the relevant charge matrix in the new physical basis is:   
\be \label{QLR_1}
Q_{L,R}^{(1)} = U_{L,R}^{(1) \dagger}\, q_{L,R}\, U^{(1)}_{L,R}\,.\ee
The invariance of gauge interactions under unitary transformations given in eq. (\ref{U1U1_trans}) indicates that $\left(Q_{L,R}^{(1)}\right)_{12}$ and $\left(Q_{L,R}^{(1)}\right)_{13}$ should equal zero. Computing explicitly $Q_{L,R}^{(1)}$, we find \cite{Mohanta:2024wcr}:
\beqa \label{Q_12}
\left(Q_L^{(1)}\right)_{12} &=& (q_{L3} - q_{L1})\, \left(U_L^{(1)}\right)^*_{31}\,\left(U_L^{(1)}\right)_{32} + (q_{L2} - q_{L1})\, \left(U_L^{(1)}\right)^*_{21}\,\left(U_L^{(1)}\right)_{22}\,,\nonumber \\
&=&\frac{(q_{L2}-q_{L1})(q_{L3}-q_{L1})}{\sqrt{N}\, (q_{L3}-q_{L2})}\,\left(\frac{\mu_{L1}}{\mu_{L3}} \left(U_L^{(1)}\right)_{32} -  \frac{\mu_{L1}}{\mu_{L2}} \left(U_L^{(1)}\right)_{22}\right)\,. \eeqa
The first line is derived by leveraging the orthogonality of the columns of $U_L^{(1)}$, whereas the subsequent line results from eq. (\ref{ev_m0}). Similarly, 
\beqa \label{Q_13}
\left(Q_L^{(1)}\right)_{13} &=&\frac{(q_{L2}-q_{L1})(q_{L3}-q_{L1})}{\sqrt{N}\, (q_{L3}-q_{L2})}\,\left(\frac{\mu_{L1}}{\mu_{L3}} \left(U_L^{(1)}\right)_{33} -  \frac{\mu_{L1}}{\mu_{L2}} \left(U_L^{(1)}\right)_{23}\right)\,, \eeqa
and
\beqa \label{Q_23}
\left(Q_L^{(1)}\right)_{23} = (q_{L3}-q_{L2})\, \left(U_L^{(1)}\right)^*_{32}\left(U_L^{(1)}\right)_{33} - (q_{L2}-q_{L1})\, \left(U_L^{(1)}\right)^*_{12}\left(U_L^{(1)}\right)_{13}\,. \eeqa
Analogous expressions for $\left(Q_R^{(1)}\right)_{ij}$ can be derived by following the procedure outlined above.

For $q_{L3} \neq q_{L2}$, it can be observed that $(Q_L^{(1)})_{12,13}$ tend to zero in the limits of either $q_{L1} \to q_{L2}$, $q_{L1} \to q_{L3}$, or $\mu_{L1} \to 0$. Thus, to ensure the breaking of the accidental $U(1)_L \times U(1)_R$ symmetry, it is essential to have fully non-degenerate flavour charges $q_{L,R}$ along with non-zero values for $(\mu_{L,R})_{1}$. This has an important consequence for inducing masses for massless states at the next leading order.

Additionally, it can be observed from the ${\cal L}_m$ that if either $\mu_{Li}$ or $\mu_{R i}$ is equal to zero, the corresponding $f^\prime_{Li}$ or $f^\prime_{Ri}$ remains unmixed with the other fermions. Since both the mass Lagrangian and gauge interactions exhibit invariance under a $U(1)_L$ or $U(1)_R$ transformation : 
\be f^\prime_{L(R) i} \to e^{i \theta} f^\prime_{L(R) i}\, , \ee 
such symmetry is preserved from quantum corrections, resulting in a fermion that remains massless at all levels. Consequently, it is essential for all $\mu_{Li}$ and $\mu_{Ri}$ to be non-zero in the current framework.

\section{UV completion of \texorpdfstring{$\mu_{L,R}$}{\mu_{L,R}} and implications}
\label{chap2:UV}
\begin{figure}[!t]
\centering
\includegraphics[width=0.6\textwidth]{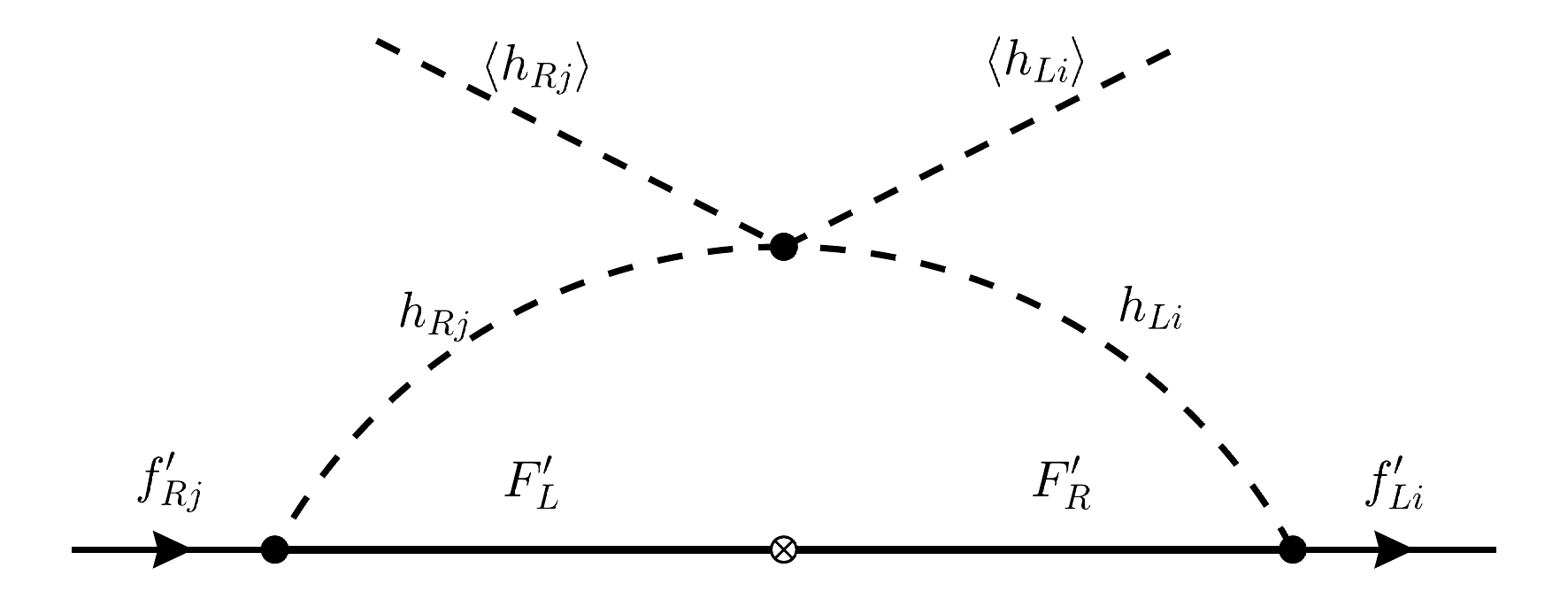}
\caption{Scalar induced fermion self-energy correction at 1-loop.}
\label{fig:1loop_scalar}
\end{figure}

It was earlier mentioned that $\mu_{L i}$ and $\mu_{R i}$ can be treated by spurions. In this section, we consider them as complex scalar fields denoted $h_{Li}$ and $h_{Ri}$ respectively. Both $h_{Li}$ and $h_{Ri}$ are non-trivially charged under the $U(1)_F$. The explicit charges can be determined from the Yukawa interaction Lagrangian:
\beqa \label{L_mass_UV}
-{\cal L}_{m} &=& y_{Li}\, \overline{f}^\prime_{Li}\,h_{L i}\, F_R^\prime +  y_{Ri}\, \overline{F}^\prime_L \,h_{R i}\, f_{Ri}^\prime + m_F\, \overline{F}^\prime_L F^\prime_R + {\rm h.c.}\,. \eeqa
Here $y_{L,Ri}$ can be identified as usual Yukawa couplings. For VL fermions as neutral under $U(1)_F$, the charges of  $h_{Li}$ and $h_{Ri}$ can be written as:
\be q_{h_{Li}}= q_{Li}\, , ~~~ ~~q_{h_{Ri}} =\,- q_{Ri}\, ,\ee
For the most general choices of $q_{L,Ri}$, the scalar potential allows a term proportional to $(h^\dagger_{L i}h_{L i})(h^\dagger_{R j}h_{R j})$. The presence of this mixing can also contribute to the mass matrix at the 1-loop level. The explicit diagram can be seen in Fig. \ref{fig:1loop_scalar}. It can be seen that, such contribution can be neglected for negligible mixing between $h_{Li}$ and $h_{Ri}$ \cite{Weinberg:2020zba,Jana:2021tlx}.

The presence of the multiplicity of the scalars and their couplings being not equal to the physical masses leads to various flavour-violating couplings among the chiral fermions. To evaluate the explicit couplings, we go to the physical basis of the scalars. Denoting $\tilde{h}_{Li}$ and $\tilde{h}_{Ri}$ as the electrically neutral components of $h_{Li}$ and $h_{Ri}$ fields respectively, their $6 \times 6$ mixing matrix can be expressed as:
\be M^2_h \,=\, \left(\ba{cc} m^2_{LL} & m^2_{LR}\\m^2_{RL} & m^2_{RR} \ea \right)\,.\ee
This is a real symmetric matrix with entries  $m^2_{PP^\prime}$ ($P,P^\prime=L,R$) each of $3\times 3$ dimension. It can be diagonalised by a real orthogonal transformation  matrix ${\cal R}$, which will have the form;
\be \label{cal R} {\cal R}\,=\, \left(\ba{cc} {\cal R}_{LL} & {\cal R}_{LR}\\ {\cal R}_{RL} & {\cal R}_{RR} \ea \right)\,.\ee
The resulting physical neutral scalars $S_a$, obtained through the aforementioned transformation, exhibit the following structure
\be S_a \,=\, {\cal R}^T_{ab}\, h_b\,, \ee
with $h_a\,=\, ( \ba{cc} \tilde{h}_{Li} & \tilde{h}_{Ri}\ea)^T$ and $a,b=1,2,..,6$.

The interaction of physical neutral scalars with the chiral fermions in the mass basis can be expressed as: 
\be -{\cal L}_Y \,=\, (\tilde{y_L})_{ia}\, \bar{f}_{Li}\, S_a\, F_R\,+\,(\tilde{y}_R)_{ia}\, \bar{F}_{L}\, S_a\, f_{Ri}\, +\, {\cal O}\left(\frac{\mu_{L}}{m_F}, \frac{\mu_{R}}{m_F}\right)\,,\ee
with, 
\beqa (\tilde{y}_L)_{ia} &=& \sum_j \left(U^{(0)\dagger}_{L}\right)_{ij}\,y_{Lj}\,({\cal R}_{L})_{ ja}\, , \nonumber \\
(\tilde{y}_R)_{ia} &=& \sum_j \left(U^{(0) T}_{R}\right)_{ij}\,y_{Rj}\,({\cal R}_{R})_{ ja}\, .
\eeqa
In the above, ${\cal R}_{L}= (\ba{cc} {\cal R}_{LL}&{\cal R}_{LR}\ea)$ and ${\cal R}_{R}= (\ba{cc} {\cal R}_{RL}&{\cal R}_{RR}\ea)$ are the $3\times6$ submatrices of ${\cal R}$. The presence of the above interactions also contributes to the self-energy corrections of the fermions at the 1-loop level, which is diagrammatically similar to  Fig. \ref{fig:1loop} with $X$ boson replaced by physical scalars $S_a$. The amplitude of this diagram has the form:
\be \sigma^{(S)}_{ij}\,=\, -\frac{m_D}{16\pi^2}\, \sum_a (\tilde{y}_L)_{ia} \,(\tilde{y}_R)_{ja} \, B_0[m_{Sa},m_F]\, ,\ee
with $B_0[m_{Sa},m_F]$ as two-point Passarino-Veltman function. Corrections to the mass matrix can be written as; 
\beqa \label{dm1:scalar}(\delta M^{(S)})_{ij}&=& \left(U^{(0)}_{L}\,\sigma^{(S)}\, U^{(0)\dagger}_{R} \right)_{ij}\,\nonumber \\
&=&\, -\frac{m_F}{16\pi^2}\,  ({y}_L)_{i} \,({y}_R)_{j} \,\sum_a  \,({\cal R}_{L})_{ ia}\,({\cal R}_{R})_{ ja} B_0[m_{Sa},m_F]\,.\eeqa
Also, these corrections, when added to the mass matrix given in eq. (\ref{chap5:M1_eff}) can potentially induce the first-family fermion masses at the 1-loop level only, and the suppression requirement compared to second-generation fermions asks for a separate mechanism. Adding the scalar contributions also introduces several other potential parameters that are unconstrained and thereby lose the computability of the novel mechanism. Therefore, it is desirable to make these contributions negligible. 

Now expanding, eq. ({\ref{dm1:scalar}}) and using eq. (\ref{cal R}), we obtain:
\beqa \label{dm1:scalar2}(\delta M^{(S)})_{ij}&=&-\frac{m_F}{16\pi^2}\,  ({y}_L)_{i} \,({y}_R)_{j} \, \left(\sum^3_{k=1}  \,({\cal R}_{LL})_{ ik}\,({\cal R}_{RL})_{ jk} B_0[m_{Sk},m_F]\,\right.\nonumber \\
& & \hspace{20mm}+\left. \sum^6_{n=3}  \,({\cal R}_{LR})_{ in}\,({\cal R}_{RR})_{ jn} B_0[m_{Sn},m_F]\,\right).\eeqa

It can be seen that both the terms in the above expression are suppressed by off-diagonal entries ${\cal R}_{LR}$ or ${\cal R}_{RL}$. The smallness of these entries can be attributed to small mixing between $h_{Li}$ and $h_{Ri}$, and/or $\langle h_{Li}\rangle \ll \langle h_{Ri}\rangle$. We assume such an arrangement in our subsequent chapters to neglect the scalar-induced contributions.

\section{Summary}
This chapter explores the general aspects of the radiative mass generation mechanism using a toy framework with three generations of chiral fermions and a pair of VL fermions. The chiral nature of the fermions, along with an extended flavour symmetry, is used to construct a scenario in which only third-generation fermions and VL states are massive. The chosen flavour symmetry is a gauged Abelian symmetry. It is shown that only flavour non-universal Abelian symmetries can accommodate the radiative mass induction mechanism. Additionally, at the one-loop level, corrections can induce masses for only second-generation fermions, while the first generation remains massless. Generating a small radiative mass for the first-generation fermion requires extending the gauge symmetry or including higher-order effects. Such scenarios are explored in the following chapters.

%% file: 30_Chapter_3/chapter_3.tex
\chapter{Flavour hierarchies from the quantum corrections in an Abelian model}
\label{chap3}
\graphicspath{{30_Chapter_3/}}

 We construct an explicit renormalisable model based on two flavour nonuniversal abelian gauge symmetries, which is shown to reproduce the observed fermion mass spectrum of the Standard Model through the radiative mass generation mechanism. As discussed in the context of the toy model, the third-generation fermions are allowed masses at zeroth order. In this chapter, we show that the framework leads to hierarchical loop-induced masses for the first and second-generation fermion masses, although both are being generated at 1-loop.  The explicit model which achieves this is constructed by taking the all-fermion generalization of the well-known leptonic $L_\mu-L_\tau$ and $L_e - L_\mu$ symmetries.

The rest of the chapter is organized as follows. An explicit model based on the findings of the toy model of the previous chapter is outlined in section \ref{chap3:model}. In section \ref{chap3:solutions}, we give example solutions which reproduce the observed fermion mass spectrum and discuss the various phenomenological implications in section \ref{chap3:pheno}. We discuss the possible flavour changing neutral currents (FCNCs) arising from the scalar sector in section \ref{chap3:scalarFCNC}.

\section{Model Implementation}
\label{chap3:model}
As outlined in the previous chapter, the suggested framework necessitates at least two $U(1)$ symmetries, wherein the three generations of Standard Model quarks and leptons are assigned distinct charges. Consequently, we extend the gauge symmetry of the SM by $G_F=U(1)_1 \times U(1)_2$ for this model. Besides the three SM generations of quarks and leptons, characterized respectively as $Q_{L i} \sim (3,2,\frac{1}{3})$, $u_{R i} \sim (3,1,\frac{4}{3})$, $d_{R i} \sim (3,1,-\frac{2}{3})$, $\Psi_{L i} \sim (1,2,-{1}{})$, and $e_{R i} \sim (1,1,-2)$, we incorporate three copies of a Higgs doublet pair: $H_{u i} \sim (1,2,-{1}{})$, $H_{d i} \sim (1,2,{1}{})$, and the SM singlets denoted by $\eta_i \sim (1,1,0)$. Additionally, each sector includes a pair of vector-like (VL) fermions: $T_{L,R} \sim (3,1,\frac{4}{3})$, $B_{L,R} \sim (3,1,-\frac{2}{3})$, and $E_{L,R} \sim (1,1,-2)$. The expressions in brackets specify their transformation characteristics under the symmetry group ${\cal G}_{SM}=$ $(SU(3)_C, SU(2)_L, U(1)_Y)$.
\begin{table}[t]
\begin{center}
\begin{tabular}{cccc} 
\hline
\hline
~~Fields~~&~~$(SU(3)_c \times SU(2)_L \times U(1)_Y)$~~&~~$U(1)_1$~~&~~$U(1)_2$~~\\
\hline
$Q_{L_i} $ & $(3,2,\frac{1}{3}) $ & \{0,1,-1\} & \{1,-1,0\}\\
$u_{R_i} $ & $(3,1,\frac{4}{3}) $ & \{0,1,-1\} & \{1,-1,0\}\\
$d_{R_i} $ & $(3,1,-\frac{2}{3}) $ & \{0,1,-1\} & \{1,-1,0\}\\
\hline
$\Psi_{L_i} $ & $(1,2,-{1}{}) $ & \{0,1,-1\} & \{1,-1,0\}\\
$e_{R_i} $ & $(1,1,-2) $ & \{0,1,-1\} & \{1,-1,0\}\\
\hline
$H_{u_i}$ & $(1,2,-{1}{}) $ & \{0,1,-1\} & \{1,-1,0\}\\
$H_{d_i}$ & $(1,2,{1}{}) $ & \{0,1,-1\} & \{1,-1,0\}\\
$\eta_{i}$ & $(1,1,0) $ & \{0,1,-1\} & \{1,-1,0\}\\
\hline 
$T_{L}, T_{R} $ & $(3,1,\frac{4}{3}) $ & 0 & 0 \\
$B_{L}, B_{R} $ & $(3,1,-\frac{2}{3}) $ & 0 & 0 \\
$E_{L}, E_{R} $ & $(1,1,-2) $ & 0 & 0 \\
\hline
\hline
\end{tabular}
\end{center}
\caption{The SM and $G_F$ charges for various fermionic and scalar particles in the model are presented. The indices $i=1,2,3$ signify the three generations, with their corresponding charges under the extended $U(1)$ given by $\{q_1,q_2,q_3\}$.}
\label{tab:fields}
\end{table}

Within the newly proposed $G_F = U(1)_1 \times U(1)_2$ symmetry, fermions and scalars from the first, second, and third generations are assigned charges of $(0,1)$, $(1,-1)$, and $(-1,0)$, respectively. Hence, $U(1)_1$ correlates with the so-called ``$2-3$ symmetry," while $U(1)_2$ is associated with ``$1-2$ symmetry." These symmetries represent generalizations of the $L_\mu - L_\tau$ and $L_e - L_\mu$ symmetries, which have been explored in the lepton sector within the literature \cite{Foot:1990mn,He:1990pn,He:1991qd}. The VL fermions are chosen neutral under $G_F$. The charges of fermions and scalar fields under the SM and $G_F$ are detailed in Table \ref{tab:fields}. Verifying that $G_F$ remains non-anomalous is straightforward, as pairs of fermions and scalars are assigned equal and opposite charges for each $U(1)$ symmetry.

The general renormalisable scalar potential of the model, which maintains invariance under the SM gauge symmetry and $G_F$, is expressed as:
\beqa \label{chap3:potential}
V &=& m_{u i}^2\, H_{u i}^\dagger H_{u i}\, +\, m_{d i}^2\, H_{d i}^\dagger H_{d i}\, +\, m_{\eta i}^2\, \eta_{i}^\dagger \eta_{i}\, \nonumber \\
& + & \left\{(m_{ud\eta})_{ijk}\, \epsilon_{ijk}\, \eta_i H_{u j} H_{d k}\,+\, (m_{\eta})_{ijk}\, \epsilon_{ijk}\, \eta_i \eta_j \eta_k + {\rm h.c.}\right\}\, \nonumber \\
& + &  (\lambda_u)_{ij}\, H_{u i}^\dagger H_{u i} H_{u j}^\dagger H_{u j}\,+\, (\lambda_d)_{ij}\, H_{d i}^\dagger H_{d i} H_{d j}^\dagger H_{d j}\,+\, (\lambda_\eta)_{ij}\, \eta_{i}^\dagger \eta_{i} \eta_{j}^\dagger \eta_{j}\, \nonumber \\
& + &  (\lambda_{ud})_{ij}\, H_{u i}^\dagger H_{u i} H_{d j}^\dagger H_{d j}\,+\, (\lambda_{u\eta})_{ij}\, H_{u i}^\dagger H_{u i} \eta_j^\dagger \eta_j\,+\, (\lambda_{d \eta})_{ij}\, H_{d i}^\dagger H_{d i} \eta_{j}^\dagger \eta_{j}\, \nonumber \\
& + &  (\tilde{\lambda}_u)_{ij}\, H_{u i}^\dagger H_{u j} H_{u j}^\dagger H_{u i}\,+\, (\tilde{\lambda}_d)_{ij}\, H_{d i}^\dagger H_{d j} H_{d j}^\dagger H_{d i}\, \nonumber \\
& + &  (\tilde{\lambda}_{ud})_{ij}\, H_{u i}^\dagger H_{u j} H_{d j}^\dagger H_{d i}\,+\, (\tilde{\lambda}_{u\eta})_{ij}\, H_{u i}^\dagger H_{u j} \eta_j^\dagger \eta_i\,+\, (\tilde{\lambda}_{d \eta})_{ij}\, H_{d i}^\dagger H_{d j} \eta_{j}^\dagger \eta_{i}\, \nonumber \\
& + & \left\{(\lambda_{ud\eta})_{ij}\, \eta_i^\dagger H_{ui} \eta_j^\dagger H_{dj} + (\tilde{\lambda}_{ud\eta})_{ij}\, \eta_i^\dagger H_{uj} \eta_j^\dagger H_{di} + {\rm h.c.}\right\}\,,\eeqa 
where, $i,j,k = 1,2,3$ are the flavour indices. The diagonal entries in each of the $\tilde{\lambda}$ matrices can be set to zero without affecting the generality of the analysis. We consider the general vacuum expectation values (VEVs) for different fields which leads to the breaking of all symmetries except for the $SU(3)_C$ and $U(1)$ associated with electromagnetism. Considering the existence of numerous parameters within eq. (\ref{chap3:potential}), we assume that such a minima can be obtained for a suitable choice of their values. Explicitly, VEVs are defined by
\be\label{chap3:Vevs}\langle {H_u}_i\rangle \equiv {v_u}_i,~~~~~ \langle {H_d}_i\rangle \equiv {v_d}_i,~~~~~\langle {\eta}_i\rangle \equiv {v_\eta}_i, \ee

The potential, in its general form, does not exhibit any enhanced global symmetry. Consequently, it does not lead to the emergence of new Goldstone bosons beyond those associated with the spontaneous symmetry breaking of the SM and $G_F$ symmetries, which are absorbed by the massive bosons $W^\pm$, $Z$, and $Z_{1,2}$. A global $SU(3)$ symmetry is present if all quadratic, cubic, and quartic couplings are flavour universal, reflecting invariance under the transformation $\Phi_i \to U_{ij} \Phi_{j}$ where $\Phi$ represents $H_u$, $H_d$, and $\eta$. Additionally, if the parameters $\tilde{\lambda}_{ud}$, $\tilde{\lambda}_{u\eta}$, $\tilde{\lambda}_{d \eta}$, $\lambda_{ud\eta}$, $\tilde{\lambda}_{ud\eta}$, $m_{ud\eta}$, and $m_\eta$ are zero, the scalar potential can have an enhanced $[U(3)]^3$ symmetry, enabling independent rotations for $H_u$, $H_d$, and $\eta$.


\subsection{Charged fermion masses}
The most general renormalisable interactions involving fermions and scalars, which remain invariant under the SM gauge symmetry and $G_F$, can be expressed as:
\beqa \label{LY}
-{\cal L}_Y &=& {y_u}_i\,\overline{Q_{L}}_i\, {H_u}_i\, T_R\, + {y_u^{\prime}}_i\,\overline{T_L}\, \eta^*_i\, u_{R i}\,+\,{y_d}_i\,\overline{Q_{L}}_i\, {H_d}_i\, B_R\, + {y_d^{\prime}}_i\,\overline{B_L}\, \eta^*_i\, d_{R i}\, \nonumber \\
& + & {y_e}_i\,\overline{\Psi_{L}}_i\, {H_d}_i\, E_R\, + {y_e^{\prime}}_i\,\overline{E_L}\, \eta^*_i\, e_{R i}\, + \, {\rm h.c.}\,.\eeqa
The interaction of two SM fermions with the Higgs fields is forbidden by the invariance under $G_F$. Vector-like fermion masses are characterized by:
\be \label{LM}
-{\cal L}_m = m_T\, \overline{T_L}\,T_R\, + \,m_B\, \overline{B_L}\,B_R\, + \,m_E\, \overline{E_L}\,E_R\, + \, {\rm h.c.}\,.\ee
The spontaneous symmetry breaking of $G_F$ via the VEVs of $\eta_i$ and ${H_{ui,di}}$ leads to $4 \times 4$ mass matrices for the charged fermions, which are structurally identical to the matrix described in eq. (\ref{M0}). Specifically,
\be \label{M_expl}
{\cal M}_{u,d,e} = \left( \ba{cc} 0 & \left(\mu_{u,d,e}\right)_{3 \times 1} \\ \left(\mu^\prime_{u,d,e}\right)_{1 \times 3} & m_{T,B,E} \ea \right)\,,\ee
where 
\be \label{mu_expl}
{\mu_u}_i = {y_u}_i {v_u}_i\,,~~{\mu_d}_i = {y_d}_i {v_d}_i\,,~~{\mu_e}_i = {y_e}_i {v_d}_i\,,\ee
\be \label{mup_expl}
{\mu^\prime_u}_i = {y^\prime_u}_i {v_\eta}_i\,,~~{\mu^\prime_d}_i = {y^\prime_d}_i {v_\eta}_i\,,~~{\mu^\prime_e}_i = {y^\prime_e}_i {v_\eta}_i\,,\ee
and VEVs ${v_u}_i,{v_d}_i$ and ${v_u}_i$ are defined in eq. (\ref{chap3:Vevs}). Here the repeated indices are not summed over. The effective $3 \times 3$ mass matrix in each charged fermion sector, analogous to eq. (\ref{M0_eff}), is
\be \label{Meff_0}
M_{u,d,e}^{(0)} \equiv - \frac{1}{m_{T,B,E}}\, \mu_{u,d,e}\, \mu^\prime_{u,d,e}\,.\ee
The above matrices are of rank one and are responsible for giving masses to the third-generation charged fermions. Thus, at the tree level, the Yukawa sector exhibits a global \(U(2)^5\) symmetry, corresponding to massless first and second generations. This symmetry doesn't commute with the flavour-dependent gauge symmetries and is completely broken if all three SM fermion generations carry distinct charges under the new symmetries, as explained in chapter \ref{chap:2}. Therefore, the gauge boson-induced loop corrections generate the masses for lighter fermions.

Using the method described in the preceding chapter, eq. (\ref{M1_eff_u1u1}), one can derive the 1-loop corrected effective $3 \times 3$ mass matrices pertinent to the charged fermions as follows:
\be \label{Mf}
M_f =  \delta M_f + M_f^{(0)}\,,\ee 
where $f=u,d,e$, with the second term expressed via eq. (\ref{Meff_0}) is the tree-level effective mass matrix. The 1-loop corrections do not impact the parameters $\mu_i$, $\mu^\prime_i$, and $m_F$, because the VL fermions are neutral under the symmetry group $G_F$. The term $\delta M_f$ comprises 1-loop corrections contributed by the gauge bosons of both the $U(1)$ symmetries. By utilizing the specified charges for $U(1)_{1,2}$ and applying eq. (\ref{M1_eff_2_u1u1}), we derive
\beqa \label{delMf}
\delta M_f &=& \frac{ g_1^2}{4 \pi^2} C_1\,\left(\ba{ccc} 0 & 0 & 0\\0 & \left(M_f^{(0)}\right)_{22} &  -\left(M_f^{(0)}\right)_{23} \\
0 & -\left(M_f^{(0)}\right)_{32} &  \left(M_f^{(0)}\right)_{33} \ea \right) \nonumber \\
& + & \frac{ g_2^2}{4 \pi^2} C_2\,\left(\ba{ccc}  \left(M_f^{(0)}\right)_{11} &  -\left(M_f^{(0)}\right)_{12} & 0 \\
-\left(M_f^{(0)}\right)_{21} &  \left(M_f^{(0)}\right)_{22} & 0 \\ 0 & 0 & 0 \ea \right)\,, \eeqa
with \be C_n =  \left(b_0[M_{Z_n},m_{f3}] - b_0[M_{Z_n},m_F]\right)\,. \ee
Here, $g_{n}$ denotes the gauge coupling, while $M_{Z_n}$ represents the mass of the gauge boson associated with $U(1)_n$. The masses of the third-generation fermion and the VL fermion in each respective sector are denoted by $m_{f3}$ and $m_F$.

The mass matrix $M_f$ for charged fermions, as defined in eq. (\ref{Mf}), together with the expressions in eq. (\ref{Meff_0}) and (\ref{delMf}), describes the 1-loop corrected mass matrix in this model. Each contribution from $\delta M_f$ generated at 1-loop is of rank one and contributes separately to the masses of the first and second generation fermions. The hierarchy between the masses of these lighter generations can be regulated by choosing $M_{Z_1} \ll M_{Z_2}$ as discussed in the previous chapter. This suppression consistently applies to all charged fermions.

In deriving eq. (\ref{delMf}), we have restricted our analysis to 1-loop corrections arising solely from the gauge boson loops. Within this framework, additional radiative corrections might result from the loops involving the emission and absorption of scalar bosons. Such contributions require mixing between the scalar fields $\eta_i$ and $H_{ui,di}$, which specifically couple to the right and left-handed chiral fermions, respectively. As discussed in section \ref{chap2:UV},  these corrections can be suppressed by assuming minimal mixing between the $\eta_i$ and $H_{ui,di}$ fields \cite{Jana:2021tlx}.
   
We have also assumed the absence of kinetic mixings among different $U(1)$ groups. Should such a mixing exist through a term like $\epsilon {F_1}_{\mu \nu} {F_2}^{\mu \nu}$ where $F_{1,2}$ denote the field strengths associated with $Z_{1,2}$ bosons, it could induce the mass of first-generation fermions via 1-loop diagrams involving the $Z_1$ boson. This is because an effective induced coupling of magnitude $\epsilon g_1$ emerges between first-generation fermions and $Z_1$, after the kinetic mixing term is appropriately transformed to obtain the physical gauge bosons (refer to \cite{Babu:1996vt} for more details). In such scenarios, both the first and second generations would derive mass from the $Z_1$ loop; however, the former would experience additional suppression by a factor of $\epsilon^2$. Consequently, if $\epsilon \ll M_{Z_1}/M_{Z_2}$, the presence of kinetic mixing is not anticipated to disrupt the hierarchical mass structure between the first two generations within this model.

\subsection{Neutrino masses}
While the primary objective is to elucidate the mass hierarchies of charged fermions, we briefly discuss the potential for neutrino mass generation in this framework. A straightforward approach is to incorporate three right-handed (RH) neutrinos, which are singlets with respect to both the SM gauge symmetry and $G_F$. These RH neutrinos can acquire Majorana masses. The gauge-invariant and renormalisable interactions in this model can be expressed as:
\be \label{L_nu}
-{\cal L}_\nu = {y_D}_{ij}\,\overline{\Psi_{L}}_i\,{H_u}_i\,{\nu_R}_j + \frac{1}{2} {M_R}_{ij}\,\nu_{R i}^T C^{-1} \nu_{R j} + {\rm h.c.}\,. \ee 
When electroweak symmetry is broken, the Dirac neutrino mass matrix is generated as ${M_D}_{ij} = {y_D}_{ij} {v_u}_i$. In instances where $M_R$ is significantly larger than $M_D$, the established type I seesaw mechanism \cite{Minkowski:1977sc,Yanagida:1979as,Mohapatra:1979ia} can be applied, resulting in the light neutrino mass matrix expressed as:
\be \label{M_nu}
M_{\nu} = - M_D\,M_R^{-1}\, M_D^T\,. \ee
In contrast to charged fermions, neutrino masses for all three generations can generally originate at the tree level. This characteristic is advantageous because the intergenerational hierarchy among neutrino masses is less pronounced compared to those of charged fermions. The Dirac neutrino Yukawa coupling matrix $y_D$ can have elements approximately of  ${\cal O}(1)$, resulting in an anarchic mass structure for $M_\nu$. Consequently, this can account for the weaker hierarchy observed in neutrino masses and the significant mixing within the lepton sector \cite{Hall:1999sn,deGouvea:2012ac}.

\section{Example solutions}
\label{chap3:solutions}
We explore the capability of the proposed framework in replicating the masses of charged fermions and the mixing of quarks by determining numerical values for the parameters $\mu_{f i}$, $\mu^\prime_{f i}$, $m_T$, $m_B$, $m_E$, and $M_{Z_{1,2}}$. From eq. (\ref{LY}), it is apparent that the parameters $y_{u i}$, $y^\prime_{u i}$, $y^\prime_{d i}$, $y_{ei}$, and $y^\prime_{ei}$ can be set to real values by eliminating their phases via redefinitions of the respective quark and lepton fields. Likewise, one of the $y_{di}$ parameters can also be rendered real. A similar approach to eq. (\ref{LM}) results in $m_T$, $m_B$, and $m_E$ being real. Additionally, we presume all the VEVs are real. Collectively, this results in 25 real parameters (including real $\mu_{u i}$, $\mu^\prime_{ui}$, $\mu_{d3}$, $\mu^\prime_{di}$, $\mu_{ei}$, $\mu^\prime_{ei}$, $m_T$, $m_B$, $m_E$, $M_{Z_1}$, $M_{Z_2}$, and complex $\mu_{d1}$, $\mu_{d2}$) to determine 13 observables (comprising 9 charged fermion masses, 3 mixing angles, and a Dirac CP phase of the quark mixing matrix). As the number of parameters exceeds the number of observables, viable solutions are anticipated. However, given that masses and mixing observables are complex non-linear functions of these input parameters, which are expected to adopt non-hierarchical values within this model, it is not immediately clear that viable solutions will materialize.

The 25 real parameters intrinsic to the framework are determined via the standard $\chi^2$ function minimization approach. For the formal definition and specifics of the $\chi^2$ function, refer to \cite{Mummidi:2021anm}. The function incorporates 13 observables, with their corresponding mean values and standard deviations detailed in Table \ref{tab:input}. 
\begin{table}[t]
\begin{center}
\begin{tabular}{cccccc} 
\hline
\hline
~~Observable~~&~~Value~~&~~Observable~~&~~Value~~\\
 \hline
$m_u$ & $1.27\pm 0.50$ MeV & $m_e$ & $0.487 \pm 0.049$ MeV \\
$m_c$ & $0.619 \pm 0.084$ GeV & $m_\mu$ &  $102.7\pm 0.103$ MeV \\
$m_t$ & $171.7\pm 3.0$ GeV & $m_\tau$ & $1.746 \pm 0.174$ GeV\\
$m_d$ & $2.90 \pm 1.24$ MeV & $|V_{us}|$ & $0.22500 \pm 0.00067$ \\
$m_s$ & $0.055 \pm 0.016$ GeV & $|V_{cb}|$ & $0.04182 \pm 0.00085$  \\
$m_b$ & $2.89 \pm 0.09$ GeV & $|V_{ub}|$ & $0.00369 \pm 0.00011$ \\
 &  & $J_{\rm CP}$ & $(3.08 \pm 0.15)\times 10^{-5}$ \\
\hline
\hline
\end{tabular}
\end{center}
\caption{Charged fermion masses and CKM parameters evaluated at $M_Z$ were employed in the fits to derive example solutions. The charged fermion masses and quark mixing parameters are sourced from \cite{Xing:2007fb} and \cite{ParticleDataGroup:2020ssz}, respectively.}
\label{tab:input}
\end{table}

To precisely determine the scale of new physics, we select three distinct values for $M_{Z_1}$ and derive a representative solution for each. The best-fit values for the remaining parameters corresponding to each solution are provided in Table \ref{tab:sol}.

\begin{landscape}
\begin{table}[t]
\begin{center}
\begin{tabular}{cccc} 
\hline
\hline
~~Parameters~~&~~Solution 1 (S1)~~&~~Solution 2 (S2)~~&~~Solution 3 (S3)~~\\
\hline
$M_{Z_1}$ & $10^4$                  & $10^6$                & $10^8$ \\
$M_{Z_2}$ & $1.0647 \times 10^{5}$  & $ 1.2687\times 10^7$  & $1.5419\times 10^9 $\\
$m_T$     & $1.1009 \times 10^{4}$  & $1.1001\times 10^6 $  & $1.1036 \times 10^8$\\
$m_B$     & $3.9983\times 10^{4}$   & $3.1298\times 10^7 $  & $3.2070\times 10^9 $\\
$m_E$     & $3.8348\times 10^{5} $  & $4.5576\times 10^7$   & $2.2698\times 10^9 $ \\
\hline
$\mu_{u1}$ & $-2.8103\times 10^2 $ & $-1.9380\times 10^2$   & $-2.6664 \times 10^2$\\
$\mu_{u2}$ & $2.1323 \times 10^2$  & $-3.0763\times 10^2 $   & $-1.2685\times 10^2 $ \\
$\mu_{u3}$ & $-3.0597\times 10^2$  & $3.0908\times 10^2$     & $3.3227 \times 10^1$\\
\hline
$\mu^\prime_{u1}$ & $-7.0782\times 10^{2}$  & $-3.7765\times 10^3 $    & $-5.9665\times 10^7 $\\
$\mu^\prime_{u2}$ & $-1.1289 \times 10^{1}$ & $2.7844\times 10^4 $     & $-6.9480\times 10^5$ \\
$\mu^\prime_{u3}$ & $3.9604\times 10^{3}$   & $3.9465\times 10^5 $     & $2.2540\times 10^7$\\
\hline
$\mu_{d1}$ & $-1.1363\times 10^{2} - i\ 2.6328 $ & $-8.7337 \times 10^{1} - i\ 3.25716  $             & $-1.3287\times 10^2 + i\ 1.6876 \times 10^1 $ \\
$\mu_{d2}$ & $8.8517 \times 10^1+ i\ 2.4820$     & $-1.4536  \times 10^{2} - i\ 1.1683 \times 10^{1}$ & $-6.6130\times 10^1 + i\ 1.5149\times 10^1 $ \\
$\mu_{d3}$ & $-1.2179 \times 10^2$               & $1.5224\times 10^{2}$                              & $1.7778 \times 10^1$ \\
\hline
$\mu^\prime_{d1}$ & $5.8106\times 10^2 $         & $3.9789\times 10^5 $      & $-5.9027\times 10^7$\\
$\mu^\prime_{d2}$ & $ -1.9029\times 10^2 $       & $-3.9048\times 10^3$      & $-7.9284\times 10^6$ \\
$\mu^\prime_{d3}$ & $-3.9601 \times 10^1 $       & $5.4490 \times 10^4$     & $-2.6330\times 10^6 $ \\
\hline
$\mu_{e1}$ & $ -1.4784 \times 10^2 $          & $-1.2824 \times 10^2 $    & $1.7061$\\
$\mu_{e2}$ & $-4.6633 \times 10^1 $           & $-0.4898 $                & $-2.8832\times10^1$ \\
$\mu_{e3}$ & $1.2534$                         & $1.3914 \times 10^2$      & $5.4213\times 10^1 $ \\
\hline
$\mu^\prime_{e1}$ & $6.4393 \times 10^2 $    & $1.5356\times 10^5$       & $8.7072\times 10^6 $\\
$\mu^\prime_{e2}$ & $1.9427 \times 10^3 $    & $1.5181 \times 10^5 $     & $-1.4819\times 10^7 $ \\
$\mu^\prime_{e3}$ & $-3.8887 \times 10^3 $   & $3.3043 \times 10^5 $      & $5.6906\times 10^7 $\\
\hline
\hline
\end{tabular}
\end{center}
\caption{The optimized values for different input parameters derived for three sample solutions are presented. All values are expressed in GeV.}
\label{tab:sol}
\end{table}
\end{landscape}

The solutions obtained align closely with the observed data, with central values matched and a total \(\chi^2 \ll 1\) for each case. Neutrino masses and lepton mixing are excluded from the fit, as they depend on a separate set of parameters (see eq. (\(\ref{M_nu}\))). These parameters can be used to obtain realistic neutrino mass and mixings.

Table \ref{tab:sol} presents example solutions which demonstrate that the model is capable of replicating a realistic charged fermion spectrum regardless of the scales at which $U(1)_{1,2}$ symmetry is broken. As previously addressed, the mass difference between $Z_1$ and $Z_2$ is dictated by the mass hierarchies between the first and second generation fermions, leading to the ratio $M_{Z_2}^2/M_{Z_1}^2 \simeq {\cal O}(10^2)$. However, the precise scale of $M_{Z_1}$ remains undetermined. For our numerical analyses, we require that $M_{Z_1} \le m_T, m_B, m_E$. It emerges that to satisfy $m_t \gg m_b, m_\tau$, it is necessary that $m_T \ll m_B, m_E$. The parameters $\mu_{fi}$, where $f=u,d,e$, arise through electroweak symmetry breaking as delineated in eq. (\ref{mu_expl}); for these, we impose the conditions $|y_{fi}|<\sqrt{4 \pi}$ and $v_{u i},v_{d i}<174$ GeV, resulting in $\mu_{fi}$ being approximately ${\cal O}(100)$ GeV or smaller. Conversely, $\mu^\prime_{fi}$ are produced by VEVs of $U(1)_{1,2}$-charged SM singlet fields and are similar in magnitude to the $U(1)_{1,2}$ breaking scale. Notably, within Table \ref{tab:sol} solutions, there are no significant hierarchical disparities among various $\mu_{fi}$ or $\mu^\prime_{fi}$ values, implying uniformity in the magnitude of the model’s dimensionless parameters. Nevertheless, the five orders of magnitude discrepancy between the first and third-generation fermion masses is successfully realised via the radiative mass generation process.


\section{Phenomenological aspects}
\label{chap3:pheno}
The model presents substantial phenomenological implications as a result of the presence of two flavourful $U(1)$ gauge symmetries and an additional vector-like set of quarks and leptons. These altogether introduce significant \gls{FCNC} interactions among the SM fermions. Such interactions can originate through three ways: (i) direct mediation by $Z_{1,2}$-bosons, which exhibit flavour-changing couplings, (ii) through the SM $Z$-boson combined with the mixing between the SM fermions and their respective VL fermions, and (iii) via mediation by neutral scalars. The first can be  parametrized in the physical basis of the quarks and charged leptons as:
\be \label{Z_couplings}
-{\cal L}_{Z_{1,2}} =  \sum_{k=1,2}\,g_{k}\, \left(\left(X^{(k)}_{f_L}\right)_{ij}\, \overline{f_{L i}}\,\gamma^\mu f_{L j} + \left(X^{(k)}_{f_R}\right)_{ij}\, \overline{f_{R i}}\,\gamma^\mu f_{R j}\right) Z_{k \mu}\,, \ee
where $f=u,d,e$. The $3 \times 3$ coupling matrices can be expressed as:
\be \label{X}
X^{(k)}_{f_L} = U_{f_L}^\dagger\, q^{(k)}_{f L}\, U_{f_L}\,,\ee
For $X^{(k)}_{f_R}$, a similar expression is derived by substituting $L$ with $R$. Within this framework, we define $q^{(1)}_{f L} = q^{(1)}_{f_R} = {\rm Diag}(0,1,-1)$ and $q^{(2)}_{f L} = q^{(2)}_{f_R} = {\rm Diag}(1,-1,0)$ for all flavours $f$, as previously outlined. The unitary matrices $U_{f_L}$ and $U_{f_R}$ are determined by diagonalising the one-loop corrected mass matrices $M_f$, where $U_{f_L}^\dagger M_f U_{f_R} = {\rm Diag.}(m_{f_1},m_{f_2},m_{f_3})$. Due to the non-universal nature of the underlying $U(1)$ symmetries, $X^{(k)}_{f_L}$ and $X^{(k)}_{f_R}$ are not typically diagonal, which can result in significant \gls{FV} within both the quark and lepton sectors.

The current associated with the $Z$-boson in the physical basis is obtained as:
\be \label{JZ_model_phys}
j^\mu_Z = \frac{-g}{2 \cos \theta_W}\, \left({\cal Y}^u_{\alpha \beta}\, \overline{u}_{L \alpha}\gamma^\mu u_{L \beta}\, -\, {\cal Y}^d_{\alpha \beta}\, \overline{d}_{L \alpha}\gamma^\mu d_{L \beta}\, -\, {\cal Y}^e_{\alpha \beta}\, \overline{e}_{L \alpha}\gamma^\mu e_{L \beta}\, -2 \sin^2 \theta_W\, J_{\rm em}^\mu \right)\,,\ee
where $J_{\rm em}^\mu$ is flavour diagonal electromagnetic current. The $4 \times 4$ coupling matrices are given by:
\be \label{Yf}
{\cal Y}^f = {\cal U}_L^{f \dagger}\, \left(\ba{cc} {\bf 1}_{3 \times 3} & 0 \\ 0 & 0 \ea \right)\,{\cal U}_L^{f}\,. \ee 
 Where ${\cal U}_{L,R}^{f}$ is the diagonalising matrix of full 1-loop corrected $4\times4$ matrix. Using the seesaw expansion, similar form as eq. (\ref{U1_ss}), one finds
\be \label{Yf_eff}
{\cal Y}^f_{ij} \simeq \delta_{ij} + {\cal O}(\frac{\mu_f\, \mu^\prime_f}{m_F^2})\,. \ee
It appears that the off-diagonal components are attenuated due to heavy-light mixing. With the specific $\mu_f$ and $m_F$ values listed in Table \ref{tab:sol}, it is observed that the FCNCs mediated by the $Z$-boson remain significantly suppressed compared to those mediated by the $U(1)_{1,2}$ boson.
 
Similarly, the FCNCs induced by the neutral scalars can be made suppressed by considering scalars heavy. This is demonstrated in detail in section \ref{chap2:UV} for the toy model, and some additional aspects will be discussed in section \ref{chap3:scalarFCNC}. Thus, if the scalar masses are approximately of ${\cal O}(M_{Z1})$, the FCNCs prompted by these scalars provide secondary contributions to those mediated by the $U(1)_1$ gauge boson. Among all the solutions outlined in the preceding section, $Z_1$ is identified as the lightest particle of the new set. Consequently, we examine various constraints on the $Z_1$ boson, specifically from the processes that include quark and lepton flavour-changing transitions. For reference, the numerical values of various $X^{(1)}_{f_L}$ and $X^{(1)}_{f_R}$ for a representative solution (S2) are provided below.
\begin{eqnarray}
\label{example:sol}
&X^{(1)}_{u_L} =\begin{bmatrix}
-0.0074 & 0 - 0.4054i & -0.0194 \\
0 + 0.4054i & 0.0573 & 0 + 0.9123i \\
-0.0194 & 0 - 0.9123i & -0.0499
\end{bmatrix}\,,  \nonumber \\
& X^{(1)}_{u_R} = \begin{bmatrix}
0.0043 & 0 - 0.0660i & -0.0002 \\
0 + 0.0660i & 0.9857 & 0 + 0.1405i \\
-0.0002 & 0 - 0.1405i & -0.9900
\end{bmatrix}\,, \nonumber\\
&X^{(1)}_{d_L} = \begin{bmatrix}
0.1760 & 0 - 0.3893i & 0 - 0.1765i \\
0 + 0.3893i & -0.0890 & 0.8911 \\
0 + 0.1765i & 0.8911 & -0.0871
\end{bmatrix}\,, \\
 & X^{(1)}_{d_R} = \begin{bmatrix}
0.9796 & 0 - 0.1996i & 0 + 0.0040i \\
0 + 0.1996i & -0.9604 & 0.1378 \\
0 - 0.0040i & 0.1378 & -0.0192
\end{bmatrix}\,,\nonumber\\
&X^{(1)}_{e_L} = \begin{bmatrix}
0.9999 & 0 - 0.0108i & 0 - 0.0007i \\
0 + 0.0108i & -0.4051 & -0.4909 \\
0 + 0.0007i & -0.4909 & -0.5948
\end{bmatrix}\, , \nonumber
\end{eqnarray}

\begin{eqnarray}
 & X^{(1)}_{e_R} = \begin{bmatrix}
0.4529 & 0 - 0.4878i & 0 + 0.1450i \\
0 + 0.4878i & 0.2124 & -0.6282 \\
0 - 0.1450i & -0.6282 & -0.6653
\end{bmatrix}\,.
\end{eqnarray}

\subsection{Quark flavour violation}
Due to the presence of flavour violating couplings with the quarks, the $Z_1$ boson mediates the meson-antimeson mixing at the tree level itself. To evaluate these contributions for $K^0-\overline{K}^0$, $B_d^0-\overline{B}_d^0$, $B_s^0-\overline{B}_s^0$ and $D^0-\overline{D}^0$, we use the effective operator-based analysis (see for example \cite{UTfit:2007eik}) and quantify the new contributions in terms of the well-known \gls{WCs}.  In the context of $M^0-\overline{M}^0$ mixing, the effective Hamiltonian corresponding to $\Delta F=2$ transitions is expressed as: 
\be \label{H_eff}  {\cal H}_{eff}\,=\, \sum_{i=1}^5 C_M^i Q^i + \sum_{i=1}^3 \tilde{C}_M^i \tilde{Q}_i\,,\ee
with $M=K,B_d,B_s,D$.
Subsequently, we use the limits on these coefficients obtained from a fit to experimental data by UTFit collaboration \cite{UTfit:2007eik} to derive constraints on the mass scale of $Z_1$ boson.
 
For $K^0-\overline{K}^0$ mixing, the detailed forms of the operators can be found in \cite{UTfit:2007eik,Ciuchini:1998ix}. By integrating out $Z_1$, the contribution to the WCs $C_K^i$ and $\tilde{C}_K^i$ at the scale $\mu=M_{Z_1}$ is determined \cite{Smolkovic:2019jow} .
\beqa \label{C_K}
C_K^1&=&\frac{g_1^2}{M_{Z_1}^2}\left[\left(X^{(1)}_{d_L}\right)_{12}\right]^2\,,~~\tilde{C}_K^1=\frac{g_1^2}{M_{Z_1}^2}\left[\left(X^{(1)}_{d_R}\right)_{12}\right]^2\,,\nonumber \\
~~C_K^5 &=&-4\frac{g_1^2}{M_{Z_1}^2}\left(X^{(1)}_{d_L}\right)_{12} \left(X^{(1)}_{d_R}\right)_{12}\,. \eeqa 
The coefficients $C_K^{2,3,4}$ and $\tilde{C}_K^{2,3,4}$ are vanishing at this energy scale, $\mu=M_{Z_1}$. To compare with experimental data, it's necessary to evolve these coefficients from $\mu=M_{Z_1}$ to $\mu=2$ GeV. This evolution is carried out using the \gls{RGE} equations outlined in \cite{Ciuchini:1998ix}. The RGE analysis reveals that it generates non-zero values for $C_K^4$, whereas $C^{2,3}_K$ and $\tilde{C}^{2,3}_K$ remain zero at the low energy scale. The values of the coefficients after RGE running are presented in Table \ref{tab:meson_WC:chap3}, where they are compared with the experimentally permissible ranges determined by the UTFit collaboration's analysis \cite{UTfit:2007eik}.
\begin{table}[t]
\begin{center}
\begin{tabular}{ccccc} 
\hline
\hline
~~W.C. ~~&~~Allowed range~~&~~S1~~&~~S2~~&~~S3~~\\
 \hline
Re$C_K^1$ & $[-9.6,9.6]\times 10^{-13}$         & \rb{$-2.7\times 10^{-9}$}    & \gb{$-1.2 \times 10^{-13} $} & \gb{$-2.5\times 10^{-18}$}\\
Re$\tilde{C}_K^1$ & $[-9.6,9.6]\times 10^{-13}$ & \rb{$-1.0\times 10^{-9}$}    & \gb{$-3.2\times 10^{-14}$}   & \gb{$-2.3\times 10^{-18}$}\\
Re$C_K^4$ & $[-3.6,3.6]\times 10^{-15}$         & \rb{$8.6\times 10^{-9}$}     & \rb{$3.3\times 10^{-13}$}    & \gb{$1.4 \times 10^{-17}$} \\
Re$C_K^5$ & $[-1.0,1.0]\times 10^{-14}$         & \rb{$7.5\times 10^{-9}$}     & \rb{$2.7\times 10^{-13}$}    & \gb{$1.1\times 10^{-17}$}\\
Im$C_K^1$ & $[-9.6,9.6]\times 10^{-13}$         & \gb{$1.5\times 10^{-23}$}    & \gb{$-5.1\times 10^{-27}$}    & \gb{$4.4\times 10^{-31}$}\\
Im$\tilde{C}_K^1$ & $[-9.6,9.6]\times 10^{-13}$ &\gb{$1.4 \times 10^{-24}$}    & \gb{$-1.8\times 10^{-27}$}   & \gb{$-2.7\times 10^{-31}$}\\
Im$C_K^4$ &  $[-1.8,0.9]\times 10^{-17}$        & \gb{$-2.9\times 10^{-23}$}   & \gb{$1.7\times 10^{-26}$}    &\gb{$-4.0\times 10^{-31}$} \\
Im$C_K^5$ & $[-1.0,1.0]\times 10^{-14}$         &\gb{$-2.6\times 10^{-23}$}    & \gb{$1.4\times 10^{-26}$}    & \gb{$-3.1\times 10^{-31}$}\\
\hline
$|C_{B_d}^1|$ & $<2.3\times 10^{-11}$           & \rb{$1.4\times 10^{-10}$}    & \gb{$2.6\times 10^{-14}$}    & \gb{$6.1 \times 10^{-19}$} \\
$|\tilde{C}_{B_d}^1|$ & $<2.3\times 10^{-11}$   & \gb{$9.7\times 10^{-13}$}    & \gb{$1.4\times 10^{-17}$}    & \gb{$7.8\times 10^{-20}$}\\
$|C_{B_d}^4|$ &  $<2.1\times 10^{-13}$          &  \rb{$2.9 \times 10^{-11}$}  & \gb{$1.5\times 10^{-15}$}    & \gb{$6.0\times 10^{-19}$}\\
$|C_{B_d}^5|$ & $<6.0\times 10^{-13}$           & \rb{$5.1 \times 10^{-11}$}   & \gb{$2.6\times 10^{-15}$}    & \gb{$9.4\times 10^{-19}$}\\
\hline
$|C_{B_s}^1|$ & $< 1.1 \times 10^{-9}$         & \rb{$4.9 \times 10^{-9}$}    & \gb{$6.6\times 10^{-13}$}    & \gb{$1.3\times 10^{-17}$}\\
$|\tilde{C}_{B_s}^1|$ & $< 1.1 \times 10^{-9}$ & \gb{$7.6\times 10^{-10}$}    & \gb{$1.6\times 10^{-14}$}    & \gb{$1.4\times 10^{-18}$}\\
$|C_{B_s}^4|$ & $< 1.6 \times 10^{-11}$        & \rb{$4.7 \times 10^{-9}$}    & \gb{$2.7\times 10^{-13}$}    & \gb{$1.2\times 10^{-17}$} \\
$|C_{B_s}^5|$ & $< 4.5 \times 10^{-11}$        & \rb{$8.2 \times 10^{-9}$}    & \gb{$4.4\times 10^{-13}$}    & \gb{$1.8 \times 10^{-17}$}\\
\hline
$|C_D^1|$ & $<7.2 \times 10^{-13}$             & \rb{$3.0\times 10^{-9}$}     & \gb{$1.3\times 10^{-13}$}    & \gb{$2.4\times 10^{-17}$}\\
$|\tilde{C}_D^1|$ & $<7.2 \times 10^{-13}$     & \rb{$8.9 \times 10^{-12}$}   & \gb{$3.5\times 10^{-15}$}    & \gb{$4.4\times 10^{-19}$}\\
$|C_D^4|$ & $<4.8\times 10^{-14}$              & \rb{$6.1\times 10^{-10}$}    & \gb{$8.5\times 10^{-14}$}    & \gb{$1.3\times 10^{-17}$} \\
$|C_D^5|$ & $<4.8 \times 10^{-13}$             & \rb{$7.1\times 10^{-10}$}    & \gb{$9.5\times 10^{-14}$}    & \gb{$1.4\times 10^{-17}$}\\
\hline
\hline
\end{tabular}
\end{center}
\caption{The estimated strength of several WCs relevant to meson-antimeson mixing is presented for three example solutions, alongside the experimentally allowed range at a $95\%$ confidence level from \cite{UTfit:2007eik}. All measurements are expressed in ${\rm GeV}^{-2}$. The coefficients highlighted in green are within the experimental limits, while those in red are outside the permissible range.}
\label{tab:meson_WC:chap3}
\end{table}

For  $B_q-\overline{B}_q^0$ $(q=d,s)$ mixing, the relevant WCs at $\mu=M_{Z_1}$ can be obtained from \cite{Smolkovic:2019jow}, and are expressed as:
\beqa \label{C_Bd}
C_{B_d}^1&=&\frac{g_1^2}{M_{Z_1}^2}\left[\left(X^{(1)}_{d_L}\right)_{13}\right]^2\,,~~\tilde{C}_{B_d}^1=\frac{g_1^2}{M_{Z_1}^2}\left[\left(X^{(1)}_{d_R}\right)_{13}\right]^2\,,\nonumber\\
~~C_{B_d}^5 &=&- 4\frac{g_1^2}{M_{Z_1}^2}\left(X^{(1)}_{d_L}\right)_{13} \left(X^{(1)}_{d_R}\right)_{13}\,,\eeqa
and 
\beqa \label{C_Bs}
C_{B_s}^1&=&\frac{g_1^2}{M_{Z_1}^2}\left[\left(X^{(1)}_{d_L}\right)_{23}\right]^2\,,~~\tilde{C}_{B_s}^1=\frac{g_1^2}{M_{Z_1}^2}\left[\left(X^{(1)}_{d_R}\right)_{23}\right]^2\,,\nonumber \\
~~C_{B_s}^5 &=&- 4\frac{g_1^2}{M_{Z_1}^2}\left(X^{(1)}_{d_L}\right)_{23} \left(X^{(1)}_{d_R}\right)_{23}\,.\eeqa
The coefficients are evolved down to $\mu = M_b = 4.6$ GeV as described in \cite{Becirevic:2001jj}.

In a similar manner, for charm quark mixing, which dictates the $D^0-\overline{D}^0$ oscillations, the Wilson coefficients at the scale $M_{Z_1}$ are specified by
\beqa \label{C_D}
C_D^1&=&\frac{g_1^2}{M_{Z_1}^2}\left[\left(X^{(1)}_{u_L}\right)_{12}\right]^2\,,~~\tilde{C}_D^1=\frac{g_1^2}{M_{Z_1}^2}\left[\left(X^{(1)}_{u_R}\right)_{12}\right]^2\,,\nonumber \\ ~~C_D^5 &=&- 4\frac{g_1^2}{M_{Z_1}^2}\left(X^{(1)}_{u_L}\right)_{12} \left(X^{(1)}_{u_R}\right)_{12}\,.\eeqa
They are also run down to the relevant low scale $\mu = 2.8$ GeV using the RGE equations given in \cite{UTfit:2007eik}.

In Table \ref{tab:meson_WC:chap3}, we present the values of the WCs at relevant hadronic scales for the three benchmark solutions and compare them against their experimental limits. It is observed that meson-antimeson mixing constraints significantly restrict the mass of the $Z_1$ boson, as it generally exhibits ${\cal O}(1)$ off-diagonal couplings with quarks, which can be seen from eq. (\ref{example:sol}). Solutions S1 and S2 are not favoured, indicating that $M_{Z_1} \gtrsim 10^5$ TeV for the solutions to remain consistent with phenomenological requirements. Consequently, the model fails to explain the neutral current $B$ anomalies, which generally demand $M_{Z_1}$ to be no more than 2 TeV, as noted in references \cite{DiLuzio:2017fdq,Allanach:2019mfl}.

\subsection{Lepton flavour violation}
The flavourful \( Z_1 \) mediates charged lepton flavour-violating processes such as \(\mu \to e\) conversion in nuclei and \( l_i \to 3l_j \) at the tree level. Additionally, processes like \( l_i \to l_j \gamma \) occur at the one-loop level, involving \( Z_1 \) and charged leptons in the loop. In this subsection, we evaluate the constraints on \( Z_1 \) imposed by these processes.

In the field of a nucleus, muons can have transition to electrons via flavour-violating couplings between \( \mu \) and \( e \) mediated by the \( Z_1 \) boson. The most stringent limit on this process is established by the SINDRUM II experiment, which utilizes a \( ^{197} \mathrm{Au} \) nucleus \cite{SINDRUMII:2006dvw}. The branching ratio for this process, as calculated in \cite{Kitano:2002mt}, is given by:
\be \label{mu2e}
{\rm BR}[\mu \to e] = \frac{2 G_F^2}{\omega_{\rm capt}}\,(V^{(p)})^2\, \left(|g^{(p)}_{LV}|^2 + |g^{(p)}_{RV}|^2\right)\,, \ee
where $V^{(p)}$ is termed as overlap integral, which involves proton distribution for a given nucleus and $\omega_{\rm capt}$ is the muon capture rate by the nucleus. The function $g^{(p)}_{LV,RV}$ is given by
\be \label{gLV}
g^{(p)}_{LV,RV} = 2 g^{(u)}_{LV,RV} + g^{(d)}_{LV,RV}\,.\ee
In the limit $M_{Z_1} \gg m_\mu$, for $Z_1$ mediated contributions, the above couplings are given by \cite{Smolkovic:2019jow}
\be \label{gLV_Z1}
g^{(q)}_{LV} \sim \frac{\sqrt{2}}{G_F} \frac{g_1^2}{M_{Z_1}^2}\,\left(X^{(1)}_{e_L}\right)_{12} \frac{1}{2}\left[\left(X^{(1)}_{q_L}\right)_{11} + \left(X^{(1)}_{q_R}\right)_{11} \right]\,\ee
with $q=u,d$. Similarly, $g^{(q)}_{RV} $ can be obtained by replacing $L \leftrightarrow R$ in the above expression. Substituting eqs. (\ref{gLV_Z1},\ref{gLV}) in (\ref{mu2e}) and using $V^{(p)}= 0.0974\, m_\mu^{5/2}$, $\omega_{\rm capt} = 13.07 \times 10^{6}\,{\rm s}^{-1}$ for $^{197}$Au from \cite{Kitano:2002mt}, we estimate ${\rm BR}[\mu \to e]$ for the solutions obtained and list them in Table \ref{tab:LFV:chap3}. For comparison purposes, we also include the most recent experimental limit on ${\rm BR}[\mu \to e]$ in the same table.
\begin{table}[t]
\begin{center}
\begin{tabular}{ccccc} 
\hline
\hline
~~LFV observable~~&~~Limit~~&~~S1~~&~~S2~~&~~S3~~\\
 \hline
${\rm BR}[\mu \to e]$    & $< 7.0 \times 10^{-13}$   &    \rb{$2.6 \times 10^{-8}$}   & \gb{$3.9 \times 10^{-15}$} & \gb{$1.8 \times 10^{-23}$}\\
\hline
${\rm BR}[\mu \to 3e]$    & $< 1.0 \times 10^{-12}$  &    \rb{$1.0 \times 10^{-9}$}   & \gb{$4.2 \times 10^{-16}$} & \gb{$2.0 \times 10^{-27}$}\\
${\rm BR}[\tau \to 3\mu]$ & $< 2.1 \times 10^{-8}$   &    \gb{$1.7\times 10^{-8}$}    & \gb{$2.8\times 10^{-17}$}  & \gb{$1.8 \times 10^{-24}$}\\
${\rm BR}[\tau \to 3 e]$  & $< 2.7 \times 10^{-8}$   &    \gb{$2.3\times 10^{-10}$}   & \gb{$6.9\times 10^{-18}$}  & \gb{$7.5 \times 10^{-30}$}\\
\hline
${\rm BR}[\mu \to e \gamma]$ & $< 4.2 \times 10^{-13}$   & \rb{$1.0\times 10^{-10}$} & \gb{$3.8\times 10^{-18}$} & \gb{$3.8 \times 10^{-27}$}\\
${\rm BR}[\tau \to \mu \gamma]$&  $< 4.4 \times 10^{-8}$ & \gb{$3.7\times 10^{-12}$} & \gb{$1.1\times 10^{-19}$} & \gb{$3.8 \times 10^{-27}$}\\
${\rm BR}[\tau \to e \gamma]$ &  $< 3.3 \times 10^{-8}$  & \gb{$2.5\times 10^{-13}$} & \gb{$2.0\times 10^{-20}$} & \gb{$8.0 \times 10^{-32}$}\\
\hline
\hline
\end{tabular}
\end{center}
\caption{Estimated values for differently charged lepton flavour violation observables are provided for the three benchmark solutions, along with current experimental limits set at a $90 \%$ confidence level. These limits are obtained from \cite{Calibbi:2017uvl}. Values highlighted in green indicate compliance with experimental constraints, whereas those in red signify violation.}
\label{tab:LFV:chap3}
\end{table}

Next, we estimate the branching ratios for the process $\mu \to 3 e$, $\tau \to 3 \mu$ and $\tau \to 3 e$ by following \cite{Heeck:2016xkh,Smolkovic:2019jow}. The relevant decay width for $l_i \to  3l_j$, estimated neglecting sub-leading terms proportional to $m_{l_j}$, is given by
\beqa \label{lto3l}
\hspace{-0.4cm}\Gamma[l_i \to 3 l_j] &\simeq& \frac{g_1^4 m_{l_i}^5}{768 \pi^3 M_{Z_1}^4}\,\left[ 4 {\rm Re}\left( \left(X_{eV}\right)_{ji} \left(X_{eA}\right)_{ji} \left(X_{eV}\right)^*_{jj} \left(X_{eA}\right)^*_{jj} \right) \right. \nonumber \\
&+& \hspace{-0.2cm}\left. 3 \left( \left| \left(X_{eV}\right)_{ji}\right|^2 + \left| \left(X_{eA}\right)_{ji}\right|^2 \right) \left( \left| \left(X_{eV}\right)_{jj}\right|^2 + \left| \left(X_{eA}\right)_{jj}\right|^2 \right) \right],
\eeqa
where
\be \label{X_VA}
X_{eV,eA} = \frac{1}{2} \left(X^{(1)}_{e_L}  \pm X^{(1)}_{e_R}\right)\,, \ee
are couplings for vector and axial-vector currents, defined in eq. (\ref{Z_couplings}), respectively.  Using the above expression, the obtained values of ${\rm BR}[l_i \to 3 l_j]$ for the example solutions are given in Table \ref{tab:LFV:chap3} along with their respective experimental constraints.

Unlike the previous flavour-violating decays, the decays like $l_i \to l_j \gamma$ arise at the 1-loop level. We calculate these decay processes by taking into account stringent constraints on ${\rm BR}[\mu \to e \gamma]$. The corresponding decay width can be written as \cite{Lavoura:2003xp}
\be \label{muegamma}
\Gamma[l_i \to l_j \gamma] = \frac{\alpha g_1^4}{4 \pi}\,\left(1-\frac{m_{l_j}^2}{m_{l_i}^2}\right)^3\,\frac{m_{l_i}^4}{M_{Z_1}^4}\,m_{l_i}\,\left( |c^\gamma_L|^2+|c_R^\gamma|^2\right)\,,\ee
 with $\alpha$ as the fine-structure constant. Also, 
 \beqa \label{c_gamma}
 c_L^\gamma & = & \sum_{k} Q_k \left[\left(X^{(1)}_{e_R}\right)^*_{jk} \left(X^{(1)}_{e_R}\right)_{ik} y_{RR} +\left(X^{(1)}_{e_L}\right)^*_{jk} \left(X^{(1)}_{e_L}\right)_{ik} y_{LL} \right. \nonumber \\
 &+& \left. \left(X^{(1)}_{e_R}\right)^*_{jk} \left(X^{(1)}_{e_L}\right)_{ik} y_{RL} +\left(X^{(1)}_{e_L}\right)^*_{jk} \left(X^{(1)}_{e_R}\right)_{ik} y_{LR} \right]\,,\eeqa
and  $c_R^\gamma$ can be obtained with replacement $L \leftrightarrow R$ in the coupling matrices appearing in  the R.H.S of eq. (\ref{c_gamma}). $Q_k$ denotes the electromagnetic charge quantum number of $l_k$ lepton. The explicit forms for the loop functions $y_{LL}$, $y_{RR}$, $y_{LR}$, and $y_{RL}$ are available in \cite{Lavoura:2003xp}. Computed values for ${\rm BR}[\mu \to e \gamma]$, ${\rm BR}[\tau \to \mu \gamma]$, and ${\rm BR}[\tau \to e \gamma]$ using these formulations are presented in Table \ref{tab:LFV:chap3} for three illustrative solutions.

Tabel \ref{tab:LFV:chap3} reveals that the most stringent constraints on the flavourful $Z_1$ interactions originate from $\mu$ to $e$ transitions and processes of the form $l_i \to 3 l_j$, both of which occur at the tree level. Additionally, the process $\mu \to e \gamma$ imposes a comparable restriction on $M_{Z_1}$. These \gls{LFV} processes exclude $M_{Z_1}$ values up to 10 TeV, rendering the benchmark solution S1 less favourable. A comparison of the values in Table \ref{tab:meson_WC:chap3} and \ref{tab:LFV:chap3} shows that LFV constraints are less severe compared to those derived from $K^0-\overline{K}^0$ oscillations.

\subsection{Direct and electroweak constraints}
\label{subsec:direct-search}
The flavour violations in the quark and lepton sector put strong lower bounds on the masses of new particles which often supersede the direct search constraints. For instance, recent findings from the LHC indicate $M_{Z_1} > 5.15$ TeV for a $Z_1$ boson with ${\cal O}(1)$ flavour-diagonal interactions with the SM fermions \cite{CMS:2021ctt}. This threshold rises to $M_{Z_1} > 7.20$ TeV if $Z_1$ has generic diquark interactions \cite{CMS:2018mgb}. Similarly, existing direct search constraints on vector-like fermions suggest $m_B > 1.57$ TeV \cite{CMS:2020ttz,ATLAS:2018mpo} and $m_T > 1.31$ TeV \cite{CMS:2018wpl,ATLAS:2018ziw}. As evident from Table \ref{tab:sol} and prior subsection results, these constraints are considerably weaker than those imposed by FCNCs.

In the present model, a different set of constraints emerges due to the mixing of $Z$ and $Z_{1,2}$, as the Higgs fields are charged under both the SM and the extended gauge symmetries. This mixing can be described, following the approach in \cite{Babu:1997st}, using mixing angles given by
\be \label{zzmixing}
\sin \theta_{1,2} = \frac{g_{1,2}}{\sqrt{g^2 + g^{\prime 2}}}\,\left(\frac{M_Z}{M_{Z_{1,2}}}\right)^2\,, \ee
where $g$ and $g^\prime$ represent the coupling strengths of the $SU(2)_L$ and $U(1)_Y$ gauge interactions, respectively. Within this framework, the hierarchy between first and second-generation fermion masses requires $M_{Z1} \ll M_{Z2}$, which in turn implies that $\theta_{2} \ll \theta_{1} \ll 1$. This indicates that the major effects arise from the $Z-Z_{1}$ mixing. This mixing introduces flavoured non-universal couplings to the SM fermions through the $Z$ boson as well, but these are attenuated by a factor of $M_Z^2/M_{Z_1}^2$ relative to the couplings of $Z_1$.

The $Z-Z_{1}$ mixing modifies the SM $\rho$ parameter which is precisely measured along with the other electroweak observables. At the leading order in $\theta_1$, the correction to the $\rho$ parameter can be obtained as \cite{Allanach:2021kzj} 
\be \label{delta_rho}
\Delta \rho = \frac{g_1^2}{g^2 + g^{\prime 2}}\,\left(\frac{M_Z}{M_{Z_{1,2}}}\right)^2\,.\ee
Based on the global fit result, $\rho = 1.00039 \pm 0.00019$ \cite{ParticleDataGroup:2020ssz}, it can be deduced that $M_{Z_1} \geq 4.5$ TeV assuming $g_1=1$. Any non-zero mixing between the $Z$ and $Z_{1}$ bosons influences the $Z$ boson's couplings with neutrinos, which can be restricted using the invisible decay width of the $Z$ boson, leading to a constraint expressed as $M_{Z_1}/g_1 \geq 0.95$ TeV \cite{Davighi:2021oel}. The mixing between $Z$ and $Z_{1}$ also results in flavour non-universal couplings of the $Z$ boson to leptons, causing lepton flavour universality violation in $Z$ boson decays. This violation is stringently constrained by LEP data, which leads to $R = 0.999 \pm 0.003$ \cite{ParticleDataGroup:2020ssz}, where $R$ represents the ratio of the partial decay widths of the $Z$ boson into an electron pair to those into a muon pair. At first order in $\theta_1$, the deviation of $R$ from unity due to new physics contributions is specified by \cite{Davighi:2021oel}:
\be \label{DeltaR}
\Delta R \simeq 4 g_1\,\sin \theta_1\, \frac{g \cos \theta_W - 3 g^\prime \sin \theta_W }{\left(g \cos \theta_W - g^\prime \sin \theta_W \right)^2 + 4 g^{\prime 2}\,\sin^2 \theta_W}\,.\ee
The LEP limit consequently implies that $M_{Z_1}/g_1\geq 1.3 $ TeV.

To summarize, the restrictions derived from direct searches and electroweak precision measurements are, at minimum, two orders of magnitude less stringent compared to those imposed by quark and lepton flavour-violating interactions. The different constraints examined in this section propose a minimum bound of $M_{Z_1}/g_1 > 10^3$ TeV for the generic feasible solutions ascertained in the current model.

\section{The scalar induced FCNCs}
\label{chap3:scalarFCNC}
The model contains scalars $H_{ui}$, $H_{di}$ and $\eta_i$, with the former two charged under both the SM and the $G_F$ gauge symmetry. The SM singlet $\eta_i$ dominantly breaks the $G_F$ symmetry. The weak doublets $H_{ui}$ and $H_{di}$ are responsible for breaking the electroweak symmetry, leaving $U(1)$ electromagnetism as a conserved symmetry. The breaking of the gauge symmetry leads to several electrically charged and neutral physical scalars. Since charged scalars explicitly couple to left-chiral fermions, they don't take part in the loop mass generation mechanism. However, the neutral scalars can contribute to the charged fermion masses through loops. Such contributions can remain relatively suppressed under reasonable conditions as discussed in section \ref{chap2:UV}. 

It is evident from the scalar potential written in eq. (\ref{chap3:potential}), the neutral scalars mix among themselves. This, along with the non-diagonal nature of the tree-level fermion mass matrix, leads to various flavour-violating couplings of scalars with the SM fermions. In this section, we obtain reasonable conditions under which these scalar-induced FCNCs will be suppressed. Identifying the electromagnetically neutral scalars present in $H_{ui}$, $H_{di}$ and $\eta_i$ as $h_{ui}$, $h_{d i}$ and $\eta_i$, respectively, their mass term can be parametrized as 
\be \label{hmass}
\frac{1}{2} \left(M^2_h \right)_{ab}\,\tilde{h}_a\,\tilde{h}_b\,, \ee
where $a=1,...,9$ and $\tilde{h} = (h_{ui},h_{dj},\eta_k)^T$. The $9 \times 9$ symmetric neutral scalar mass matrix, denoted by $M^2_h$, can be expressed using $3 \times 3$ block matrices as
\be \label{M9}
M^2_h = \left(\ba{ccc} m_{u u}^2 & m^2_{ud} &  m^2_{u\eta} \\ (m^2_{ud})^T & m_{d d}^2 & m^2_{d\eta}\\ (m^2_{u\eta})^T & (m^2_{d\eta})^T & m^2_{\eta \eta} \ea \right)\,.\ee
These parameters are derivable from the scalar potential and are functions of various parameters, present in eq. (\ref{chap3:potential}), as well as the VEVs of $H_{ui}$, $H_{di}$, and $\eta_i$. Typically, one can determine the physical neutral scalar states $h_a$ by performing diagonalisation on $M^2_h$. Specifically, we define
\be \label{R}
\tilde{h}_a = (R_h)_{ab}\, h_b\,,\ee
where $R_h$ is an orthogonal matrix such that $R_h^T M^2_h R_h = {\rm Diag.}(m_{h_1}^2,...,m^2_{h_9})$. 
 While the typical mass scale for scalars is ${\cal O}(M_X)$, one scalar must account for the detected SM-like Higgs boson with a mass of 125 GeV \cite{ATLAS:2015yey}. Minimally, the lightest of
the neutral scalars, $h_1$, can fulfil this purpose. This scenario requires ${\rm Det.} M_h^2 \ll M_X^{18}$, demanding precise tuning of the parameters within the scalar potential. Considering the numerous indeterminate parameters in the scalar potential, we presume that achieving this configuration is feasible. 

As mentioned earlier, the scalar mixing and the feature that Yukawa couplings are not proportional to the masses for the SM fermions in the underlying framework lead to flavour-changing transitions mediated by the neutral scalars. For example, consider the interactions with the up-type quarks in eq. (\ref{LY}).In the physical basis, they lead to the following flavour changing interactions between the SM up-type quarks and neutral scalars at the leading order.
\beqa \label{hup}
&&\left[\left(U^{u \dagger}_{L } \right)_{mi} {y_u}_i (R_u)_{ia} \left(\rho_R^\dagger U^u_R \right)_n+  \left(U^{u\dagger}_{L} \rho_L \right)_m {y_u^{\prime}}_k (R_\eta)_{k a} \left(U^u_{R} \right)_{kn} \right] \overline{u}_{Lm} {h}_a u_{Rn}\nonumber\\
&&+\, {\rm h.c.}\,,   \eeqa
where $U^u_{L,R}$ are unitary matrices which relate the mass and interaction basis of the up-type quarks. $R_u$ ($R_\eta$) is $3 \times 9$ matrix made up of the first (last) three rows of the orthogonal matrix $R_h$.  In comparison to eq. (\ref{Z_couplings}), the flavour violating couplings in eq. (\ref{hup}) are suppressed by heavy-light fermion mixing. Clearly, if the masses of the scalars are of ${\cal O}(M_X)$, then the induced FCNC are suppressed compared to the ones mediated by the gauge boson. The existing experimental constraints, therefore, only require that the flavour-changing couplings of the light Higgs,$(R_{h})_{1a}$, be suppressed. Considering the large number of parameters in the scalar potential, we assume that this suppression is feasible, albeit with some fine-tuning.

\section{Conclusions and Outlook}
Based on the findings presented in the previous chapter, the realization of a radiative mass generation mechanism in an Abelian extension necessitates that the SM fermions possess non-universal charges under the associated symmetry. To explore this, we construct an explicit model based on a \( U(1)_1 \times U(1)_2 \) symmetry, which generalizes the well-known leptonic \( L_\mu - L_\tau \) and \( L_e - L_\mu \) symmetries, respectively. Using the invariance under these symmetries, at the tree level, the Yukawa sector is designed to exhibit a global \(U(2)^5\) symmetry, corresponding to massless first and second generations. The breaking of \( U(1)_1 \) induces radiative masses for the second-generation fermions, while the breaking of \( U(1)_2 \) contributes to the first-generation masses. The mass hierarchy between the first two fermion generations is naturally linked to the hierarchy of the symmetry-breaking scales of \( U(1)_1 \) and \( U(1)_2 \). 

Although the model presented here reproduces the observed hierarchical pattern of fermion masses, it still involves a significant number of free parameters. However, unlike the SM, the fundamental parameters of the proposed model do not vary across a wide range of magnitudes.

We provide three representative numerical solutions that successfully reproduce the observed charged fermion masses and quark mixing parameters. These solutions are analysed under various experimental constraints, including those arising from quark and lepton flavour violations, direct searches for new particles, and electroweak precision observables. Although the radiative mechanism does not strictly determine the absolute scale of new physics, current constraints strongly suggest that the masses of new particles must be of the order of \( 10^5 \, \mathrm{TeV} \). Furthermore, the requirement to achieve a viable fermion mass spectrum places significant constraints on the relative mass scales of the new vector bosons and VL fermions in the model. However, the experimental verification of such a high scale seems not feasible in the near future. So, a natural question arises: can we further lower the scale of the radiative mass mechanism? The answer lies in the identification of which process gives the stringent bound and in suppressing such a coupling that induces the process. It can be seen that, for this model, the meson-antimeson oscillations, which involve the first two generation fermions, give the strongest bound. So, the new physics scale can be further lowered by postulating a radiative mass mechanism model in which such couplings are naturally suppressed. In the next chapter, we identify a set of non-universal charges which viably lower the scale of new physics by at least two orders.

%% file: 40_Chapter_4/chapter_4.tex
\chapter{Optimising the flavour violation in the Abelian frameworks}
\label{chap4}
It is noted from the previous chapter that the flavour non-universality of the new gauge interactions leads to flavour-changing neutral currents for the SM fermions, and its constraints dictate the lower limit on the new physics scale in these types of frameworks. The magnitude of flavour changing neutral current couplings, namely $Q_{ij}$, between the $i^{\rm th}$ and $j^{\rm th}$ generations,  primarily depends on the structure of $G_F$. For example, the choices of $G_F$ in the previous chapter lead to more or less universal values of $Q_{ij}$. On the other hand, the present experimental limits from quark and lepton flavour violations are more stringent in the $1$-$2$ sector than in the $2$-$3$ or $1$-$3$ sectors. This suggests that it would be desirable to have $|Q_{12}|<|Q_{23}|,|Q_{13}|$ to somewhat relax the lower limit on the new gauge boson mass mentioned in section \ref{chap3:pheno}. This has phenomenological and technical advantages in terms of observability and naturalness, respectively. 

The chapter attempts to investigate these questions systematically. The specific nature of $G_F= U(1)_F$, which can lead to the desired ordering for $Q_{ij}$, is obtained by using the symmetry analysis given in chapter \ref{sec:sym_dec}. It will be shown that optimal $G_F$ charges which lead to minimal flavour violation, can also induce the non-vanishing and small first-generation fermion masses. It is found that an interesting correlation exists between the loop-induced first-generation mass and flavour violation in the $1$-$2$ sector, such that the latter vanishes completely in the limit of a massless first-generation. All these aspects are discussed in section \ref{chap4:optimal}. We implement the optimal flavour-violating framework in the standard model in a phenomenologically viable manner in section \ref{sec:SM} and constraints arising mainly from flavour violations are discussed in section \ref{sec:constraints}. We also discuss possibilities to accommodate massive neutrinos within this framework in section \ref{sec:neutrino}.  Before concluding in section \ref{chap4:summary}, we discuss the possible origin of new gauge charges chosen in section \ref{chap4:gaugecharges} .

\section{Optimal gauge charges and 2-loop masses}
\label{chap4:optimal}
In this section, we use the symmetry deconstruction discussion of the toy model framework discussed in Chapter \ref{chap:2} to identify the set of optimal charges. It is explicitly shown that the 1-loop corrected mass Lagrangian has a residual symmetry $U(1)_L \times U(1)_R$, which can be defined by the transformations:
\be \label{U1U1_trans:1}
f_{L 1} \to e^{i \alpha_L} f_{L 1}\,,~~f_{R 1} \to e^{i \alpha_R} f_{R 1}\,.\ee
The above symmetry is not respected by the 1-loop corrected gauge Lagrangian for non-vanishing $(Q_{L,R}^{(1)})_{12,13}$. Explicitly, the breaking
\be \label{2loop_symm}
U(1)_L \times U(1)_R\, \xrightarrow{\text{at 2-loop}}\, U(1)_{\rm fn}\,,\ee
induces a small yet positive mass for the light fermion. Here $U(1)_{\rm fn}$ is a global fermion number, which remains as a conserved quantity, at least at the perturbative level in this setup.

For our convenience, we rewrite the following elements $Q_{L,R}^{(1)}$ from eqs. (\ref{Q_12},\ref{Q_13}) and (\ref{Q_23}):
\beqa \label{Q_1}
\left(Q_L^{(1)}\right)_{12} &=&\frac{(q_{L2}-q_{L1})(q_{L3}-q_{L1})}{\sqrt{N}\, (q_{L3}-q_{L2})}\,\left(\frac{\mu_{L1}}{\mu_{L3}} \left(U_L^{(1)}\right)_{32} -  \frac{\mu_{L1}}{\mu_{L2}} \left(U_L^{(1)}\right)_{22}\right)\,. \nonumber\\
\left(Q_L^{(1)}\right)_{13} &=&\frac{(q_{L2}-q_{L1})(q_{L3}-q_{L1})}{\sqrt{N}\, (q_{L3}-q_{L2})}\,\left(\frac{\mu_{L1}}{\mu_{L3}} \left(U_L^{(1)}\right)_{33} -  \frac{\mu_{L1}}{\mu_{L2}} \left(U_L^{(1)}\right)_{23}\right)\,, \\
\left(Q_L^{(1)}\right)_{23} &=& (q_{L3}-q_{L2})\, \left(U_L^{(1)}\right)^*_{32}\left(U_L^{(1)}\right)_{33} - (q_{L2}-q_{L1})\, \left(U_L^{(1)}\right)^*_{12}\left(U_L^{(1)}\right)_{13}\,.\nonumber \eeqa
From a phenomenology, it is preferable to satisfy the condition $\left|\left(Q_L^{(1)}\right)_{12}\right| < \left|\left(Q_L^{(1)}\right)_{23}\right|$, due to the more stringent experimental limitations on flavour-violating processes within the $1$-$2$ sector. Eq. (\ref{Q_12}) demonstrates that this arrangement is feasible if 
\be \label{cond_1_qs}
|q_{L2}-q_{L1}| \ll |q_{L3}-q_{L2}|~~~{\rm or}~~~ |q_{L3}-q_{L1}| \ll |q_{L3}-q_{L2}|\,,\ee
and 
\be \label{cond_2_mus}
|\mu_{L 1}| < |\mu_{L2}|,\,|\mu_{L3}|\,.\ee 
The same holds true for parameters where $L$ is replaced by $R$ in the subscripts. 
These observations are crucial in determining the precise nature of $U(1)_F$, which can implement the radiative mass mechanism with minimal flavour violation. 

To maintain flexibility in selecting flavour-specific charges, a straightforward approach to guarantee the $U(1)_F$ symmetry remains anomaly-free is to consider $U(1)_F$ as a vector-like symmetry, implying $q_{L i} = q_{R i}$ for every $i$. This configuration is adequate to cancel both $[U(1)_F]^3$ and mixed gauge-gravity anomalies. The vector-like nature of $U(1)_F$ alone does not prevent the occurrence of bare mass terms such as $\overline{f}^\prime_{L i} f^\prime_{R i}$, which, if present in ${\cal L}_m$ as indicated in eq. (\ref{L_mass}), can undermine the mechanism of radiative mass generation. Consequently, the introduction of an additional chiral symmetry may be necessary. In the case of the Standard Model, the inherent gauge symmetry of the model can be leveraged to fulfill this role.

After selecting $q_{Li} = q_{Ri}$, fixing one of the three charges to a non-zero value is permissible without any loss of generality. With this consideration and the constraints given in eq. (\ref{cond_1_qs}), we choose:
\be \label{gauge_charges}
q_{L1} = q_{R1} = 1 - \epsilon\,,~~~q_{L2} = q_{R2} = 1 + \epsilon\,,~~~q_{L3} = q_{R3} = -2\,, \ee
with $0< \epsilon \le 1$. Moreover, we selected the charges to ensure that ${\rm Tr}(q_{L,R}) = 0$. This simplifies the cancellation of certain mixed gauge anomalies when extending this toy model to incorporate the Standard Model. When $\epsilon \ll 1$, it is possible to achieve a relative suppression of flavour violation strength in the $1$-$2$ sector, as discussed in section \ref{sec:sym_dec}. Specifically, under this assumption, eqs. (\ref{Q_12},\ref{Q_23}) yield the expression:
\beqa \label{Q_rat}
\left| \frac{\left(Q_L^{(1)}\right)_{12}}{\left(Q_L^{(1)}\right)_{23}} \right| & \simeq & \frac{2 |\epsilon|}{3 \sqrt{N}}\, \left| \frac{\frac{\mu_{L1}}{\mu_{L3}} \left(U_L^{(1)}\right)_{32} -  \frac{\mu_{L1}}{\mu_{L2}} \left(U_L^{(1)}\right)_{22}}{ \left(U_L^{(1)}\right)^*_{32}\left(U_L^{(1)}\right)_{33}}\right|\,. \eeqa
When the condition, eq. (\ref{cond_2_mus}), is applied, $|\epsilon| \ll 1$ leads to $\left|\left(Q_L^{(1)}\right)_{12}\right| \ll \left|\left(Q_L^{(1)}\right)_{23}\right|$. Nevertheless, when $\epsilon=0$, it results in a massless generation of fermion, requiring appropriate optimisation of $\epsilon$.
\subsection{2-loop fermion masses}
Following the methodology applied in the section \ref{subsec:1loop}, we proceed to evaluate the mass of the first generation, which emerges through the subsequent order correction to ${\cal M}^{(1)}$. This is depicted by the diagram in Fig. \ref{fig:2loop}. The correction can be expressed as:
\begin{figure}[t]
\centering
\includegraphics[width=0.6\textwidth]{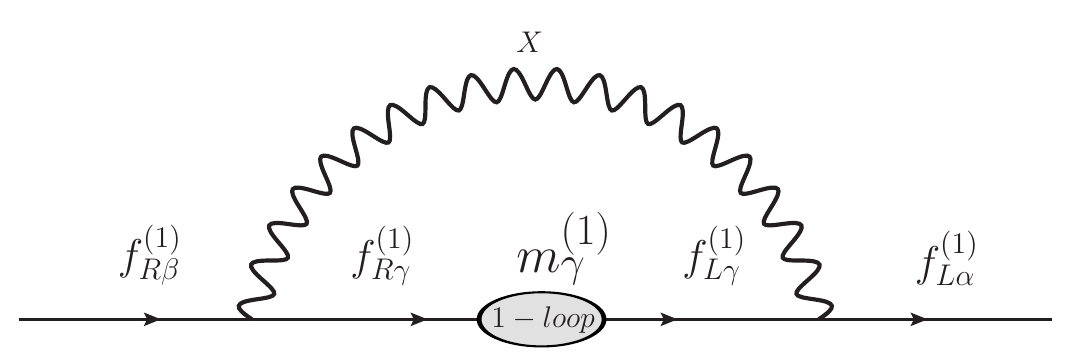}
\caption{Next-to-leading order correction to the 1-loop corrected mass matrix.}
\label{fig:2loop}
\end{figure}

\be \label{M2}
{\cal M}^{(2)} = {\cal M}^{(1)} + \delta {\cal M}^{(1)}\,, \ee
with
\be \label{dM1}
\delta {\cal M}^{(1)} = {\cal U}_L^{(1)}\,\sigma^{(1)}\,{\cal U}_R^{(1) \dagger}\,,\ee
and $\sigma^{(1)}$ has form similar to $\sigma^R$ of eq. (\ref{sigma_LR}). Explicitly,
\be \label{sigma_1}
\sigma^{(1)}_{\alpha \beta} = \frac{g_X^2}{4 \pi^2}\,\sum_\gamma ({\cal Q}^{(1)}_L)_{\alpha \gamma} ({\cal Q}^{(1)}_R)_{\gamma \beta}\, m_\gamma^{(1)}\, B_0[M_X,m_\gamma^{(1)}]\,,\ee
where
\be \label{Q1}
{\cal Q}_{L,R}^{(1)} = {\cal U}_{L,R}^{(1) \dagger}\,\left(\ba{cc} q_{L,R} & 0\\0 & 0 \ea \right)\,{\cal U}_{L,R}^{(1)}\,,\ee
is the 1-loop corrected charge matrix in the new physical basis. Here, for simplification of further analysis, we assume VL fermions as neutral.

From eq. (\ref{dM1}), one finds $\delta {\cal M}^{(1)}_{\alpha 4} = \delta {\cal M}^{(1)}_{4 \alpha} = 0$, which results from the neutral behaviour of vector-like states under $U(1)_F$. This can be anticipated by comparing with eq. (\ref{del_calM_44}). The correction to $3\times 3$ upper left block of $\delta {\cal M}^{(1)} $ is explicitly computed as 
\beqa \label{dM1_33}
\delta{\cal M}^{(1)}_{ij} &=& \frac{g_X^2}{4 \pi^2}\, q_{Li}  q_{Rj}\, \left({\cal U}^{(1)}_L\right)_{i \gamma} \left({\cal U}^{(1)}_R\right)^*_{j \gamma}\,m_{\gamma}^{(1)}\,B_0[M_X,m_\gamma^{(1)}]\,.\eeqa
Altogether, ${\cal M}^{(2)}$ can be written as
\be \label{M2_1}
{\cal M}^{(2)} = \left( \ba{cc} \left(\delta M^{(1)}\right)_{3 \times 3} & \mu_L\\ \mu_R & m_F \ea \right)\,,\ee
with 
\be \label{dM1_eff}
\delta M^{(1)}_{ij} = \delta M^{(0)}_{ij} + \delta{\cal M}^{(1)}_{ij}\,.\ee
Further simplification is attainable in the seesaw approximation. By inserting eq. (\ref{U1_ss}) into eq. (\ref{dM1_33}) and performing some basic algebraic simplifications, we obtain
\beqa \label{dM1_33_1}
\delta{\cal M}^{(1)}_{ij} &=&  \frac{g_X^2}{4 \pi^2}\, q_{Li}  q_{Rj}\, \left(\sum_{k=2,3} (U_L^{(1)})_{ik} (U_R^{(1)})^*_{jk} m_k^{(1)} B_0[M_X,m_k^{(1)}] - M^{(0)}_{ij} B_0[M_X,m_F] \right)\,. \nonumber \\
\eeqa
This leads to
\beqa \label{dM1_eff_1}
\delta M^{(1)}_{ij} &=& \frac{g_X^2}{4 \pi^2}\, q_{Li} q_{Rj} \Big(M^{(0)}_{ij} (B_0[M_X,m_3^{(0)}]-2 B_0[M_X,m_F]) \Big. \nonumber \\
&+& \Big. \sum_{k=2,3}\,(U_L^{(1)})_{ik} (U_R^{(1)})^*_{jk} m_k^{(1)} B_0[M_X,m_k^{(1)}]\Big)\,,\eeqa
for the $3 \times 3$ matrix appearing in ${\cal M}^{(2)}$.

The diagonalisation of ${\cal M}^{(2)}$ is carried out by 
\be \label{M2_diag}
{\cal U}_L^{(2) \dagger}\,{\cal M}^{(2)}\,{\cal U}_R^{(2)} \equiv {\cal D}^{(2)} = {\rm Diag}.\left( m_1^{(2)},m_2^{(2)},m_3^{(2)},m_4^{(2)}\right)\,,\ee
resulting in a non-zero mass for the lightest fermion induced at the two-loop level. As before, the unitary matrices can be represented as
\be \label{U2_ss}
{\cal U}^{(2)}_{L,R} \approx \left(\ba{cc} U_{L,R}^{(2)} & - \rho_{L,R}^{(2)} \\
 \rho_{L,R}^{(2) \dagger} U_{L,R}^{(2)} & 1 \ea\right)\,. \ee
Comparing the form of ${\cal M}^{(2)}$ in eq. (\ref{M2_1}) with eq. (\ref{M0}) implies $\rho^{(2)}_{L,R}=\rho^{(0)}_{L,R}$. Using the above in eq. (\ref{M2_diag}), the effective $3\times 3$ mass matrix at 2-loop is
\be \label{M2_eff}
M^{(2)}_{ij} = M^{(0)}_{ij} + \delta M^{(1)}_{ij}\,,\ee
leading to
\be \label{M2_eff_diag}
U_L^{(2) \dagger}\, M^{(2)}\, U_R^{(2)} = {\rm Diag.}\left(m_1^{(2)},m_2^{(2)},m_3^{(2)}\right)\,.\ee

From eqs. (\ref{dM1_eff_1},\ref{M2_eff}) and some further simplification using eq. (\ref{M1_eff_diag}), we finally obtain the following effective chiral fermion mass matrix:
\beqa \label{M2_eff_fnl}
M^{(2)}_{ij} &=& M^{(0)}_{ij} \left(1+\frac{g_X^2}{4 \pi^2}\, q_{Li}\, q_{Rj}\, (B_0[M_X,m_3^{(1)}]- B_0[M_X,m_F]) \right) \nonumber \\
&+& \delta M^{(0)}_{ij} \left(1+\frac{g_X^2}{4 \pi^2}\, q_{Li}\, q_{Rj}\, B_0[M_X,m_3^{(1)}] \right)  \\
&+& \frac{g_X^2}{4 \pi^2}\, q_{Li}\, q_{Rj} \,(U_L^{(1)})_{i2} (U_R^{(1)})^*_{j2}\, m_2^{(1)}\, (B_0[M_X,m_2^{(1)}] - B_0[M_X,m_3^{(1)}])\,.\nonumber\eeqa
The initial term mentioned above represents the typical tree-level contribution. The subsequent second and third terms characterize the next-to-leading order contribution, being proportional to $q_{L i} q_{R j} M^{(0)}_{ij}$. Combined with the tree-level contribution, these terms result in only two non-zero masses. The next-to-next-to-leading order effects emerge from the fourth and fifth terms in eq. (\ref{M2_eff_fnl}), which generates the mass for the first generation.

The singular part of $M^{(2)}$, as can be read from eq. (\ref{M2_eff_fnl}), can be expressed as: 
\beqa \label{div_M2}
{\rm Div.}\left(M^{(2)}_{ij}\right) & \propto & q_{Li}\, q_{Rj}\, \delta M^{(0)}_{ij} \propto q^2_{Li}\, q^2_{Rj}\, \left(U^{(0)}_{L}\right)_{i3}\,\left(U^{(0)}_{R}\right)^*_{j3}\,m_3^{(0)}\,. \eeqa
This term is a rank-1 contribution and is proportional to $m_3^{(0)}$. The divergence can be absorbed by renormalising $m_3^{(0)}$, as this parameter is inherently part of the theory at the tree level. Consequently, the masses of the first and second generations, which result from eq. (\ref{M2_eff_fnl}), are finite and can be explicitly calculated.


\begin{figure}[t]
\centering
\subfigure{\includegraphics[width=0.48\textwidth]{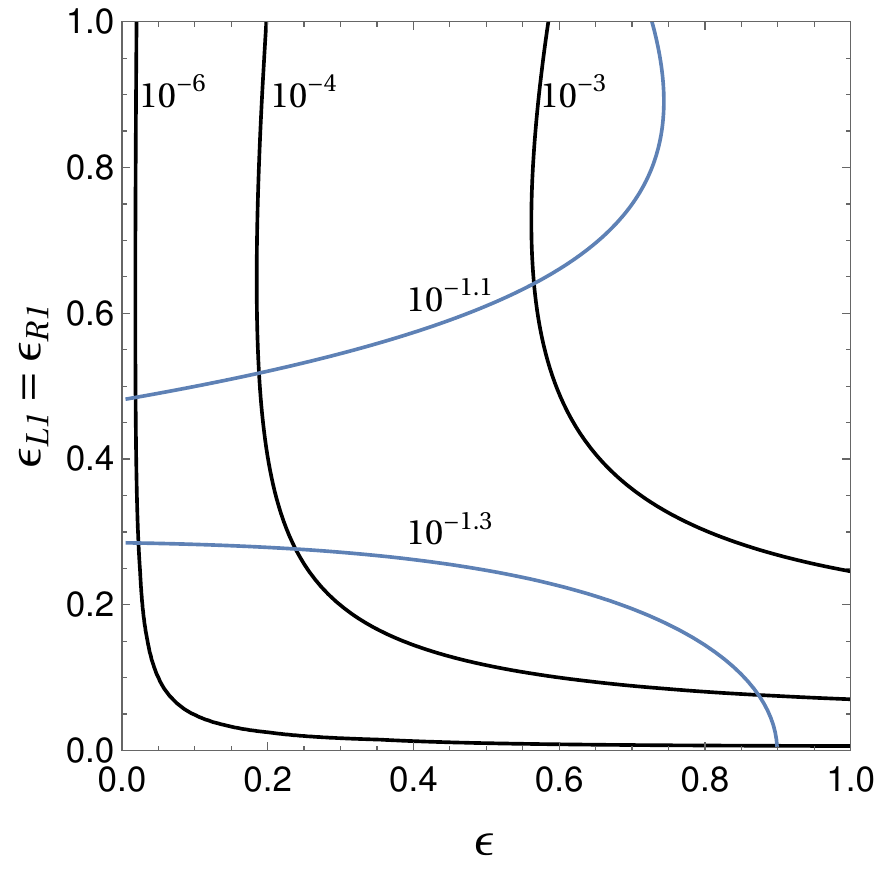}}\hspace*{0.5cm}
\subfigure{\includegraphics[width=0.48\textwidth]{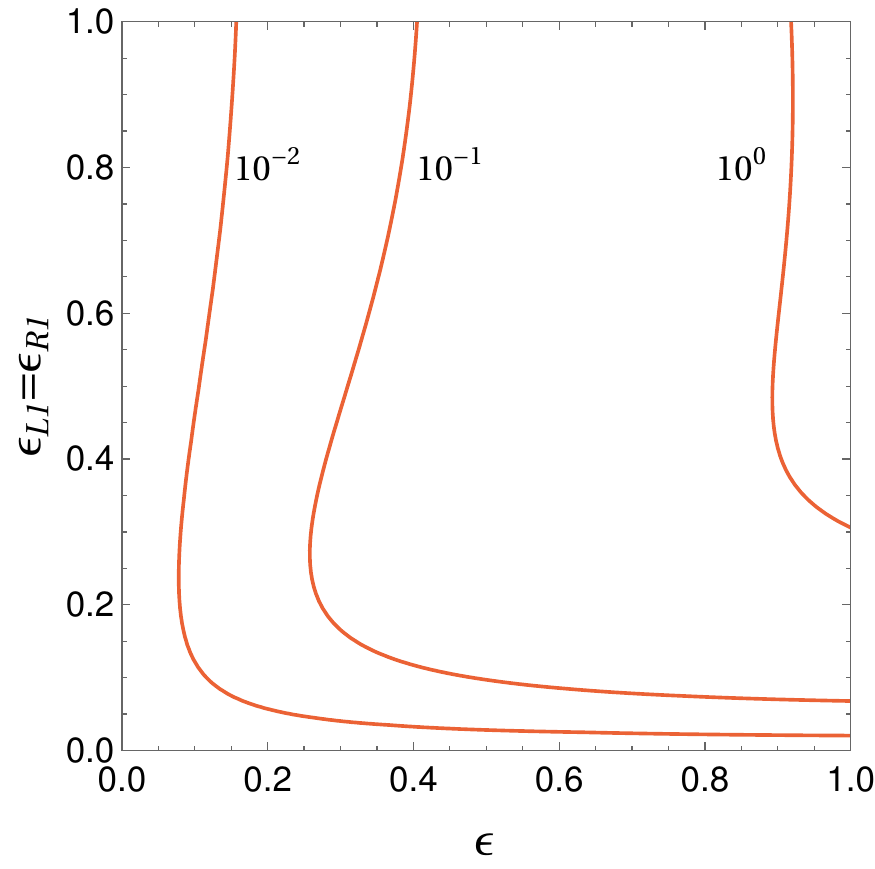}}
\caption{Left sub-figure: contours of $m_1/m_3 = 10^{-6}$, $10^{-4}$, $10^{-3}$ (black) and $m_2/m_3 = 10^{-1.1}$, $10^{-1.3}$ (blue). Right sub-figure: contours of $\left|Q^{(2)}_{12}\right|^2/\left|Q^{(2)}_{23}\right|^2 = 10^{-2}$, $0.1$ and $1$.}
\label{fig1}
\end{figure}
In order to clearly illustrate how the choice of gauge charges in eq. (\ref{gauge_charges}) results in the desired fermion mass hierarchy and flavour-violating interactions, we perform a numerical evaluation of the two-loop corrected mass matrix $M^{(2)}$ as derived from eq. (\ref{M2_eff_fnl}) using a representative set of input parameters. The parameters chosen include $g_X=0.5$, $M_X=10$ TeV, $m_F=10\, M_X$, $\mu_{L}= (\epsilon_{L1}, 0.3, 1)\, \frac{v}{\sqrt{2}}$, $\mu_{R}= (\epsilon_{R1}, 0.3, 1)$ TeV, where $v=246$ GeV and $Q=M_Z$. The relationships between the masses $m_1/m_3$ and $m_2/m_3$ are analyzed for various $\epsilon$ values with $\epsilon_{L 1} = \epsilon_{R 1} \equiv \epsilon_1$. These relationships are illustrated in the left panel of Fig. \ref{fig1}. As expected, the ratio $m_1/m_3$ diminishes with decreasing $\epsilon$ or $\epsilon_1$. Conversely, $m_2/m_3$ is relatively insensitive to $\epsilon$, as its hierarchy with respect to $m_3$ is predominantly influenced by the loop factor. Similarly, contour plots for the ratio $|Q^{(2)}_{12}|^2/|Q^{(2)}_{23}|^2$ are presented in the right panel of Fig. \ref{fig1}.

\begin{figure}[t]
\centering
\subfigure{\includegraphics[width=0.46\textwidth]{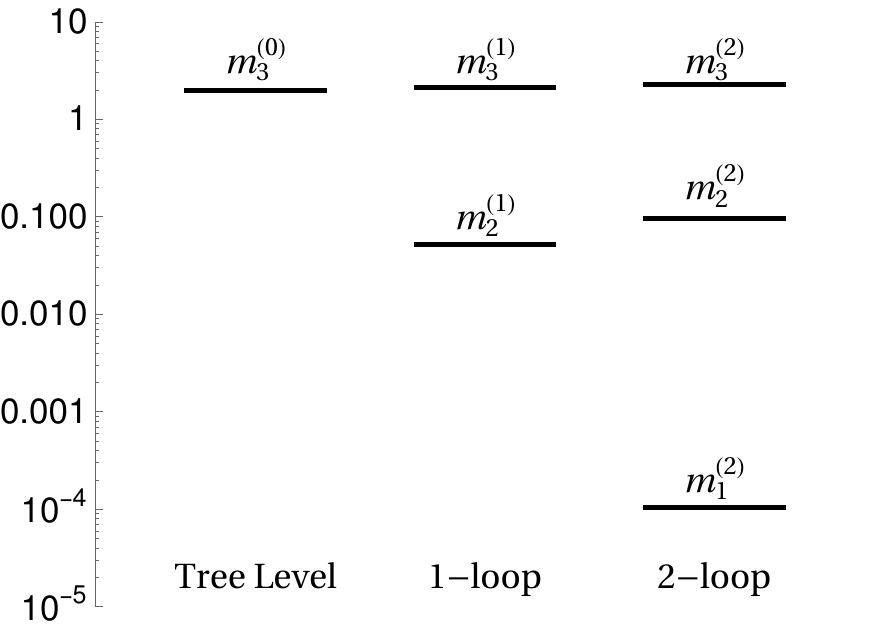}}\hspace*{0.5cm}
\subfigure{\includegraphics[width=0.46\textwidth]{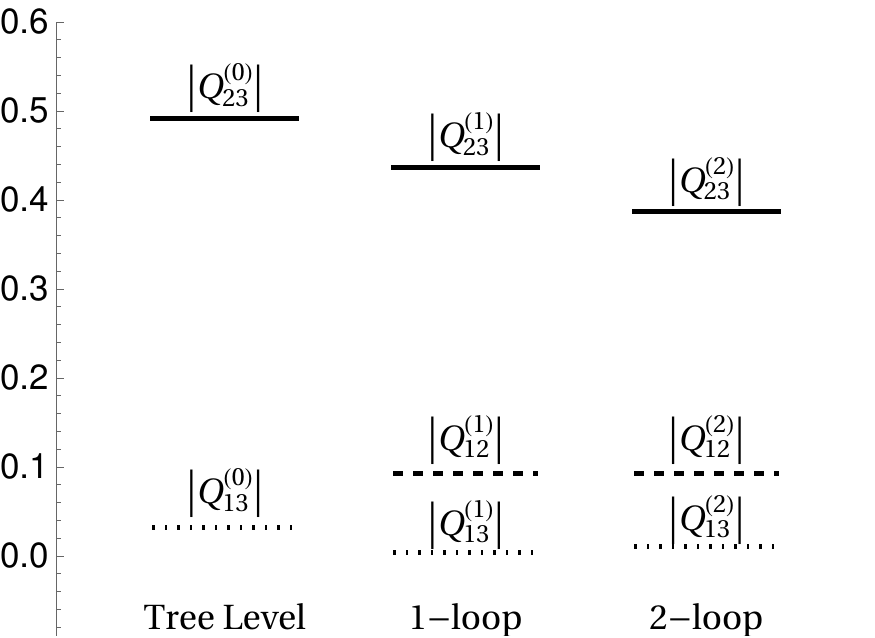}}
\caption{Left sub-figure: Charged fermion masses at tree level, 1-loop, and 2-loop for \(\epsilon = \epsilon_{L1} = \epsilon_{R1} = 0.2\), with the remaining parameters as defined in the text.  
Right sub-figure: The magnitudes of \(Q^{(n)}_{ij}\) for the same parameter values.}
\label{fig2}
\end{figure}
As demonstrated, the typical range $10^{-6} \gtrsim m_1/m_3 \gtrsim 10^{-4}$ usually favors $\epsilon$ values between $0.05$ and $0.5$ for arbitrary $\epsilon_1$. Consequently, this implies $|Q^{(2)}_{12}|^2 \lesssim 0.1\,|Q^{(2)}_{23}|^2$, which is anticipated. However, $|Q^{(2)}_{12}|$ cannot be reduced excessively, as it necessitates extremely small values for both $\epsilon$ and $\epsilon_1$. As detailed in the preceding section and depicted in Fig. \ref{fig1}, this situation pertains to a nearly massless first generation. By setting $\epsilon = \epsilon_1 = 0.2$, we also present the absolute values of the three masses and the flavour-violating couplings $|Q^{(n)}_{ij}|$, $n=1,2$ in Fig. \ref{fig2} at the $n^{\rm}$ loop level. At the leading order, due to the degeneracy between $m_1$ and $m_2$, $|Q^{(0)}_{12}|$ remains undefined, but it becomes physical at the 1-loop level. Fig. \ref{fig2} illustrates that the masses and $|Q^{(n)}_{ij}|$ only experience minor corrections at the subsequent order, and their relative hierarchy is preserved.

\section{Integration with the standard model}
\label{sec:SM}
This section discusses the integration of the minimal flavour violating radiative mechanism framework with the SM and its viability in reproducing the observed charged fermion mass spectrum. We proceed by utilising the gauge charges presented in eq. (\ref{gauge_charges}) and consistently apply them to all the SM fermions.

\subsection{Implementation}
\label{subsec:implementation}
The underlying mechanism for SM fermions is readily implementable, facilitating loop-induced mass hierarchies for the charged fermions. The left-handed chiral fields are extended to include the electroweak quark and lepton doublets, denoted as $Q_{L i}$ and $\Psi_{L i}$. Correspondingly, the right-handed chiral SM fields comprise $u_{R i}$, $d_{R i}$, and $e_{R i}$. As done in the previous chapter, in each sector, the VL fermions are represented by $T_{L,R}$, $B_{L,R}$, and $E_{L,R}$, which transform under the SM gauge symmetry similarly to $u_{R i}$, $d_{R i}$, and $e_{R i}$, respectively. The entities $\mu_{L i}$ of the toy model are replaced by three pairs of electroweak doublet scalars $H_{u i}$ and $H_{d i}$, and terms proportional to $\mu_{R i}$ are derived from three SM singlet scalars $\eta_{i}$. The three generations of chiral fermions and scalars possess flavour non-universal charges as specified in eq. (\ref{gauge_charges}). The field contents of the model and their charges under the SM and $U(1)_F$ gauge symmetries are detailed in Table \ref{tab:fields:u1}.
\begin{table}[!t]
\begin{center}
\begin{tabular}{ccc} 
\hline
\hline
~~Fields~~&~~$SU(3)_C \times SU(2)_L \times U(1)_Y$~~&~~$U(1)_F$~~~~\\
\hline
$Q_{L i} $ & $(3,2,\frac{1}{3}) $ & $(1-\epsilon, 1+\epsilon, -2)$ \\
$u_{R i} $ & $(3,1,\frac{4}{3}) $ & $(1-\epsilon, 1+\epsilon, -2)$ \\
$d_{R i} $ & $(3,1,-\frac{2}{3}) $ & $(1-\epsilon, 1+\epsilon, -2)$ \\
$\Psi_{L i} $ & $(1,2,-{1}{}) $ & $(1-\epsilon, 1+\epsilon, -2)$ \\
$e_{R i} $ & $(1,1,-2) $ & $(1-\epsilon, 1+\epsilon, -2)$ \\
$\nu_{R i} $ & $(1,1,0) $ & $(1-\epsilon, 1+\epsilon, -2)$ \\
\hline
$T_{L,R}$ & $(3,1,\frac{4}{3}) $ & 0\\
$B_{L,R}$ & $(3,1,-\frac{2}{3}) $ & 0\\
$E_{L,R}$ & $(1,1,-2) $ & 0\\
\hline
$H_{u i}$ & $(1,2,-{1}{}) $ & $(1-\epsilon, 1+\epsilon, -2)$ \\
$H_{d i}$ & $(1,2,{1}{}) $ & $(1-\epsilon, 1+\epsilon, -2)$ \\
$\eta_i$  & $(1,1,0) $ & $(1-\epsilon, 1+\epsilon, -2)$ \\
\hline
\hline
\end{tabular}
\end{center}
\caption{Field contents alongside their flavour-independent SM gauge charges and flavour-specific $U(1)_F$ quantum numbers.}
\label{tab:fields:u1}
\end{table}

The inclusion of three generations of $\nu_R$, the right-handed neutrino partners, with $U(1)_F$ charges specified in Table \ref{tab:fields:u1}, is crucial to maintaining the vector structure of the $U(1)_F$ symmetry. As discussed in the previous section, this structure ensures the cancellation of both the cubic $U(1)_F$ and mixed gauge-gravity anomalies. Additionally, it guarantees a vanishing $U(1)_Y \times U(1)_F^2$ anomaly. The condition ${\rm Tr}(q_{L,R}) = 0$ further ensures the cancellation of anomalies associated with $SU(2)_L^2 \times U(1)_F$ and $U(1)_Y^2 \times U(1)_F$. Consequently, the field content and gauge charges outlined in Table \ref{tab:fields:u1} establish a theoretically consistent and anomaly-free framework.

The Yukawa and mass Lagrangian, which is both renormalisable and gauge invariant, can be formulated for the fermions as
\beqa \label{LY_SM}
-{\cal L}_Y &=& {y_u}_i\,\overline{Q}_{Li}\, {H_u}_i\, T_R\, +\,{y_d}_i\,\overline{Q}_{Li}\, {H_d}_i\, B_R\,+\,  {y_e}_i\,\overline{\Psi}_{Li}\, {H_d}_i\, E_R\,  \nonumber \\
& + & {y_u^{\prime}}_i\,\overline{T}_L\, \eta^*_i\, u_{R i}\, +\, {y_d^{\prime}}_i\,\overline{B}_L\, \eta^*_i\, d_{R i}\, +\, {y_e^{\prime}}_i\,\overline{E}_L\, \eta^*_i\, e_{R i}\, 
\nonumber \\
& + & m_T\, \overline{T}_L\,T_R\, + \,m_B\, \overline{B}_L\,B_R\, + \,m_E\, \overline{E}_L\,E_R\,+ \, {\rm h.c.}\,.\eeqa
As in standard scenarios, direct mass terms for the three generations of fermions are prohibited by the chiral structure of the SM gauge symmetry, even if such terms are permitted by the $U(1)_F$ symmetry. Additionally, the non-trivial transformation properties assigned to the Higgs fields under the new gauge symmetry prevent direct Yukawa couplings between the left- and right-chiral fields of the SM. 

The fermionic current associated with the flavour non-universal gauge interactions is expressed as \({\cal L}_X = j_X^\mu X_\mu\), where \(j_X^\mu\) represents the corresponding current.
\be \label{JX_model}
j^\mu_X = g_X\, \left( \sum_{f = Q,L} q_{L i}\, \overline{f}_{Li}\,\gamma^\mu \, f_{L i}\, +\,  \sum_{f = u,d,e} q_{R i}\, \overline{f}_{R i}\,\gamma^\mu \, f_{R i} \right)\,,\ee
with the choice of $q_{L i}$ and $q_{R i}$ as listed in Table \ref{tab:fields:u1}.

The scalar sector of this model is identical to the one described in the previous chapter. The renormalisable scalar potential takes the same form as in eq. (\ref{chap3:potential}). Accordingly, we assume that the potential yields a suitable vacuum configuration that breaks both the flavour symmetry \( G_F \) and the electroweak symmetry. The most general non-vanishing vacuum configuration of \(\eta_i\) leads to the complete breaking of \( U(1)_F \). As is customary, the electroweak symmetry is broken by VEVs of \( H_{ui} \) and \( H_{di} \), which preserve \( U(1)_{\rm em} \) while also contributing to the breaking of \( U(1)_F \). These VEVs are parametrized as follows:
\be \label{vev_def}
y_{fi}\,\langle H_{fi} \rangle \equiv \mu_{f i}\,,~~ y^\prime_{fi}\,\langle \eta^*_{i} \rangle \equiv \mu^\prime_{f i}\,,\ee
with $f = u,d,e$ and $\langle H_{ei}\rangle = \langle H_{di}\rangle$. The criterion $\sum_i \left(|\langle H_{ui}\rangle|^2 + |\langle H_{di}\rangle|^2\right) = (246\,)^2$ ${\rm GeV}^2$ generally suggests that the $\mu_{fi}$ values do not exceed the electroweak scale, whereas no similar restriction is imposed on the magnitudes of $\mu_{fi}^\prime$.

Substituting eq. (\ref{vev_def}) into eq. (\ref{LY_SM}) transforms the latter into the exact form of the interactions shown in eq. (\ref{L_mass}). The matrix \({\cal M}^{(0)}\) defined in eq. (\ref{M0}) is reproduced for \(f = u, d, e\) with the replacements \(\mu_{Li} \to \mu_{fi}\) and \(\mu_{Ri} \to \mu^\prime_{fi}\). utilising the results from eq. (\ref{M2_eff_fnl}), the two-loop corrected effective \(3 \times 3\) mass matrix for the charged fermions is found to be:
\beqa \label{M2_eff_fnl_f}
\left(M^{(2)}_f\right)_{ij} &=& \left(M^{(0)}_f\right)_{ij} \left(1+\frac{g_X^2}{4 \pi^2}\, q_{Li}\, q_{Rj}\, ({b}_0[M_X,m_{f3}^{(1)}]- {b}_0[M_X,m_F]) \right) \nonumber \\
&+& \left(\delta M_f^{(0)}\right)_{ij} \left(1+\frac{g_X^2}{4 \pi^2}\, q_{Li}\, q_{Rj}\, {{b}_0}[M_X,m_{f3}^{(1)}] \right) \\
&+& \frac{g_X^2}{4 \pi^2}\, q_{Li}\, q_{Rj} \,(U_{fL}^{(1)})_{i2} (U_{fR}^{(1)})^*_{j2}\, m_{f2}^{(1)}\, ({b}_0[M_X,m_{f2}^{(1)}] - {b}_0[M_X,m_{f3}^{(1)}])\,,\nonumber\eeqa
for $f=u,d,e$. The corresponding mass matrices at the tree-level and one-loop are expressed as 
\be \label{M0_eff_f}
\left(M^{(0)}_f\right)_{ij} = -\frac{1}{m_F}\, \mu_{f i}\, \mu_{f j}^\prime\,,\ee
and 
\be \label{M1_eff_f}
\left(M^{(1)}_f\right)_{ij} = \left(M^{(0)}_f\right)_{ij}\,\left(1+ C_f\, q_{Li}\, q_{Rj} \right)\,, \ee
respectively, with 
$C_f= \frac{g_X^2}{4 \pi^2} ({b}_0[M_X,m_{f3}^{(0)}] - {b}_0[M_X,m_F])$.

Eqs. (\ref{M2_eff_fnl_f}, \ref{M0_eff_f}, \ref{M1_eff_f}) are straightforward generalizations of eqs. (\ref{M2_eff_fnl}, \ref{M0_eff}, \ref{M1_eff}). The term \(m_{fi}^{(n)}\) in these expressions represents the \(i^{\text{th}}\) eigenvalue of \(M_f^{(n)}\), which can be computed using the provided expressions. Similarly, \(U^{(1)}_{fL}\) and \(U^{(1)}_{fR}\) are obtained by diagonalising \(M_f^{(1)}\), following the standard definition in eq. (\ref{M1_eff_diag}). The parameter \(\tilde{b}_0\), given in eq. (\ref{b0}), denotes the finite part of the function \(B_0\) as defined in the \(\overline{\rm MS}\) scheme.

\subsection{Test of viability}
\label{subsec:numerical}
To show that \( M^{(2)}_f \), as derived in eq. (\ref{M2_eff_fnl_f}), can yield realistic charged fermion masses for \( f = u, d, e \) and quark mixing parameters, we perform a numerical analysis using the method developed and detailed in the previous chapter.

In our analysis, we fix \( g_X = 0.5 \) and optimise the \(\chi^2\) function for several values of \(\epsilon\) and select example values of \( M_X \). The remaining 23 parameters are optimised while adhering to standard constraints, as discussed in Section \ref{chap3:solutions}. This process is conducted for the following two scenarios:
\begin{itemize}
\item Case A: Ordered $\mu_{fi}$ and $\mu_{fi}^\prime$, i.e. for $f=u,d,e$, we impose
\be \label{caseA}
|\mu_{f1}| < |\mu_{f2}| < |\mu_{f3}|\,,~~{\rm and}~~~|\mu_{f1}^\prime| < |\mu_{f2}^\prime| < |\mu_{f3}^\prime|\,.\ee
\item Case B: Strongly ordered $\mu_{di}$ and $\mu_{di}^\prime$. In this scenario, besides the constraints specified in eq. (\ref{caseA}), we also impose
\be \label{caseB}
\frac{|\mu_{d1}|}{ |\mu_{d2}|} < 0.1\,,~~{\rm and}~~~\frac{|\mu_{d1}^\prime|}{ |\mu_{d2}^\prime|}<0.1\,.\ee
\end{itemize}
Both conditions contribute to achieving a phenomenologically favorable hierarchy in flavour violations, as discussed in Section \ref{chap4:optimal}, and establish a clear link between flavour violation in the 1-2 sector and the parameter \(\epsilon\). Notably, Case B highlights a region of parameter space where the stringent constraints from \(K^0\)-\(\overline{K}^0\) oscillations can be more effectively avoided, as demonstrated in the next section.

\begin{figure}[t] 
\centering
\subfigure{\includegraphics[width=0.48\textwidth]{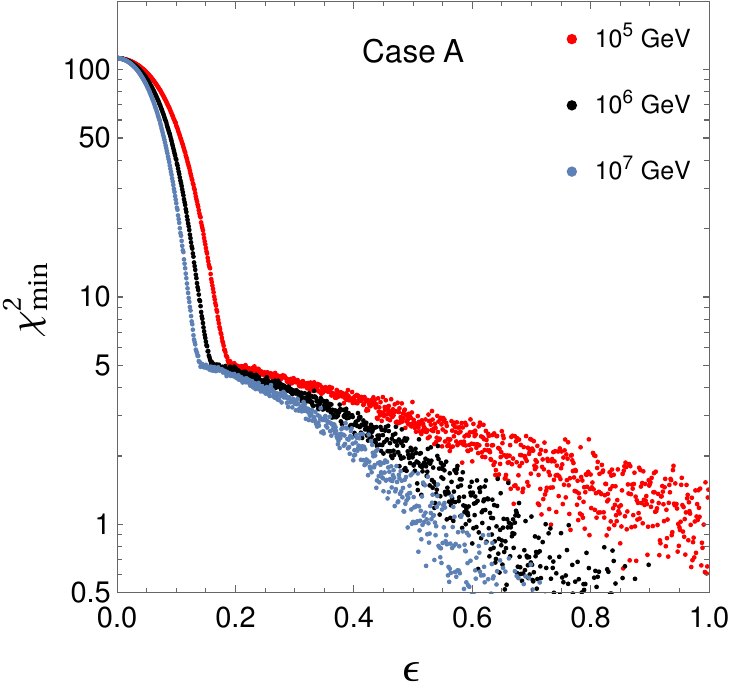}}\hspace*{0.5cm}
\subfigure{\includegraphics[width=0.48\textwidth]{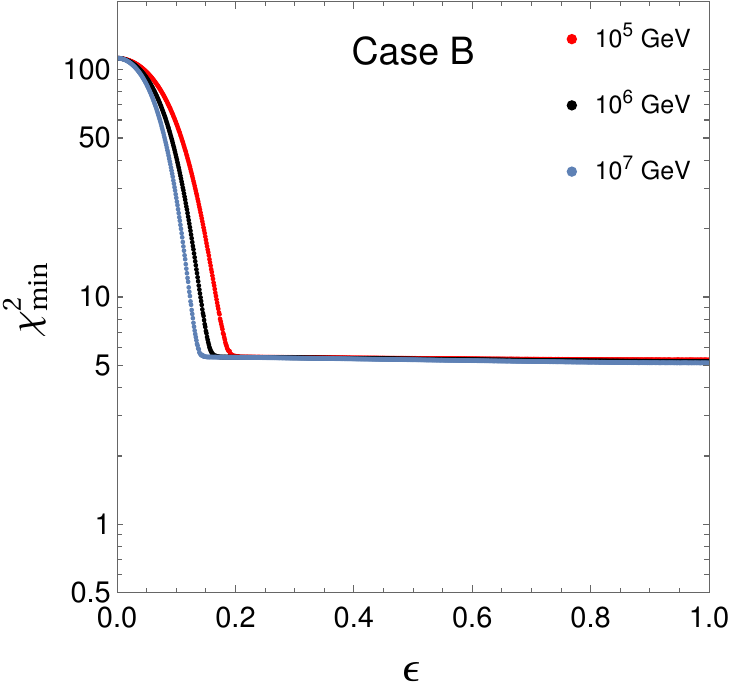}}
\caption{The calculated $\chi^2_{\rm min}$ as a function of $\epsilon$ for three exemplary values of $M_X$, with scenarios classified as case A (left panel) and case B (right panel). }
\label{fig:chi2}
\end{figure}

The minimised \(\chi^2\) results for different values of \(\epsilon\) and \(M_X\) in both scenarios are presented in Fig. \ref{fig:chi2}.
Fits corresponding to \(\chi^2_{\rm min} \leq 9\) are considered acceptable, as no observable deviates by more than \(3\sigma\) from its mean value. With this criterion in mind, the key observations from Fig. \ref{fig:chi2} are as follows: 
\begin{itemize}
    \item \textbf{Fits disfavor \(\epsilon < 0.15\):} For smaller values of \(\epsilon\), the first-generation masses fail to fit their experimentally determined values within their \(3\sigma\) ranges. This aligns with expectations discussed in Section \ref{chap4:optimal}.
    \item \textbf{Case A:} The fit improves with larger values of \(\epsilon\). Specifically, for \(M_X \geq 10^6\) GeV, excellent fits are achieved with \(\chi^2_{\rm min} \lesssim 1\) when \(\epsilon > 0.5\), indicating that all observables are within their \(1\sigma\) ranges. Additionally, for a given \(\epsilon\), larger values of \(M_X\) lead to further improvement in the fits.
    \item \textbf{Case B:} In contrast, \(\chi^2_{\rm min} < 5\) is not attainable for any combination of \(\epsilon\) and \(M_X\). Under the conditions specified in eq. (\ref{caseB}), the down quark mass remains \(2.3\sigma\) away from its central value, regardless of the other parameters.
\end{itemize}

These observations indicate that the mechanism can only be implemented in the Standard Model in a phenomenologically consistent manner if \(\epsilon > 0.15\). This lower bound implies a finite level of flavour violation in the 1-2 sector, which is analyzed in detail in the next section.

In Table \ref{tab:fit}, we present two specific benchmark examples to illustrate the effectiveness of the fits and the behavior of the fitted parameters.
\begin{table}[t]
\begin{center}
\begin{math}
\begin{tabular}{cccccccc}
\hline
\hline
 &   &
      \multicolumn{2}{c}{\bf Solution 1} &
      \multicolumn{2}{c}{\bf Solution 2} \\
~~~Observable~~~ & $O_{\rm exp}$  & $O_{\rm th}$ & Pull & $O_{\rm th}$ & Pull  \\
\hline
$m_u$\, [MeV]  &  $1.27 \pm 0.5$   &$ 1.26$    & $ -0.02$ & $ 1.27$     &  $ 0$ \\
$m_c$\, [GeV]   & $0.619 \pm 0.084$    &$ 0.614$    & $ -0.06$ & $ 0.617$     &  $-0.02$ \\
$m_t$\, [GeV]   & $171.7 \pm 3.0$     & $ 171.8$  & $ 0.03$ & $ 171.7 $   & $ 0$  \\
$m_d$\, [MeV]  & $2.90 \pm 1.24$& $ 0.16$  & $ -2.21$  & $0.02$ &$ -2.32$\\
$m_s$\, [GeV]  & $0.055 \pm 0.016$      & $ 0.056$  & $0.06 $ & $ 0.055$    & $0$ \\
$m_b$\, [GeV]  & $2.89 \pm 0.09$         & $ 2.89$     & $0$ & $ 2.89$       & $0 $ \\
$m_e$\, [MeV]  & $0.487 \pm 0.049$        & $ 0.489$   & $0.04$ & $ 0.487$     & $0$  \\
$m_{\mu}$\,[GeV] &$0.1027 \pm 0.0103$      & $ 0.1025$  & $-0.02$  & $ 0.1025$     & $-0.02 $ \\
$m_{\tau}$ \,[GeV] & $1.746 \pm 0.174$       & $ 1.746$   & $0$ & $ 1.742$      & $-0.02$ \\
$|V_{us}|$     & $0.22500 \pm 0.00067$      & $0.22499$    & $-0.01 $ & $0.22500$     & $0$  \\
$|V_{cb}|$     & $0.04182 \pm 0.00085$          & $0.04182$   & $0$& $0.04182$    & $0$  \\
$|V_{ub}|$     & $0.00369 \pm 0.00011$          & $0.00369 $   & $0 $ & $0.00369$    & $0$  \\
$J_{\rm CP}$   & $3.08\times10^{-5}$   & $3.08\times10^{-5}$  & $0 $ & $3.08\times10^{-5}$ & $0$ \\
\hline
    $\chi^2_{\rm min}$    & & & $4.9$ & & $5.4$ \\
\hline
\hline
\end{tabular}
\end{math}
\end{center}
\caption{Two optimal solutions from benchmark cases A and B are presented. \(O_{\rm exp}\) denotes the extrapolated values of the observables at the renormalization scale \(Q = M_Z\). The computed values of the observables at \(\chi^2_{\rm min}\) are listed under \(O_{\rm th}\), with corresponding pulls indicating the deviation from the mean value \(O_{\rm exp}\).}
\label{tab:fit}
\end{table}
Solution 1 (\( S1 \)) is classified under case A, while Solution 2 (\( S2 \)) pertains to a scenario with strong ordering of \(\mu_{di}\) and \(\mu_{di}^\prime\) within case B. These solutions are selected with values of \(\epsilon\) and \(M_X\) that have been optimised considering flavour violation, as discussed in the forthcoming section. It is evident that all observables, with the exception of \(m_d\), show excellent agreement. However, \(m_d\) exceeds a deviation of 2\(\sigma\) in both scenarios. Hence, reducing the uncertainty in \(m_d\) in the future could substantially impact the feasibility of these solutions.

The values of \(\epsilon\) and \(M_X\) for the selected benchmark solutions, along with the numerical values of the remaining 23 real parameters at the \(\chi^2\) minimum, are listed in Table \ref{tab:sol}. A significant observation in both cases is that all \(\mu_{fi}\) and \(\mu_{fi}^\prime\) are confined to a range spanning at most two orders of magnitude. Interpreted in the context of eq. (\ref{vev_def}), this suggests that all fundamental Yukawa couplings can naturally be of \({\cal O}(1)\) in this model. The observed hierarchies in masses are thus attributed to the intricate and carefully arranged structure of the theory. Furthermore, it is evident that the VL fermions must remain close to the \(U(1)_F\) breaking scale to prevent excessive seesaw suppression of the third-generation fermion masses. This is particularly apparent for the top quark mass, which necessitates \(m_U \simeq \mu_{u3}^\prime\). In contrast, the relatively lighter \(m_b\) and \(m_\tau\) masses are achieved through \(m_D > \mu_{d3}^\prime\) and \(m_E > \mu_{e3}^\prime\), respectively.
\begin{table}[t]
\begin{center}
\begin{tabular}{ccc} 
\hline
\hline
~~Parameters~~&~~\textbf{Solution 1} ~~&~~\textbf{Solution 2}~~\\
\hline
$M_X$           & $3\times 10^6$   & $10^6$             \\
$\epsilon$      & $ 0.178$         & $ 0.285 $        \\
\hline
$m_U$   & $5.6653\times 10^{6 }$  & $ 6.2838\times 10^{6 } $   \\
$m_D$   & $4.9336 \times 10^{8}$  & $ 1.5524\times 10^{ 8} $   \\
$m_E$   & $ 8.2489\times 10^{6}$  & $ 3.9884\times 10^{ 6} $   \\
\hline
$\mu_{u1}$ & $-9.6518$            & $ 1.2821 $          \\
$\mu_{u2}$ & $ 9.9336 $           & $ -7.2968$          \\
$\mu_{u3}$ & $ -3.7757\times10^2$ & $ 4.6534\times10^2$ \\
\hline
$\mu^\prime_{u1}$ & $6.9209\times 10^5 $    & $1.1248\times 10^{6}$   \\
$\mu^\prime_{u2}$ & $-1.6434 \times 10^6 $  & $-1.4161 \times 10^{6}$ \\
$\mu^\prime_{u3}$ & $-1.8288\times 10^6 $   & $1.4585\times 10^{6}$   \\
\hline
$\mu_{d1}$ & $-1.8003 \times 10^{1} + i\ 2.4815  $  & $1.9233 + i\ 1.9146 $           \\
$\mu_{d2}$ & $2.9553  \times 10^{1} + i\ 1.6954$    & $-3.2718+ i\ 1.9150\times 10^1$ \\
$\mu_{d3}$ & $-2.2993 \times 10^2$                  & $2.2572\times 10^2$             \\
\hline
$\mu^\prime_{d1}$ & $-7.1617\times 10^5 $ & $-4.8001\times 10^4 $ \\
$\mu^\prime_{d2}$ & $7.2368\times 10^5$   & $ 4.8498\times 10^5 $ \\
$\mu^\prime_{d3}$ & $4.4255 \times 10^6$  & $1.4044 \times 10^6 $ \\
\hline
$\mu_{e1}$     & $2.5748\times 10^1 $     & $ -1.3349 \times 10^1$   \\
$\mu_{e2}$     & $-2.6008\times 10^{1} $  & $-1.5363\times 10^1 $    \\
$\mu_{e3}$     & $-4.7027 \times10^1$     & $2.1028\times 10^1 $     \\
\hline
$\mu^\prime_{e1}$ & $-1.3925\times 10^5$    & $3.6639 \times 10^4 $   \\
$\mu^\prime_{e2}$ & $-1.4093 \times 10^5 $  & $-7.9852 \times 10^4 $  \\
$\mu^\prime_{e3}$ & $-1.4252 \times 10^5 $  & $-2.1704 \times 10^5 $  \\
\hline
\hline
\end{tabular}
\end{center}
\caption{These are the optimised values of different input parameters for two exemplary solutions of Table \ref{tab:fit}. All dimension-full parameters are expressed in GeV.}
\label{chap4:sol}
\end{table}

\section{Phenomenological constraints}
\label{sec:constraints}
In this section we discuss the constraints arising from the flavour violations mainly induced by the $X$ boson. The other sources of flavour violations, and their suppression relative to the $X$ boson induced, are discussed in detail in section \ref{chap3:pheno}. Assuming FCNCs are dominantly controlled by the $X$ boson, which is parametrized by eq. (\ref{JX_model}), its flavour-violating couplings in the physical basis of the quarks and charged leptons are given by:
\be \label{JX_model_phys}
j^\mu_X = g_X\, \sum_{f = u,d,e} \left(\left(X_{fL}\right)_{ij}\, \overline{f}_{Li}\,\gamma^\mu \, f_{L j}\, +\,  \left(X_{fR}\right)_{ij}\, \overline{f}_{R i}\,\gamma^\mu \, f_{R j} \right)\,,\ee
with 
\be \label{Xf}
X_{fL,R} = U^{f \dagger}_{L,R}\, q_{L,R}\, U^f_{L,R}\,,\ee
and $U^{f}_{L,R}$ being the $3 \times 3$ unitary matrices that diagonalise the corresponding 2-loop corrected $M^{(2)}_f$ given in eq. (\ref{M2_eff_fnl_f}) satisfying the relation $U^{f \dagger}_{L}\, M^{(2)}_f\, U^f_{R} = {\rm Diag.}(m^{(2)}_{f1},m^{(2)}_{f2},m^{(2)}_{f3})$. In the previous chapter, section \ref{chap3:pheno}, we computed various flavour violating observables by considering flavour-violating charges as $X^{(1)}_{fL,R}$, given in eq. (\ref{X}). The same expressions for those observables will be applicable for the present model phenomenological analysis with $X^{(1)}_{fL,R}$ replaced by $X_{fL,R}$ given in eq. (\ref{Xf}).

\subsection{Meson-antimeson oscillation}
The most stringent constraints on quark sector FCNC couplings arise from the neutral meson-antimeson oscillations, namely $M^0$-$\overline{M}^0$ transitions, where $M=K,\, B_d,\, B_s,\, D$. At $Q=M_X$ scale, the nonvanishing WCs for $K^0$-$\overline{K}^0$, $B_d^0$-$\overline{B_d}^0$,$B_s^0$-$\overline{B_s}^0$ and $D^0$-$\overline{D}^0$ transition are given in eq. (\ref{C_K}, \ref{C_Bd}, \ref{C_Bs}) and (\ref{C_D}) respectively, with $X^{(1)}_{fL,R}$ replaced by $X_{fL,R}$. Subsequently, all coefficients are run down by appropriate renormalisation group equations (RGEs) from $Q=M_X$ to $Q = 2$ GeV for the $K^0$-$\overline{K}^0$ system \cite{Ciuchini:1998ix}, to $Q = 4.6$ GeV for the $B_{d,s}^0$-$\overline{B}_{d,s}^0$ system \cite{Becirevic:2001jj}, and to $Q = 2.8$ GeV for the $D^0$-$\overline{D}^0$ system \cite{UTfit:2007eik}. It is observed that this running results in non-zero values for $C^4_M$, while $\tilde{C}^{2,3}_M$ and $C^{2,3}_M$ persist at zero. The non-vanishing Wilson coefficients at their relevant low-energy scales for each point are displayed in Fig. \ref{fig:chi2}. Subsequently, these values are compared with the current experimental boundaries as determined by the UTFit collaboration\cite{UTfit:2007eik}.

\begin{figure}[t!]
\centering
\subfigure{\includegraphics[width=0.48\textwidth]{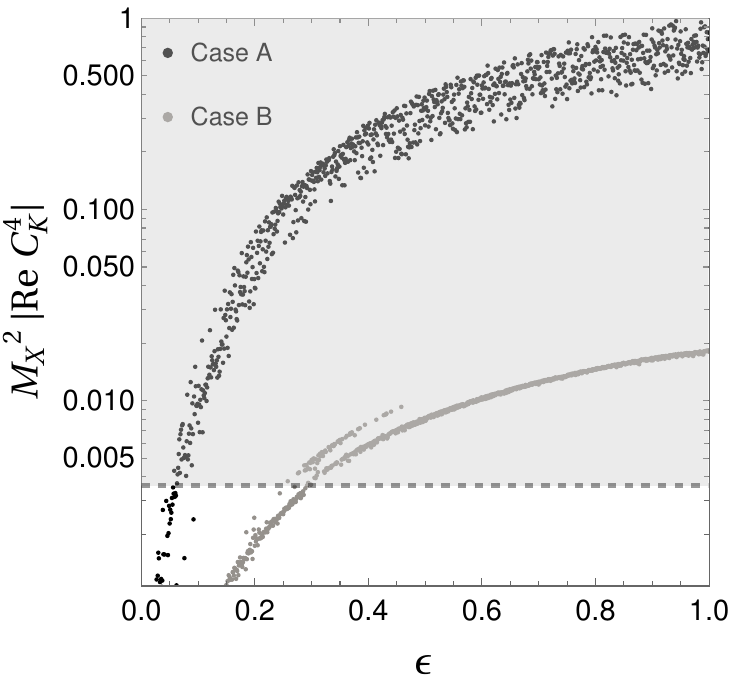}}\hspace*{0.5cm}
\subfigure{\includegraphics[width=0.48\textwidth]{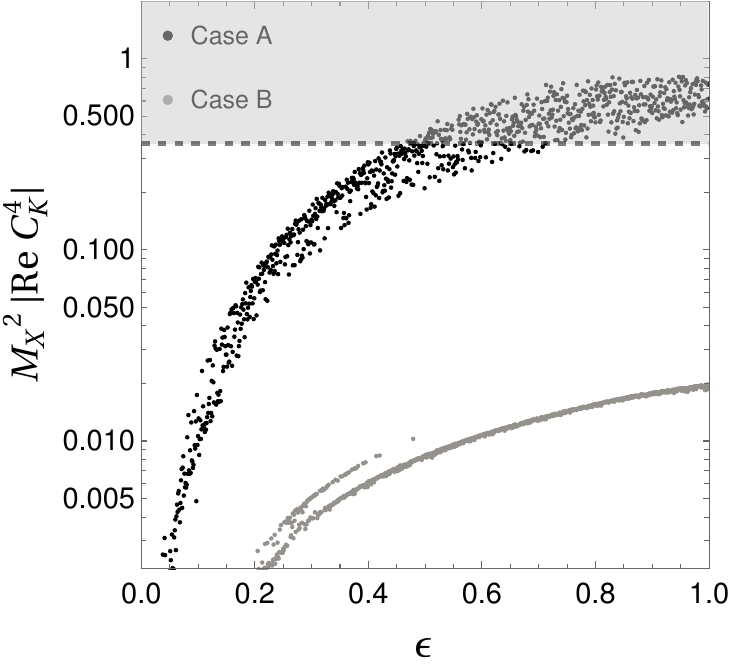}}
\caption{Magnitude of \(\text{Re}C^{4}_K\) calculated for the best-fit points with different values of \(\epsilon\), for \(M_X = 10^6\) GeV (left sub-figure) and \(M_X = 10^7\) GeV (right sub-figure). The shaded regions indicate exclusions based on current limits at the 95\% confidence level.}
\label{fig:C4}
\end{figure}

\begin{figure}[t]
\centering
\subfigure{\includegraphics[width=0.48\textwidth]{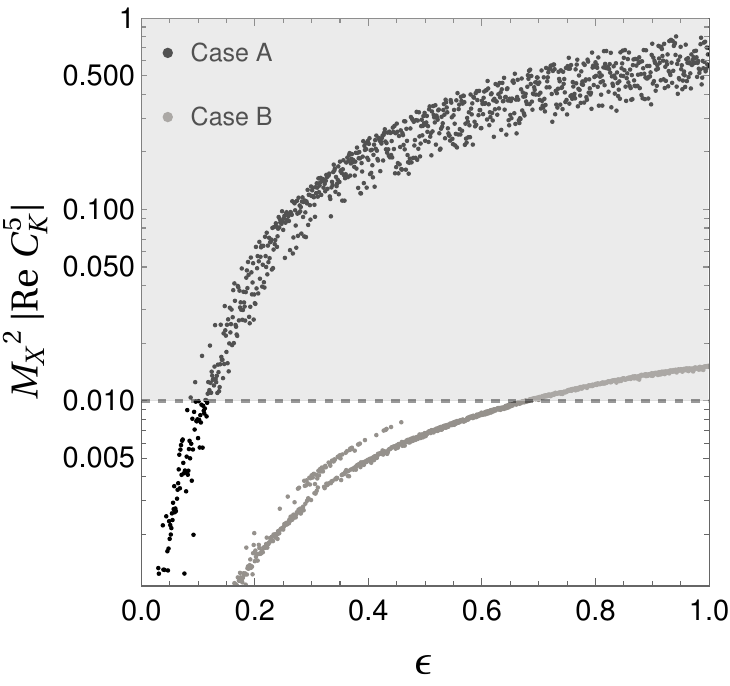}}\hspace*{0.5cm}
\subfigure{\includegraphics[width=0.48\textwidth]{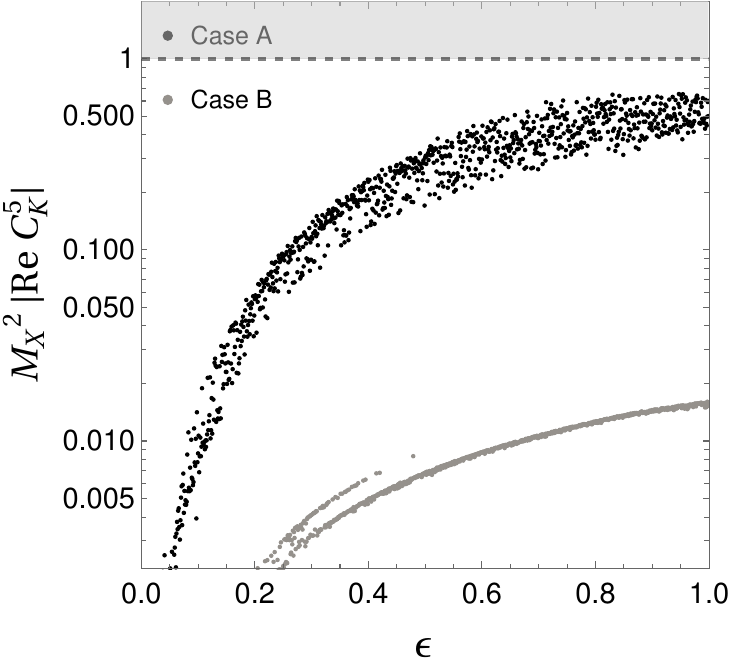}}
\caption{Magnitude of ${\rm Re}C^{5}_K$ computed for the best-fit points for different values of $\epsilon$ and for $M_X=10^6$ GeV (left sub-figure) and $M_X=10^7$ GeV (right sub-figure).  The shaded regions are excluded by the present limits at $95\%$ confidence level.}
\label{fig:C5}
\end{figure}

Among all the $C^i_M$ and $\tilde{C}^i_M$ computed in the present model, we find that the strongest limits on $M_X$ are predominantly influenced by  ${\rm Re} C^{4,5}_K$. This is illustrated by presenting their values as functions of $\epsilon$ for $M_X = 10^6$ and $M_X = 10^7$ GeV, shown in Figs. \ref{fig:C4} and \ref{fig:C5}, respectively. 

We also present these estimations for Cases A and B, as discussed in the previous section. From both figures, it is evident that flavour constraints can be effectively avoided for smaller values of \(\epsilon\) in all cases. Notably, Case B, characterized by strongly ordered \(\mu_{di}\) and \(\mu_{di}^\prime\), significantly reduces the magnitudes of \(C^4_K\) and \(C^5_K\). For Case A with \(M_X = 10^6\) GeV, the current limits on \(C^4_K\) only allow \(\epsilon < 0.15\), which is already disfavored due to the large \(\chi^2_{\rm min}\) value, as shown in Fig. \ref{fig:chi2}. In contrast, Case B with the same \(M_X\) yields solutions with acceptable \(\chi^2_{\rm min}\) and permitted values for \({\rm Re}C^{4,5}_K\). For \(M_X = 10^7\) GeV, both Cases A and B produce viable solutions, as illustrated in Figs. \ref{fig:chi2}, \ref{fig:C4}, and \ref{fig:C5}.

Our analysis shows that the minimum mass required for the \(X\)-boson to evade constraints from meson-antimeson oscillations, without relying on strongly ordered \(\mu_{di}\) and \(\mu_{di}^\prime\), is \(M_X = 3 \times 10^6\) GeV. This ensures a viable spectrum of charged fermion masses and quark mixing parameters. Strong ordering of \(\mu_{di}\) and \(\mu_{di}^\prime\) can further reduce this to \(M_X = 10^6\) GeV. Example solutions for each scenario are provided in Tables \ref{tab:fit} and \ref{chap4:sol}. 

For these solutions, we calculate all non-vanishing \(C^i_M\) and \(\tilde{C}^i_M\), which are listed in Table \ref{tab:FV}. All values comply with current experimental limits. Additionally, it is observed that the Wilson coefficients associated with flavour violation in the 1-2 sector are generally smaller than those in the 2-3 sector, as intended by the small \(\epsilon\) parameter in this framework.
\begin{table}[!ht]
\begin{center}
\begin{tabular}{cccc} 
\hline
\hline
~~Flavour observables~~&~~Experimental limit~~&~~~{\bf Solution 1}~~~&~~~{\bf Solution 2}~~~\\
 \hline
Re$C_K^1$ & $[-9.6,9.6]\times 10^{-13}$     &$-5.0\times 10^{-16}$ & $-8.4\times 10^{-16}$  \\        Im$C_K^1$ & $[-9.6,9.6]\times 10^{-13}$    & $-1.7\times 10^{-30}$& $1.5\times 10^{-29}$  \\
Re$\tilde{C}_K^1$ & $[-9.6,9.6]\times 10^{-13}$ &  $-7.2\times 10^{-16}$ & ${-4.7\times 10^{-16}}$  \\
  Im$\tilde{C}_K^1$ & $[-9.6,9.6]\times 10^{-13}$  & $2.5\times 10^{-30}$ &{$3.3 \times 10^{-30}$}\\
Re$C_K^4$ & $[-3.6,3.6]\times 10^{-15}$  & $-3.2\times 10^{-15}$  & ${-3.4\times 10^{-15}}$\\ Im$C_K^4$ &  $[-1.8,0.9]\times 10^{-17}$   & $8.3\times 10^{-32}$ & {$4.1\times 10^{-29}$}\\
Re$C_K^5$ & $[-1.0,1.0]\times 10^{-14}$ &  $-2.7\times 10^{-15}$  & ${-2.8\times 10^{-15}}$\\ Im$C_K^5$ & $[-1.0,1.0]\times 10^{-14}$  & $6.8\times 10^{-32}$ &{$3.4\times 10^{-29}$}\\
\hline
$|C_{B_d}^1|$ & $<2.3\times 10^{-11}$  &  $6.6\times 10^{-18}$ & $3.3\times 10^{-18}$\\
  $ |\tilde{C}_{B_d}^1|$ & $<2.3\times 10^{-11}$   & $2.0\times 10^{-17}$ & $2.9\times 10^{-17}$\\
$|C_{B_d}^4|$ &  $<2.1\times 10^{-13}$ & $3.0\times 10^{-17}$  &  $2.5\times 10^{-17}$\\
  $|C_{B_d}^5|$ & $<6.0\times 10^{-13}$  &  $5.0\times 10^{-17}$ & $4.2 \times 10^{-17}$ \\

\hline
$|C_{B_s}^1|$ & $< 1.1 \times 10^{-9}$  &  $1.3\times 10^{-15}$ & $3.9 \times 10^{-15}$ \\
 $|\tilde{C}_{B_s}^1|$ & $< 1.1 \times 10^{-9}$  & $2.8\times 10^{-15}$ & $5.2\times 10^{-14}$\\
$|C_{B_s}^4|$ & $< 1.6 \times 10^{-11}$  &   $5.0\times 10^{-15}$ & $3.7 \times 10^{-14}$\\
 $|C_{B_s}^5|$ & $< 4.5 \times 10^{-11}$  & $8.1\times 10^{-15}$ & $6.2\times 10^{-14}$\\
\hline
$|C_D^1|$ & $<7.2 \times 10^{-13}$  &   $6.9\times 10^{-16}$  & $1.3\times 10^{-15}$ \\
 $|\tilde{C}_D^1|$ & $<7.2 \times 10^{-13}$  & $6.9\times 10^{-16}$ & $3.9 \times 10^{-14}$\\
$|C_D^4|$ & $<4.8\times 10^{-14}$   &   $2.8\times 10^{-15}$  & $2.8\times 10^{-14}$ \\
 $|C_D^5|$ & $<4.8 \times 10^{-13}$  &  $3.0\times 10^{-15}$ & $3.1\times 10^{-14}$\\
\hline
 ${\rm BR}[\mu \to e]$ & $< 7.0 \times 10^{-13}$  ~~&~$5.1 \times 10^{-17}$ & $3.2 \times 10^{-15}$\\
\hline
${\rm BR}[\mu \to 3e]$ & $< 1.0 \times 10^{-12}$   & $2.9\times 10^{-19}$ & $2.0 \times 10^{-17}$\\ 
${\rm BR}[\tau \to 3\mu]$ & $< 2.1 \times 10^{-8}$  & $5.3\times 10^{-19}$ & $2.8\times 10^{-17}$\\ 
${\rm BR}[\tau \to 3 e]$ & $< 2.7 \times 10^{-8}$  & $6.4\times 10^{-22}$ & $7.2\times 10^{-20}$ \\
\hline 
${\rm BR}[\mu \to e \gamma]$ & $< 4.2 \times 10^{-13}$  &  $6.1\times 10^{-21}$ & $2.9\times 10^{-19}$\\
${\rm BR}[\tau \to \mu \gamma]$ &  $< 4.4 \times 10^{-8}$  & $6.1\times 10^{-22}$ & $2.4\times 10^{-19}$\\
 ${\rm BR}[\tau \to e \gamma]$ &  $< 3.3 \times 10^{-8}$  & $7.3\times 10^{-24}$ & $1.9\times 10^{-22}$\\
\hline
\hline
\end{tabular}
\end{center}
\caption{The magnitudes of the various WCs (in units of GeV\(^{-2}\)) for \(\Delta F = 2\) processes in the quark sector, along with the branching ratios for the lepton flavour-violating process, are computed for solutions 1 and 2. The respective experimental limits are also provided.}
\label{tab:FV}
\end{table}

\subsection{\texorpdfstring{$\mu$}{\mu} to \texorpdfstring{$e$}{e} conversion}
\begin{figure}[t]
\centering
\subfigure{\includegraphics[width=0.48\textwidth]{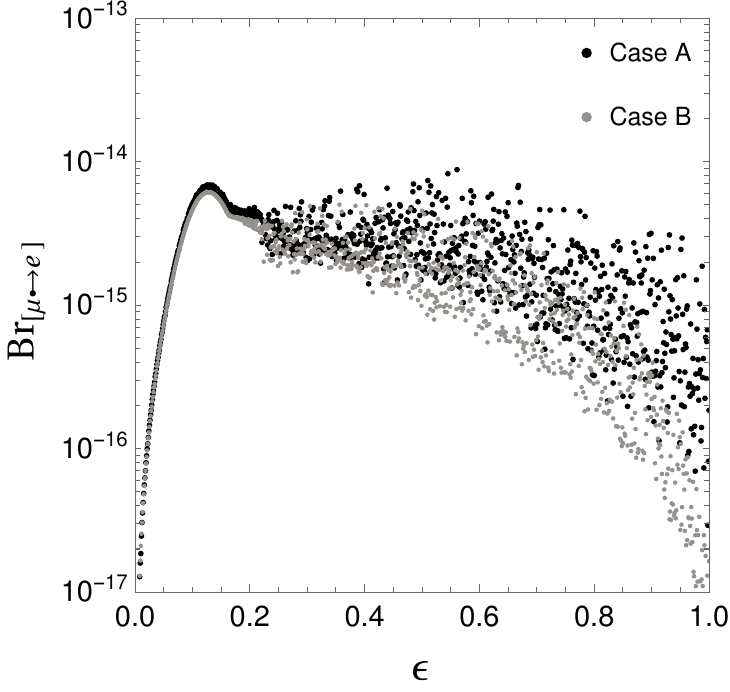}}\hspace*{0.5cm}
\subfigure{\includegraphics[width=0.48\textwidth]{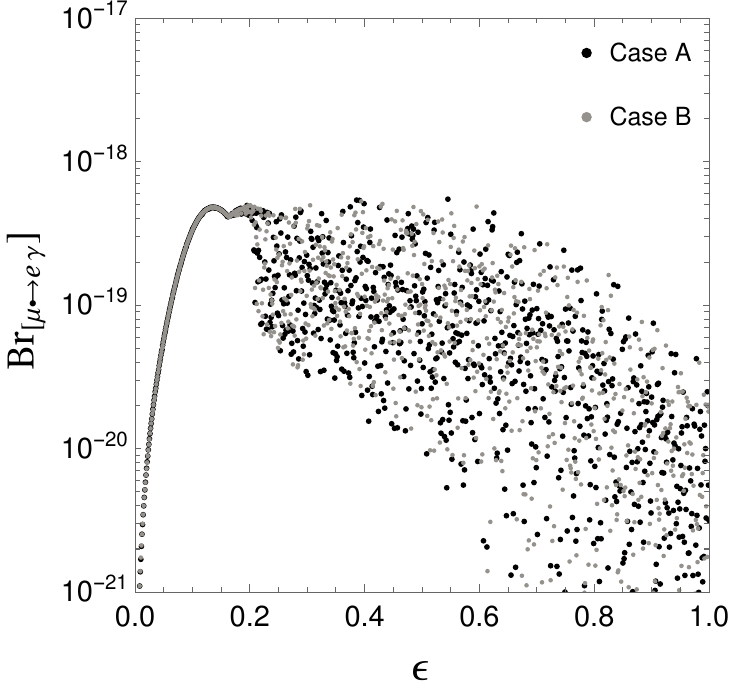}}
\caption{Left sub-figure: ${\rm BR}[\mu \to e]$ in $^{197}$Au nucleus for various values of $\epsilon$ and for $M_X=10^6$ GeV estimated from the best-fit solutions. Right sub-figure: The same for the lepton flavour violating observable ${\rm BR}[\mu \to e\, \gamma]$.}
\label{fig:mu2e}
\end{figure}

The \(X\)-boson, through its flavour-conserving couplings with \(u\) and \(d\) quarks and flavour-violating couplings with \(e\) and \(\mu\) leptons, can mediate \(\mu \to e\) conversion in nuclei at tree level. The branching ratio for this transition, \({\rm BR}[\mu \to e]\), can be calculated using eq. (\ref{mu2e}) for each parameter point shown in Fig. \ref{fig:chi2}, corresponding to \(M_X = 10^6\) GeV. The results are presented in the left sub-figure of Fig. \ref{fig:mu2e}. Additionally, we compute the branching ratios for the two benchmark solutions and provide them in Table \ref{tab:FV}.

It can be seen that, in both scenarios A and B, the predicted magnitude of ${\rm BR}[\mu \to e]$ is quite similar. This similarity arises because the factors distinguishing these scenarios do not alter the flavour-violating interactions in the lepton sector. With $\epsilon > 0.15$, ensuring feasible solutions in both scenarios, the implication is that ${\rm BR}[\mu \to e] \le 7 \times 10^{-15}$, which remains two orders of magnitude below the current limit set by SINDRUM II. Consequently, the present framework does not appear to be restricted by limits from $\mu$ to $e$ conversion.

\subsection{\texorpdfstring{$l_i \to 3 l_j$ }{l_i \to 3 l_j}and \texorpdfstring{$l_i \to  l_j \gamma$ }{l_i \to  l_j \gamma}}
Other flavour-violating processes in the lepton sector include \(l_i \to 3 l_j\) and \(l_i \to l_j \, \gamma\). Using eqs. (\ref{lto3l}) and (\ref{muegamma}), the branching ratios for \(\mu \to 3 e\), \(\tau \to 3 \mu\), \(\tau \to 3 e\), \(\mu \to e \gamma\), \(\tau \to \mu \gamma\), and \(\tau \to e \gamma\) can be estimated in the present model by appropriately replacing \(X^{(1)}_{eL,R}\) with \(X_{eL,R}\). For \(M_X \geq 10^6\) GeV, none of these processes produce branching ratios large enough to impose meaningful constraints on the model. The estimated values for the benchmark solutions are listed in Table \ref{tab:FV}. Notably, due to the optimised arrangement of flavour-violating couplings, flavour violation in the \(1\)-\(2\) sector is typically smaller or comparable to that in the \(2\)-\(3\) sector. This represents an improvement over the model discussed in the previous chapter, where the opposite trend was observed. Additionally, the evaluated values of \({\rm BR}[\mu \to e \gamma]\) for \(M_X = 10^6\) GeV and various \(\epsilon\) values are shown in Fig. \ref{fig:mu2e}. 

In summary, the most stringent phenomenological constraint on this model arises from \(K^0\)-\(\overline{K}^0\) oscillations, which necessitate a new gauge boson mass of at least \(10^3\) TeV or higher. This scale is nearly two orders of magnitude smaller than the lowest scale obtained in the $U(1)_1\times U(1)_2$ model. As discussed in the previous section, the vector-like quarks and leptons also tend to cluster near this scale. Under these conditions, other constraints, such as those from direct searches and electroweak precision tests, are easily satisfied, as discussed in the previous chapter.

\section{Neutrino masses}
\label{sec:neutrino}
While the primary focus of this study has been on the charged fermion mass spectrum, we also explore how the framework can be extended to incorporate neutrino masses. The main observational features that distinguish neutrinos from charged fermions \cite{Capozzi:2017ipn, deSalas:2020pgw, Esteban:2020cvm} are: (a) The overall mass scale of neutrinos is several orders of magnitude smaller than that of charged fermions, and (b) Neutrinos exhibit a relatively weaker mass hierarchy, with \(m_{\nu 2}/m_{\nu 3} \simeq 0.2\) for \(m_{\nu 1} = 0\) and \(m_{\nu 2}/m_{\nu 3} \simeq 1\) in the case of quasi-degenerate neutrinos. As outlined below, the current framework can be extended in two qualitatively distinct ways to explain neutrino masses while accounting for these features. These approaches align with common extensions of the Standard Model designed to address neutrino mass generation.

\subsection{Majorana option}
Neutrinos can be of Majorana nature, introducing lepton number violation into the framework. If the scale associated with this violation, denoted as \(\Lambda_{\rm LN}\), is higher than both the electroweak and \(U(1)_F\) breaking scales, neutrino masses can be effectively described using the familiar dimension-5 operator \cite{Weinberg:1979sa}. In the current framework, these operators take the form:
\be \label{dim5}
{\cal L}_{\rm dim-5} \supset \frac{c^{(1)}_{ij}}{2 \Lambda_{\rm LN}}\, \left(\overline{\Psi}_{Li} H_{u i}\right)\left(H^T_{u j} \Psi^c_{Lj}\right)\,+\,\frac{c^{(2)}_{ij}}{2 \Lambda_{\rm LN}}\, \left(\overline{\Psi}_{Li} H_{u j}\right)\left(H^T_{u i} \Psi^c_{Lj}\right)\,+\,{\rm h.c.}\,.
\ee
The first operator arises from a complete theory by integrating out fermions neutral under both the SM and \(U(1)_F\) gauge symmetries, corresponding to the type I seesaw mechanism \cite{Minkowski:1977sc, Yanagida:1979as, Mohapatra:1979ia, Schechter:1980gr}. In contrast, the second operator requires UV completions involving either \(U(1)_F\)-neutral fermions charged under the electroweak symmetry (type III seesaw \cite{Foot:1988aq}) or complex scalars charged under both the electroweak and \(U(1)_F\) gauge symmetries (type II seesaw \cite{Lazarides:1980nt, Magg:1980ut, Mohapatra:1980yp}). Importantly, none of these extensions introduces additional contributions to gauge or mixed gauge-gravity anomalies.

After electroweak symmetry breaking, the neutrino masses are expressed as:
\be
\left(m_\nu\right)_{ij} = \frac{c_{ij}}{\Lambda_{\rm LN}}\, \mu_{ui}\,\mu_{uj}\,,
\ee
where \(c_{ij}\) are the coefficients from eq. (\ref{dim5}), appropriately scaled by the Yukawa couplings \(y_{ui}\). With general \(c_{ij}\), all three neutrinos acquire tree-level masses that are suppressed by \(\Lambda_{\rm LN}\), consistent with the observed features (a) and (b) outlined earlier in this section. Furthermore, the coefficients \(c_{ij}\) provide sufficient flexibility to reproduce viable leptonic mixing parameters.To achieve a more predictive and insightful understanding of the leptonic mixing parameters, additional symmetries or specific structures in the UV completion would be required. Developing such a framework would involve detailed model-building efforts.

\subsection{Dirac option}
Alternatively, neutrino masses can be introduced analogously to how charged fermion masses are generated in this framework. This requires three Weyl fermions, \(\nu_{Ri}\), with \(U(1)_F\) charges \((1-\epsilon, 1+\epsilon, -2)\), and a neutral vector-like pair \(N_{L,R}\). The \(\nu_{Ri}\) are already present in the original model as a requirement for anomaly cancellation (see Table \ref{tab:fields:u1}). All these fields are singlets under the SM gauge symmetries. At leading order, the Dirac neutrino mass matrix is given by:  
\be
\left(M^{(0)}_\nu\right)_{ij} = -\frac{1}{m_N}\, \mu_{\nu i}\, \mu_{\nu j}^\prime\,,
\ee
where \(\mu_{\nu i} = y_{\nu i} \langle H_{u i} \rangle\), \(\mu^\prime_{\nu i} = y^\prime_{\nu i} \langle \eta^*_i \rangle\), and \(m_N\) is the Dirac mass of the \(N_{L,R}\) pair. This structure leads to one massive light neutrino state.  

In this framework, the universal seesaw-like structure allows the smallness of the neutrino mass to be explained by a large \(m_N\). By identifying the massive neutrino state with the atmospheric neutrino oscillation scale \cite{Esteban:2020cvm}, the Dirac mass \(m_N\) is estimated as:  
\be
m_N \approx 2\times 10^{17}\, {\rm GeV}\,\left(\frac{0.05\,{\rm eV}}{m_{\nu 3}}\right)  \left(\frac{\langle \eta \rangle}{100\,{\rm TeV}}\right)  \left(\frac{\langle H_u \rangle}{100\,{\rm GeV}}\right)\,,
\ee
assuming all dimensionless parameters are of \({\cal O}(1)\). Here, the issue of unnaturally small neutrino Yukawa couplings in the standard SM extension with Dirac neutrinos is reframed as a hierarchy problem, \(m_N \gg m_{U,D,E}\), in this model.  

The solar neutrino mass scale can arise when higher-order corrections are introduced to \(M^{(0)}_\nu\). In this case, the same expression as eq. (\ref{M2_eff_fnl_f}) applies, with appropriate modifications. However, some fine-tuning may be necessary, as the desired magnitude of \(m_{\nu 2}/m_{\nu 3}\) is larger than the typical loop suppression factor. This tuning can be avoided by introducing two or more copies of \(N_{L,R}\) and multiple \(\nu_{Ri}\) multiplets, which would allow the tree-level neutrino mass matrix to have a rank greater than one. In such a scenario, both the solar and atmospheric mass scales could be generated at tree level and would naturally exhibit less hierarchy.

\section{On the origin of the gauge charges}
\label{chap4:gaugecharges}
In this section we show that the choice made in eq. (\ref{gauge_charges}) for the charges of underlying Abelian flavour symmetry can simply be obtained from kinetic mixing. Consider two $U(1)$ symmetries characterized by gauge bosons $X_{1,2}^\mu$, alongside their respective gauge interactions
\be \label{LG_app}
- {\cal L}_G = g^{(1)}\, q^{(1)}_{ii}\, \overline{f}^\prime_{i}\,\gamma^\mu\,f^\prime_{i}\, X^{(1)}_\mu + g^{(2)}\, q^{(2)}_{ii}\, \overline{f}^\prime_{i}\,\gamma^\mu\,f^\prime_{i}\, X^{(2)}_{\mu}\,,  \ee
where $f^\prime$ stands for both $f^\prime_L$ and $f^\prime_R$. A suitable choice for the charge matrices is
\be \label{q1-q2}
q^{(1)} = {\rm Diag.}(1,1,-2)\,,~~q^{(2)} = {\rm Diag.}(1,-1,0)\,.\ee
Subsequently, consider the kinetic terms of these gauge bosons with non-zero kinetic mixing \cite{Holdom:1985ag}
\be \label{kin}
-{\cal L}_{\rm kin} = \frac{1}{4} F^{(1)}_{\mu \nu} F^{(1) \mu \nu} + \frac{1}{4} F^{(2)}_{\mu \nu} F^{(2) \mu \nu} + \frac{\chi}{2} F^{(1)}_{\mu \nu} F^{(2) \mu \nu}\,,\ee
with $F^{(\alpha)}_{\mu \nu}$ represents the field strength of the gauge boson $X^{(\alpha)}_\mu$. The kinetic terms can be made diagonal by applying a linear transformation
\be \label{LT_kin}
\left(\ba{c} X^{(1)}_\mu \\ X^{(2)}_\mu \ea \right) \to \left(\ba{cc} 1 & 0  \\ -\chi & 1 \ea \right)  \left(\ba{c} X^{(1)}_\mu \\ X^{(2)}_\mu \ea \right)\,.  \ee
This process eliminates the kinetic mixing from the kinetic terms, which subsequently re-emerge within the gauge interactions. By inserting eq. (\ref{LT_kin}) into eq. (\ref{LG_app}), we obtain
\be \label{LG_app_1}
- {\cal L}_G = g^{(1)} \left(q^{(1)}_{i i} - \chi \frac{g^{(2)}}{g^{(1)}}\, q^{(2)}_{i i} \right)\, \overline{f}^\prime_{i}\,\gamma^\mu\,f^\prime_{i}\, X^{(1)}_{\mu} + g^{(2)} q^{(2)}_{i i}\, \overline{f}^\prime_{i}\,\gamma^\mu\,f^\prime_{i}\, X^{(2)}_{\mu}\,.  \ee
Setting $g^{(1)}=g_X$, $\chi g^{(2)}/g^{(1)} \equiv \epsilon$ and $X^{(1)}_\mu = X_\mu$, we find 
\be \left(q^{(1)}_{i i} - \epsilon\, q^{(2)}_{i i} \right)=\lbrace 1-\epsilon, 1+\epsilon, -2\rbrace \ee for $i=1,2,3$ respectively. These are the desired charges given in eq. (\ref{gauge_charges}).  By setting $M_{X_2} \gg M_{X}$, the effects of the gauge interactions associated with $X^{(2)}_\mu$ gauge boson can be decoupled from the theory, and the effective theory would contain only one new $U(1)$ symmetry with charges $\lbrace 1-\epsilon, 1+\epsilon, -2\rbrace$ for three generations of the SM fermions as considered in section \ref{subsec:implementation}.

The two $U(1)$ symmetries displaying flavour-dependent charges, as detailed in eq. (\ref{q1-q2}), combined with any flavour universal $U(1)$ symmetry (such as hypercharge within the current framework), can be reorganized into $U(1)_1 \times U(1)_2 \times U(1)_3$. In this configuration, solely the $i^{\rm th}$ generation fermion carries a charge under $U(1)_i$. This can be understood by examining the gauge interactions:
\be \label{three_u1}
\sum_{\alpha = 1}^3\,g^{(\alpha)}\, q^{(\alpha)}_{ii}\, \overline{f}^\prime_{i}\,\gamma^\mu\,f^\prime_{i}\, X^{(\alpha)}_{\mu}\,,  \ee
with $q^{(1),(2)}$ as already given in eq. (\ref{q1-q2}) and $q^{(3)} = {\rm Diag.}(1,1,1)$. A transformation 
\be \label{rot_gb}
X_\mu^{(\alpha)} \to \tilde{X}_\mu^{(\alpha)} = {\cal R}_{\alpha \beta}\,X_\mu^{(\beta)}\,,\ee
can be applied to redefine the gauge bosons so that in the new basis, the gauge interactions are represented as 
\be \label{three_u1_2}
\sum_{\alpha = 1}^3\,\tilde{g}^{(\alpha)}\, \tilde{q}^{(\alpha)}_{ii}\, \overline{f}^\prime_{i}\,\gamma^\mu\,f^\prime_{i}\, \tilde{X}^{(\alpha)}_{\mu}\,.  \ee
The new couplings are 
\be \label{rot_gc}
\tilde{g}^{(\alpha)} \tilde{q}_{ii}^{(\alpha)} = {\cal R}_{\alpha \beta}\,g^{(\beta)} q_{ii}^{(\beta)} \,,\ee
where ${\cal R}$ is an $3 \times 3$ orthogonal matrix.

For a specific choice, $\sqrt{6} g^{(1)}=\sqrt{2} g^{(2)}=\sqrt{3} g^{(3)} \equiv \tilde{g}$, and 
\be \label{R_gb}
{\cal R} = \left(
\begin{array}{ccc}
 \frac{1}{\sqrt{6}} & \frac{1}{\sqrt{2}} & \frac{1}{\sqrt{3}} \\
 \frac{1}{\sqrt{6}} & -\frac{1}{\sqrt{2}} & \frac{1}{\sqrt{3}} \\
 -\sqrt{\frac{2}{3}} & 0 & \frac{1}{\sqrt{3}} \\
\end{array}
\right)\,,\ee
one finds,
\beqa \label{q_tilde}
 \tilde{g}^{(1)} \tilde{q}^{(1)} &=& \tilde{g}\,{\rm Diag.}(1,0,0)\,,\nonumber\\
 ~~\tilde{g}^{(2)} \tilde{q}^{(2)} &=& \tilde{g}\,{\rm Diag.}(0,1,0)\,,\nonumber\\
 ~~\tilde{g}^{(3)}\tilde{q}^{(3)} &=& \tilde{g}\,{\rm Diag.}(0,0,1)\,.\eeqa
Thus, the three Abelian symmetries in the underlying theory: two flavour non-universal new Abelian symmetries and hypercharge, can be arranged such that each generation is exclusively charged under only one \(U(1)\), with equal interaction strengths. This framework has recently been proposed under the names "tri-hypercharge" \cite{FernandezNavarro:2023rhv} and "deconstructed hypercharge" \cite{Davighi:2023evx}.

\section{Conclusion}
\label{chap4:summary}
It is demonstrated that extending the SM with a single Abelian gauge symmetry is sufficient to provide a viable framework for radiatively generating the masses of the lighter generations. The second-generation masses are generated at one loop, while the first-generation masses arise at two loops. The two-loop corrected mass matrix can be fully expressed in terms of the tree-level mass matrix, gauge charges, and the gauge boson mass, as shown in eq. (\ref{M2_eff_fnl}). This represents a significant improvement over the previous $U(1)_1\times U(1)_2$ model, where the masses of both lighter generations were induced at one loop, and the intergenerational hierarchy was attributed to the specific ordering of gauge boson masses.

The framework inherently introduces large FCNCs mediated by the new gauge boson, which impose the strongest constraints on the scale of new physics. As demonstrated in this study, the flavour-changing couplings can be organized such that their strength in the 1-2 sector is minimised. This is phenomenologically advantageous, as the most stringent constraints on flavour violations arise from the 1-2 sector, particularly from \(K^0\)-\(\overline{K}^0\) mixing and \(\mu\)-\(e\) conversion in nuclei. The suppression of flavour-changing couplings in the 1-2 sector is proportional to the difference between the charges of the first- and second-generation fermions under the new gauge symmetry. However, these charges cannot be made arbitrarily close, as complete degeneracy would result in a strictly massless first generation. 

This interplay between charge differences and flavour violation is systematically explored through numerical analysis. It is found that the down-type quark sector plays a dominant role in this optimisation. If the down quark mass is to be reproduced within its \(3\sigma\) range, as determined from lattice computations, the mass of the \(U(1)_F\) gauge boson must be at least \(10^3\) TeV, while satisfying all flavour constraints. This lower bound is nearly two orders of magnitude smaller than those derived in the previous $U(1)_1\times U(1)_2$ model.

%% file: 50_Chapter_5/chapter_5.tex
\chapter{Radiative mass mechanism in Non-Abelian Gauge theory}
\label{chap5}
\graphicspath{{50_Chapter_5/}}
The most advantageous feature of non-abelian symmetries is that it naturally provides larger irreducible representations for fermion multiplets. This feature has significant implications in the radiative mass generation framework as the unification of three families of fermions into a single multiplet leads to a smaller number of Yukawa couplings as compared to the Abelian frameworks discussed in previous chapters. Also, another essential feature of non-abelian $G_F=SU(3)_F$ is that it naturally accommodates gauge bosons with flavour non-diagonal couplings, which play a crucial role in the underlying mechanism. The same symmetry can also be effectively utilised in order to ensure that only the third generations receive mass at the tree level. Moreover, being a simple group, it minimally modifies the SM gauge structure and can lead to a predictive scenario.
The horizontal $SU(3)$ symmetry was previously proposed in \cite{HernandezGaleana:2004cm,Hernandez-Galeana:2015zap} for similar purposes. However, a systematic and comprehensive analysis of loop-induced fermion masses, mixing parameters, and the associated phenomenological constraints on the flavour symmetry breaking scale has not been performed. Another non-abelian alternative, $G_F = SO(3)_L \times SO(3)_R$, was recently investigated in \cite{Weinberg:2020zba}, but it was found to yield an inconsistent flavour spectrum. This chapter addresses these gaps by presenting a complete and realistic model for radiatively induced quark and lepton masses based on non-abelian flavour symmetry. 

We demonstrate that a viable implementation of the $SU(3)_F$ framework within the SM necessitates multiple electroweak Higgs doublets and the presence of VL fermions. The latter is crucial for achieving the observed mass spectrum while satisfying flavour violation constraints. Moreover, the hierarchy between the first and second-generation masses can naturally arise when the flavour symmetry is broken in a specific pattern. These features, combined with improved predictivity, make this model less arbitrary compared to the abelian symmetry-based approaches discussed earlier.

The remainder of this chapter is organized as follows: In the next section, we outline the general framework of $SU(3)_F$ and the mechanism for radiative mass generation. Section \ref{sec:GBmass} details the breaking of horizontal symmetry and the resulting gauge boson mass spectrum. In section \ref{sec:model}, we present the implementation of this scheme within the SM. Numerical solutions validating the viability of the model are discussed in section \ref{sec:numerical}. Section \ref{sec:fv} explores some phenomenological implications of the framework, and conclusions are presented in section \ref{sec:concl}.

\section{\texorpdfstring{$SU(3)_F$ }{SU(3)_F} and fermion mass generation}
\label{sec:su3fm}
We compute the loop-corrected fermion masses induced by the gauge bosons of $SU(3)_F$. We begin with the toy model chiral fermions $f_{L i}^\prime$ and $f_{R i}^\prime$ and a pair of VL fermions, $F^\prime_{L,R}$ and write the arranged tree-level mass matrix ${\cal M}^{0}$, same as eq. (\ref{M0}), in the basis $f^\prime_{ L \alpha} \equiv (f^\prime_{L i}, F^\prime_L)$ and $f^\prime_{R \alpha} \equiv (f^\prime_{R i}, F^\prime_R)$. The mass matrix ${\cal M}^{0}$ leads to massive third-generation fermions and vector-like states, as discussed in section \ref{toy:treelevel}.

We also assume that $f^\prime_{L i}$ and $f^\prime_{R i}$  as fundamental representations under the horizontal gauged symmetry $SU(3)_F$, while the VL fermions are treated as singlets with respect to the same symmetry. Consequently, the non-vanishing $\mu_L$ and $\mu_R$ in the mass Lagrangian mentioned above lead to the breaking of $SU(3)_F$. It is possible to derive a vanishing $3\times 3$ sub-matrix through the chiral nature of $f^\prime_{L i}$ and $f^\prime_{R i}$ within the framework of the SM gauge symmetry, allowing it to remain zero even when $SU(3)_F$ is broken. Based on the SM charges assigned to $F^\prime_L$ or $F^\prime_R$, either $\mu_L$ or $\mu_R$ is also protected by the chiral symmetry.

 Within this framework, the relatively small masses of the first two generations can arise due to quantum corrections. To quantify these corrections, we examine the $SU(3)_F$ gauge interactions that involve fermions and gauge bosons $A_\mu^a$ as described by
\be \label{L_gauge:su3}
-{\cal L}_{\rm gauge} = g_F  \left( \overline{f}^\prime_{L i} \gamma^\mu A^a_\mu \left(\frac{\lambda^a}{2}\right)_{ij}f^\prime_{L j } +  \overline{f}^\prime_{R i} \gamma^\mu A^a_\mu \left(\frac{\lambda^a}{2}\right)_{ij}f^\prime_{R j } \right)\,, \ee
where $a =1,..,8$ and $\lambda^a$ are the Gell-Mann matrices. For the latter, we use the expressions on a different basis than the conventional one, and they are listed in Appendix \ref{app:GM} for clarity. The above can be extended to include the VL fermions as
\be \label{L_gauge:su3_2}
-{\cal L}_{\rm gauge} = \frac{g_F}{2}  \left( \overline{f}^\prime_{L \alpha} \gamma^\mu A^a_\mu \left( \Lambda^a \right)_{\alpha \beta}f^\prime_{L \beta} +  \overline{f}^\prime_{R \alpha} \gamma^\mu A^a_\mu \left( \Lambda^a \right)_{\alpha \beta}f^\prime_{R \beta}  \right)\,, \ee
where $\Lambda^a$ are $4 \times 4$ matrices given by
\be \label{Lambda}
{ \Lambda}^a= \, \left(\ba{cc}\lambda^a& 0\\0 &0\ea \right)\,.\ee

The physical basis of fermions, denoted by $f_{L,R}$, can be obtained from the canonical basis using the unitary transformations $f^\prime_{L,R}= {\cal U}_{L,R}\, f_{L,R}$, as in eq. (\ref{M0_diag}).

In a similar manner, the physical gauge bosons $B_\mu^a$ can be derived from $A_\mu^a$ by using an $8 \times 8$ real orthogonal matrix ${\cal R}$, described by the equation 
\be \label{AtoB}
A_{a \mu}\,=\, {\cal R}_{ab}\, B_{b \mu} \,.\ee
The matrix ${\cal R}$ is explicitly determined by diagonalising the gauge-boson mass matrix, which is real and symmetric.

Thus, the gauge interactions in the physical basis of fermions and gauge bosons are obtained as
\be \label{L_gauge:su3_3}
-{\cal L}_{\rm gauge} = \frac{g_F}{2}  \left(\overline{f}_{L \alpha} \gamma^\mu  \left({\cal U}_L^\dagger{\Lambda^a} {\cal U}_L \right)_{\alpha \beta}f_{L \beta} + \overline{f}_{R \alpha} \gamma^\mu  \left({\cal U}_R{}^\dagger{\Lambda^a} {\cal U}_R \right)_{\alpha \beta}f_{R \beta} \right) {\cal R}_{ab}B^b_\mu\,. \ee
Due to the non-commutative nature of $\Lambda^a$ matrices for every $a$, it is impossible to simultaneously diagonalise the matrices ${\cal U}_{L,R}^\dagger{\Lambda^a} {\cal U}_{L,R}$. Therefore, there always exists a set of gauge bosons which has flavour-changing interactions with fermions, which is necessary for the generation of masses for the first and second family fermions through radiative corrections.

The fermion mass matrix, corrected by the $SU(3)_F$ gauge interactions at 1-loop, similar to eq. (\ref{M_corr}), can be expressed as
\be \label{1loopmass}
{\cal M}= {\cal M}^{0}+ \delta {\cal M}\,, \ee 
where 
\be \label{dm}
\delta {\cal M} = {\cal U}_{L}\, \Sigma(0)\, {\cal U}^{ \dagger}_{R}\, .\ee
\begin{figure}[!t]
    \centering
    \includegraphics[width=8cm]{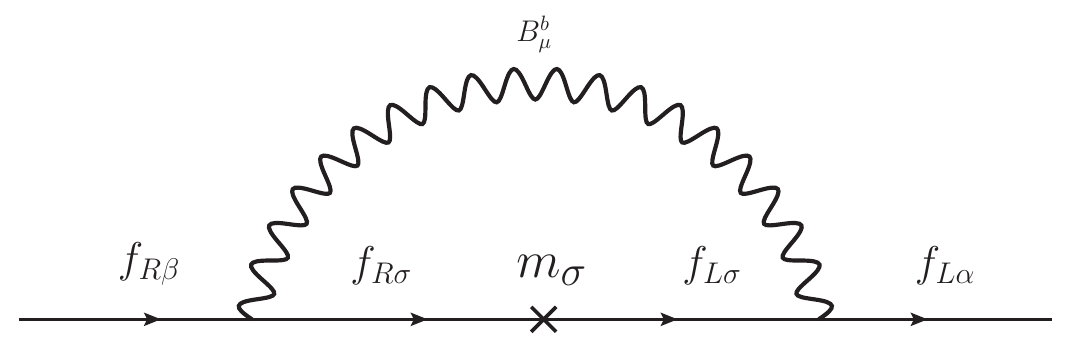}
    \caption{Self-energy correction induced by $SU(3)_F$ gauge boson at 1-loop.}
    \label{fig:loop}
\end{figure}
Using eq. (\ref{L_gauge:su3_3}), the 1-loop correction (see Fig. \ref{fig:loop}) can be computed as 
\beqa \label{sigma:su3}
-i(\Sigma(p))_{\alpha \beta} &=& \int \frac{d^4 k}{(2\pi)^4}\left(-i\frac{g_F}{2} {\cal R}_{ab} ({\cal U}_L^{ \dagger} {\Lambda^a} {\cal U}_L)_{\alpha \sigma}\right) \gamma^\mu \frac{i m_{\sigma}}{ (k+p)^2-m_{\sigma}^2 + i\epsilon} \, \nonumber \\
& & \hspace{5mm}\left(-i\frac{g_F}{2} {\cal R}_{c b} ({\cal U}_R^{ \dagger}{\Lambda^c}{\cal U}_R)_{\sigma \beta}\right) \gamma^\nu\, \Delta_{\mu \nu}(k)\,,\eeqa
with 
\be \label{propa}
\Delta_{\mu \nu}(k) = \frac{-i}{k^2-{M_{b}}^2 + i\epsilon} \left( \eta_{\mu \nu}-(1-\zeta)\frac{k_\mu k_\nu}{k^2-\zeta M_{b}^2}\right)\,.\ee
Here $M_{b}$ represents the  mass of  the  gauge boson $B^b_{\mu}$. In the Feynmann-'t Hooft gauge, the above integral can be evaluated as:
\be  \label{sigma_2}
\left(\Sigma_f(0)\right)_{\alpha \beta} = \frac{{g_F}^2}{16\pi^2}\, {\cal R}_{ab} ({\cal U}_L^\dagger{\Lambda^a}{\cal U}_L)_{\alpha \sigma} 
{\cal R}_{c b}({\cal U}_R^\dagger{\Lambda^c}{\cal U}_R)_{\sigma \beta}\, m_{\sigma}\, B_0[M_{b},m_{\sigma}]\,, \ee 
with $B_0[M_{b},m_{\sigma}]=\Delta_\epsilon+b_0[M_{b},m_{\sigma}]$ as the 2-point PV function defined in eq. (\ref{B0}).

 Equation (\ref{sigma_2}) clearly demonstrates that the $m_\sigma$-independent term, the divergent part , in $\delta {\cal M}$ arising from $B_0[M_b,m_\sigma]$ is vanishing. Explicitly,
\be
\delta {\cal M}_{\rm div} = {\cal U}_{L}\, \Sigma(0)_{\rm div}\, {\cal U}^{ \dagger}_{R}\,,\ee
where $\Sigma(0)_{\rm div}$ contains the terms proportional to  $\Delta_\epsilon$ from eq. (\ref{sigma_2}). Explicitly,
\beqa 
(\delta {\cal M}_{\rm div})_{\rho \kappa}&=&  
  \frac{{g_F}^2}{16\pi^2}\,({\cal R R^T})_{ac}\,{\cal U}_{L}{}_{\rho \alpha}  ({\cal U}_L^\dagger{\Lambda^a}{\cal U}_L)_{\alpha \sigma} 
({\cal U}_R^\dagger{\Lambda^c}{\cal U}_R)_{\sigma \beta}\, m_{\sigma}\, \Delta_\epsilon\, {\cal U}^{ \dagger}_{R}{}_{\beta \kappa}\, , \nonumber\\
&=& \frac{{g_F}^2 \Delta_\epsilon}{16\pi^2}\,  \,{\cal U}_{L}{}_{\rho \alpha}\, ({\cal U}_L^\dagger{\Lambda^a}{\cal U}_L)_{\alpha \sigma}\, {\cal D}_{\sigma \sigma}\,({\cal U}_R^\dagger{\Lambda^a}{\cal U}_R)_{\sigma \beta}\,{\cal U}^{ \dagger}_{R}{}_{\beta \kappa}\,. \eeqa 
The last equality is obtained by using the orthogonality of ${\cal R}$, and it can be further simplified by using eq. (\ref{M0_diag}) and unitarity of ${\cal U}_{L,R}$, as
\be \label{DM_div:su3}
\delta {\cal M}_{\rm div} = \frac{{g_F}^2 \Delta_\epsilon}{16\pi^2}\, \Lambda^a {\cal M}^{0}\Lambda^a = 0\,,
\ee
where the form of ${\cal M}^{0}$ and $\Lambda^a$ given in eqs. (\ref{M0},\ref{Lambda}), is used to get the last equality. As mentioned earlier, the vanishing of $\delta {\cal M}_{\rm div}$ is in accordance with the renormalisability \cite{Barr:1978rv,Weinberg:1972ws} of the theory as there are no corresponding counterterms to renormalise.

Expanding the finite part of eq. (\ref{sigma_2}) and substituting eqs. (\ref{Lambda}), (\ref{U0_ss}), and (\ref{Meff_diag_comp}), the one-loop correction to the effective $3 \times 3$ mass matrix simplifies to:
\be \label{dm_ij}
(\delta M)_{ij} \simeq \frac{{g_F}^2}{16 \pi^2} {\cal R}_{ab} {\cal R}_{cb} (\lambda^a M^0\lambda^c)_{ij}\, \Delta b_0[M_b^2]\,, \ee
where $M^0$ is the effective $3\times3$ mass matrix which has the form as given in eq. (\ref{M0_eff}), and
\be \label{delta_b0} 
\Delta b_0[M_b^2]  = -  \frac{M_b^2 \ln \frac{M_b^2}{\mu^2} - m_3^2 \ln \frac{m_3^2}{\mu^2}}{M_b^2 - m_3^2} + \frac{M_b^2 \ln \frac{M_b^2}{\mu^2} - m_4^2 \ln \frac{m_4^2}{\mu^2}}{M_b^2 - m_4^2} \, .\ee 
Using eq. (\ref{Lambda}), we also find that $\delta{\cal M}_{4 \alpha} = \delta{\cal M}_{\alpha 4} = 0$. The absence of observable new gauge bosons or vector-like states suggests the hierarchy $m_3 \ll M_b, m_4$. In this regime, the loop function in eq. (\ref{dm_ij}) can be approximated as  
\be \label{loop_approx}  
\Delta b_0[M_b^2] \simeq \frac{m_4^2}{M_b^2 - m_4^2} \ln\left(\frac{M_b^2}{m_4^2}\right)\,.  
\ee

Using eq. (\ref{loop_approx}) and the explicit expressions for the $\lambda^a$ provided in Appendix \ref{app:GM}, the following key features of the loop-corrected fermion mass matrix can be inferred from eq. (\ref{dm_ij}): 
\begin{itemize}
    \item The loop-induced masses are suppressed by the loop factor $g_F^2/(16 \pi^2)$ when $m_4 > M_b$. If $m_4 < M_b$, an additional suppression by a factor of $m_4^2/M_b^2$ occurs. 
    \item For a generic choice of gauge boson masses and the orthogonal matrix ${\cal R}$, eq. (\ref{dm_ij}) predicts that the first and second-generation fermions acquire masses of the same order at 1-loop.  
    \item A desirable scenario is one where only the second-generation fermions gain mass at 1-loop, while the first generation remains massless and acquires mass at higher orders. However, from the inspection of eq. (\ref{dm_ij}) and the Gell-Mann matrices, no configuration is found that strictly prevents the first-generation fermions from obtaining mass at 1-loop.
\end{itemize}

These results suggest that while the loop-suppressed masses for the first and second generations naturally emerge, achieving a hierarchy between them requires additional mechanisms. Utilising the first feature, consider first $\alpha$ gauge bosons, $1 <\alpha < 8 $, have masses as $M_{\alpha} < m_4$, and the rest have masses $M_n > m_4$ with $n=\alpha+1,...,8$, then eq. (\ref{dm_ij}) leads to:
\be \label{dm_ij_2}
(\delta M)_{ij} \simeq \frac{{g_F}^2}{16 \pi^2} (\lambda^a M^0\lambda^c)_{ij}\,\left(\sum^\alpha_{\alpha=1} {\cal R}_{a\alpha} {\cal R}_{c\alpha} \Delta b_0[M_\alpha^2]+\sum^8_{n=\alpha+1} {\cal R}_{a n} {\cal R}_{c n} \Delta b_0[M_n^2]\right). \ee
The relative suppression of the second term w.r.t the first term is $\frac{m_4^2}{M_n^2}$. If such a scenario is achieved along with restricting the values of $ {\cal R}_{ab}$ such that the first term of eq. (\ref{dm_ij_2}) would generate second family masses and the second term the first family, then the framework will account for the hierarchical spectrum of the SM fermions. In the next section, we provide explicit realisation of this proposal. 
 
\section{ Gauge-boson mass hierarchy}
\label{sec:GBmass}
Consider a two-step breaking of the \( SU(3)_F \) symmetry as follows:  
\be \label{su3:breaking}
SU(3)_F \xrightarrow{\langle \eta_1 \rangle} SU(2)_F \xrightarrow{\langle \eta_2 \rangle} \text{nothing}\,, 
\ee  
where \( \langle \eta_1 \rangle \gg \langle \eta_2 \rangle \). In this setup, the gauge bosons associated with the \( SU(2)_F \) generators are expected to be significantly lighter than the remaining five gauge bosons. This hierarchy in the gauge boson masses naturally translates into the mass hierarchy between the first- and second-generation fermions, as we demonstrate below.

In our chosen basis for the Gell-Mann matrices, it is convenient to associate the intermediate \( SU(2)_F \) symmetry with the generators \( \lambda^\alpha \) (\( \alpha = 1, 2, 3 \)). The remaining indices are labeled as \( m = 4, \dots, 8 \). In the basis \( A^\mu_a = \left(A^\mu_\alpha, A^\mu_m \right) \), the gauge boson mass term can be expressed as:  
\be \label{LGB}  
-{\cal L}^{M}_{\rm GB} = \frac{1}{2} {\cal M}^2_{ab} A^\mu_a A_{b \mu},  
\ee  
with  
\be \label{MGB_block}  
{\cal M}^2 = \begin{pmatrix} M^2_{(33)} & M^2_{(35)} \\ (M^2_{(35)})^T & M^2_{(55)} \end{pmatrix}.  
\ee  
Here, \( M^2_{(AB)} \) are sub-blocks of dimensions \( A \times B \) in the gauge boson mass matrix. The two-step breaking of the \( SU(3)_F \) symmetry implies a hierarchy in the mass blocks, with \( M^2_{(33)}, M^2_{(35)} \ll M^2_{(55)} \). This hierarchical structure allows the gauge boson mass matrix to be diagonalised using a see-saw-like procedure. At leading order, the resulting orthogonal transformation matrix can be written as:  
\be \label{R_block}  
{\cal R} = \begin{pmatrix} R_3 & -\rho R_5 \\ \rho^T R_3 & R_5 \end{pmatrix} + {\cal O}(\rho^2),  
\ee  
where \( \rho = -M^2_{(35)} (M_{(55)}^2)^{-1} \). In this expression, \( R_3 \) and \( R_5 \) are real orthogonal matrices of dimensions \( 3 \times 3 \) and \( 5 \times 5 \), respectively. The explicit form of $R_3$ ($R_5$) can be obtained by diagonalising the real symmetric matrix $M^2_{(33)}$ ($M^2_{(55)}$) given in eq. (\ref{MGB_block}).

By substituting eq. (\ref{R_block}) into eq. (\ref{dm_ij}) and focusing on the leading order terms of the seesaw expansion parameter $\rho$, we obtain
\beqa \label{dmij2}
({\delta M})_{ij} = \frac{{g_F}^2}{16 \pi^2} &\Big[& (R_3)_{\alpha \beta}(R_3)_{\gamma \beta} \left(\lambda^\alpha M^0 \lambda^\gamma \right)_{ij} \Delta b_0[M_\beta^2] \Big. \nonumber \\
& + & (R_3)_{\alpha \beta}(\rho^T R_3)_{m\beta} \left(\lambda^\alpha M^0\lambda^m + \lambda^m M^0\lambda^\alpha \right)_{ij}  \Delta b_0[M_\beta^2] \nonumber \\
& + & (R_5)_{mn}(R_5)_{pn} \left(\lambda^m M^0\lambda^p \right)_{ij} \Delta b_0[M_n^2]  \\
& - & \Big. (R_5)_{mn}(\rho R_5)_{\alpha n} \left(\lambda^m M^0\lambda^\alpha + \lambda^\alpha M^0\lambda^m \right)_{ij}  \Delta b_0[M_n^2]\, +\, {\cal O}(\rho^2) \Big]\,. \nonumber\eeqa
Recall that the indices \( \alpha, \beta, \dots = 1, 2, 3 \) correspond to the lighter gauge bosons, while \( m, n, \dots = 4, \dots, 8 \) correspond to the heavier ones. The first term in the expression for \( \delta M \) provides the dominant contribution since \( M_\alpha^2 < M_m^2 \). Due to the structure of \( \lambda^\alpha \), which has a vanishing first row and first column, this contribution is rank-one, generating mass only for the second-generation fermion. For \( M_\alpha < m_4 \), this second-generation mass is suppressed only by the loop factor relative to the third-generation mass. The masses of the first-generation fermions, however, arise from the remaining terms in eq. (\ref{dmij2}) and are further suppressed by factors of \( M_\alpha^2/M_m^2 \) or \( m_4^2/M_m^2 \) compared to the second-generation mass. This hierarchy ensures that the 1-loop-induced corrections produce the desired mass pattern for fermions if the scales satisfy:  
\be \label{scales}  
M_\alpha^2 \lesssim m_4^2 \lesssim M_m^2\,.  
\ee  
This relationship implies that the scale of \( SU(3)_F \) symmetry breaking and the mass scale of VL fermions must be comparable. However, the overall scale of these new states remains unconstrained by the fermion mass considerations alone, as the finite corrections depend only on the ratios of \( m_4 \) and \( M_a \).

\section{An explicit Model}
\label{sec:model}
Building on the general conditions required to generate a fermion mass hierarchy through quantum corrections induced by $SU(3)_F$ gaguge bosons, we present a specific and minimal implementation of the framework where these aspects are explicitly realized. As anticipated, we assume that the three generations of the SM fermions transform as fundamental representations under the horizontal gauged symmetry \( SU(3)_F \). Additionally, we include \( N_R \), a triplet of three SM singlet fermions under \( SU(3)_F \), which is essential for anomaly cancellation.  

The SM Higgs doublet is replaced by two Higgs doublets, each appearing in three copies to form triplets of \( SU(3)_F \). Furthermore, two SM singlet scalar fields, \( \eta_s \), transforming as triplets under \( SU(3)_F \), are introduced. These fields are necessary to achieve consistent gauge symmetry breaking and to generate the desired fermion mass matrices at the tree level.  

As outlined earlier in Section \ref{sec:su3fm}, the framework also requires VL fermions, which are assumed to be singlets under the new symmetry. The complete set of matter and scalar fields, along with their transformation properties under the SM and \( SU(3)_F \), is summarized in Table \ref{tab:fields:su3}.
\begin{table}[!t]
\begin{center}
\begin{tabular}{ccc} 
\hline
\hline
~~Fields~~&~~$(SU(3)_c \times SU(2)_L \times U(1)_Y)$~~&~~$SU(3)_F$~~~~\\
\hline
$Q_{L} $ & $(3,2,\frac{1}{3}) $ & 3 \\
$u_{R} $ & $(3,1,\frac{4}{3}) $ & 3 \\
$d_{R} $ & $(3,1,-\frac{2}{3}) $ & 3\\
\hline
$\Psi_{L} $ & $(1,2,-{1}{}) $ & 3 \\
$e_{R} $ & $(1,1,-2) $ & 3 \\
$N_{R} $ & $(1,1,0) $ & 3 \\
\hline
$H_{u}$ & $(1,2,-{1}{}) $ & 3 \\
$H_{d}$ & $(1,2,{1}{}) $ & 3 \\
{$\eta_1$,\,$\eta_2$}  &{ $(1,1,0) $} & ${3}$ \\
\hline
$T_L, \, T_R $ & $(3,1,\frac{4}{3}) $ & 1\\
$B_L, \, B_R $ & $(3,1,-\frac{2}{3}) $ & 1\\
$E_L, \, E_R $ & $(1,1,-2) $ & 1\\
\hline
\end{tabular}
\end{center}
\caption{The SM and \(G_F\) quantum numbers of different fermions and scalars in the model.}
\label{tab:fields:su3}
\end{table}

\subsection{\texorpdfstring{$SU(3)_F$ }{SU(3)_F} breaking and gauge boson mass ordering}
\label{subsec:su3breaking}
The absence of \( SU(3)_F \) gauge bosons in experimental observations so far suggests that the breaking scale of this symmetry is significantly higher than the weak scale. As a result, the primary breaking of the new gauge symmetry must be driven by the SM singlet fields \( \eta_{1,2} \), with contributions from the electroweak doublets expected to be negligible. Based on this reasoning, we focus on \( SU(3)_F \) breaking exclusively driven by \( \eta_{1,2} \).

The most general and renormalisable potential involving $\eta_{1,2}$ can be written as
\beqa \label{potential:su3}
V(\eta_1,\eta_2) &=& m_{11}^2\, \eta_1^\dagger \eta_1 + m_{22}^2\, \eta_2^\dagger \eta_2 - \left\{ m_{12}^2\, \eta_1^\dagger \eta_2 + {\rm h.c.} \right\} \nonumber \\
 &+& \frac{\xi_1}{2}\, (\eta_1^\dagger \eta_1)^2 +  \frac{\xi_2}{2}\, (\eta_2^\dagger \eta_2)^2 + \xi_3\,  (\eta_1^\dagger \eta_1) (\eta_2^\dagger \eta_2) + \xi_4\,  (\eta_1^\dagger \eta_2) (\eta_2^\dagger \eta_1) \nonumber \\
 &+& \left\{\frac{\xi_5}{2} (\eta_1^\dagger \eta_2)^2 + \xi_6\,  (\eta_1^\dagger \eta_1) (\eta_1^\dagger \eta_2) + \xi_7\, (\eta_2^\dagger \eta_2) (\eta_1^\dagger \eta_2) + {\rm h.c.} \right\}\,.\eeqa
 Here, all the parameters except $\xi_{5,6,7}$ and $m_{12}^2$ are real. 
In order to minimise the potential w.r.t $\eta_{1,2}$, we define the VEVs as:  
\be \label{eta_vevs}  
\langle \eta_1 \rangle = (v_F, 0, 0)^T\,, \quad \langle \eta_2 \rangle = (0, 0, \epsilon v_F)^T\,.  
\ee  
 The preferred way of the breaking of $SU(3)_F$, as given in eq. (\ref{su3:breaking}), requires \( \epsilon < 1 \). We will show that such a minima exists for a suitable choice of parameters. Also, without losing generality, one of these VEVs can always be chosen in this form using an \( SU(3)_F \) rotation. A single field in this configuration does not completely break the gauge symmetry, leaving an unbroken \( SU(2) \) subgroup. Therefore, at least two scalar fields with VEVs in different directions are required to fully break the \( SU(3)_F \) symmetry ( for an overview of $SU(N)$ breaking with vector representations, see \cite{Li:1973mq}).  
 
For the VEV configuration of \( \eta_{1,2} \) given in eq. (\ref{eta_vevs}), the minimisation of the potential leads to  
\beqa \label{minimum}
v\, (m_{11}^2 + v^2 \xi_1 + \epsilon^2 v^2 \xi_3) &=& 0 \,, \nonumber \\
\epsilon v\, (m_{22}^2 + \epsilon^2 v^2 \xi_2 + v^2 \xi_3) &=& 0\,.
\eeqa
The non-trivial solutions of these equations correspond to  
\be \label{vevs_expl}
v^2 = \frac{-m_{11}^2 \xi_2 + m_{22}^2 \xi_3}{\xi_1 \xi_2 - \xi_3^2}\,,~~~(\epsilon v)^2 = \frac{-m_{22}^2 \xi_1 + m_{11}^2 \xi_3}{\xi_1 \xi_2 - \xi_3^2}\,.
\ee
The VEVs are expressed in terms of real parameters \( m_{11}^2 \), \( m_{22}^2 \), and \( \xi_{1,2,3} \), with the latter constrained by the stability conditions of the potential:  
\be \label{stability}
\xi_{1,2} \geq 0\,,~~\xi_3 \geq - \sqrt{\xi_1 \xi_2}\,.  
\ee
For \( 0 > \xi_3 \geq -\sqrt{\xi_1 \xi_2} \), it follows that \( \xi_1 \xi_2 - \xi_3^2 \geq 0 \). Further assuming \( |m_{22}^2| \ll |m_{11}^2| \), \( \xi_1 \ll \xi_2 \), and \( m_{11}^2 < 0 \), the VEVs in eq. (\ref{vevs_expl}) remain real. Their ratio is then determined as  
\be \label{ratapp}
\epsilon^2 \approx -\frac{\xi_3}{\xi_2} \leq \sqrt{\frac{\xi_1}{\xi_2}} \ll 1\,.
\ee
Moreover, for \( \xi_3 \approx -\sqrt{\xi_1 \xi_2} \), the VEVs in eq. (\ref{vevs_expl}) correspond to the global minima of the potential among the available solutions of eq. (\ref{minimum}). In conclusion, the desired VEV configurations for \( \eta_{1,2} \) can be achieved while satisfying stability constraints for an appropriate choice of parameters.


Following the spontaneous breaking of $SU(3)_F$, the kinetic terms associated with $\eta_{1,2}$ result in the gauge boson mass matrix defined in eq. (\ref{LGB}):
\be \label{MGB}
{\cal M}^2_{a b} =\frac{g_F^2}{2}\,\sum_{s=1,2} \langle\eta_s\rangle^\dagger \lambda^{a \dagger} \lambda^{b} \langle\eta_s\rangle\,.\ee
Using eq. (\ref{eta_vevs}) and explicit forms of the generators $\lambda^a$, the above mass matrix can be written  in the notation of eq. (\ref{MGB_block}) as 
\beqa \label{MGB_block2}
M^2_{(33)} &=& \frac{g_F^2 v_F^2}{2}\,{\rm Diag.}\left(\epsilon^2, \epsilon^2, \epsilon^2\right)\,, \nonumber \\
M^2_{(55)} &=& \frac{g_F^2 v_F^2}{2}\,{\rm Diag.}\left(1, 1, 1+\epsilon^2, 1+\epsilon^2, \frac{1}{3}(4 + \epsilon^2) \right)\,, \nonumber \\
M^2_{(35)} &=& \frac{g_F^2 v_F^2}{2}\,\left(\ba{ccccc} 0 & 0 & 0 & 0 & 0 \\
0 & 0 & 0 & 0 & 0 \\
0 & 0 & 0 & 0 & -\frac{\epsilon^2}{\sqrt{3}} \\  \ea \right) \eeqa
The structure of the gauge boson mass matrix, in this case, is extremely simple, with mixing occurring solely between the states $A^\mu_3$ and $A^\mu_8$, both associated with diagonal generators.

Due to the simple form of the matrix ${\cal M}^2$,  the diagonalising matrix, parametrized by eq. (\ref{R_block}),  has following explicit forms for its block elements:

\be \label{R_res}
R_3 = \mathbb{I}_{3 \times 3}\,,~~R_5 = \mathbb{I}_{5 \times 5}\,,~~\rho = \left(\ba{ccccc} 0 & 0 & 0 & 0 & 0 \\
0 & 0 & 0 & 0 & 0 \\
0 & 0 & 0 & 0 & -\frac{\sqrt{3}}{4}\epsilon^2 \\  \ea \right)\,.\ee
The diagonal gauge boson mass matrix then has the form:
\be \label{MGB_diag}
{\cal D}^2 = \frac{g_F^2 v_F^2}{2}\,{\rm Diag.}\left(\epsilon^2, \epsilon^2, \epsilon^2, 1, 1, 1+\epsilon^2, 1+\epsilon^2, \frac{4}{3}+\frac{1}{3}\epsilon^2 \right)\, +\,  {\cal O}(\epsilon^4)\,. \ee
Therefore, the hierarchical gauge boson masses \( M^2_{1,2,3} \ll M^2_{4,\dots,8} \) are naturally achieved, as required to generate the mass gaps between the first and second-generation fermions.

Substituting eq. (\ref{R_res}) in the $(\delta M)_{ij}$, we find
\beqa \label{dmij3}
\frac{16 \pi^2}{g_F^2}(\delta M)_{ij} &= & \sum_{\alpha=1}^3 \left(\lambda^\alpha M^{0} \lambda^\alpha \right)_{ij} \Delta b_0 [M_\alpha^2] + \sum_{m=4}^8 \left(\lambda^m M^{0} \lambda^m \right)_{ij} \Delta b_0[M_m^2]  \nonumber \\
& + & \frac{\sqrt{3}}{4}\epsilon^2 \left(\lambda^3 M^{0} \lambda^8 + \lambda^8 M^{0} \lambda^3 \right)_{ij}  \left(\Delta b_0[M_3^2] - \Delta b_0[M_8^2] \right)\,\nonumber \\
&+&\,{\cal O}(\epsilon^4)\,. \eeqa
An apparent degeneracy of some of the gauge bosons allows further simplification. Denoting,
\beqa \label{MZ1-MZ2}
M_1^2 \simeq M_2^2 \simeq M_3^2 \equiv M_{Z_1}^2\,,\nonumber\\
~~ M_4^2 \simeq ... \simeq M_7^2 \simeq \frac{3}{4} M_8^2 \equiv M_{Z_2}^2\,,\eeqa
and using $\epsilon^2 = M_{Z_1}^2/M_{Z_2}^2$, we find
\beqa \label{dmij4}
\frac{16 \pi^2}{g_F^2}\,{\delta M} & \simeq & \left(\ba{ccc} 0 & 0 & 0\\ 0 & M^0_{22} + 2 M^0_{33} & -M^0_{23} \\ 0 & - M^0_{32} & 2 M^0_{22} + M^0_{33} \ea \right)\, \Delta b_0[M_{Z_1}^2]\nonumber \\
& + &  \left(\ba{ccc} 2(M^0_{22} + M^0_{33}) & 0 & 0\\ 0 & 2 M^0_{11}  & 0 \\ 0 & 0 & 2 M^0_{11}  \ea \right)\, \Delta b_0[M_{Z_2}^2]\nonumber \\
& + & \frac{1}{3} \left(\ba{ccc} 4 M^0_{11}   & -2 M^0_{12}  &  -2 M^0_{13} \\ -2 M^0_{21}  & M^0_{22}   &  M^0_{23}  \\ -2 M^0_{31}  & M^0_{32}  & M^0_{33}   \ea \right)\, \Delta b_0\Big[\frac{4}{3}M_{Z_2}^2\Big]  \\
& + & \frac{M_{Z_1}^2}{2 M_{Z_2}^2} \left(\ba{ccc} 0   & - M^0_{12}  &   M^0_{13} \\ - M^0_{21}  & M^0_{22}   & 0  \\  M^0_{31}  & 0  & -M^0_{33}   \ea \right) \left(\Delta b_0[M_{Z_1}^2] - \Delta b_0\Big[\frac{4}{3}M_{Z_2}^2\Big] \right). \nonumber\eeqa  
As expected, the first term exhibits a vanishing row and column, leading to mass generation for only the second-generation fermions. The remaining terms are suppressed by either \( m_4^2/M_{Z_2}^2 \) or \( M_{Z_1}^2/M_{Z_2}^2 \) relative to the second-generation mass, resulting in a small mass for the first-generation fermions.


\subsection{Charged fermion masses}
\label{subsec:CFM}
Using the defined set of fields and their transformation properties listed in Table \ref{tab:fields:su3}, the most general renormalisable Yukawa Lagrangian of the model can be expressed as
\beqa \label{yukawa}
-{\cal L}_{Y}&=& {y}_u \, \overline{Q}_{Li} H^i_u  T_{R}\, +   {y}_d \, \overline{Q}_{L i} H_d^i  B_R\, + {y}_e \, \overline{\Psi}_{Li} H_d^i  E_R\,  \nonumber \\
&+& {y}'^{(s)}_u \,\overline{T}_{L} \eta^\dagger_{s i}\,  u^{i}_{R} \,+ {y}'^{(s)}_d \,\overline{B}_L \eta^\dagger_{s i}\,  d^{ i}_R\, +  {y}'^{(s)}_e \,\overline{E}_L \eta^\dagger_{s i}\,  e^{ i}_R \, \nonumber \\
& + & m_T \overline{T}_L T_R+\, m_B \overline{B}_L B_R+\, m_E \overline{E}_L E_R\,+{\rm h.c.}\, \eeqa
where $ i=1,2,3$ is an $SU(3)_F$ index and $s= 1,2$ denotes multiplicity of $\eta$ fields. Here, all the fields are written in the flavour basis.

It is evident that following the breaking of $SU(3)_F$ and electroweak symmetries, the Yukawa interactions presented in eq. (\ref{yukawa}) result in tree-level mass matrices in the specified structure of eq. (\ref{M0_eff}) with
\be \label{mumup} 
\mu_f\,=\,\left(\ba{ccc}y_f v^f_1 & y_f v^f_2&y_f v^f_3 \ea \right)^T\,~ \text{and }\,~ \mu^\prime_f\,=\,\left(\ba{ccc} y^{\prime (1)}_f\, v_F & 0 &y^{\prime (2)}_f \, \epsilon v_F \ea \right)\,,\ee
where $f=u, d, e$ corresponds to the three types of charged fermions. Furthermore, $v^u_i = \langle H_u^i \rangle$, and $v^d_i = v^e_i = \langle H_d^i \rangle$. The VEVs of $\eta_{1,2}$ are given in eq. (\ref{eta_vevs}). The tree-level effective mass matrix, after integrating the heavy vector-like states, can be expressed as
\be \label{Meff_0:su3}
M_{u,d,e}^{0} \equiv - \frac{1}{m_{T,B,E}}\, \mu_{u,d,e}\, \mu^\prime_{u,d,e}\,.\ee
The specific forms of \( \mu_f \) and \( \mu^\prime_f \) given in eq. (\ref{mumup}) result in the matrix \( M_f^{0} \) having a vanishing second column.


At the 1-loop level, the mass matrices for the charged fermions are expressed as follows
\be \label{Meff_1}
M_f = M_f^0 + \delta M_f\,,\ee  
and the value of $\delta M_f$ can be deduced by applying the general formula, eq. (\ref{dmij4}). By incorporating the structure of the tree-level mass matrices, we obtain
\beqa \label{dm:model}
\delta M_f & \simeq & \frac{g_F^2}{16 \pi^2} \left[\left(\ba{ccc} 0 & 0 & 0\\ 0 &  2 (M^0_f)_{33} & -(M^0_f)_{23} \\ 0 & 0 &   (M^0_f)_{33} \ea \right)\, \Delta b_0[M_{Z_1}^2] \right. \nonumber \\
& + & 2 \left(\ba{ccc} (M^0_f)_{33} & 0 & 0\\ 0 & (M^0_f)_{11}  & 0 \\ 0 & 0 & (M^0_f)_{11}  \ea \right)\, \Delta b_0[M_{Z_2}^2]\nonumber \\
& + & \frac{1}{3} \left(\ba{ccc} 4 (M^0_f)_{11}   & 0  &  -2 (M^0_f)_{13} \\ -2 (M^0_f)_{21}  & 0   &  (M^0_f)_{23}  \\ -2 (M^0_f)_{31}  & 0  & (M^0_f)_{33}   \ea \right)\, \Delta b_0\Big[\frac{4}{3}M_{Z_2}^2\Big]  \\
& + & \left. \frac{\epsilon^2}{2} \left(\ba{ccc} 0   & 0  &   (M^0_f)_{13} \\ - (M^0_f)_{21}  & 0   & 0  \\  (M^0_f)_{31}  & 0  & -(M^0_f)_{33}   \ea \right)\, \left(\Delta b_0[M_{Z_1}^2] - \Delta b_0\Big[\frac{4}{3}M_{Z_2}^2\Big] \right) \right] \,. \nonumber\eeqa
In the following section, we show that the aforementioned $M_f$ is capable of reproducing the observed mass spectrum of charged fermions as well as the quark mixing patterns. 

\subsection{Neutrino masses}
\label{subsec:Neutrino}
As mentioned earlier, the anomaly-free nature of the model requires the inclusion of SM singlet fermions \( N_R \). Given the field content and symmetry of the model, it is clear that there is no Dirac Yukawa coupling between \( L_L \) and \( N_R \), nor is there a Majorana mass term for \( N_R \) at the renormalisable level. This precludes \( N_R \) from contributing to the light neutrino masses via the conventional type-I seesaw mechanism.

However, the model's symmetry allows for the following Weinberg operators:  
\beqa \label{nu_op}  
&&\frac{c_1}{\Lambda} \left(\overline{\Psi_L^c}^i H_{u i}^* \right) \left(\Psi_L^j  H_{u j}^\dagger \right)  
\,+ \, \frac{c_2}{\Lambda} \left(\frac{\lambda^a}{2}\right)^k_i \left(\frac{\lambda^a}{2}\right)^l_j \left(\overline{\Psi_L^c}^i  H_{u k}^* \right) \left(\Psi_L^j   H_{u l}^\dagger \right)  \nonumber\\
&&+ \frac{c_3}{\Lambda} \left(\overline{\Psi_L^c}^i \Psi_L^j  \right) \left(H_{u i}^\dagger H_{u j}^* \right),  
\eeqa
which can generate suppressed neutrino masses relative to the charged fermions. Extending the model to achieve an ultraviolet completion of these operators is straightforward. 

For instance, the simplest approach involves introducing two or more fermions, \( \nu_{R k} \), which are singlets under the full gauge symmetry and hence do not contribute to anomalies. These singlets can couple to the \( SU(3)_F \) triplet \( L_L^i \) and anti-triplet \( H_{u i}^\dagger \), producing a standard Dirac Yukawa term, and can also have a Majorana mass term unrestricted by the gauge symmetry of the model. Integrating out \( \nu_{R k} \) generates the first operator in eq. (\ref{nu_op}). Likewise, the second and third operators can arise from integrating out heavy \( SU(3)_F \) adjoint fermions and sextet scalars, respectively.

Unlike charged fermions, neutrinos can acquire masses at tree level through dimension-5 operators, bypassing the constraints imposed by the underlying flavour symmetries. This feature is particularly advantageous since the inter-generational mass hierarchy among neutrinos is significantly weaker compared to that of charged fermions. As a result, the neutrino masses and mixing parameters remain largely unconstrained within the framework of the effective theory. However, it is important to note that specific constraints may arise depending on the particular UV completion chosen.

\section{Numerical Solutions}
\label{sec:numerical}
To validate the model and understand the structure of its parameters, we perform a numerical analysis to find example solutions that reproduce the observed charged fermion masses and quark mixing parameters. By redefining various matter fields in eq. (\ref{yukawa}) to remove unphysical phases, we ensure that the parameters \( y_f \), \( y_f^{\prime (1)} \), and \( m_{T,B,E} \) are real. Additionally, we assume all VEVs are real, leaving only three complex parameters in the model: \( y_{u,d,e}^{\prime (2)} \).  

From eq. (\ref{mumup}), this results in real values for \( (\mu_f)_i \) (for \( i=1,2,3 \)), \( (\mu_f^\prime)_1 \), and a complex \( (\mu_f^\prime)_3 \). Furthermore, we impose the relation  
\be \label{mue-mud}  
\mu_e =\frac{y_e}{y_d} \mu_d \equiv r\, \mu_d\,,  
\ee  
where \( r \) is a real parameter. In total, the model contains 21 real parameters, including real values for \( \mu_{ui} \), \( \mu_{di} \), \( r \), \( \mu_{u1}^\prime \), \( \mu_{d1}^\prime \), \( \mu_{e1}^\prime \), \( m_T \), \( m_B \), \( m_E \), \( M_{Z_1} \), and \( M_{Z_2} \), along with complex \( \mu_{u3}^\prime \), \( \mu_{d3}^\prime \), and \( \mu_{e3}^\prime \). These parameters must account for 13 observables: 9 charged fermion masses, 3 quark mixing angles, and a CP phase. Despite having more parameters than observables, it is not immediately clear whether the model can successfully reproduce the experimental values due to various constraints and correlations among parameters, as we describe below.

\begin{table}[t]
\begin{center}
\begin{tabular}{cccc} 
\hline
\hline
~~P~~&~~Solution 1 (S1)~~&~~Solution 2 (S2)~~&~~Solution 3 (S3)~~\\
 \hline
$M_{Z_1}$         & $10^4$ GeV     & $10^6$ GeV      & $10^8$ GeV \\
$\epsilon$        & $ 0.0284 $     & $ 0.0365 $      & $ 0.0239 $ \\
$\epsilon_T$      & $ 0.0123 $     & $ 0.0155   $    & $ 0.0106 $ \\
$\epsilon_B$      & $ 0.5742  $    & $ 0.9587   $    & $0.7666  $ \\
$\epsilon_E$      & $ 0.8626  $    & $ 0.9019   $    & $ 0.6797 $ \\
\hline
$\epsilon_{u1}$        & $  -0.2351  $    & $0.3936  $   & $ -0.1797 $ \\
$\epsilon_{u2}$        & $  0.0272  $    & $ -0.3794  $   & $ 0.1144 $ \\
$\epsilon_{u3}$        & $ 0.2581 $       & $ -0.8025  $   & $ 0.2438 $ \\
\hline
$\epsilon^\prime_{u1}$ & $ -0.5324 \times 10^{-4} $      & $0.3015 \times 10^{-3}$     & $ -0.2084\times 10^{-3} $ \\
$\epsilon^\prime_{u3}$ & $ -0.0341  + i\, 0.0066 $     & $ 0.0144  - i\,0.0063$ & $ -0.0160- i\,0.0279 $ \\
\hline
$\epsilon_{d1}$        & $-0.3782  $    & $0.2035   $    & $ -0.2885  $ \\
$\epsilon_{d2}$        & $0.0675  $    & $ -0.1759 $    & $ 0.2067  $ \\
$\epsilon_{d3}$        & $ 0.4154 $    & $ -0.4239 $    & $ 0.3919  $ \\
\hline
$\epsilon^\prime_{d1}$ & $ 0.0168  $                         & $ 0.0317  $                         & $0.0240  $ \\
$\epsilon^\prime_{d3}$ & $ 0.0016  + i\,0.43 \times 10^{-3}$ & $ 0.0022 + i\,0.0008  $ & $ 0.0022+ i\,0.0010 $ \\
\hline
$r$                    & $ -0.2676 $         & $-0.2293 $          & $0.3748 $ \\
\hline
$\epsilon^\prime_{e1}$ & $0.0118 $            & $-0.0325 $          & $-0.0083 $\\
$\epsilon^\prime_{e3}$ & $0.0545 +i\, 0.0017$ & $-0.0708-i\,0.0027  $ & $ -0.0341   - i\, 0.11 \times 10^{-3} $\\
\hline
\hline
\end{tabular}
\end{center}
\caption{Three different set of the optimised values of the model parameters (P) for different $M_{Z_1}$ which lead to viable charged fermion masses and quark mixing.}
\label{tab:sol:su3}
\end{table}


To simplify the discussion, different dimension-full parameters can be represented using the mass scales and dimensionless variables. Since, it is desirable to have $M_{Z_1}^2 \lesssim m_{T,B,E}^2 \lesssim M_{Z_2}^2$, as suggested in eq. (\ref{scales}), to obtain the hierarchical spectrum, we introduce
\be \label{dimless1}
m_T = \epsilon_T M_{Z_2},\, m_B = \epsilon_B M_{Z_2},\, m_E = \epsilon_E M_{Z_2}, \ee
with $\epsilon_{T,B,E}$ as dimensionless parameters.
Additionally, the $\epsilon$ parameter is obatined from $M_{Z_1} = \epsilon M_{Z_2}$. 

Since $\mu_{fi}$ are solely originated from the electroweak symmetry breaking, we define
\be \label{dimless3}
\mu_{ui} = \epsilon_{ui}  v,\, ~~\mu_{di} = \epsilon_{di}  v, \ee
where $v=174$ GeV.

Similarly, for flavour symmetry-breaking parameters  $\mu_{fi }^\prime$, we write
\be \label{dimless2}
\mu_{f1 }^\prime = \epsilon_{f 1}^\prime M_{Z_2},\, ~~\mu_{f3}^\prime = \epsilon_{f 3}^\prime M_{Z_2}\,.\ee 
 In this way, various dimensionless parameters $\epsilon$ and $\epsilon^\prime$ can preferably take values less than unity.

For a given value of \( M_{Z_1} \), the remaining dimensionless parameters are determined using the \( \chi^2 \) optimisation technique, as detailed in chapter \ref{chap3}. Three benchmark solutions obtained through this method are presented in Table \ref{tab:sol:su3} for different \( M_{Z_1} \). The minimised \( \chi^2 \) values for these solutions are 6.97, 6.90, and 6.46 for S1, S2, and S3, respectively. Additionally, Table \ref{tab:res} provides the resulting charged fermion masses and quark mixing parameters for each solution, alongside their corresponding experimental values for comparison.

\begin{table}[t]
\begin{center}
\begin{tabular}{ccccc} 
\hline
\hline
~Observable~& Value &~~S1~~&~~S2~~&~~S3~~\\
\hline
$m_u$ [MeV] & $1.27 \pm 0.5$       & $1.27$    &$1.30$     & $1.30$\\
$m_c$ [GeV] & $0.619 \pm 0.084$    &$0.618$    &$0.623$    & $0.609$\\
$m_t$ [GeV] & $171.7 \pm 3.0$      & $171.7$   &$171.6$    & $171.7$\\
\hline
$m_d$ [MeV] & $2.90 \pm 1.24$      & $3.17$    &$2.39$    & $3.46$\\
$m_s$ [GeV] & $0.055 \pm 0.016$    & $0.005$   &$0.005$   & $0.009$\\
$m_b$ [GeV] & $2.89 \pm 0.09$      & $2.89$    &$2.89$    & $2.90$\\
\hline
$m_e$ [MeV] & $0.487 \pm 0.049$    & $0.486$   &$0.488$   &$0.486$ \\
$m_\mu$ [GeV]& $0.1027 \pm 0.0103$ & $0.1027$  &$0.100$   & $0.100$\\
$m_\tau$ [GeV] & $1.746 \pm 0.174$ & $1.738$   &$1.762$   & $1.806$\\
\hline 
$|V_{us}|$ & $0.22500 \pm 0.00067$ & $0.22396$  &$0.22739$  & $0.21922$\\
$|V_{cb}|$ & $0.04182 \pm 0.00085$ & $0.04175$  &$0.04251$  & $0.04099$\\
$|V_{ub}|$ & $0.00369 \pm 0.00011$ & $0.00368$  &$0.00375$  & $0.00364$\\
$J_{CP}$ & $(3.08 \pm 0.15)\times 10^{-5}$ & $3.09\times 10^{-5}$ & $3.03\times 10^{-5}$ & $3.1\times 10^{-5}$\\
\hline
\hline
\end{tabular}
\end{center}
\caption{The fitted values of the charged fermion masses and quark mixing parameters at the minimum \( \chi^2 \) value for the three example solutions are presented in Table \ref{tab:sol:su3}. The second column shows the experimentally measured values of the corresponding observables, extrapolated to \( M_Z \), which were used in the \( \chi^2 \) function.}
\label{tab:res}
\end{table}

Several key features of the model can be inferred from Table \ref{tab:sol:su3}. The minimised \( \chi^2 \) values remain nearly the same for different choices of \( M_{Z_1} \), indicating that the ability to reproduce realistic flavour hierarchies depends on the relative masses of the new gauge bosons and VL states rather than the overall flavour symmetry breaking scale. This aligns with the expectation that flavour hierarchies are technically natural. All \( \epsilon_{fi} \) values are of \( \mathcal{O}(10^{-1}) \), suggesting that the fundamental Yukawa couplings \( y_f \) are of the same order, with no significant hierarchy among the VEVs \( v^{u,d}_i \). Additionally, as expected from eq. (\ref{mumup}), the fitted values show \( \epsilon^\prime_{f3} < \epsilon^\prime_{f1} \) for \( f = u, d \). However, achieving the fitted values of \( \epsilon^\prime_{e1} \) requires a two-order-of-magnitude separation between \( y_e^{\prime(1)} \) and \( y_u^{\prime(1)} \). Overall, the fundamental Yukawa couplings in this model span only two orders of magnitude, in contrast to the SM, where they vary across at least five orders. Since third-generation fermions acquire their masses at the tree level, the hierarchy between \( m_t \) and \( m_{b,\tau} \) does not naturally emerge and instead requires \( \epsilon_T \ll \epsilon_{B,E} \).  

From Table \ref{tab:res}, it can be noticed that all observables, except for \( m_s \), are fitted within the \( \pm 1\sigma \) range of their reference values across all solutions. The fitted value of \( m_s \) deviates by approximately \( 3\sigma \) from its experimental value evolved at $M_Z$ scale. Despite the model containing more parameters than observables, its inability to precisely match the central value of \( m_s \) suggests the presence of non-trivial correlations among observables arising from the predictive nature of the non-Abelian flavour symmetry. Notably, a more precise measurement of the strange quark mass could potentially falsify the model, irrespective of the \( SU(3)_F \) breaking scale.

\section{Flavour violation}
\label{sec:fv}
As we have done in previous chapters, we discuss the phenomenological implications due to the inherent presence of flavour-changing neutral currents induced by the gauge bosons of $SU(3)_F$ only. We study them in detail in this section by first deriving the general dimension-6 effective operators and then estimating various relevant quark and lepton flavour transitions.

Rewriting eq. (\ref{L_gauge:su3_2}) in the physical basis of fermions and gauge bosons, one finds
\be \label{Lm_guage}
-{\cal L}_{\rm gauge} = \frac{g_F}{2}  \left( \overline{f}_{L i} \gamma^\mu   \left(\tilde{\lambda}^a_{f}{}_L \right)_{ij}  f_{L  j}\, +\,  \overline{f}_{R i} \gamma^\mu  \left(\tilde{\lambda}^a_{f}{}_R \right)_{ij}  f_{R j} \right)\, {\cal R}_{ab}B^b_\mu\,, \ee
where
\be \label{}
 \tilde{\lambda}^a_{f}{}_{L,R} = U^{f}{}^\dagger_{L,R}\, \lambda^a\, U^{f}{}_{L,R}\,,\ee
and $U^{f}{}_{L,R}$ are the unitary matrices that diagonalise the 1-loop corrected mass matrix $M_f$. From above,  integrating out the gauge bosons, we find the effective dimension-6  four fermion operators as
\beqa \label{d6_ops}
{\cal L}_{\rm eff} =&& C^{(ff^\prime)LL}_{ijkl}\, \overline{f}_{Li} \gamma^\mu f_{L j}\,\overline{f^\prime}_{L k} \gamma_\mu f^\prime_{L l} \nonumber \\ &+& C^{(ff^\prime)RR}_{ijkl}\, \overline{f}_{R i} \gamma^\mu f_{R j}\,\overline{f^\prime}_{R k} \gamma_\mu f^\prime_{R l}\,\nonumber \\
 &+& C^{(ff^\prime)LR}_{ijkl}\, \overline{f}_{L i} \gamma^\mu f_{L j}\,\overline{f^\prime}_{R k} \gamma_\mu f^\prime_{R l}\nonumber \\ & +& C^{(ff^\prime)RL}_{ijkl}\, \overline{f}_{R i} \gamma^\mu f_{R j}\,\overline{f^\prime}_{L k} \gamma_\mu f^\prime_{L l}\,,\eeqa
where
\be \label{C}
C^{(ff^\prime)P P^\prime}_{ijkl} = \frac{g_F^2}{8 M_b^2}\,{\cal R}_{ab} {\cal R}_{cb}\, \left(\tilde{\lambda}^a_{f}{}_P \right)_{ij} \left(\tilde{\lambda}^c_{f^\prime}{}_{P^\prime} \right)_{kl}\,,\ee
and $P,P^\prime = L,R$ and $f, f^\prime = u, d, e$. For a hierarchical mass spectrum of gauge bosons, the coefficients of the effective operators can be further simplified. By applying eqs. (\ref{MGB_block},\ref{R_block}), we obtain at the leading order in $\rho$ as
\beqa \label{C_1}
\frac{8}{g_F^2} C^{(ff^\prime)P P^\prime}_{ijkl} &\simeq & \frac{1}{M_\alpha^2} \left( (R_3)_{\beta \alpha} (R_3)_{\gamma \alpha}  \left(\tilde{\lambda}^\beta_f{}_P \right)_{ij} \left(\tilde{\lambda}^\gamma_{f^\prime}{}_{P^\prime} \right)_{kl} \right. \nonumber \\
&&\hspace{6mm}+ \left.  (R_3)_{\beta \alpha} (\rho^T R_3)_{m \alpha}  \left(\tilde{\lambda}^\beta_{f}{}_P \right)_{ij} \left(\tilde{\lambda}^m_{f^\prime}{}_{P^\prime} \right)_{kl} \nonumber \right. \\
&&\hspace{6mm}+\left.  (\rho^T R_3)_{m \alpha} (R_3)_{\beta \alpha}  \left(\tilde{\lambda}^m_{f}{}_P \right)_{ij} \left(\tilde{\lambda}^\beta_{f^\prime}{}_{P^\prime} \right)_{kl}  \right) \nonumber \\
& + & \frac{1}{M_m^2} \left( (R_5)_{nm} (R_5)_{pm}  \left(\tilde{\lambda}^n_{f}{}_P \right)_{ij} \left(\tilde{\lambda}^p_{f^\prime}{}_{P^\prime} \right)_{kl} \right. \nonumber \\
&&\hspace{6mm}- \left.  (\rho R_5)_{\alpha m} (R_5)_{nm}  \left(\tilde{\lambda}^\alpha_{f}{}_P \right)_{ij} \left(\tilde{\lambda}^n_{f^\prime}{}_{P^\prime} \right)_{kl} \nonumber \right.\\
&&\hspace{6mm}- \left. (R_5)_{nm} (\rho R_5)_{\alpha m}  \left(\tilde{\lambda}^n_{f}{}_P \right)_{ij} \left(\tilde{\lambda}^\alpha_{f^\prime}{}_{P^\prime} \right)_{kl}  \right)\,.\eeqa
It can be seen that the first three terms are proportional to $M^{-2}_{\alpha}$. Out of these, $2$nd and $3$rd terms have a relative suppression $\rho$ compared to the first. Similarly, the other three terms with proportionality $M^{-2}_{n}$ follow the same trend for relative suppression among them.

Further simplification is possible within the explicit model by utilising eqs. (\ref{R_res}) and (\ref{MZ1-MZ2}). Substituting these into the above equation, we obtain:
\beqa \label{C_2}
\frac{8 M_{Z_1}^2}{g_F^2}\, C^{(ff^\prime) P P^\prime}_{ijkl} &\simeq &  \sum_{\alpha=1}^3 \left(\tilde{\lambda}^\alpha_{f}{}_P \right)_{ij} \left(\tilde{\lambda}^\alpha_{f^\prime}{}_{P^\prime} \right)_{kl} + \epsilon^2 \sum_{m=4}^7 \left(\tilde{\lambda}^m_{f}{}_P \right)_{ij} \left(\tilde{\lambda}^m_{f^\prime}{}_{P^\prime} \right)_{kl} \nonumber \\
& - & \frac{\sqrt{3}}{4} \epsilon^2 \left( \left(\tilde{\lambda}^3_{f}{}_P \right)_{ij} \left(\tilde{\lambda}^8_{f^\prime}{}_{P^\prime} \right)_{kl}  + \left(\tilde{\lambda}^8_{f}{}_P \right)_{ij} \left(\tilde{\lambda}^3_{f^\prime}{}_{P^\prime} \right)_{kl} \right) \nonumber \\
& + & \frac{3}{4} \epsilon^2 \left(\tilde{\lambda}^8_{f}{}_P \right)_{ij} \left(\tilde{\lambda}^8_{f^\prime}{}_{P^\prime} \right)_{kl} + {\cal O} (\epsilon^4)\,.\eeqa

At leading order, flavour violation in the model is dictated by the coupling matrices \(\tilde{\lambda}^\alpha_f{}_{L,R}\). Since these matrices do not commute, it is not possible to simultaneously diagonalise all of them for any choice of \(U_{L,R}\). As a result, the most significant flavour-violating effects arise from the coefficients  
\beqa \label{C_3}
C^{(ff^\prime) P P^\prime}_{ijkl} &\simeq & \frac{g_F^2}{8 M_{Z_1}^2}  \sum_{\alpha=1}^3 \left(\tilde{\lambda}^\alpha_{f}{}_P \right)_{ij} \left(\tilde{\lambda}^\alpha_{f^\prime}{}_{P^\prime} \right)_{kl}\,.
\eeqa  
This expression can be used to estimate the leading contributions to various flavour-violating processes in both the quark and lepton sectors.

\subsection{Quark sector}
The strongest constraints on various $C^{(ff^\prime) P P^\prime}_{ijkl}$ primarily arise from meson-antimeson oscillations such as $K^0-\overline{K}^0$, $B^0_d-\overline{B^0}_d$, $B^0_s-\overline{B^0}_s$ and $D^0-\overline{D}^0$. To quantify these constraints, we closely follow the procedure adopted in the previous chapters.  For $M^0-\overline{M^0}$ transitions, the effective Wilson coefficients $C^i_M$, defined in eq. (\ref{H_eff}), can obtained at $\mu=M_{Z_1}$ as:
\begin{align}
C^1_K \,=\, -C^{(dd)LL}_{1212} &,& \tilde{C}^1_K= -C^{(dd)RR}_{1212} &,&{C}^5_K= -4\, C^{(dd)LR}_{1212}\,,\\
C^1_{B_d} \,=\, -C^{(dd)LL}_{1313} &,& \tilde{C}^1_{B_d}= -C^{(dd)RR}_{1313} &,&{C}^5_{B_d}= -4\, C^{(dd)LR}_{1313}\,,\\
C^1_{B_s} \,=\, -C^{(dd)LL}_{2323} &,& \tilde{C}^1_{B_s}= -C^{(dd)RR}_{2323} &,&{C}^5_{B_s}= -4\, C^{(dd)LR}_{2323}\,,\\
C^1_{D} \,=\, -C^{(uu)LL}_{1212} &,& \tilde{C}^1_D= -C^{(uu)RR}_{1212} &,& {C}^5_D= -4\, C^{(uu)LR}_{1212}\,.
\end{align} 
The remaining $C^i_M$ and $\tilde{C}^i_M$ are zero at this high scale.

Using the renormalisation group equations, we evolve all the coefficients from \( \mu = M_{Z_1} \) to their respective mesonic scales, as discussed in previous chapters. This running induces a non-vanishing \( C^4_M \), while \( C^{2,3}_M \) and \( \tilde{C}^{2,3}_M \) remain zero. The evolved Wilson coefficients are computed using eq. (\ref{C_3}) for the three benchmark solutions are listed in Table \ref{tab:sol:su3} and are compared with the corresponding experimental limits provided by the UTFit collaboration \cite{UTfit:2007eik}. The results are summarized in Table \ref{tab:meson_WC:su3}. 
\begin{table}[t]
\begin{center}
\begin{tabular}{ccccc} 
\hline
\hline
~~WC ~~&~~Allowed range~~&~~S1~~&~~S2~~&~~S3~~\\
 \hline
Re$C_K^1$ & $[-9.6,9.6]\times 10^{-13}$         & \rb{$ 3.8\times 10^{-12}$}  & \gb{$-2.5 \times 10^{-15} $} & \gb{$ 3.4 \times 10^{-19}$}\\
Re$\tilde{C}_K^1$ & $[-9.6,9.6]\times 10^{-13}$ & \gb{$ 6.1\times 10^{-15}$} & \gb{$ 2.1\times 10^{-19}$}  & \gb{$-9.9 \times 10^{-23}$}\\
Re$C_K^4$ & $[-3.6,3.6]\times 10^{-15}$         & \rb{$ 3.7 \times 10^{-10}$} & \rb{$-3.7 \times 10^{-15}$}  & \gb{$ 2.0 \times 10^{-18}$} \\
Re$C_K^5$ & $[-1.0,1.0]\times 10^{-14}$         & \rb{$ 3.3 \times 10^{-10}$} & \gb{$ -3.2 \times 10^{-15}$} & \gb{$ 1.6 \times 10^{-18}$}\\
Im$C_K^1$ & $[-9.6,9.6]\times 10^{-13}$         & \rb{$ 2.1 \times 10^{-11}$} & \gb{$ 1.4\times 10^{-15}$}   & \gb{$ 8.6 \times 10^{-20}$}\\
Im$\tilde{C}_K^1$ & $[-9.6,9.6]\times 10^{-13}$ &\gb{$ 4.2 \times 10^{-15}$} & \gb{$3.1 \times 10^{-19}$}  & \gb{$-1.5 \times 10^{-22}$}\\
Im$C_K^4$ &  $[-1.8,0.9]\times 10^{-17}$        & \rb{$ 5.8 \times 10^{-10}$}  & \rb{$ 1.5 \times 10^{-14}$} &\gb{$ -2.8 \times 10^{-18}$} \\
Im$C_K^5$ & $[-1.0,1.0]\times 10^{-14}$         &\rb{$  5.1\times 10^{-10}$} & \rb{$ 1.3 \times 10^{-14}$} & \gb{$ -2.2 \times 10^{-18}$}\\
\hline
$|C_{B_d}^1|$ & $<2.3\times 10^{-11}$          & \rb{$ 2.4\times 10^{-10}$}  & \gb{$ 1.4\times 10^{-14}$}  & \gb{$ 2.0 \times 10^{-18}$} \\
$|\tilde{C}_{B_d}^1|$ & $<2.3\times 10^{-11}$  & \gb{$ 8.5\times 10^{-12}$}  & \gb{$ 4.1\times 10^{-16}$}  & \gb{$ 7.8\times 10^{-20}$}\\
$|C_{B_d}^4|$ &  $<2.1\times 10^{-13}$         &  \rb{$ 1.3 \times 10^{-10}$}   & \gb{$ 7.7\times 10^{-15}$} & \gb{$1.6 \times 10^{-18}$}\\
$|C_{B_d}^5|$ & $<6.0\times 10^{-13}$          & \rb{$ 2.2 \times 10^{-10}$}  & \gb{$ 1.3\times 10^{-14}$} & \gb{$ 2.5\times 10^{-18}$}\\
\hline
$|C_{B_s}^1|$ & $< 1.1 \times 10^{-9}$        & \gb{$ 2.0  \times 10^{-11 }$} & \gb{$ 6.4\times 10^{-16 }$} & \gb{$ 1.7\times 10^{ -19}$}\\
$|\tilde{C}_{B_s}^1|$ & $< 1.1 \times 10^{-9}$ & \gb{$ 9.5\times 10^{-13}$}  & \gb{$ 9.8\times 10^{-17}$}  & \gb{$2.4 \times 10^{-20}$}\\
$|C_{B_s}^4|$ & $< 1.6 \times 10^{-11}$       & \gb{$ 5.8 \times 10^{-12}$}  & \gb{$ 9.4\times 10^{-16}$}  & \gb{$ 1.6\times 10^{-19}$} \\
$|C_{B_s}^5|$ & $< 4.5 \times 10^{-11}$       & \gb{$ 1.0 \times 10^{-11}$}  & \gb{$ 1.6\times 10^{-15}$}  & \gb{$  2.4\times 10^{-19}$}\\
\hline
$|C_D^1|$ & $<7.2 \times 10^{-13}$            & \rb{$ 1.5\times 10^{-12 }$} & \gb{$ 2.4\times 10^{-15 }$} & \gb{$2.8 \times 10^{ -19}$}\\
$|\tilde{C}_D^1|$ & $<7.2 \times 10^{-13}$  & \gb{$ 9.4 \times 10^{ -17}$} & \gb{$ 8.2\times 10^{-18}$}  & \gb{$8.0\times 10^{-23}$}\\
$|C_D^4|$ & $<4.8\times 10^{-14}$           & \gb{$ 4.4\times 10^{-14}$}   & \gb{$ 5.5\times 10^{-16}$}  & \gb{$ 2.0\times 10^{-20}$} \\
$|C_D^5|$ & $<4.8 \times 10^{-13}$          & \gb{$ 5.2\times 10^{-14}$}   & \gb{$ 6.1\times 10^{-16}$}  & \gb{$2.1 \times 10^{-20}$}\\
\hline
\hline
\end{tabular}
\end{center}
\caption{Numerical values of WCs (in \({\rm GeV}^{-2}\) units) for operators contributing to meson-antimeson oscillations, estimated for three example solutions. The experimentally allowed ranges are taken from \cite{UTfit:2007eik}. Values highlighted in red exceed the respective limits and are therefore excluded.}
\label{tab:meson_WC:su3}
\end{table}

\subsection{Lepton sector}

As noted in the previous section, the dominant contribution to the flavour violation process is governed by the first three gauge bosons $B^\alpha_\mu $. The exchange of $B^\alpha_\mu $ mediate lepton flavour violating process like $\mu \to e$ conversion in nuclei, $l_i \to 3 l_j $ and $l_i \to l_j \gamma$. The first two processes arise at the tree level, whereas the latter is at the one-loop level in the present model. 

The branching ratio for the process $\mu \to e$ conversion in the field of the nucleus can be computed using eq. (\ref{mu2e}) and (\ref{gLV}). The couplings $g^{(q)}_{LV, RV}$, $q=u,d$, appearing in these expressions,  can be parametrized for the present theory as
\beqa \label{gLV_su3}
g^{(q)}_{LV} &\sim & \frac{\sqrt{2}}{G_F} \, \frac{1}{2}
\left[C^{(eq) L L}_{2111}\,+\,C^{(eq) L R}_{2111} \right]\,\nonumber \\
g^{(q)}_{RV} &\sim & \frac{\sqrt{2}}{G_F} \, \frac{1}{2}
\left[C^{(eq) R R}_{2111}\,-\,C^{(eq) R L}_{2111} \right]\, .\eeqa
The computed branching fraction for the three benchmark solutions are given in Table \ref{tab:lfv}.  It can be seen that the present experimental limit disfavours S1  in this case.
\begin{table}[t]
\begin{center}
\begin{tabular}{ccccc}
\hline
\hline
~~LFV observable~~&~~Limit~~&~~S1~~&~~S2~~&~~S3~~\\
\hline
${\rm BR}[\mu \to e]$ & $< 7.0 \times 10^{-13}$    & \rb{$ 4.6 \times 10^{-9}$} & \gb{$ 1.2 \times 10^{-17 }$}   & \gb{$ 3.1 \times 10^{-25 }$}\\
\hline
${\rm BR}[\mu \to 3e]$ & $< 1.0 \times 10^{-12}$   & \rb{$1.2 \times 10^{-11 }$} & \gb{$8.7  \times 10^{-20 }$}  & \gb{$ 5.6 \times 10^{-27 }$}\\
${\rm BR}[\tau \to 3\mu]$ & $< 2.1 \times 10^{-8}$ & \gb{$8.5\times 10^{-12 }$} & \gb{$1.2 \times 10^{-19 }$}    & \gb{$ 5.3 \times 10^{-27 }$}\\
${\rm BR}[\tau \to 3 e]$ & $< 2.7 \times 10^{-8}$  & \gb{$1.7 \times 10^{-10}$} & \gb{$ 1.5\times 10^{-19 }$}    & \gb{$ 9.4 \times 10^{-27 }$}\\
\hline
${\rm BR}[\mu \to e \gamma]$ & $< 4.2 \times 10^{-13}$    & \rb{$ 3.6\times 10^{-10 }$} & \gb{$ 1.8\times 10^{-17 }$}  & \gb{$ 3.5 \times 10^{-25 }$}\\
${\rm BR}[\tau \to \mu \gamma]$ &  $< 4.4 \times 10^{-8}$ & \gb{$5.6\times 10^{-13 }$}& \gb{$5.1 \times 10^{-20 }$}    & \gb{$1.1  \times 10^{-27 }$}\\
${\rm BR}[\tau \to e \gamma]$ &  $< 3.3 \times 10^{-8}$   & \gb{$ 2.5\times 10^{ -12}$} & \gb{$ 1.3\times 10^{-20 }$}  & \gb{$2.3  \times 10^{-28}$}\\
\hline
\hline
\end{tabular}
\end{center}
\caption{Branching ratios evaluated for various charged lepton flavour violating processes for the three example solutions listed in Table \ref{tab:sol:su3}. The corresponding experimental limits are extracted from \cite{Calibbi:2017uvl}.  The values excluded by the limits are highlighted in red.}
\label{tab:lfv}
\end{table}

In the present model, the exchange of new gauge boson $B^\alpha_\mu$  dominantly contributes to the trilepton decay \( l_i \to 3 l_j \) at the leading order. Following the approach outlined in \cite{Ramond:1999vh}, the decay width for this process can be systematically estimated. In the limit \( m_i \gg m_j \), the expression takes the form
\beqa \Gamma[l_i \to 3 l_j] &=&  \frac{4 \, m_i^5}{1536}\, \sum_{P,P'}\left|
C^{(ee) P P^\prime}_{jijj}\right|^2\,,\eeqa
which, using eq. (\ref{C_3}), takes the following form
\beqa
\label{l_i3lj}\Gamma[l_i \to 3 l_j] \,=\, \frac{g_F^4}{16}\frac{ m_i^5}{1536}\, \sum_{\alpha,\beta=1}^3 &{}& \left[ \left(\tilde{\lambda}^\alpha_{e}{}_L \right)_{ji} \left(\tilde{\lambda}^\beta_{e}{}_L \right)_{ji} +  \left(\tilde{\lambda}^\alpha_{e}{}_R \right)_{ji} \left(\tilde{\lambda}^\beta_{e}{}_R \right)_{ji} \right] \, \nonumber\\
& & \times \left[ \left(\tilde{\lambda}^\alpha_{e}{}_L \right)_{jj} \left(\tilde{\lambda}^\beta_{e}{}_L \right)_{jj} +  \left(\tilde{\lambda}^\alpha_{e}{}_R \right)_{jj} \left(\tilde{\lambda}^\beta_{e}{}_R \right)_{jj} \right]\, . 
\eeqa
The branching ratios for  $\mu \to 3 e $, $\tau \to 3 e $ and $\tau \to 3 \mu $  calculated using the above expression are given in Table \ref{tab:lfv} for three solutions along with their corresponding experimental limits.

The contribution arising from the exchange of $B^\alpha_\mu$ to the process $l_i \to l_j \gamma$ can be estimated by following \cite{Lavoura:2003xp}. The computed decay width in the approximation $M_{Z_1} \gg m_i, m_j$ can be parametrized as
\be \label{i2j_gamma} \Gamma[l_i \to l_j \gamma] \,\simeq\, \frac{\alpha {g_F}^4}{64} \left(1-\frac{m_j^2}{m_i^2}\right)^3\, {m_i^5}\,\left( \left|\sigma_L \right|^2\,+\, \left|\sigma_R \right|^2\right)\,,\ee
where
\be \sigma_L \,=\, \sum_{\alpha=1}^{3} \sum_{k=1}^{3} \left[Y_1\, \frac{m_j}{m_i}\left(\tilde{\lambda}^\alpha_L \right)_{jk}\left(\tilde{\lambda}^\alpha_L \right)_{ki} \, +\,Y_2\,\left(\tilde{\lambda}^\alpha_R \right)_{jk}\left(\tilde{\lambda}^\alpha_R \right)_{ki}\,
- 4 Y_3\,\frac{m_k}{m_i}\,\left(\tilde{\lambda}^\alpha_R \right)_{jk}\left(\tilde{\lambda}^\alpha_L \right)_{ki}\right]\, . \ee
In a similar manner, $\sigma_R$ can be derived by swapping $L$ and $R$ in the above expression. The loop functions $Y_1, Y_2$, and $Y_3$ are defined as follows:
\be Y_1 = Y_2 = 2a+6c+3d\, ~~\,~~
Y_3 = a + 2c\,, \ee
and the explicit expressions of $a, c$, and $d$ are given in \cite{Lavoura:2003xp}. Using eq. (\ref{i2j_gamma}), the estimated branching ratios for $\mu \to e  \gamma$, $\tau \to \mu \gamma$ and $\tau  \to e \gamma $  are listed in Table \ref{tab:lfv} for the three example solutions.

By comparing the estimated magnitudes of various flavour-violating observables in Tables \ref{tab:lfv} and \ref{tab:meson_WC:su3} with their corresponding experimental limits, we find that among the three benchmark solutions, only \( S3 \) remains viable. This suggests that the lowest possible scale for new physics is approximately \( 10^3 \) TeV. However, a slightly lower scale could be allowed if we impose a specific ordering of the parameters \( \mu_{fi} \) and \( \mu^\prime_{fi} \), as discussed in Section \ref{subsec:numerical}. Since our primary goal is to demonstrate the existence of viable solutions, we do not explore such cases where fine-tuning of parameter values might permit a marginally lower phenomenologically allowed scale for new physics. Additionally, even in such cases, the predicted scale would still be higher than the estimates presented in Chapter \ref{chap4}, which considers optimal charge assignments for a flavour-nonuniversal \( U(1)_F \).

\section{Conclusion}
\label{sec:concl}
We have demonstrated that a gauged horizontal \( SU(3)_F \) symmetry can be effectively utilised to ensure that only the third-generation fermions acquire mass at the tree level, while the masses of the lighter generations arise from self-energy corrections. Our analysis shows that radiative corrections typically generate masses for both the second and first-generation fermions at the one-loop level, necessitating a separate explanation for the hierarchy between them. We establish that this hierarchy can be achieved if \( SU(3)_F \) undergoes a two-step breaking with an intermediate \( SU(2) \) symmetry. This results in a mild hierarchy among the gauge bosons of the local flavour group, which is then transferred to the fermion sector through quantum corrections. To illustrate this mechanism, we construct an explicit model and demonstrate how it successfully reproduces the observed hierarchical structure of quark and lepton masses.

Phenomenologically, the breaking scale of \( SU(3)_F \) is primarily constrained by flavour violation. The new gauge bosons exhibit \( \mathcal{O}(1) \) flavour-changing couplings with the SM fermions as also observed in the previous models. This leads to significant rates for flavour-violating processes. Constraints on such processes imply that the lowest viable new physics scale must be approximately \( \mathcal{O}(10^3) \) TeV. A similar conclusion is reached in Chapter \ref{chap3}, which explores a flavour-nonuniversal \( U(1) \times U(1) \) symmetry. As previously noted, this predicted scale is nearly two orders of magnitude higher than the lowest allowed new physics scale for the \( U(1)_F \) symmetry with optimal charge assignments, as discussed in Chapter \ref{chap4}.  

Although the current framework does not achieve the scenario of optimal flavour violation, it introduces two key improvements over previous models. First, the gauge boson mass hierarchy in Chapter \ref{chap3}, or the smallness of the \( \epsilon \) parameter in Chapter \ref{chap4}, both introduced as ad hoc assumptions, now emerge naturally from a sequential breaking of a single gauge group, \( SU(3)_F \). Second, the use of a single non-Abelian flavour group results in a more predictive framework by reducing the number of free Yukawa couplings in theory. After removing unphysical phases, the number of independent Yukawa couplings is reduced from 20 in the previous two chapters to 12 in the present model. This reduction introduces correlations among the masses of various quarks and charged leptons. One example of this is evident in the model’s preference for a strange quark mass that is \( 3\sigma \) lower than the current experimental value. Therefore, the present scenario can also be tested by precise measurements of the light quark masses.

%% file: 60_Chapter_6/chapter_6.tex
\chapter{Left-Right symmetric extension and Accommodating the strong CP solution}
\label{chap6}
\graphicspath{{60_Chapter_6/}}
We briefly discussed the strong CP puzzle in chapter \ref{strong CP} and mentioned that the physical CP violating phase in the strong interaction sector is $\bar{\theta } =\, \theta_{QCD}\, +\, \arg (\det(M_u M_d))\,$. In the radiative mass models described in chapter \ref{chap3} and \ref{chap4}, the first two generation fermions remain massless at tree level. This, in turn, makes \( \bar{\theta} \) unphysical and can be fully rotated away by redefining the massless chiral fields similar to massless quark solutions of strong CP. However, this doesn't hold true when the quantum corrections are added to induce the lighter generation masses. In other words, large CP-violating phases are induced when all of the fermion masses are generated. So, the radiative mass models based on the SM extension frameworks don't solve the strong CP puzzle automatically.

In this chapter, we show that the radiative mass mechanism, when implemented in a left-right (L-R) framework, finds a possible explanation for a strong CP puzzle. An additional requirement is the imposition of parity invariance, which makes the mass matrices to possess Hermitian structure. One crucial difference between this framework and Babu-Mohapatra solutions \cite{Babu:1988mw,Babu:1989rb} is that here, the lighter generation fermion masses are induced radiatively as opposed to their model, where masses were obtained through the universal seesaw mechanism \cite{Davidson:1987mh}. A similar framework is proposed in \cite{Cherchiglia:2024ssz}, which explains the quark mass hierarchies through the seesaw mechanism and solves the strong CP puzzle by incorporating the so-called Neslon-Barr mechanism. A key characteristic of our mechanism is that, since the masses in our scenario are partially computable parameters of the theory, the CP phase \(\bar{\theta}\) also becomes a computable quantity.

The remainder of the chapter is structured as follows: Section \ref{chap5:model} presents the explicit model incorporating radiative mass generation and a strong CP solution. Section \ref{sec:masses-theta} examines fermion mass generation and the strong CP phase across different orders in perturbation theory. Section \ref{sec:qualitative} discusses the qualitative features of the framework along with some phenomenological analysis. Finally, Section \ref{sec:summary} summarizes our findings. 

\section{Overview of the Model}
\label{chap5:model}
The gauge symmetry of the model is  
\(
SU(3)_C \times SU(2)_L \times SU(2)_R \times U(1)_{B-L} \times G_F
\)
with parity invariance imposed. Here, \( G_F = U(1)_{2-3} \) is a generalized version of the well-known \( L_\mu - L_\tau \) symmetry, extended to include all fermions. The particle content of the model is summarized in Table \ref{tab:particle}, where \( i = 1,2,3 \) represents the three generations of SM fermions. Under the new flavour non-universal symmetry \( G_F \), the second and third-generation fermions carry charges, while the first generation remains neutral. In addition to the SM fermions, each sector of charged fermions includes a pair of vector-like fermions, which are taken neutral under the flavour symmetry. The model also introduces three copies of scalar fields \( H_{Li} \) and \( H_{Ri} \), which transform as doublets under \( SU(2)_L \) and \( SU(2)_R \), respectively.

\begin{table}[!ht]
\begin{center}
\begin{tabular}{ccc}
\hline
\hline
~~Particles~~ &~~ ${\cal G}_{LRSM}$~~&~~ $G_F$\\
\hline
$Q_{Li}=\left(\ba{c}u\\d \ea \right)_{Li}$ & $ \left(3,2,1,\frac{1}{3} \right)$ & ~~ $ \{0,1,-1 \}$ \\
$Q_{Ri}=\left(\ba{c}u\\d \ea \right)_{Ri}$ & $\left(3,1,2,\frac{1}{3} \right) $& ~~ $ \{0,1,-1 \}$ \\
$\Psi_{Li}=\left(\ba{c}\nu\\e \ea \right)_{Li}$ & $ \left(1,2,1,-{1} \right)$& ~~ $ \{0,1,-1 \}$ \\
$\Psi_{Ri}=\left(\ba{c}\nu\\e \ea \right)_{Ri}$ & $\left(1,1,2,-{1}\right)$ & ~~ $ \{0,1,-1 \}$ \\
\hline 
$H_{Li}$ & $\left(1,2,1,1 \right)$ & ~~ $ \{0,1,-1 \}$ \\
$H_{Ri}$ & $\left(1,1,2,1 \right)$& ~~ $ \{0,1,-1 \}$ \\
\hline
$U_{L,R}$ & $\left(3,1,1,\frac{4}{3} \right)$ & 0\\
$D_{L,R}$  & $\left(3,1,1,-\frac{2}{3} \right)$ & 0\\
$E_{L,R}$ & $\left(1,1,1,-2 \right)$ & 0 \\
\hline
\end{tabular}
\end{center}
\caption{Particle contents of the model.}
\label{tab:particle}
\end{table}

The transformation properties of the fermions and scalars under the parity are defined below:
 \begin{align}
 \label{parity}
Q_{Li} & \longleftrightarrow\,  Q_{Ri} ,&\, \Psi_{Li}\longleftrightarrow  \,  \Psi_{Ri}\,, \nonumber\\
F_{L}&  \longleftrightarrow  \,  F_{R} ,&\, H_{Li} \longleftrightarrow  \,  H_{Ri} \,,
\end{align}
with $F= U, D, E$ denoting three types of vector-like fermions with electromagnetic charges $\frac{2}{3}, -\frac{1}{3}, -1$ respectively .

The covariant derivative for the gauge symmetry can  written as: 
\be D_\mu = \begin{cases}
    \ba{c} \partial_\mu \,+\, i g W^i_{\mu L} \frac{\sigma^i}{2} \,+\, i g_1 B_\mu \frac{Y_1}{2}\, +\, i g_X X_\mu \frac{X}{2} ~~~~~~\text{For LH fields}\\ \partial_\mu \,+\, i g W^i_{\mu R} \frac{\sigma^i}{2} \,+\, i g_1 B_\mu \frac{Y_1}{2}\, +\, i g_X X_\mu \frac{X}{2}     
 ~~~~~~\text{For RH fields}\ea  
\end{cases}\,.\ee
The couplings \( g \), \( g_1 \), and \( g_X \) correspond to the gauge bosons of \( SU(2)_{L,R} \), \( U(1)_{B-L} \), and \( U(1)_F \), respectively. To ensure invariance under the parity transformation \( W_{\mu L} \leftrightarrow W_{\mu R} \), the coupling \( g \) is taken to be the same for both \( SU(2)_L \) and \( SU(2)_R \).  The quantum numbers \( Y_1 \) and \( X \) represent the charges under \( B-L \) and \( G_F \), respectively, for the fields on which the covariant derivative acts. When the scalar fields acquire VEVs, the full gauge symmetry undergoes spontaneous breaking, leaving \( SU(3)_C \times U(1)_{EM} \) as the unbroken subgroup. Diagrammatically, the symmetry-breaking pattern is:
\beqa \label{breaking}SU(3)_C \times SU(2)_L \times SU(2)_R \times U(1)_{B-L} \times G_F & \xrightarrow[]{\langle H_{Ri}\rangle} &\, {SU(3)_C \times SU(2)_L \times U(1)_Y} \,\nonumber \\  & \xrightarrow[]{\langle H_{Li}\rangle}& \,  SU(3)_C \times U(1)_{EM}\, . \eeqa
In this framework, the hyper-charge $Y$ and the electromagnetic charge $Q$ are defined as follows:
\be \frac{Y}{2} = T^3_R \,+\, \frac{Y_1}{2}\, , ~~~~~~~ Q\,=\,T^3_L \,+\, \frac{Y}{2}\, . \ee
The VEVs mentioned in eq. (\ref{breaking}) are denoted as follows:
\be \langle H_{Li} \rangle = v_{Li} \, ~~~ \text{and}~~~ \langle H_{Ri} \rangle = v_{Ri}\, , \label{vevs}\ee
and it will be subsequently shown that both $v_{Li}$ and $v_{Ri}$ can be chosen as real without losing generality.\\

The renormalisable scalar potential, which is both gauge and parity invariant, can be expressed as:
\beqa \label{potential}
V &=& \mu_{L i}^2\, H_{L i}^\dagger H_{L i}\, +\, \mu_{Ri}^2\, H_{R i}^\dagger H_{R i}\, +\,   (\lambda)_{ij}\,\left[ (H_{L i}^\dagger H_{L i})\,( H_{L j}^\dagger H_{L j})\,+\, (H_{R i}^\dagger H_{R i})\,( H_{R j}^\dagger H_{R j})\,\right] \nonumber \\
& + &  (\tilde{\lambda})_{ij}\,\left[ (H_{L i}^\dagger H_{L j})\,( H_{L j}^\dagger H_{L i})\,+\, (H_{R i}^\dagger H_{R j})\,( H_{R j}^\dagger H_{R i})\,\right] \,+\,   (\lambda^\prime)_{ij}\, (H_{L i}^\dagger H_{L i})\,( H_{R j}^\dagger H_{R j})\nonumber \\
&+&  (\tilde{\lambda^\prime})_{ij} \,(H_{L i}^\dagger H_{L j})\,( H_{R j}^\dagger H_{R i}) \,,\eeqa 
with $(\,)$ bracket indicating singlet combination. To softly break the parity symmetry, the condition \( \mu_{L i}^2 \neq \mu_{R i}^2 \) is imposed. Additionally, the matrices \( \tilde{\lambda} \) and \( \tilde{\lambda}^\prime \) can have vanishing diagonal elements without any loss of generality.  

Moreover, it can be verified that all the parameters in the potential are real.
\begin{itemize}
    \item $\mu_{L,R i}^2, \lambda, \tilde{\lambda} $ and ${\lambda^\prime}$  are real parameters as  corresponding operators are self-conjugate.
    \item $(\tilde{\lambda^\prime})_{ij}$ is real as the conjugate of the associative operator is the same as its parity-transformed one.
\end{itemize}   The vacuum expectation values are entirely determined by the potential parameters. Since all the parameters in the potential are real, we assume that the resulting vacuum expectation values are also real.


For the choice of vevs given in eq. (\ref{vevs}), the masses of the charged gauged bosons can be written as; 
\be M^2_{W^{\pm}_L} \,=\, \frac{1}{4} g^2 \sum_i v^2_{Li}, \,~~~\text{and}~~M^2_{W^{\pm}_R} \,=\, \frac{1}{4} g^2 \sum_i v^2_{Ri}.\ee
At the leading order, the charged gauge bosons do not mix. However, the neutral gauge bosons do, and their mass-squared matrix, expressed in terms of \((W^3_L{}_\mu, W^3_R{}_\mu, B_\mu, X_\mu)\), can be parameterised as:
{\small
\be \label{mass:gauge} {\mathbb{M}}^2 \,=\,\frac{1}{4} 
\left(\ba{cccc} g^2 \sum_i v^2_{Li} & 0 & -g\,g_1\,   \sum_i v^2_{Li} & g\,g_X (v^2_{L3}-v^2_{L2}) \\
0& g^2 \sum_i v^2_{Ri}  & -g\,g_1\,   \sum_i v^2_{Ri} & g\,g_X (v^2_{R3}-v^2_{R2}) \\
-g\,g_1 \sum_i v^2_{Li} & -g\,g_1 \sum_i v^2_{Ri} & g^2_1\,   \sum_i (v^2_{Li}\,+\,v^2_{Ri}) & g_1\,g_X \sum_P(v^2_{P2}-v^2_{P3}) \\
 g\,g_X (v^2_{L3}-v^2_{L2}) & g\,g_X (v^2_{R3}-v^2_{R2}) & g_1\,g_X \sum_{P}(v^2_{P2}-v^2_{P3})  & g^2_X (v^2_{R2}+v^2_{R3}) \ea \right)\,.
\ee}
Here, the summation \(\sum_P\) includes terms with \(P = L, R\). It follows that the \(4 \times 4\) mass matrix, \(\mathbb{M}^2\), has a vanishing determinant, with its upper-left \(3 \times 3\) block also having a determinant zero. The corresponding massless eigenstate can be identified as the photon \( A_\mu \), with its coupling constant defined as:
\be \frac{1}{e^2}\,=\,\frac{2}{g^2}\,+\, \frac{1}{g_1^2}\, .\ee
Additionally, the submatrix:  
\be  
\left( \ba{cc}  
g^2 \sum_i v^2_{Ri}  & -g\,g_1\,   \sum_i v^2_{Ri} \\  
-g\,g_1 \sum_i v^2_{Ri} & g^2_1\,   \sum_i (v^2_{Li}\,+\,v^2_{Ri})  
\ea\right)  
\ee  
has a vanishing eigenvalue in the limit \( v_{Li} \to 0 \). For small but nonzero \( v_{Li} \), the lighter gauge boson, with a mass proportional to \( v_{Li} \), can be identified as the SM \( Z \) boson, while the heavier state, with a mass proportional to \( v_{Ri} \), corresponds to the \( Z_R \) boson.  The direct search limits on  \( W_R \) and  \( Z_R \) boson imply $v_{Ri} \gg v_{Li}$. It can be demonstrated that such a fine-tuned minima can be obtained from the scalar potential defined in eq. (\ref{potential}) (for details, see \cite{Mohanta:2024wmh}).

In this minimal framework, the VEVs of the \( H_{Ri} \) fields break both L-R symmetry and \( U(1)_F \). Since these symmetries share a common breaking source, the right-handed sector and the \( U(1)_F \) gauge boson acquire masses of similar magnitude. Consequently, for further analysis, the masses of \( X \), \( Z_R \), and \( W^\pm_R \) are taken to be of the same order as \( M_X \), the mass of the \( X \) boson.

From eq. (\ref{mass:gauge}), the mixing between \( X \) and \( Z_R \) is suppressed by a factor of \( \mathcal{O}\left(\frac{v^2_{R3}-v^2_{R2}}{v^2_{R3}+v^2_{R2}}\right) \). As the $X$ boson couplings are flavour non-universal in nature, the symmetry breaking introduces flavour-violating couplings as discussed in chapter \ref{chap:2}. The aforementioned mixing then induces FCNC couplings for the $Z_R$ boson. In a similar fashion, the mixing of $X$ boson and the SM $Z$ boson introduces FCNC couplings for the $Z$ boson. However, for a multi-TeV \( X \) boson, its mixing with the \( Z \) boson is further suppressed by \( \mathcal{O}(\frac{m^2_Z}{M^2_X}) \) ensuring compatibility with all electroweak constraints (see, for example, section \ref{subsec:direct-search}).

\section{Fermion masses and  \texorpdfstring{$\bar{\theta}$ }{\bar{\theta}} }
\label{sec:masses-theta}
\subsection{Tree level}
\label{chap5:subsec:1loop}
The enforcement of a well-defined parity symmetry ensures that the QCD vacuum angle, \( \theta_{QCD} \), is absent from the Lagrangian. 
%
%
As a result, the contribution to \( \bar{\theta} \) originates solely from the second term of eq. (\ref{theta-bar}), which arises from the phases of the quark mass matrices.  

Given the set of fields listed in Table \ref{tab:particle}, the most general renormalisable fermionic mass Lagrangian that is both gauge and parity invariant can be expressed as:
  \beqa 
  -{\cal L}_y &=&  y^d_{i} \left(\bar{Q}_{Li}\,  {H}_{Li}\, D_{R} \, +\,\bar{Q}_{Ri} \, {H}_{Ri}\,  D_{L} \right) \,+\,y^e_{i} \left(\bar{\Psi}_{Li}\,  {H}_{Li} \, E_{R} \,+\,\bar{\Psi}_{Ri} \, {H}_{Ri}\,  E_{L} \right)\, \nonumber \\
  &+& y^u_{1} \left(\bar{Q}_{L1}\,  \tilde{H}_{L1} \, U_{R} \,+ \,\bar{Q}_{R1} \, \tilde{H}_{R1}\,  U_{L} \right)\, +\, y^u_{2} \left(\bar{Q}_{L2}\,  \tilde{H}_{L3} \, U_{R} \,+ \,\bar{Q}_{R2} \, \tilde{H}_{R3}\,  U_{L} \right)\, \nonumber \\ 
  &+& y^u_{3} \left(\bar{Q}_{L3}\,  \tilde{H}_{L2} \, U_{R} \,+ \,\bar{Q}_{R3} \, \tilde{H}_{R2}\,  U_{L} \right)\, \,+\, m_U\, \bar{U}_{L} U_{R } \, +\, m_D\, \bar{D}_{L} D_{R } \nonumber\\ &+& m_E\, \bar{E}_{L} E_{R }  \, +\, H.c\,.
  \eeqa
The mass terms for vector-like fermions, \( m_{U,D,E} \), are real due to parity invariance. Additionally, field redefinitions allow the absorption of certain phases in the Yukawa couplings \( y^{u,d,e} \). Specifically, it can be shown that \( y^{u,e}_i \) and \( y^d_3 \) can be chosen to be real, leaving \( y^d_{1,2} \) as the only complex parameters in the theory. These complex parameters are responsible for generating the weak CP phase, which manifests in the CKM matrix. Furthermore, an accidental CP symmetry exists in the up-quark and charged lepton sectors at the tree-level Lagrangian.

When both the flavour symmetry and the LR symmetry are spontaneously broken down to \( SU(3)_C \times U(1)_{EM} \), the mass matrices for the charged fermions take the following form:
  \beqa {\cal M}_u^{(0)} &=&  \left( \ba{cc} 0_{3\times 3} & \ba{c}y^u_1\, v_{L1}\\y^u_2\, v_{L3}\\y^u_3\, v_{L2}\ea \\\ba{ccc}y^u_1 v_{R1}& y^u_2 v_{R3} & y^u_3 v_{R2}\ea & m_U \ea\right),\,\nonumber\\  {\cal M}_{d}^{(0)} &=&  \left( \ba{cc} 0_{3\times 3} & \ba{c}y^{d}_1\, v_{L1}\\y^{d}_2\, v_{L2}\\y^{d}_3\, v_{L3}\ea \\\ba{ccc}y^{d}_1{}^* v_{R1}& y^{d}_2{}^* v_{R2} & y^{d}_3 v_{R3}\ea & m_{D} \ea\right),  \label{Mf_0_1}\eeqa
  with $ v_{Li}, v_{Ri}$ as the VEVs defined in eq. (\ref{vevs}) which are real. The mass matrix \( {\cal M}_{e}^{(0)} \) for charged leptons can be obtained by replacing all \( y^d_i, y^d_i{}^* \) with real \( y^e_i \) values. 
  
  It follows that the diagonal elements of the mass matrices in eq. (\ref{Mf_0_1}) are real, while the phases of the off-diagonal elements are equal and opposite to those of their respective transposed counterparts. This structure ensures a real determinant (in this case, it is exactly zero), as discussed in Appendix \ref{real-determinant}. Due to this property, these matrices are referred to as "Hermitian Type." Additionally, two out of the four eigenvalues vanish, implying that at tree level, a massless quark solution holds, leading to an unphysical \( \bar{\theta} \).

The mass matrices presented in eq. (\ref{Mf_0_1}) can be expressed concisely as: 
   \be \label{M0:LR} {\cal M}_f^{(0)} =  \left( \ba{cc} 0 & \mu_f \\ \mu_f^\prime & m_F \ea\right)\, ,\ee
  with  $m_F$ as the is the VL fermion mass term for $F$ type, and
  \beqa \label{chap6:munmup} \mu_u &=& \left( \ba{ccc}y^u_1 v_{L1}& y^u_2 v_{L3} & y^u_3 v_{L2}\ea \right)^T\,, ~~  \mu^\prime_u = \left( \ba{ccc}y^u_1 v_{R1}& y^u_2 v_{R3} & y^u_3 v_{R2}\ea \right)\,, \nonumber \\
  \mu_{d} &=& \left( \ba{ccc}y^{d}_1 v_{L1}& y^{d}_2 v_{L2} & y^{d}_3 v_{L3} \ea \right)^T\,, ~~  \mu^\prime_{d} = \left( \ba{ccc}y^{d}_1{}^* v_{R1}& y^{d}_2{}^* v_{R2} & y^{d}_3 v_{R3}\ea  \right) \,,\nonumber\\
  \mu_{e} &=& \left( \ba{ccc}y^{e}_1 v_{L1}& y^{e}_2 v_{L2} & y^{e}_3 v_{L3} \ea \right)^T\,, ~~  \mu^\prime_{e} = \left( \ba{ccc}y^{e}_1 v_{R1}& y^{e}_2 v_{R2} & y^{e}_3 v_{R3}\ea  \right)\,.\eeqa 

As previously mentioned, the matrix \( {\cal M}_f^{(0)} \) is of the Hermitian type, as the complex part of \( \mu_{fi} \) is proportional to \( y^d_i \), while that of \( \mu^\prime_{fi} \) is proportional to \( y^{d*}_i \). 

The structure of \( {\cal M}_f^{(0)} \) in eq. (\ref{M0:LR}) closely resembles that of eq. (\ref{M0}). As a result, it also possesses two nonzero eigenvalues, which correspond to the masses of the third-generation SM fermions and their associated vector-like fermion states.  

In the seesaw limit, the effective \( 3 \times 3 \) mass matrices for the SM fermions take a form similar to eq. (\ref{M0_eff}):
\be \label{chap5:M0} M^{(0)}_f = -\frac{1}{m_F}\mu_f \mu_f^\prime\, .\ee
The Hermitian-type structure of \( M^{(0)}_f \) remains intact, as the complex parts of \( \mu_f \) and \( \mu^\prime_f \) are conjugates of each other. Additionally, this mass matrix generates masses only for the third-generation fermions, while the masses of the lighter-generation fermions arise through self-energy corrections. These corrections are induced by the flavour non-universal gauge boson and may contribute to the second term of eq. (\ref{theta-bar}).  

Notably, higher-order corrections to \( \bar{\theta} \) need to be evaluated only for the down-quark sector, as the up-quark mass matrix given in eq. (\ref{chap5:M0}) is real and does not introduce complex parameters, even when gauge corrections are included.  

Quantum gravity effects are expected to violate all global symmetries, including parity. However, for the proposed solution to remain valid, such gravity-induced corrections must be sufficiently suppressed. The leading dimension-five operators that could contribute are:  
\be  
{\cal L}^{d=5} \,=\, \frac{{\cal O}(1)}{M_{Pl}}\, \bar{Q}_{Li}Q_{Rj} H^\dagger_{Rj}H_{Li} + \dots  
\ee  
These operators introduce non-Hermitian corrections to eq. (\ref{chap5:M0}). However, such contributions are suppressed by a factor of \( m_F/M_{Pl} \). For \( m_F \lesssim 10^{-6} M_{Pl} \), these terms are too small to even play a role in inducing first-generation fermion masses. Consequently, such corrections can be safely neglected.

\subsection{1-loop masses and \texorpdfstring{$\bar{\theta}$ }{\bar{\theta}}}
\label{sec:modelB}
As already discussed in chapter \ref{chap:2}, the tree-level mass matrices get corrections induced by diagrams involving massive fermions and the $X$-gauge boson in the loop (see Fig. \ref{fig:1loop}). Such corrections can be evaluated by following the procedure outlined in section \ref{subsec:1loop}, and the resulting 1-loop corrected mass matrix can be written as 
\be \label{chap5:M1_1}
{\cal M}_f^{(1)} = \left( \ba{cc} \left(\delta M_f^{(0)}\right)_{3 \times 3} & \mu_{f}\\ \mu^\prime_{f} & m_F \ea \right)\,.\ee
where
\be \label{chap5:dM0_eff} \left(\delta M_f^{(0)}\right)_{ij} \simeq \frac{g_X^2}{16 \pi^2}\, q_{Li}\, q_{Rj}\, \left(M_f^{(0)}\right)_{ij}\, \left(b_0[M_X,m_3^{(0)}] - b_0[M_X,m_4^{(0)}] \right)\,.\ee
Here, $b_0$ is the finite part of the Passarino-Veltmann function defined in eq. (\ref{b0}), and $q_{Li}$ ($q_{Ri}$) is the $U(1)_F$ charges of the SM chiral field $f_{Li}$ ($f_{Ri}$). Explicitly,
\beqa
q_{Li}&=& q_{Ri} ~\cong~ q_{i}\,=\, \lbrace 0,1,-1\rbrace\,.
\eeqa

In the seesaw limit, the unitary matrices that block-diagonalise \( {\cal M}^{(1)} \) from eq. (\ref{chap5:M1_1}) can be approximated as  
\be \label{chap5:U1_ss}  
{\cal U}_f^{(1)}{}_{L,R} \approx \left(\ba{cc} U_f{}_{L,R}^{(1)} & - \rho_f{}_{L,R}^{(1)} \\  
 \rho_f{}_{L,R}^{(1) \dagger} U_f{}_{L,R}^{(1)} & 1 \ea\right)\,,  
\ee  
where \( \rho_f^{(1)}{}_{L,R} = \rho_f^{(0)}{}_{L,R} \) represent the seesaw expansion parameters. In this model, they take the form  
\be  
\rho_f^{(0)}{}_{L} = -\frac{\mu_f}{m_F}, \quad \rho_f^{(0) \dagger}{}_{R} = -\frac{\mu^\prime_f}{m_F}.  
\ee  
The matrices \( U_f{}_{L,R}^{(1)} \) are the \( 3\times 3 \) unitary matrices that diagonalise the effective one-loop corrected \( 3\times 3 \) mass matrix:  
\beqa \label{chap5:M1_eff}  
\left(M_f^{(1)}\right)_{ij} &=& \left(M_f^{(0)}\right)_{ij} + \left(\delta M_f^{(0)}\right)_{ij} \nonumber \\  
&=&  \left(M_f^{(0)}\right)_{ij} \left( 1+ q_{Li} q_{Rj} C \right)\,,  \eeqa  
where the correction factor is given by  
\be  
C = \frac{g_X^2}{16\pi^2} \left(b_0[M_X,m_3^{(0)}] - b_0[M_X,m_4^{(0)}]\right).  
\ee  
It follows from eq. (\ref{chap5:M1_eff}) that the determinant of this mass matrix remains zero, ensuring the presence of at least one massless state. For flavour non-universal charges, second-generation fermion masses are generated, and the diagonalisation of the effective mass matrix \( M^{(1)} \) takes the form:  
\be \label{chap5:M1_eff_diag}  
U_f{}_L^{(1) \dagger} M_f^{(1)} U_f{}_R^{(1)} = {\rm Diag.}\left(0,m_{f2}^{(1)},m_{f3}^{(1)}\right)\,,  
\ee  
where \( m_{fi}^{(n)} \) represents the mass of the \( i^{\rm th} \) generation fermion at the \( n^{\rm th} \) order. Since one of the eigenvalues of \( {\cal M}^{(1)} \) (given in eq. (\ref{chap5:M1_1})) is zero, \( \bar{\theta} \) remains an unphysical parameter at 1-loop level also. Additionally, from eq. (\ref{chap5:M1_eff}), it is evident that the Hermitian structure of the mass matrix is also preserved.

In addition to corrections induced by the \( X \)-boson, the mass matrix also receives contributions from diagrams involving physical neutral scalar mixings. These corrections can be computed following the procedure outlined in Section \ref{chap2:UV} and are given by  
\be \label{dm1:scalar3}  
(\delta M^{(S)}_f)_{ ij}\,=\, -\frac{m_F}{16\pi^2}\, {y^f_{i}y^{f*}_{j}}\sum_a  \,({\cal R}_{L})_{ ia}\,({\cal R}_{R})_{ ja} B_0[m_{Sa},m_F]\, .  
\ee  
This contribution is proportional to \( {y^d_{i}y^{d*}_{j}} \), similar to \( (M^{(0)}_f)_{ij} \), ensuring that the phases of \( (\delta M^{(S)}_f)_{ ij} \) and \( (M^{(0)}_f)_{ij} \) remain identical. As a result, the scalar corrections at the one-loop level are also of the Hermitian type and do not induce a strong CP phase.


However, when these corrections are added to the mass matrix given in eq. (\ref{chap5:M1_eff}), they can potentially generate first-generation fermion masses exclusively at the one-loop level. Since the masses of first-generation fermions are significantly suppressed compared to those of the second generation, an additional mechanism is required to account for this hierarchy. To avoid this scenario as well as improve the computability power of the mechanism, it is preferable to minimise these contributions by considering small scalar mixings, as discussed in Section \ref{chap2:UV}.

It is worth to note that, a possible contribution to the neutron electric dipole moment, \( d^e_n \) which is also a CP violating parameter, arises at 1-loop level when a photon line is attached to the internal fermion lines in the left panel of Fig. \ref{fig:1loop}. This contribution is proportional to the imaginary part of \( \delta M^d_{11} \). For the X-boson exchange diagram, \( \delta M^d_{11} \) remains real at the one-loop level (see eq. (\ref{chap5:dM0_eff})), and the same holds for the scalar-mediated contribution given in eq. (\ref{dm1:scalar3}). Consequently,  
\be  
d^e_n ~\propto ~ {\rm Im}\left((\delta M^{(X)}_d)_{11}\,+\,(\delta M^{(S)}_d)_{11}\right) = 0 \,.  
\ee  
This result is a distinctive feature of our model, stemming from the fact that vector-like fermion masses are real. In contrast, other models that employ the universal seesaw mechanism generate fermion masses by treating vector-like fermion mass terms as complex parameters. As a consequence, those models generally predict a nonzero \( d^e_n \) at the one-loop level.

\subsection{2-loop masses and \texorpdfstring{$\bar{\theta}$ }{\bar{\theta}}}
\label{sec:2loop}
As mentioned earlier, the one-loop corrections induced by the flavoured gauge boson generate masses only for second-generation fermions. In this subsection, we demonstrate that two-loop corrections can produce viable first-generation fermion masses. Following \cite{Mohanta:2024wcr}, the two-loop corrected mass matrix, incorporating \( X \)-boson-induced effects (similar to  of Fig. \ref{fig:2loop}), can be parameterised as: 
\beqa \label{M2_eff_fnl_f:scp}
\left(M^{(2)}_f\right)_{ij} &=& \left(M^{(0)}_f\right)_{ij} \left(1+\frac{g_X^2}{16 \pi^2}\, q_{Li}\, q_{Rj}\, (b_0[M_X,m_{f3}^{(1)}]- b_0[M_X,m_F]) \right) \nonumber \\
&+& \left(\delta M_f^{(0)}\right)_{ij} \left(1+\frac{g_X^2}{16 \pi^2}\, q_{Li}\, q_{Rj}\, b_0[M_X,m_{f3}^{(1)}] \right) \\
&+& \frac{g_X^2}{16 \pi^2}\, q_{Li}\, q_{Rj} \,\left(U_{fL}^{(1)}\right)_{i2} \left(U_{fR}^{(1)}\right)^*_{j2}\, m_{f2}^{(1)}\, (b_0[M_X,m_{f2}^{(1)}] - b_0[M_X,m_{f3}^{(1)}])\,\nonumber,\eeqa
where $U^{(1)}_{L,R}$ are the unitary matrices which are defined in eq. (\ref{chap5:M1_eff_diag}) and $b_0$ is the finite part of $B_0$ defined in eq. (\ref{b0}). The first two terms of the above expression are proportional to \( M_f^{(0)} \) and \( \delta M_f^{(0)} \), respectively, meaning they cannot introduce non-Hermitian entries in the mass matrix. Therefore, only the third term remains as a potential source of non-Hermitian contributions, which we now analyze. Also, the parameter \( m^{(1)}_{f2} \) in the third term is real, which results from the fact that the mass matrix \( M_f^{(1)} \) is of Hermitian type.

In the up-quark and charged lepton sectors, the matrices \( M_{u,e}^{(0)} \)and  \( \delta M_{u,e}^{(0)} \) are all real, which implies \( U^{(1)}_{L,R} \) is also real. This ensures that \( M_{u}^{(2)} \) does not contribute to \( \bar{\theta} \) at the two-loop level. This result is expected, as an accidental CP symmetry exists in the up sector at leading order, as discussed in Subsection \ref{chap5:subsec:1loop}. Moreover, the gauge interactions of the \( X \) boson with up quarks do not violate this symmetry, further preventing any contributions to \( \bar{\theta} \) at this order.

In the down-quark sector, although the third term, proportional to  \( (U_{dL}^{(1)})_{i2} \,\times (U_{dR}^{(1)})^*_{j2} \), appears to introduce complex entries in the mass matrix, it can be shown that these entries remain Hermitian. The unitary matrices \( U_{dL}^{(1)} \) and \( U_{dR}^{(1)} \) diagonalise \( M_d^{(1)}M_d^{(1)}{}^{\dagger} \) and \( M_d^{(1)}{}^{\dagger}M_d^{(1)} \), respectively. Due to the Hermitian nature of \( M_d^{(1)} \), the corresponding elements of \( M_d^{(1)}M_d^{(1)}{}^{\dagger} \) and \( M_d^{(1)}{}^{\dagger}M_d^{(1)} \) share the same phase. 

The above argument can be understood as follows: defining \( \theta_i \) as the phase of \( y^d_i \), then from eq. (\ref{chap6:munmup}) phases of \( \mu_{di} \) and \( \mu^\prime_{di} \) are \( \theta_i \) and \( -\theta_i \), respectively, with \( \theta_3 = 0 \). Consequently, the elements \( \left(M_d^{(1)}M_d^{(1)}{}^{\dagger}\right)_{ij} \) and \( \left(M_d^{(1)}{}^{\dagger}M_d^{(1)}\right)_{ij} \) can be expressed as:
\beqa 
\label{m2-m2}
\left(M_d^{(1)}M_d^{(1)}{}^{\dagger}\right)_{ij} &=& \frac{|\mu_{di}|\,|\mu_{dj}|}{m_F^2}\,e^{i \theta_{ij}}\,\sum_k |\mu^\prime_{dk}|^2\,\left(1+C q_{i}q_{k}\right)\,\left(1+C q_{j}q_{k}\right)\, ,\nonumber \\
\left(M_d^{(1)}{}^{\dagger}M_d^{(1)}\right)_{ij}&=& \frac{|\mu^\prime_{di}|\,|\mu^\prime_{dj}|}{m_F^2}\,e^{i \theta_{ij}}\,\sum_k |\mu_{dk}|^2\,\left(1+C q_{i}q_{k}\right)\,\left(1+C q_{j}q_{k}\right)\, ,
\eeqa
with $q_{Li} = q_{Ri} \cong q_{i}$ and 
\be \label{theta_ij}\theta_{ij}=\theta_i-\theta_j\,.\ee 
The constant factor \( C \) in equations is defined in eq. (\ref{chap5:M1_eff}). From eq. (\ref{m2-m2}), it is evident that both \( \left(M_d^{(1)}M_d^{(1)}{}^{\dagger}\right)_{ij} \) and \( \left(M_d^{(1)}{}^{\dagger}M_d^{(1)}\right)_{ij} \) share the same phase, and their elements are related by the interchange \( |\mu_i| \leftrightarrow |\mu^\prime_i| \). Since these two matrices have identical phases, the elements \( \left(U_{dL}^{(1)}\right)_{ik} \) and \( \left(U_{dR}^{(1)}\right)_{ik} \) also share the same phases (see Appendix \ref{app:phases} for details), and the phase factor can be written as: 
\be \left(U_{dL,R}^{(1)}\right)_{ik}~ \sim ~e^{i\theta_{ik}}\,.\ee 

As a result, the product \( \left(U_{dL}^{(1)}\right)_{i2} \left(U_{dR}^{(1)}\right)^*_{j2} \) in eq. (\ref{M2_eff_fnl_f:scp}) carries a phase factor \( e^{i\theta_{ij}} \). Furthermore, the phase factors of \( (M_d^{(0)})_{ij} \) and \( (\delta M_d^{(0)})_{ij} \) are also \( e^{i\theta_{ij}} \) since both are proportional to \( \mu_{di} \mu^\prime_{dj} \). Consequently, from eq. (\ref{M2_eff_fnl_f:scp}), the phase factor of \( \left(M^{(2)}_f\right)_{ij} \) is \( e^{i\theta_{ij}} \), while for \( \left(M^{(2)}_f\right)_{ji} \), it is \( e^{-i\theta_{ij}} \). This ensures that the Hermitian nature of the mass matrix is preserved and its determinant remains real, similar to eq. (\ref{real-det}). Therefore,  
\be \label{theta:2}  
\bar{\theta}_{\rm 2-loop}^{(X)}\,=\,\text{Arg} \det (M^{(2)}_d) \,=\, 0\, .  
\ee



\section{Qualitative analysis}
\label{sec:qualitative}
The absence of gauge-induced corrections to \(\bar{\theta}\) at both the one-loop and two-loop levels can be understood from the fact that the gauge interactions of the \(X\) boson do not violate parity symmetry. This becomes particularly evident in the flavour basis. For instance,  
\be \label{chap5:L_gauge}  
-{\cal L}_{\rm gauge} = \frac{g_X}{2} X_\mu \left(q_{L i}\, \overline{f}_{L i} \gamma^\mu f_{L i} + q_{R i}\, \overline{f}_{R i} \gamma^\mu f_{R i} \right)\,,  
\ee  
with \( q_{L i} = q_{R i} \approx q_i \). It is straightforward to verify that the interaction given in eq. (\ref{chap5:L_gauge}) remains invariant under the parity transformations defined in eq. (\ref{parity}).

In the mass basis of the fermions, when all the masses are induced perturbatively, the eq. (\ref{chap5:L_gauge}) can be expressed as: 
\be \label{chap5:L_gauge2}
-{\cal L}_{\rm gauge} = \frac{g_X}{2} X_\mu \left((Q_f^{(2)}{}_{L})_{ ij}\, \overline{\textbf{f}}_{L i} \gamma^\mu {\textbf{f}}_{L j} + (Q_f^{(2)}{}_{R})_{ ij}\, \overline{\textbf{f}}_{R i} \gamma^\mu {\textbf{f}}_{R j} \right)\,, \ee
with $\mathbf{f}_{L,R i}$ as the physical states obtained at 2-loop when masses of all the masses are induced, and 
\be \label{Q2}
{ Q}_f^{(2)}{}_{L,R} = { U_f^{(2)}}_{L,R}^\dagger\,q\,{U_f^{(2)}}_{L,R}\,.\ee
Here ${U_f^{(2)}}_{L,R}$ are the unitary matrices that diagonalise the 2-loop corrected effective mass matrix given in eq. (\ref{M2_eff_fnl_f:scp}). In the up-quark sector, these matrices are orthogonal and real, thus ${ Q}_u^{(2)}{}_{L,R}$ does not introduce complex parameters into the Lagrangian. For the down-quark sector: following the argument given in the last paragraph of section \ref{sec:2loop} and appendix \ref{app:phases}, it can be shown that $\left({U_d^{(2)}}_{L}\right)_{ik}$ and $\left({U_d^{(2)}}_{R}\right)_{ik}$ also share a common phase factor of $e^{i \theta_{ik}}$. With this, eq. (\ref{chap5:L_gauge2}) can be rewritten for the down-quark sector as:
\be \label{chap5:L_gauge3}
-{\cal L}_{\rm gauge} = \frac{g_X}{2} X_\mu \left(\lvert(Q_d^{(2)}{}_{L})_{ ij}\rvert\,e^{i\theta_{ij}} \,\overline{\textbf{d}}_{L i} \gamma^\mu {\textbf{d}}_{L j} + \lvert(Q_d^{(2)}{}_{R})_{ ij}\rvert\,e^{i\theta_{ij}} \, \overline{\textbf{d}}_{R i} \gamma^\mu {\textbf{d}}_{R j} \right)\,. \ee
It shows that the couplings of $X$ in the left-handed sector and the right-handed sector have the same phase. As the mass terms require the mixing of both the chiralities and using the fact that the mass insertion $m^{(2)}_f$ is real, the product is always hermitian type. For example, consider a general three-loop diagram as shown in Fig. \ref{fig:xloops}, the complex part of the diagram corresponds to:
\begin{figure}[t]
\centering
\includegraphics[width=0.8\textwidth]{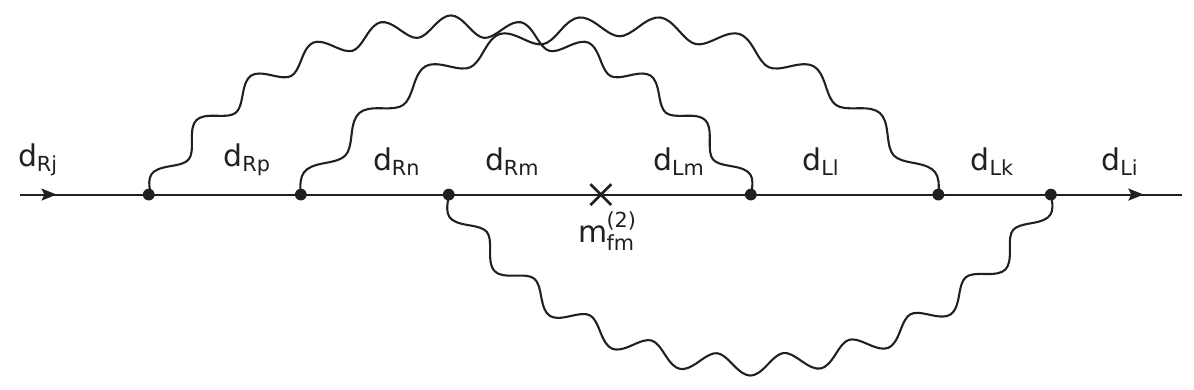}
\caption{Gauge boson induced next order correction at 3-loop.}
\label{fig:xloops}
\end{figure}
\beqa &{}& (Q_d^{(2)}{}_{L})_{ ik}(Q_d^{(2)}{}_{L})_{ kl}(Q_d^{(2)}{}_{L})_{ lm}(Q_d^{(2)}{}_{R})_{ mn}(Q_d^{(2)}{}_{R})_{ np}(Q_d^{(2)}{}_{R})_{ pj}\nonumber\\
&{}& \propto {\rm Constant }\times\,e^{i\theta_{ik}}\,e^{i\theta_{kl}}\,e^{i\theta_{lm}}\,e^{i\theta_{mn}}\,e^{i\theta_{np}}\,e^{i\theta_{pj}} \nonumber \\
&{}&= {\rm Constant }\times\,e^{i\theta_{ij}}\eeqa 
The last equality is obtained by using eq. (\ref{theta_ij}). It can be seen that the phase factor of the $ij^{,th}$ element of the loop corrected mass matrix will be proportional to $e^{i\theta_{ij}}$ as $M^{(2)}_{ij}$ of eq. (\ref{M2_eff_fnl_f:scp}). Thus, up to an arbitrary order of loops, the corrected mass matrix is of hermitian type, and the determinant is real, as shown in eq. (\ref{real-det}).  Therefore, we conclude that the gauge corrections of the $X$ boson at any order of perturbation theory do not introduce the strong CP phase.

The above argument can also be understood by comparing the structure of this model with that of \cite{Babu:1989rb}. Expressing the Lagrangian in the form  
\be  
{\cal L} \,\supset\, {\cal L}_{\rm Yukawa} \,+\, {\cal L}_{X} \,+\, {\cal L}_{\rm scalar} \,,  
\ee  
it is evident that the combined terms \( {\cal L}_{\rm Yukawa} \,+\, {\cal L}_{X} \) respect parity symmetry and effectively play the same role as \( {\cal L}_{\rm Yukawa} \) in the universal seesaw mechanism of \cite{Babu:1989rb}. As a result, the corrections introduced by \( {\cal L}_{X} \) do not contribute to \( \bar{\theta} \) at any order in perturbation theory.  

Since parity symmetry is softly broken in the scalar sector, radiative effects from scalar interactions can introduce a small but nonzero strong CP phase. Following \cite{Babu:1989rb}, the leading scalar-induced two-loop contribution to \( \bar{\theta} \) can be approximated as  
\be  \label{2loop:theta-bar}
\bar{\theta} \,\sim\, \left(\frac{1}{16\pi^2}\right)^2 \, \left(\frac{v_L}{v_R}\right)^2\, \phi^2 \,.  
\ee  
Here, the phase \( \phi \) originates from the mixing of various \( \mu \) and \( \mu^\prime \) parameters, while the factor \( v_L/v_R \) arises due to the mixing between \( H_{Li} \) and \( H_{Rj} \). The phase \( \phi \) can be chosen \(\sim {\cal O}(1) \) for the most general case.  \( v_L \) at the electroweak scale, and \( v_R \) corresponding to the \( U(1)_F \) breaking scale. The  \( U(1)_F \) breaking scale is basically determined from the mass scale of gauge boson $X$, which is mainly constrained from flavour violations, as we have seen in previous chapters.

The flavour-violating interactions of the \( X \) boson naturally induce meson-antimeson oscillations at tree level. In the lepton sector, its exchange leads to flavour-violating processes such as \( \mu \to e \) conversion in nuclei and trilepton decays \( l_i \to 3 l_j \) at leading order, while processes like \( l_i \to l_j \gamma \) emerge at the one-loop level. A detailed phenomenological analysis of these effects was previously carried out in chapter \ref{chap3}, indicating a new physics scale of approximately \( M_X \approx 10^8 \) GeV, which remains consistent with all flavour constraints. A similar bound can also be found in chapter \ref{chap4} for the case of \( \epsilon = 1 \) in the model discussed there. 
 Now for \( v_R \sim M_X \sim 10^8 \) GeV, and $\phi, v_L$ as their specified order, eq. (\ref{2loop:theta-bar}) reads
 \[ \bar{\theta} \sim 10^{-14}\,. \]  
This result is three orders of magnitude below the current experimental limit \cite{Abel:2020pzs}, demonstrating that the model provides a viable solution to the strong CP problem. Also, it can be tested in future ultracold neutron experiments, which will have the sensitivity of $ \sim 10^{-11}$ for the strong CP phase \cite{Alarcon:2022ero}.

\section{Summary}
\label{sec:summary}
In this chapter, we demonstrate that the radiative mass generation mechanism, when implemented in a parity-invariant L-R symmetric framework, offers a potential solution to the strong CP problem. This mechanism is realized by extending the gauge sector with a flavour-dependent Abelian gauge symmetry, \( G_F \).  

In this framework, the third-generation SM fermions acquire tree-level seesaw masses through the introduction of an additional generation of vector-like fermions. Meanwhile, the masses of the first two generations are generated at the one- and two-loop levels via gauge corrections, as discussed in Chapter \ref{chap4}. The required flavour non-universal symmetry, \( G_F \), corresponds to \( U(1)_{2-3} \), which can be thought of as an all-fermion version of the \( L_\mu - L_\tau \) symmetry. Additionally, the mechanism exploits the progressive increase in the rank of the mass matrices at different loop levels and shows that the strong CP phase becomes physical at the 2-loop level when all the quark masses are generated.  

A key result is that gauge corrections induced by \( {\cal L}_X \) alone do not break parity symmetry, ensuring that the induced strong CP phase remains zero at all orders of perturbation theory. However, scalar-induced corrections lead to a nonzero contribution to \( \bar{\theta} \), estimated to be around \( 10^{-14} \) at the two-loop level. A precise measurement of \( \bar{\theta} \) in the future could potentially test and falsify the model, depending on the observed value.  

The minimal version of this model suggests that the flavour symmetry-breaking scale and the \( SU(2)_R \) (parity) breaking scale are of similar magnitude. Given that the flavour symmetry-breaking scale is constrained to be around \( 10^3 \) TeV, any detectable parity-breaking signatures are expected to be negligible.

%% file: 70_Conclusion/Conclusion.tex
\chapter{Summary and Outlook}\label{chap:summary}
\label{chap7}
This thesis studies the application of the radiative mass generation mechanism to explain the hierarchical structure of fermion masses observed in the SM. In this mechanism, the tree-level masses are allowed only for the third-generation charged SM fermions, and the masses for the first and second-generation charged fermions are induced through the quantum corrections. Since the latter can be fully or partially expressed in terms of quantities that can be measured independently in experiments, at least in principle, the loop-induced masses become calculable. This is what distinguishes the framework of radiative fermion masses from other models addressing the flavour hierarchies. 

However, the SM alone cannot accommodate the radiative mass mechanism as for only massive third-generation cases, the Yukawa sector at the tree level exhibits a global \( U(2)^5 \) symmetry. Since the full action remains invariant under this symmetry, it results in massless first- and second-generation fermions for all orders of perturbation theory. Thus, the successful implementation of such a mechanism requires an extension of the SM gauge symmetry.

We systematically analyze the mechanism in various gauged extensions of the SM, involving both Abelian and non-Abelian flavour symmetries. The extended gauge symmetry can also be utilised to prevent tree-level mass terms for all the SM fermions, and the third-generation fermions can acquire masses with the help of an additional vector-like family through the seesaw-like mechanism. Subsequently, the radiative corrections induced by the spontaneous breaking of extended gauge symmetry can give rise to masses for the remaining fermions, explaining their hierarchical spectrum. 

The successful realisation of a radiative mass generation mechanism in an Abelian extension necessitates non-universal charges for the SM fermions under the associated symmetry. At the 1-loop level, the first-generation fermion remains massless. The simplest way to induce the lightest generation fermion masses is to consider two abelian symmetries, i.e., \( U(1)_1 \times U(1)_2 \). The breaking of \( U(1)_1 \) induces radiative masses for the second generation, while \( U(1)_2 \) breaking contributes to the first generation. The mass hierarchy between the two generations is naturally linked to the mild hierarchy of these symmetry-breaking scales.  

In chapter \ref{chap3}, we constructed a model based on a \( U(1)_{2-3} \times U(1)_{1-2} \) symmetry, which is the all-fermion generalisation of the well-known leptonic \( L_\mu - L_\tau \) and \( L_e - L_\mu \) symmetries. The renormalisable framework contains 25 real parameters that appear in the expressions of loop-induced masses. The viability of the model is checked by obtaining three representative numerical solutions that successfully reproduce the observed charged fermion masses and quark mixing parameters while satisfying experimental constraints from flavour violations, direct searches, and electroweak precision observables. The most stringent constraints arise from meson-antimeson oscillations, mainly when the first two generation fermions are involved, requiring the new physics scale to be at least \( 10^5 \) TeV. This high scale poses a challenge for experimental verification as well as being unpleasant from a naturalness point of view. To have a further lower new physics scale, a refined model is needed, where specific couplings responsible for these oscillations are naturally suppressed. The lowering can't be arbitrarily made small as such couplings also take part in generating masses for first-generation fermions.

The correlation between the flavour violation in the $1$-$2$ sector and the loop-induced masses for first-generation fermion masses is studied in chapter \ref{chap4}. We identify a set of optimal charges for a single Abelian $U(1)_F$, which leads to the induction of first-generation fermion masses at the 2-loop level. This setup improves upon the previous two-\( U(1) \) model by providing a more direct explanation of the inter-generational mass hierarchy. Also, it naturally introduces optimal flavour-changing couplings in $1 $-$2$ sector, allowing the new physics scale to be reduced to \( 10^3 \) TeV, nearly two orders of magnitude lower than in the previous model. 

While this single-\( U(1) \) model significantly improves the phenomenology, it still involves a large number of Yukawa couplings as the two-\( U(1) \) case of chapter \ref{chap3}. A promising direction is to unify all three fermion generations under a single representation of non-Abelian flavour symmetry, thereby reducing the number of free parameters. This motivates the introduction of a gauged horizontal \( SU(3)_F \) symmetry under which three generations of fermions transform as the fundamental representation. This flavour symmetry can be utilised to ensure that only the third-generation fermions acquire mass at tree level, while the first two generations obtain their masses through radiative corrections. In this setup, both first- and second-generation fermions typically receive masses at the one-loop level, making it necessary to introduce an additional mechanism to explain their mass hierarchy. This is achieved by breaking \( SU(3)_F \) in two steps, first to an intermediate \( SU(2) \) symmetry and then to nothing. This sequential breaking introduces a mild hierarchy among the new gauge bosons, which is transferred to the fermion masses through loop corrections.  

The breaking scale of \( SU(3)_F \) is primarily constrained by flavour violation, as the new gauge bosons mediate significant FCNCs. The resulting constraints require the symmetry-breaking scale to be at least \( 10^5 \) TeV, comparable to the bounds derived in the previous Abelian extension in chapter \ref{chap3}. However, this framework introduces two key improvements over the latter: first, the hierarchy among new gauge boson masses arises naturally from sequential symmetry breaking rather than being imposed arbitrarily; second, the use of a single non-Abelian flavour group reduces the number of independent Yukawa couplings from 20 to 12, making the model more predictive. This leads to correlations among quark and lepton masses, with the model, for instance, preferring a strange quark mass somewhat lower than the current experimental central value.  

Beyond explaining the fermion mass hierarchy, the radiative mass generation mechanism, when implemented in an L-R symmetric framework, can also address the strong CP problem. This is achieved by extending the gauge sector of the L-R framework with an additional flavour-dependent Abelian gauge symmetry, \( G_F \). Likewise, in the SM extension case, in this setup also, third-generation fermions acquire seesaw masses at tree level through an additional generation of vector-like fermions, while first- and second-generation masses arise via one- and two-loop gauge corrections. The required flavour symmetry, \( U(1)_{2-3} \), is an all-fermion version of the well-known \( L_\mu - L_\tau \) symmetry. Additionally, the flavour symmetry-breaking scale and the \( SU(2)_R \) (parity) breaking scale are found to be comparable. Given that the former is constrained to be around \( 10^8 \) GeV, any observable parity-breaking signatures are expected to be negligible.    

A key result is that gauge boson-induced corrections alone do not break parity symmetry, ensuring that the strong CP phase remains zero at all orders of perturbation theory. However, scalar-induced corrections can lead to a nonzero \( \bar{\theta} \), estimated to be of order \( 10^{-14} \) at the two-loop level. A precise future measurement of \( \bar{\theta} \) could potentially falsify this model if the observed value is significantly larger. 

Overall, this study presents a systematic exploration of the radiative mass generation mechanism in different gauge extensions of the SM, demonstrating its potential to explain fermion mass hierarchies while addressing key flavour and CP violation challenges. The findings highlight the trade-offs between complexity, predictiveness, and experimental testability, suggesting that a non-Abelian extension provides the most promising framework for a predictive theory of flavour. However, there are areas where the radiative mass framework requires further improvement. For example,
\begin{itemize}
\item The $SU(3)_F$ framework, which accounts for the minimal number of  Yuakawa couplings and a more predictive setup, does not account for the salient features of the optimal flavour violating gauge sector presented by $U(1)_F$ of chapter \ref{chap4}. One possible way to incorporate the optimal flavour violation in $SU(3)_F$ framework is by suitably breaking the symmetry which lead to $U(1)_F$ as an intermediate symmetry. However, in such cases, obtaining the hierarchy between first- and second-generation fermions will require special treatment, and this has yet to be checked. 
\item The number of Yukawa couplings for the $U(1)_F$ case can also be reduced to some extent by implementing this scheme in models that provide quark-lepton unification \cite{Georgi:1974sy,Fritzsch:1974nn,Pati:1974yy} at high energies.
\item Although the framework accounts for the hierarchical nature of charged fermion masses, it does not explain the smallness of mixing in the quark sector. This is because the underlying symmetries and the field content of the model allow different and arbitrary tree-level Yukawa couplings for the up-type and down-type quarks. Frameworks with an improved degree of calculability can address this issue.
\item The lowest scale of new physics remains at $10^3$ TeV. While this scale, or even a larger one, does not hinder the core mechanism discussed here, the significant separation between it and the electroweak scale poses a challenge. This separation can be managed at the expense of fine-tuning in the generic case. However, in more ambitious theories where the parameters of the scalar potential are also calculable, this becomes a concrete problem.
\item  In all discussed Abelian or non-Abelian extension scenarios, we have neglected the scalar-induced corrections to the mass matrix induced by considering small mixing and/ or heavy scalars. However, it is interesting to check whether such corrections can lead to viable first-generation fermion masses. Then, optimal $U(1)_F$ with $\epsilon=0$ case would lead to a new physics scale much lower than $10^3$ TeV. This is because non-universal charges $\lbrace 1,1,-2\rbrace$ would lead to a $U^{(f_1)}(1)_L\times U^{(f_1)}(1)_R$ invariant Lagrangian up to 1-loop for gauge boson induced corrections. When scalar corrections are included, the comparatively small FV couplings for stringent $K^0-\overline{K^0}$ oscillation process can be induced, and thereby, a relatively lower new physics scale can be phenomenologically allowed.
\item The feasibility of these frameworks is indirectly inferred by analysing the various flavour-violating processes. However, the precision measurements of light quark masses can potentially give direct verification. Also, the computation of Higgs couplings to the fermions in these theories and comparing them with their respective experimentally obtained values can give another potential area to test these specific models.
\end{itemize}
We believe that the above issues require a set of systematic investigations to establish further the validity and robustness of the radiative mass generation scheme.  

%% file: 100_Appendices/appendix_1.tex
\chapter{\texorpdfstring{$SU(3)_{F}$}{SU(3)_{F}}: The generators}
\label{appendix_1}
\graphicspath{{100_Appendices/}}
\label{app:GM}
In this study, we utilize a specific representation for the $SU(3)_F$ generators $\lambda^a$, with $a=1,2,...,8$, which is explicitly given as follows:
\beqa
\label{su(3)_lambda}
\lambda^1 &=& \left( \ba{ccc} 0 & 0 & 0 \\ 0 & 0 & 1\\0&1&0 \ea\right)\,,\,~~~ \lambda^2 \,=\, \left( \ba{ccc} 0 & 0 & 0 \\ 0 & 0 & -i\\0&i&0 \ea\right)\,,\, ~ \lambda^3 \,=\, \left(\ba{ccc} 0 & 0& 0\\ 0& 1 &0\\0 & 0 & -1 \ea \right)\,,\, \nonumber \\
\lambda^4 &=& \left(\ba{ccc} 0 & 1& 0\\ 1& 0 &0\\0 & 0 & 0 \ea \right)\,,~~~
 \lambda^5 \,=\, \left(\ba{ccc} 0 & i& 0\\ -i& 0 &0\\0 & 0 & 0 \ea \right)\,,\, ~
\lambda^6 \,=\, \left(\ba{ccc} 0 & 0& 1\\ 0& 0 & 0\\1 & 0 & 0 \ea \right)\,,\, \nonumber \\
\lambda^7 &=& \left(\ba{ccc} 0 & 0& i\\ 0& 0 &0\\-i & 0 & 0 \ea \right)\,,\,  ~\lambda^8 \,=\,\frac{1}{\sqrt{3}} \left(\ba{ccc} -2 & 0& 0\\ 0& 1 &0\\0 & 0 & 1 \ea \right)\,.\eeqa
These are expressed in a basis where the first three generators correspond to gauge bosons that do not couple to the first generation in the canonical basis. An \( SU(2) \) subgroup, associated with these three generators of the full flavour symmetry group, remains unbroken by the vacuum expectation value of \( \eta_1 \) in eq. (\ref{eta_vevs}).  

The matrix representation of \( SU(3)_F \) generators in eq. (\ref{su(3)_lambda}) can be derived from the conventional Gell-Mann matrices \( \lambda^a_G \) using the following transformations:
\be\lambda^a = P\cdot \lambda^a_{G}\cdot P^{-1} \ee 
where $P=P_{23}\cdot P_{13}$ with 
\be P_{23}=\left( \ba{ccc} 1 & 0 & 0 \\ 0 & 0 & 1\\0&1&0 \ea\right)\,,~~P_{13}=\left( \ba{ccc} 0 & 0 & 1 \\ 0 & 1 & 0\\1&0&0 \ea\right)\,. \ee
The explicit form of $\lambda^a_{G}$ matrices is written below.

\[
\lambda_G^1 =
\left(\ba{ccc}
0 & 1 & 0 \\
1 & 0 & 0 \\
0 & 0 & 0
\ea\right), \quad
\lambda_G^2 =
\left(\ba{ccc}
0 & -i & 0 \\
i & 0 & 0 \\
0 & 0 & 0
\ea\right), \quad
\lambda_G^3 =
\left(\ba{ccc}
1 & 0 & 0 \\
0 & -1 & 0 \\
0 & 0 & 0
\ea\right),
\]

\[
\lambda_G^4 =
\left(\ba{ccc}
0 & 0 & 1 \\
0 & 0 & 0 \\
1 & 0 & 0
\ea\right), \quad
\lambda_G^5 =
\left(\ba{ccc}
0 & 0 & -i \\
0 & 0 & 0 \\
i & 0 & 0
\ea\right), \quad
\lambda_G^6 =
\left(\ba{ccc}
0 & 0 & 0 \\
0 & 0 & 1 \\
0 & 1 & 0
\ea\right),
\]

\[
\lambda_G^7 =
\left(\ba{ccc}
0 & 0 & 0 \\
0 & 0 & -i \\
0 & i & 0
\ea\right), \quad
\lambda_G^8 =
\frac{1}{\sqrt{3}}
\left(\ba{ccc}
1 & 0 & 0 \\
0 & 1 & 0 \\
0 & 0 & -2
\ea\right).
\]

%% file: 100_Appendices/appendix_2.tex
\chapter{Notes on Hermitian Type matrices}
\label{appendix_2}
\graphicspath{{100_Appendices/}}
\section{The real determinant}
\label{real-determinant}
The simplest form of hermitian type matrix $X$ can be formed if the elements $X_{ij}$ has the phase factor $e^{i\theta_{ij}}$ with $\theta_{ij}=\theta_i -\theta_j$. Let us define a hermitian-type matrix 
\be\label{hermitian-type} M_{ij}\,=\, r_{ij}\, e^{i\theta_{ij}} \ee
with all $r_{ij}$ real. Considering as $3\times3$ matrix, the determinant of $M$ can be written as 
\beqa
\det M &=& \epsilon_{ijk}\, M_{1i}\,M_{2j}\,M_{3k}\, \nonumber \\
&=& \epsilon_{ijk}\, r_{1i}\,r_{2j}\,r_{3k}\,e^{i(\theta_1+\theta_2+\theta_3)}\,\,e^{-i(\theta_i+\theta_j+\theta_k)}\eeqa
Here we have used the fact $\theta_{ij}=\theta_i\,-\,\theta_j$. It can be seen that the factor $(\theta_i+\theta_j+\theta_k)$ is always $(\theta_1+\theta_2+\theta_3)$ for $i\neq j\neq k$. This implies the determinant is always real, and 
\be \label{real-det} \det M\,=\, \epsilon_{ijk}\,r_{1i}\,r_{2j}\,r_{3k}\,.\ee
It is also straightforward to see that any $n\times n$ hermitian type matrices has real determinant.

\subsection*{Note:} 
\begin{itemize}
    \item The sum of two hermitian type matrices $X$ and $Y$ is also a hermitian type matrix only when the phase factor of the elements $X_{ij}$ and $Y_{ij}$ are the same. 
\end{itemize}


\section{Phase of the elements of  \texorpdfstring{$U_{L,R}$}{U_{L,R}}}
\label{app:phases}
Using eq. (\ref{hermitian-type}) we can write: 
\beqa \label{m2-m2:2} 
\left(M M^{\dagger}\right)_{ij} &=& c_{ij}\, e^{i\theta_{ij}}\, , \nonumber\\
\left(M^{\dagger}M\right)_{ij} &=& d_{ij}\, e^{i\theta_{ij}}\,,\eeqa
where $c_{ij}$ and $d_{ij}$ are the real constants, and $\theta_{ij}$ is defined in eq. (\ref{theta_ij}). It can be seen that $c_{ij}$ and $d_{ij}$ are the elements of real symmetric matrices. A similar form can be obtained for $M^{(0,1)}M^{(0,1)\dagger}$ and $M^{(0,1)\dagger}M^{(0,1)}$. Now, inverting the bi-unitary diagonalization equations, we get
\beqa\label{m2-m2:3} 
\left(M M^{\dagger}\right)_{ij} &=& \left(U_L \,D^2 \,U^\dagger_L\right)_{ij} \,,  \nonumber\\\left(M^{\dagger}M\right)_{ij} &=& \left(U_R\, D^2 \,U^\dagger_R\right)_{ij}  \,,\eeqa
with $D= {\rm Diag.} (m_k)$. Now, putting eq. (\ref{m2-m2:2}) in eq. (\ref{m2-m2:3}) and using the the fact $\theta_{ij}=\theta_i -\theta_j$, we can write 
\beqa\label{m2-m2:4}  c_{ij} &=& \left(e^{-i\theta_{ik}}\,(U_L)_{ik}\right)\, m_k^2 \,\left(e^{i\theta_{jk}}\,(U_L)^*_{jk}\right)\,  \,, \nonumber\\
d_{ij} &=& \left(e^{-i\theta_{ik}}\,(U_R)_{ik}\right)\, m_k^2 \,\left(e^{i\theta_{jk}}\,(U_R)^*_{jk}\right) \,,\eeqa
Defining: \be\left(S_{L,R}\right)_{ik}= e^{i\theta_{jk}}\,(U_{L,R})_{jk}\,,\ee the above eq. (\ref{m2-m2:4}) can be rewritten as: 
\beqa\label{m2-m2:5}  c_{ij} &=& \left(S_L\right)_{ik}\, m_k^2 \,\left(S_L\right)^*_{jk}\,  \,, \nonumber\\
d_{ij} &=& \left(S_R\right)_{ik}\, m_k^2 \,\left(S_R\right)^*_{jk}\,,\eeqa
Since  $m_k^2$ is real and $c_{ij}$ and $d_{ij}$ are elements of a real symmetric matrix, already mentioned after eq. (\ref{m2-m2:2}), the diagonalising matrix $S_{L,R}$ has to be real orthogonal matrix. Therefore, 
\beqa
(U_L)_{ik}&=& a^L_{ik}\,e^{i\theta_{ik}}\,,\nonumber\\
(U_R)_{ik}&=& a^R_{ik}\,e^{i\theta_{ik}}
\eeqa
with $a^{L,R}_{ik}$ as any real constants. Hence it is shown that the phase of $(U_L)_{ik}$ and $(U_R)_{ik}$ is equal. 